%% file: main.tex
\documentclass[b5paper,12pt]{book}
\pdfoutput=1
\usepackage[b5paper,hmargin={2.03cm,2.03cm},vmargin={2.0cm,2.0cm}]{geometry}
\usepackage{amsmath,amssymb,lmodern}
\usepackage{packages/amsthm}
\usepackage[numbers,sort&compress]{natbib}
\usepackage{bigints}
\usepackage{epsfig}
\usepackage{braket}
\usepackage{slashed}
\usepackage{bm}
\usepackage{bbold}
\usepackage{graphicx}
\graphicspath{{figures/}}
\usepackage{txfonts} 
\usepackage{bigints}
\usepackage{caption}
\usepackage{titling}
\usepackage{emptypage}
\usepackage{titlesec}
\titleformat*{\section}{\large\bfseries\centering}
\titleformat*{\subsubsection}{\normalsize\bfseries\centering}
\usepackage{fancyhdr}
\makeatletter
\def\@makechapterhead#1{%
  \vspace*{50\p@}%
  {\parindent \z@ \raggedright \normalfont
    \ifnum \c@secnumdepth >\m@ne
      \if@mainmatter
        \huge\centering\bfseries\thechapter\mdseries
        \par\nobreak
        \vskip 10\p@
      \fi
    \fi
    \interlinepenalty\@M
    \centering\Large  #1\par\nobreak
    \vskip 90\p@
  }}
\makeatother
\renewcommand{\Re}{\operatorname{Re}}
\renewcommand{\Im}{\operatorname{Im}}

\DeclareMathOperator{\Tr}{Tr}
\renewcommand{\v}[0]{\bm}
\DeclareMathOperator{\arcsinh}{arcsinh}
\begin{document}
\newcommand{\head}[1]{\textbf{#1}} \title{High energy scattering and emission
  in QED$\&$QCD media} \author{X.G. Feal\\\\[3\baselineskip] IGFAE/Instituto
  Galego de F\'isica de Altas Enerx\'ias
  \\\\ [\baselineskip]\footnotesize{This work has been done as a PhD thesis for the}
  \\ \footnotesize{University of Santiago de Compostela and is partially}
  \\ \footnotesize{based on already published research by}
  \\ \footnotesize{the author.}}
\frontmatter
\date{}
\maketitle 
\tableofcontents 
\include{preface/preface}

\include{acknowledgments/acknowledgments}

\include{introduction/introduction} \nocite{apsrev41Control}
\mainmatter
\label{partpage}\part{QED}
\include{chapter1/chapter1}
\include{chapter2/chapter2}
\include{chapter3/chapter3}
\label{partpage}\part{QCD}
\include{chapter4/chapter4}
\include{chapter5/chapter5}

\include{conclusions/conclusions}

\listoffigures
\appendix
\include{appendices/appendix1}
\backmatter
\bibliography{bibliography/bibliography}
\bibliographystyle{packages/apsrev4-1}
\end{document}

%% file: preface/preface.tex
\chapter*{Preface}

We certify that all the possible mistakes have been made, without exception,
in the course of this little work. The reader will be surprised to find that
some of them have been randomly corrected in the final version. The resulting
work is a mess. Put the book it in a bookshelf. Keep in a cold and dry place
and please do not touch it anymore. Keep out of reach of researchers and then
try to be happy.

%% file: acknowledgments/acknowledgments.tex
\chapter*{Acknowledgments}

I am particularly grateful to Ricardo V\'azquez for the very valuable help and
the time he spent in revising in detail the final manuscript, in supervising
the work behind and in double checking most of the results. Also a very
important part, if not all, of the underlying ideas present here, belong in
fact to him.

I am grateful also to Carlos Salgado and N\'estor Armesto for making us to
address the question, long time ago, if there was a way of evaluating
bremsstrahlung spectra, whether for photons or gluons, circumventing some
limitations widely present in the classic works. Indeed it was possible, and
from that anecdote arised this little work. Together with Carlos Pajares,
their explanations and discussions were very helpful. I thank also Joaqu\'in
S\'anchez-Guillen and Jaime \'Alvarez-Mu\~niz for their suggestions after
carefully reading the papers.

The effort behind this little work goes beyond oneself, it extends to the
people I must admire for their supporting role. In what concerns to make this
work see the light of day, they deserve the same credit as the one writing
these lines. To, Bea, Alba, Gema, Mart\'in, Irene, Susana, David, Dani... in
commutative order, and so many others whose distractions were also very
valuable. I am also grateful to the colleagues at my office, for always
keeping an eye on my back, to my home town friends, old friends, including
some beers, and such kind of very important things.

%% file: introduction/introduction.tex
\chapter*{Introduction}

The study of coherence phenomena for the emission of quanta in a multiple
interaction scenario dates back to the seminal works of Ter-Mikaelian
\cite{termikaelian1953}, Landau and Pomeranchuk \cite{landau1953a,landau1953b}
and Migdal \cite{migdal1956} on the bremsstrahlung suppression due to
condensed media, the well known Landau-Pomeranchuk-Migdal (LPM) effect. The
existence of phases in the elastic amplitudes, which keep track of the
placement and momentum change of each of the interactions, lead to
interference effects in the squared sum of all the involved amplitudes. At
high energies these phases set a coherence length, in the dominant direction
of the traveling particle, in which the individual amplitudes can be
considered to coherently add in the squared amplitude, but incoherently
interfere with the rest of the processes. A manifestation of the coherence
between Feynmann diagrams could acquire a more familiar form in analogy with
the soft photon theorem \cite{weinberg1965}. For sufficiently low energy of
the emitted photon, the diagrams representing emissions from the internal legs
cancel and the total amplitude restricts to the photons coming from the first
and last legs. Thus in that limit radiation can be understood as if all the
internal interactions coherently emit as a single entity. This immediately
addresses the question on how the intensity of photons behaves for larger
photon energies and how this energy and the medium modulate the coherence. In
other words, when the medium still emits as a single entity, and when, if it
occurs, a regime of maximal incoherence between its constituents may be found,
in which the total intensity is just the sum of the single \cite{bethe1934}
intensities.

Predictions and observations of this coherence effect have been extensively
carried in the past for quantum electrodynamic interactions (QED) and
they have suscited a renewed interest in the era of the hadron colliders as a
way to indirectly observe the hadronic matter produced in heavy ion
collisions. Soon after the birth of quantum chromodynamics (QCD) as the theory
of the strong interactions \cite{fritzsch1973a,gross1973a,politzer1973a}, it
was assumed that a new state of the hadronic matter, consisting in a
deconfined phase of quarks and gluons at extremely high temperatures, should
exist in nature
\cite{hagedorn1965a,huang1970a,collins1975a,shuryak1978a,susskind1978a,polyakov1978a,bjorken1982b}.
At temperatures well above the QCD critical temperature $\sim$ 170 MeV and/or
at densities well above the ordinary nuclear density $\sim$ 1 GeV/fm$^3$,
collisions involving large momentum change are required in order to consider
the interactions between particles significative. The weak value of the
coupling at such high gluon momenta leads, however, to a relatively small
amplitude of these kind of hard interactions to occur. This asymptotically
free behavior of QCD at high energies suggests, then, a picture of quasi free,
softly interacting quarks and gluons under extreme conditions. In analogy with
the weakly coupled QED plasmas, this new state of the matter at such extreme
conditions was quickly coined \cite{shuryak1980a} with the name of Quark Gluon
Plasma (QGP). The study of the QGP properties results of interest not only as
a way of understanding the complex nature and rich phenomenology of QCD, but
it also becomes crucial for the understanding of the initial stages of the
universe in the quark epoch just after the Big Bang, for example. In the
search of evidences of QGP formation at the RHIC and LHC hadron colliders
several indirect probes can be considered. Among these QGP signatures, like
the observation of anisotropies in the charged particle distributions, related
to an hydrodynamical behavior of the QGP, or the suppression of heavy quark
mesons, a quantitative prediction of the radiative energy loss of the
underlying jet particles, while traveling through the formed QCD medium,
should provide an indirect evidence of the QGP formation and its
characteristics. For that purpose an accurate study of the LPM effect for the
gluon bremsstrahlung spectrum in a multiple collision scenario with the QCD
medium becomes, then, indispensable.

The structure of this work is approximately guided by the historical
developments. We also put some focus in understanding the coherence effects,
both in the elastic and the emission processes, following a conventional
quantum field theory (QFT) description with amplitudes. In Chapter 1 we lay
the foundations for the treatment of high energy scattering under multiple
interactions by defining a high energy integration of the fermion state under
a general, static and classical external field \cite{glauber1959}. In Chapter
2 we present a formalism for multiple scattering at arbitrary perturbative
order, and define the medium averaged intensities and observables leading to
transverse and longitudinal coherency effects. In Chapter 3 we simply use the
former multiple scattering results to build a QFT formalism for high energy
emission which can be evaluated for general interactions with mediums of
finite size. Under the Fokker-Planck approximation and the semi-infinite
medium limit we are able to recover the Migdal result \cite{migdal1956} as a
particular case. In Chapter 4 we extend the QED multiple scattering results
for the QCD scenario and finally, in Chapter 5, we evaluate the gluon
bremsstrahlung intensity with these tools. Under the Fokker-Planck
approximation, Wiedemann result \cite{wiedemann1999} is found and within some
length or mass approximations the well known results of Migdal/Zakharov
\cite{migdal1956,zakharov1996a} or the BDMPS group \cite{baier1995,baier1996}
are also recovered.

%% file: chapter1/chapter1.tex
\chapter{High energy fermions in an external field}

The purpose of this chapter is to introduce the Glauber high energy
integration of the wave \cite{glauber1959} for particles obeying the Dirac
equation. These wave solutions will be used in the following chapters to
evaluate the QED and QCD scattering and emission amplitudes in a multiple
collision scenario. Consider a high energy particle traveling and interacting
with a medium, classically characterized as a set of sources distributed in
some region of the space. Since the state of the particle is expected to
contain an arbitrary number of interactions with the medium, an
exact/non-perturbative solution for the traveling wave is required. Processes
typically involving the creation or annihilation of other particles will
emerge as a result of the interaction with the medium. Their amplitudes, when
expressed in terms of the exact wave integrations, correspond to a sum of all
the perturbative orders in $g$, the medium coupling parameter, or, in other
words, all the ways in which the process can be represented by means of
elementary interactions with medium constituents. This superposition of
Feynman diagrams will lead to a quantum fluctuation of the observables in the
number of collisions and they may be expressed, on average, as functions of
the number of particles, the geometry and the nature and strength of the
interaction with the medium.

Early efforts in computing amplitudes in this exact approach exist
\cite{bethe1954}. For photon bremsstrahlung due to a single Coulomb field, for example, an exact Dirac 
or Klein-Fock-Gordon solution for the electron can be obtained in polar coordinates
\cite{darwin1928,sommerfeld1935}, but at the expense of making the
emission intensity calculation an almost impossible task
\cite{bethe1954}. It is obvious then, that if we try to extend the problem to 
multiple external fields randomly distributed, some kind of 
approximations have to be done. Fortunately,
as we will see later, two approximations will prove to simplify our problem. 
First, as already mentioned under the conditions of our interest the medium is
considered a classical and static source. The validity of this approximation
and the form of the interaction is briefly explained in 
Section \ref{section_1_1}. Second, the high energy solution of the Dirac
equation can be given as an operator acting on a Schrodinger
like state which is easier to solve in the high energy approximation \cite{glauber1959}. A derivation of this result and a discussion of the 
conditions for the approximation are given in Section
\ref{section_1_2}. In Section \ref{section_1_3} we 
prove that the perturbative expansion of the found wave agrees with the
standard Born approximation, and in Section \ref{section_1_4} we 
compute the propagator in momentum and light cone space, showing that except
for a negligible spinorial correction, the well known Glauber scattering
amplitude \cite{glauber1959} is found. This correspondence is a 
direct consequence of the Schrodinger behavior, except for the spin structure, 
of the Dirac solution at high energies.
\section{The medium as a classical source}
\label{section_1_1}
We consider a medium of ordinary solid matter composed, then, of $n$ static
nuclei of charge $Ze$ and mass $m_n$. When a high energy electron of mass
$m$ and charge $e$ passes through this medium a photon field emanates from
the induced current, which is going to be considered classical and static. By
classical we mean that the nuclei or sources originating the field are
completely localized in coordinate space, and by static that these sources
have a negligible recoil. Both conditions read for the 4-current
\begin{align}
J_0^{(n)}(x)=Ze\sum_{i=1}^{n}
\delta^{(3)}(\v{x}-\v{r}_i),\medspace\medspace\medspace\medspace
\v{J}^{(n)}(x)=0,
\label{classical_source}
\end{align}
where $\v{r}_i$ is the position of each source. In order to take 
into account screening effects of the electron shell, we 
will consider that the photon field originated in the interaction
 has an effective mass $\mu_d$ canceling photon propagation to distances 
much larger than $r_d=1/\mu_d$. This screening mass of the nuclei can be
estimated as $\mu_d=\alpha^2m Z^{1/3}$, being $\alpha$ the fine structure constant. 
In the Lorenz gauge \cite{jackson1962}, now required for current conservation, the Maxwell equation for
the massive photon $A_\mu^{(n)}(x)$ due to the $n$ sources reads
\begin{equation}
\left(\partial_\nu\partial^\nu +\mu_d^2\right) A_\mu^{(n)}(x)=4\pi
J_\mu^{(n)}(x).
\label{photon_field_equation}
\end{equation}
To get the form of $A_\mu^{(n)}(x)$ we simply propagate the 
current \eqref{classical_source} as
\begin{equation}
A_\mu^{(n)}(x)=\int d^4 y
D^F_{\mu\nu}(x-y)J_\nu^{(n)}(y),
\label{photon_field_solution}
\end{equation}
by means of a Feynman-Stueckelberg propagator. By direct inspection of equations
\eqref{photon_field_equation} and \eqref{photon_field_solution}, the
propagator, then, has to satisfy the equation  
\begin{equation}
\left(\partial_\eta\partial^\eta +\mu_d^2\right) D^F_{\mu\nu}(x-y)=4\pi
\delta^{(4}(x-y)g_{\mu\nu}.
\label{photon_propagator}
\end{equation}
By Fourier transforming equation \eqref{photon_propagator} we easily find
\begin{equation}
\hat{D}^F_{\mu\nu}(q^2)=g_{\mu\nu}\frac{4\pi}{-q^2+\mu_d^2},
\label{photon_propagator_momentum}
\end{equation}
With the above equation, and using \eqref{classical_source} and
\eqref{photon_field_solution} it is straightforward to show that 
the field does not carry energy, as result
of time translation invariance, and it has only one 
non-vanishing component, namely $A_0^{(n)}(x)$, given by
\begin{align}
A_\mu^{(n)}(x) = g_{\mu0}Ze\sum_{i=1}^n \int\frac{d^3\v{q}}{(2\pi)^3} \frac{4\pi 
  }{\v{q}^2+\mu_d^2}e^{-i\v{q}\cdot (\v{x}-\v{r}_i)}.
\end{align}
Here we notice that the inverse screening radius $\mu_d$ is related to
the typical momentum change $\v{q}$ in a single collision with one nuclei.
Finally, the last integral can be accomplished by using Cauchy's theorem,
\begin{align}
\int
\frac{d^3\v{q}}{(2\pi)^3}\frac{4\pi}{\v{q}^2+\mu_d^2}e^{-i\v{q}\cdot\v{x}}=\frac{1}{\pi
|\v{x}|} \Im \int_{-\infty}^{+\infty} dq\frac{q}{q^2+\mu_d^2}e^{iq|\v{x}|}=\frac{1}{|\v{x}|}e^{-\mu_d|\v{x}|},
\end{align}
so we find the interaction
\begin{align}
A_0^{(n)}(x)=Ze \sum_{i=1}^{n} D^F_{00}(\v{x}-\v{r}_i)=\sum_{i=1}^{n}\frac{Ze}{|\v{x}-\v{r}_i|}
e^{-\mu_d|\v{x}-\v{r}_i|}.
\label{yukawa_external_field}
\end{align}
This is known as a Yukawa/Debye interaction with screening $\mu_d$. In the
limit $\mu_d\to 0$ the field equations transform into the gauge
covariant Maxwell equations and correspondingly we find the Coulomb
interaction. This simple result derived under the classicity 
of the current can be justified by doing the pertinent approximations
in the quantum current. For a single nuclei we would have,
\begin{equation}
J_\mu(x) = \sqrt{\frac{m_n}{p_b^0}}\bar{u}_{s_b}(p_b)\gamma_\mu
u_{s_a}(p_a)\sqrt{\frac{m_n}{p_a^0}} e^{i(p_b-p_a)\cdot x},\label{quantum_current1}
\end{equation}
where we used spinor conventions given in Appendix \ref{appendix1}.
The amplitude of finding the photon carrying momentum $q=p_a-p_b$ 
is, using \eqref{photon_field_solution}, \eqref{photon_propagator} and \eqref{quantum_current1}
\begin{equation}
A_\mu(x) = \frac{4\pi Ze}{-(p_a-p_b)^2+\mu_d^2}\sqrt{\frac{m_n}{p_a^0}}\bar{u}_{s_a}(p_a)\gamma_\mu
u_{s_b}(p_b)\sqrt{\frac{m_n}{p_b^0}}e^{+i(p_a-p_b)x}.
\label{photon_field_quantum}
\end{equation}
As usual, this photon has to be joined with the electron current and
then the resulting quantity integrated in $x$. Due to both external constraints
of the other nuclei reducing its recoil and because it is 
much heavier than the typical momentum
change produced into a single collision $\mu_d$, we have 
$\v{p}_a=0$ and $\v{p}_b\approx \v{p}_a$, so $p_b^0\approx m_n$,
\begin{align}
J_0&=\sqrt{\frac{m_n}{p_a^0}}\bar{u}_{s_a}(p_a)\gamma_0
u_{s_b}(p_b)\sqrt{\frac{m_n}{p_b^0}}\approx
1\nonumber\\
 J_k&=\sqrt{\frac{m_n}{p_a^0}}\bar{u}_{s_a}(p_a)\gamma_k
u_{s_b}(p_b)\sqrt{\frac{m_n}{p_b^0}}\approx \frac{(p_b)_k}{m_n}\ll 1.
\label{quantum_current2}
\end{align}
Similar conclusions can be placed for the electron in the high energy limit.
Using then these slow dependences of the nuclei and electron currents on the momentum, using 
\eqref{quantum_current2} integration in time
and momentum can be carried independently at \eqref{photon_field_quantum} and we obtain
\begin{equation}
\langle A_\mu(x) \rangle_{q,t} \equiv \delta_{\mu 0}Ze\int\frac{d^4q}{(2\pi)^4} \int dt  \frac{4\pi 
  }{-q^2+\mu_d^2}e^{+iq\cdot x} =\delta_{\mu 0} \frac{Ze}{|\v{x}|}e^{-\mu_d|\v{x}|},
\end{equation}
in such a way that the coherent superposition of photon amplitudes
recovers the classical field limit. The classical approximation is allowed, 
then, if the inertias of the nuclei and the electron (the
rest mass and the energy, respectively) greatly exceed the typical momentum 
change $\mu_d$ of a single interaction. For moving constituents, like the ones 
in an ideal gas, it is also sufficient to guarantee that the typical energy of each constituent is much larger than $\mu_d$, so the medium can be considered
classic. And if the traveling fermion energy greatly exceeds also the typical
energy of the constituents, the medium can be considered static. 
\section{High energy limit of the Dirac equation}
\label{section_1_2}
We depart, then, from a medium characterized by $n$ sources of the form
\begin{align}
A_0^{(n)}(x)=\sum_{i=1}^n
A_0^{(1)}(\v{x}-\v{r}_i),\medspace\medspace\medspace\medspace \partial_0
A_0^{(n)}(x)=0.\label{general_external_field}
\end{align}
The state $\psi^{(n)}(x)$ of the electron is coupled with
strength $g=e$, its charge, to these $n$ fields. 
Since these fields carry themselves the
charge of each target $Ze$, let us relabel the overall coupling to $g=Ze^2$ and
leave the interaction $A_0^{(n)}(x)$ coupling independent. The electron obeys
the Dirac equation under this field, so it has also to obey the squared Dirac equation, and we write
\begin{align}
\left(i\gamma^\mu\frac{\partial}{\partial x^\mu}+m-g\gamma^0
A_0^{(n)}(x)\right)\left(i\gamma^\mu\frac{\partial}{\partial x^\mu}-m-g\gamma^0
A_0^{(n)}(x)\right)\psi^{(n)}(x)=0.
\end{align}
Energy operator represents the $x_0$ evolution of the states
, $i\partial_{0}\equiv \pm p_0$. Accordingly energy positive solutions are the
ones represented by
\begin{align}
\psi^{(n)}(x)=\psi^{(n)}(0,\v{x})e^{-ip_0x_0}.
\end{align}
Using the relations $\{\gamma^\mu,\gamma^\nu\}=2g^{\mu\nu}$,
$\alpha^i=\gamma^i\gamma^0$ and taking into account the condition $\partial^0 A_0^{(n)}(x)=0$, we find an eigenvalue equation in $p_0$ 
\begin{align}
\bigg(p_0^2-m^2+\v{\nabla}^2&-2gp_0A_0^{(n)}(\v{x})\bigg)\psi^{(n)}(0,\v{x})\nonumber\\&=
\left(ig\v{\alpha}\cdot\v{\nabla}A_0^{(n)}(\v{x})-\left(gA_0^{(n)}\right)^2(\v{x})\right)\psi^{(n)}(0,\v{x}).
\label{psiequation}
\end{align}
We take first the infinite momentum limit of \eqref{psiequation}.
To do that, we will assume that the energy $p_0$ of the fermion greatly exceeds the average
magnitude of the interaction $\braket{gA_0^{(n)}(x)}$, a condition which reads
\begin{align}
g\braket{A_0^{(n)}(x)}\equiv g\int d^3\v{x}\bar{\psi}^{(n)}(0,\v{x})\gamma_0
A_0^{(n)}(\v{x})\psi^{(n)}(0,\v{x})\ll p_0.
\end{align}
Under this condition, in \eqref{psiequation} we can drop off the terms linear in
$g$ and $g^2$ but not the term in $gp_0$, leading to
\begin{align}
\left(p_0^2-m^2+\v{\nabla}^2-2gp_0A_0^{(n)}(\v{x})\right)\varphi_s^{(n)}(0,\v{x})=0,
\end{align}
where we denoted the solution in this limit $\varphi_s^{(n)}(x)$. We state that a free 
solution with 4-momentum $p$ must be a good approximation up to a modified wave, 
which we call $\phi_s^{(n)}(\v{x})$. Keeping this in mind we write down the ansatz 
\begin{align}
\varphi_s^{(n)}(0,\v{x})=\sqrt{\frac{m}{p_0}}e^{-i\v{p\cdot
    x}}u(p)\phi_s^{(n)}(\v{x}).
\end{align}
Solutions in the pure high energy limit $\varphi_s(x)$ and $\phi_s(x)$ 
have been labeled with a ($s$) because they are solutions to an effective Schrodinger
equation, as we will see below. Here $u(p)$ stands for a free spinor and
$\mathcal{N}(p)=\sqrt{m/p_0}$ the normalization (see Appendix \ref{appendix1}). With this
ansatz we find the equation for $\phi_s^{(n)}(\v{x})$ 
\begin{equation}
\v{\nabla}^2\phi_s^{(n)}(\v{x})-2i\v{p}\cdot\v{\nabla}\phi_s^{(n)}(\v{x})-2gp_0 A_0^{(n)}(\v{x})\phi_s^{(n)}(\v{x})=0.
\label{phiequation}
\end{equation}
For multiple external fields the
above equation does not have a closed exact solution. We investigate, by now, the
momentum change induced by $A_0^{(n)}(\v{x})$. The modified wave $\phi_s(\v{x})$ changes
asymptotic momentum, which was $\v{p}$, to
\begin{align}
\braket{\varphi_s^{(n)}|\tilde{\v{p}}\varphi_s^{(n)}}=\v{p}+\int d^3
\v{x}\medspace \phi_s^{(n),*}(\v{x})\left(+i\v{\nabla}
\phi_s^{(n)}(\v{x})\right)=\v{p}+\delta\v{p}.
\end{align}
In order to see how the modification $\delta \v{p}$ compares to $\v{p}$, one multiplies the
last equation by $\v{p}$ and uses \eqref{phiequation}, finding
\begin{align}
\v{p}\braket{\varphi_s^{(n)}|\tilde{\v{p}}\varphi_s^{(n)}}=\v{p^2}+\frac{1}{2}\int & d^3
\v{x}\medspace \phi_s^{(n),*}(\v{x})\v{\nabla}^2\phi_s^{(n)}(\v{x})\nonumber\\
&-gp_0\int d^3
\v{x}\medspace \phi_s^{(n),*}(\v{x})A_0^{(n)}(\v{x})\phi_s^{(n)}(\v{x}),
\end{align}
and, by using the energy momentum relation, we see that, in order to have
$\delta\v{p}\ll \v{p}$, we find the condition
\begin{equation}
g\braket{A_0^{(n)}(\v{x})}\equiv g\int d^3\v{x}\phi_s^{(n),*}(\v{x})A_0^{(n)}(\v{x})\phi_s^{(n)}(\v{x})\ll p_0,
\label{glaubercondition1}
\end{equation}
which guarantees that the initial energy greatly exceeds the averaged 
potential, together with
\begin{equation}
\v{\nabla}^2\phi_s^{(n)}(\v{x})\sim 0,
\label{glaubercondition2}
\end{equation}
which guarantees that the wave function varies slowly in a wavelength,
as expected if the particle wavelength is much smaller than the spread
of the interaction. Conditions \eqref{glaubercondition1} and
\eqref{glaubercondition2} are 
essentially the same as the ones required for the Glauber high
energy approximation of Schrodinger particles \cite{glauber1959}. They
guarantee that the quotient $\delta p/p$ is very small. If
we place the initial particle direction along $\v{z}$ the transverse $\v{q}_t$
and longitudinal $\v{q}_z$ momentum changes verify
\begin{align}
|\v{q}_t| = \beta p_0
\sin\theta \sim \beta p_0
\theta, \medspace \medspace \medspace \medspace|\v{q}_z| =
\beta p_0\left(1-\cos\theta\right) \sim  \beta p_0 \theta^2,
\end{align}
since $\theta$ can be made arbitrarily small. Under these conditions the longitudinal
momentum change can be neglected and the paraxial  
limit for the electron propagation is achieved.

We now introduce the corrections to 
$\varphi_s^{(n)}(x)$ resulting of not neglecting the $g$ and $g^2$ terms in
\eqref{psiequation}. The full solution $\psi^{(n)}(x)$ to
\eqref{psiequation} can be given with a modified ansatz \cite{bess1950} 
\begin{align}
\psi^{(n)}(0,\v{x})=\mathcal{N}(p)e^{-i\v{p\cdot
    x}}(1+{\Lambda}(p))u(p)\phi_s^{(n)}(\v{x})\label{modified_ansatz},
\end{align}
where ${\Lambda}(p)$ is defined as an operator commuting with $\v{\nabla}^2$,
$\v{\nabla}$ and thus $\v{p}$, and $\phi_s(\v{x})$ is given by
\eqref{phiequation}. Inserting \eqref{modified_ansatz} in \eqref{psiequation} we get
\begin{align}
2gp_0\bigg(\Lambda(p) &A_0^{(n)}(\v{x})\bigg)u(p)\phi_s^{(n)}(\v{x})
\nonumber \\
&=\left\{ig\v{\alpha}\cdot\v{\nabla}A_0^{(n)}(\v{x})-
  \left(gA_0^{(n)}\right)^2(\v{x})\right\}(1+{\Lambda}(p))u(p)\phi_s^{(n)}(\v{x}).
\end{align}
Inspecting this equation it's clear that, for positive energy solutions, the
operator in the ansatz has to be defined as\footnote{The wave in the form \eqref{modified_ansatz} with the operator defined as
\eqref{operator_definition} corresponds to a $g^2/l^2$ truncation of Darwin's solution \cite{darwin1928} of the Dirac
equation under a Coulomb field, given as an infinite series in spherical harmonics, so
it remains also valid at low energies. Further truncations
\cite{furry1934,sommerfeld1935} neglecting also terms of order $1/p^0$
in Darwin's solution had already been made, a restriction later shown
\cite{bethe1954} unnecessary.} 
\begin{align}
{\Lambda}(p)=\frac{i}{2p_0}\v{\alpha}\cdot\v{\nabla},
\label{operator_definition}
\end{align}
so the first term on the right hand side is canceled. The approximation is
good up to the neglection of the remaining term, of squared order $(gA_0^{(n)})^2$. 
Its contribution to certain amplitudes, like the bremsstrahlung one, however, 
is negligible \cite{bethe1954}. The correction introduced by the
operator seems
of order $1/p_0$, but its matrix structure may introduce terms of order $p_0$ 
coming from the spinors in the amplitudes. It results convenient to move the 
operator completely to the left, acting over $\varphi_s^{(n)}(x)$ instead of over
$\phi_s^{(n)}(x)$,
\begin{equation}
\psi^{(n)}(0,\v{x})=\left(1+\frac{i}{2p_0}\v{\alpha}\cdot\v{\nabla}+\frac{1}{2p_0}\v{\alpha}\cdot\v{p}\right)\varphi_s^{(n)}(0,\v{x}).
\label{states}
\end{equation}
Same equation can be put in terms of the propagators. By noticing
\begin{equation}
\psi^{(n)}(x)=i\int d^3\v{y} \medspace S_F^{(n)}(x,y)\gamma^0\psi^{(n)}(y),\medspace\medspace\medspace
\varphi_s^{(n)}(x)=\int d^3\v{y}\medspace G_s^{(n)}(x,y)\varphi_s^{(n)}(y)
\label{propagationrelation}
\end{equation}
and using \eqref{states}, the propagator at high energies for
$\psi^{(n)}(x)$ is related with the propagator for $\varphi_s^{(n)}(x)$ as
\begin{equation}
S_{F}^{(n)}(x,y)=\left(1+\frac{i}{2p_0}\v{\alpha}\cdot\v{\nabla}+\frac{1}{2p_0}\v{\alpha}\cdot\v{p}\right)\gamma_0 G_s^{(n)}(x,y).
\label{diracpropagator}
\end{equation}
We are now in position to solve for $\varphi_s^{(n)}(x)$. To do that we notice that 
equation \eqref{psiequation} can be arranged 
in a Schrodinger form by solving for the eigenvalue $p_0$. 
In this way the time evolution operator $H_s^{(n)}$ is found to be
\begin{equation}
H_s^{(n)}=-\frac{1}{2(p_0/2)}\v{\nabla}^2+2gA_0^{(n)}(\v{x}),
\label{hamiltonian}
\end{equation}
so the solution at arbitrary time $x_0$ reads
\begin{align}
\varphi_s^{(n)}(x_0,\v{x})=\exp\left(-i x_0
H_s^{(n)}\right)\varphi_s^{(n)}(0,\v{x}).
\end{align}
Notice that $\varphi_s^{(n)}(x)$ has spinorial structure but it evolves like a
scalar with a Hamiltonian where the mass role is played by the energy, meaning
that in the infinite momentum limit energy becomes the only inertia. In order
to solve it, we can write the Schrodinger propagator \cite{feynman1948} as a
path integral
\begin{align}
G_s^{(n)}(x_f,x_i)\equiv \int \mathcal{D}^3\v{x}(t) \medspace\exp\left[i\int^{t_f}_{t_i}
  dt\medspace \left(\frac{p_0}{4}\dot{\v{x}}^2(t)-2gA_0^{(n)}(\v{x}(t))\right)\right],
\end{align}
and take the limit $t_i\to -\infty$ and $p_0\gg 1$. In order to do that
we go to the reference frame of the asymptotic particle, of
initial momentum $p$, given by the coordinate transformation
\begin{align}
\v{w}(t)=\v{x}(t)-2\frac{\v{p}}{p_0}t=\v{w}(t)+2\v{\beta}t.
\label{transformed_paths}
\end{align}
In this Galilean transformation the role of mass is played by the energy and
the Lagrangian transforms to
\begin{align}
\frac{p_0}{4}\dot{\v{x}}^2(t)-2gA_0^{(n)}(\v{x}(t)):=\frac{p_0}{4}\dot{\v{w}}^2(t)+\frac{\v{p}^2}{p_0}+\v{\dot{w}}(t)\cdot\v{p}-2gA_0^{(n)}(\v{w}(t)+2\v{\beta}t),
\end{align}
where the two new non path-dependent terms can be directly integrated out, yielding
\begin{align}
G_s^{(n)}(x_f,x_i)&=\exp\left[+i\v{p}\cdot (\v{w}(t_f)-\v{w}(t_i))+i\frac{\v{p}^2}{p_0}(t_f-t_i)\right]\\
&\cdot\int \mathcal{D}^3\v{w}(t) \exp\left[i\int^{t_f}_{t_i}
  dt\left(\frac{p_0}{4}\dot{\v{w}}^2(t)-2gA_0^{(n)}(\v{w}(t)+2\frac{\v{p}}{p_0}t)\right)\right].
\end{align}
We can see that the interaction is evaluated over the deformed paths \eqref{transformed_paths}.
Path fluctuations weighted by the kinetic term, however, affect only to
$\v{w}(t)$, so paths in the interaction term start at
$\v{x}(t_i)+2\v{\beta}t_i$ and end at $\v{x}(t_f)+2\v{\beta}t_f$. When
the large limit of $\beta$ is taken and $t_i\to-\infty$, the deformed
paths in the interaction term are almost straight lines parallel to
$\v{p}$, ending at a transverse distance $\v{w}_t(t_f)$
with respect to $\v{p}$ and at longitudinal distance
$\v{w}_l(t_f)=2\v{\beta} t_f$. By replacing them by straight lines the path integral in the interaction
term can be taken approximately constant for all the fluctuations so it factorizes as
\begin{align}
&\int \mathcal{D}^3\v{w}(t) \exp\bigg[i\int^{t_f}_{t_i}
  dt\medspace \bigg(\frac{p_0}{4}\dot{\v{w}}^2(t)-2gA_0^{(n)}(\v{w}(t)+2\v{\beta}t)\bigg)\bigg]\\
&\approx\exp\left[-2ig\int^{t_f}_{-\infty}dt\medspace
A_0^{(n)}(\v{w}_t(t_f)+2\v{\beta}t)\right]\left(\int \mathcal{D}^3\v{w}(t) \exp\left[i\int^{t_f}_{t_i}
  dt\medspace \bigg(\frac{p_0}{4}\dot{\v{w}}^2(t)\bigg)\right]\right)\nonumber.
\end{align}
and, by adding the phase terms, defining $\v{s}=2\v{\beta}t$, placing the
$\v{p}$ along the z-axis and undoing the change, one finds
\begin{equation}
G_s^{(n)}(x_f,x_i)=G_s^{(0)}(x_f,x_i)\exp\left[-i\frac{g}{\beta}\chi_0^{(n)}(\v{x}_f)\right]\equiv
G_s^{(0)}(x_f,x_i)W^{(n)}(\v{x}_f,p).\label{schrodingerpropagator}
\end{equation}
where we have introduced the free propagator $G_s^{(0)}(x_f,x_i)$ for
\eqref{hamiltonian} and defined the shorthand notation
\begin{align}
\chi_0^{(n)}(\v{x})\equiv \int^{\hat{\v{p}}\cdot
    \v{x}}_{-\infty}ds \medspace A_0^{(n)}(\hat{\v{p}}\times(\v{x}\times\hat{\v{p}})+s\hat{\v{p}}).
\end{align}
In terms of an asymptotic free wave $\psi^{(0)}(x)$ 
of momentum $p$ we finally find, using equations \eqref{propagationrelation},
 \eqref{diracpropagator} and \eqref{schrodingerpropagator},
\begin{align}
\psi^{(n)}(x)=\left\{\left(1+i\frac{\gamma_k\gamma_0}{2p_0^0}\partial_k\right)W^{(n)}(\v{x},p)\right\}\psi^{(0)}(x),\label{wave_integration}
\end{align}
sum in repeated indices assumed. We see that, except for the operator correction, as expected, the above result is the generalization for
spin-$1/2$ particles of the Glauber's high energy waves
\cite{glauber1959}. Finally, an equation satisfied by the high energy limit of
the wave can be written. Indeed, by simple derivation we find
\begin{align}
\frac{\partial \varphi_s^{(n)}(0,\v{x})}{\partial x_3} = \bigg(ip_3-\frac{i}{\beta}gA_0^{(n)}(\v{x})\bigg)\varphi_s^{(n)}(0,\v{x}).\label{highenergy_equation}
\end{align}
This result, or alternatively \eqref{schrodingerpropagator}, says us that the
medium, in the high energy limit, only adds a phase to a free propagation,
the phase being just the integration of the static component of the external
field along the direction of propagation of the fermion till the position
$x_3$. Furthermore, second derivatives can be neglected in virtue of \eqref{glaubercondition2}.
\section{Perturbative expansion}
\label{section_1_3}
Glauber's integration \cite{glauber1959} of the wave at high
energies do reproduce the Born approximation at high energies of a Schrodinger wave. 
Therefore, equation \eqref{wave_integration} must agree with the
series expansion in $g$ of a spin-1/2 particle. When expanded in the
external field $A_0^{(n)}(\v{x})$ we get
\begin{equation}
\psi^{(n)}(x)=\psi^{(0)}(x)-i\frac{g}{\beta}\left(1+\frac{i\v{\alpha}\cdot\v{\nabla}}{2p_0}+\frac{\v{\alpha}\cdot\v{p}}{2p_0}\right)\chi_0^{(n)}(\v{x},p)\psi^{(0)}(x)+(...).
\label{perturbationeikonal}
\end{equation}
This series expansion must agree with the standard perturbative series for
$\psi(x)$, this is, the Born series given by
\begin{align}
\psi^{(n)}(x)=\psi^{(0)}(x)+g\int dy\medspace S_F^{(0)}(x-y)\gamma^0A_0^{(n)}(y)\psi^{(0)}(y) + \ldots.
\label{perturbationfeynman1}
\end{align}
In order to check if the eikonal integration 
in \eqref{perturbationeikonal} equals the leading order term of
\eqref{perturbationfeynman1}, we write the external field as the Fourier transform of some $A_0^{(n)}(\v{q})$,
\begin{align}
A_0^{(n)}(y)=\int \frac{d^4q}{(2\pi)^4}\medspace 2\pi\delta(q_0)e^{-iq\cdot
  y}A_0^{(n)}(q),
\end{align}
and use the Feynman-Stueckelberg propagator for the Dirac equation
\begin{align}
S_F^{(0)}(x-y) =\int \frac{d^4k}{(2\pi)^4}\frac{\slashed{k}+m}{k^2-m^2}
e^{-ik\cdot(x-y)}.
\label{electron_propagator}
\end{align}
By inserting these two expressions in equation \eqref{perturbationfeynman1}
and integrating in $y$ and $k$ one arrives to
\begin{align}
\psi^{(n)}(x)
=\psi^{(0)}(x)+ g&\int \frac{d^4q}{(2\pi)^4} \medspace 2\pi \delta(q_0)
\frac{\slashed{q}+\slashed{p}+m}{(q+p)^2-m^2}\gamma^0A_0^{(n)}(\v{q})e^{-iq \cdot x}\psi^{(0)}(x).
\end{align}
Here we have to express the numerator as
\begin{align}
(\slashed{q}+\slashed{p}+m)\medspace\gamma_0\psi^{(0)}(x)
&=-q_i\gamma_i\gamma_0\psi^{(0)}(x)+(i\gamma_u\partial^u+m)\gamma_0\psi^{(0)}(x)\nonumber\\
&=-(\v{q}\cdot\v{\alpha})\psi^{(0)}(x)+2i\partial_0\psi^{(0)}(x)-\gamma_0(\slashed{p}-m)\psi^{(0)}(x)\nonumber\\
&=-(\v{q}\cdot\v{\alpha})\psi^{(0)}(x)+2p_0\psi^{(0)}(x),
\end{align}
and replace the term $\v{q}$ with a gradient in order to move it out of the
integral sign
\begin{align}
-(\v{q}\cdot\v{\alpha})e^{-iq\cdot
  x}\psi^{(0)}(x)=\left(i\v{\alpha}\cdot\v{\nabla}+\v{\alpha}\cdot\v{p}\right)e^{-iq\cdot
  x}\psi^{(0)}(x).
\end{align}
Since the denominator becomes 
\begin{align}
(p+q)^2-m^2=m^2+q^2+2p\cdot
q-m^2=-\v{q}^2-2\v{q}\cdot\v{p},
\end{align}
one finds
\begin{align}
\psi^{(n)}(x)=\psi^{(0)}(x)-g&\left(1+\frac{i\v{\alpha}\cdot\nabla}{2p_0}+\frac{\v{\alpha}\cdot\v{p}}{2p_0}\right)\int
\frac{d^3\v{q}}{(2\pi)^3}e^{i\v{q\cdot x}}\frac{2p_0}{\v{q}^2+2\v{q\cdot
    p}}A_0^{(n)}(\v{q})\psi^{(0)}(x).
\end{align}
When $\v{p}\gg 1$, the integral is dominated by the term $\v{q\cdot p}$ and we
can neglect $\v{q}^2$. Notice that $\v{q}\cdot \v{p}$ is the momentum change
in the initial direction, which we will take as the $z$ direction. Now, using
the decomposition
$\v{x}=\hat{\v{p}}\times(\v{x}\times\hat{\v{p}})+\hat{\v{p}}(\v{x}\cdot\hat{\v{p}})$
we write the following trick
\begin{align}
\frac{e^{i\v{q\cdot x}}}{\v{q\cdot p}}\equiv
  \frac{i}{|\v{p}|}\int_{-\infty}^{\v{x\cdot \hat{p}}} ds\medspace 
  \exp\bigg[i\v{q}\cdot
    (\hat{\v{p}}\times(\v{x}\times\v{\hat{p}})+s\hat{\v{p}})\bigg].
\end{align}
By using the Fourier transform of the field we identify now
\begin{align}
\int
\frac{d^3\v{q}}{(2\pi)^3}e^{i\v{q}\cdot\v{x}}\frac{p_0}{\v{q}\cdot\v{p}}A^{(n)}_0(\v{q})&=\frac{i}{\beta}\int_{-\infty}^{\v{x}\cdot\hat{\v{p}}}
ds
\int\frac{d^3\v{q}}{(2\pi)^3}e^{i\v{q}\cdot(\hat{\v{p}}\times(\v{x}\times\hat{\v{p}})+s\hat{\v{p}})}A_0^{(n)}(\v{q})\nonumber\\
&=\frac{i}{\beta} \int_{-\infty}^{\v{x}\cdot\hat{\v{p}}} ds \medspace
A_0^{(n)}(\hat{\v{p}}\times(\v{x}\times\hat{\v{p}})+s\hat{\v{p}}),
\end{align}
which enables us to recover the first term of the series in \eqref{perturbationeikonal}
\begin{equation}
\psi^{(n)}(x)=\psi^{(0)}(x)-i\frac{g}{\beta}\left(1+\frac{i\v{\alpha}\cdot\v{\nabla}}{2p_0}+\frac{\v{\alpha}\cdot\v{p}}{2p_0}\right)\chi^{(n)}(\v{x},p)\psi^{(0)}(x)+\ldots.
\label{perturbationfeynman2}
\end{equation}
The high energy integration of the wave, thus, reproduces the perturbative
expansion of the scattered wave.
\section{Propagators in momentum and light cone spaces}
\label{section_1_4}
Computing the propagator for the Schrodinger wave in momentum space 
follows the same procedure taken to evaluate the propagator in position space. 
As we will see in the next chapter, the scattering amplitude for $\psi^{(n)}(x)$, 
omitting the spin and the operator correction, is essentially given
by the following Schrodinger propagator in momentum space. In order to find
it we simply Fourier transform
\begin{align}
G_s^{(n)}(p_f,p_i)\equiv\int d^3\v{x}_f\int
d^3\v{x}_i\medspace 
e^{-i\v{p}_f\cdot\v{x}_f}G_s^{(n)}(x_f,x_i)e^{i\v{p}_i\cdot
  \v{x}_i}.
\end{align}
Now, define the momentum change $\v{q}=\v{p}_f-\v{p}_i$ and rewrite the above
expression as
\begin{align}
G_s^{(n)}(p_f,p_i)
&=\int d^3\v{x}_f\int
d^3\v{x}_i\medspace e^{-i\v{q}_f\cdot\v{x}_f}\\
\cdot &\int \mathcal{D}^3\v{x}(t)\exp\left[i\int^{t_f}_{t_i} dt\left(
  \frac{p_0}{4}\dot{\v{x}}^2(t)-\v{p}_i\cdot\dot{\v{x}}(t)-2gA_0^{(n)}(\v{x}(t))\right)\right].\nonumber
\end{align}
Here, in the last step, the term $i\v{p}_i\cdot(\v{x}_i-\v{x}_f)$ has been
introduced back in the integral as a total derivative. Now, as before, let us
introduce the change \eqref{transformed_paths} leading to 
\begin{align}
G_s^{(n)}(p_f,p_i)&=\exp\left[-i\frac{\v{p}_i^2}{p_i^0}(t_f-t_i)-2i\frac{\v{q}\cdot\v{p_i}}{p_i^0}t_f\right]\int
d^3\v{w}_f\int
d^3\v{w}_i\medspace e^{-i\v{q}_f\cdot\v{w}_f}\nonumber\\
\cdot &\int \mathcal{D}^3\v{w}(t)\exp\bigg[
i\int_{t_i}^{t_f}
dt\medspace\bigg(\frac{p_i^0}{4}\dot{\v{w}}^2(t)-2gA_0^{(n)}(\v{w}(t)+2\v{\beta}t)\bigg)\bigg],
\end{align}
so
\begin{align}
G_s^{(n)}(p_f,p_i)\approx &\exp\bigg[-i\frac{\v{p}_i^2}{p_i^0}(t_f-t_i)-2i\frac{\v{q}\cdot\v{p_i}}{p_i^0}t_f\bigg]\int
d^3\v{w}_f\int
d^3\v{w}_i\medspace e^{-i\v{q}_f\cdot\v{w}_f}\nonumber\\
&\cdot \bigg(\int \mathcal{D}^3\v{w}(t)\exp\bigg[
i\int_{t_i}^{t_f}
dt\medspace\bigg(\frac{p_i^0}{4}\dot{\v{w}}^2(t)\bigg)\bigg]\bigg)\nonumber\\
&\cdot\exp\bigg[-2ig \int_{t_i}^{t_f} dt\medspace
  A_0^{(n)}\bigg(\v{w}_t(t_f)+2\v{\beta}t\bigg)\bigg],
\end{align}
where the path integration in the interaction term has been factorized out of
all the path fluctuations due to the limit $t_i\to-\infty$ and $\beta\to
1$, and the longitudinal and perpendicular coordinates refer to $\v{p}_i$. 
The integral in $\v{w}(t_i)$ only affects to the free propagator and
consequently, using the normalization condition 
\begin{align}
\int d^3\v{w}_i \medspace G_s^{(0)}(w_f,w_i)=1,
\end{align}
and transforming back to the original system and denoting
$\hat{\v{p}}_is=2\v{\beta}t$ one finds
\begin{align}
G_s^{(n)}(p_f,p_i)=\int d^3\v{x}_f\medspace
e^{-i\v{q}\cdot\v{x}_f}\exp\left[-i\frac{g}{\beta}
  \int_{-\infty}^{x_f^l} ds\medspace 
  A_0^{(n)}\left(\v{x}_f^t+\hat{\v{p}}_is\right)\right].
\end{align}
The particular limit $t_f\to\infty$ corresponds to the scattering
amplitude if the spinorial part and the low energy corrections of the state
are omitted. In this case one finds a conservation of longitudinal
momentum change  $2\pi\delta(q_l)$ and the integral restricts to
the transverse plane to $\v{p}_i$. For completeness we may compute 
the propagator in light-cone variables too. We define light cone variables as customary
\begin{equation}
x_+=\frac{x_0+x_3}{\sqrt{2}},\medspace\medspace\medspace\medspace
x_-=\frac{x_0-x_3}{\sqrt{2}},\medspace\medspace\medspace\medspace \v{x}_t=(x_1,x_2),
\end{equation}
in such way that the scalar product of two 4-vectors reads
\begin{align}
x_\mu y^\mu =
x_0y_0-x_3y_3-x_1x_1-x_2x_2=x_+y_-+x_-y_+-\v{x}_t\cdot\v{y}_t.
\end{align}
We depart from a free fermion $\psi^{(0)}(x)$ satisfying the Dirac equation and, thus, the
squared Dirac equation
\begin{align}
(\slashed{p}-m)\psi^{(0)}(x)=0 \rightarrow
  (\slashed{p}+m)(\slashed{p}-m)\psi^{(0)}(x)=(i\partial_\mu\partial_\nu-m^2)\psi^{(0)}(x)=0.
\end{align}
Now, since the spinorial structure in the squared equation is factorizable, by
using $\psi^{(0)}(x)=\sqrt{m/p_0}u(p)\varphi^{(0)}(x)$ and the identity
\begin{align}
\partial_0\partial_0-\partial_3\partial_3=\frac{1}{2}\left(\frac{\partial}{\partial
x_+}+\frac{\partial}{\partial
  x_-}\right)^2-\frac{1}{2}\left(\frac{\partial}{\partial
  x_+}-\frac{\partial}{\partial x_-}\right)^2=2\frac{\partial}{\partial
  x_+}\frac{\partial}{\partial x_-},
\end{align}
then the squared Dirac equation for $\varphi^{(0)}(x)$ can be rewritten in the light cone variables as 
\begin{equation}
\left(2\frac{\partial}{\partial
  x_+}\frac{\partial}{\partial x_-}-\frac{\partial^2}{\partial^2
  \v{x}_t}-m^2\right)\varphi^{(0)}(x)=0.
\label{kleingordonlightcone}
\end{equation}
Let us use $p_-$ as the generator of the evolution in $x_+$ of the states and
define the Fourier transform of $\varphi^{(0)}(x)$ at time point $x_+$
\begin{align}
 \varphi^{(0)}(x_+,x_-,\v{x}_t)=\int \frac{dp_+}{2\pi} \int
\frac{d^2\v{p}_t}{(2\pi)^2}
e^{-ip_+x_-+i\v{p}_t\cdot\v{x}_t}\tilde{\varphi}^{(0)}(x_+,p_+,\v{p}_t).
\end{align}
By inserting this expression in equation \eqref{kleingordonlightcone} one
finds
\begin{align}
\int \frac{dp_+}{2\pi} \int
\frac{d^2\v{p}_t}{(2\pi)^2}e^{-ip_+x_-+i\v{p}_t\cdot\v{x}_t}\left(-2ip_+\frac{\partial}{\partial
  x_+}+\v{p}_t^2-m^2\right)\tilde{\varphi}^{(0)}(x_+,p_+,\v{p}_t)=0,
\end{align}
from which it derives the equation 
\begin{equation}
\left(-2ip_+\frac{\partial}{\partial
  x_+}+\v{p}_t^2-m^2\right)\tilde{\varphi}^{(0)}(x_+,p_+,\v{p}_t)=0,
\end{equation}
or
\begin{equation}
i\frac{\partial}{\partial x_+}\tilde{\varphi}^{(0)}(x_+,p_+,\v{p}_t)=\left(\frac{\v{p}_t^2}{2p_+}-\frac{m^2}{2p_+}\right)\tilde{\varphi}^{(0)}(x_+,p_+,\v{p}_t),
\end{equation}
which is a Schrodinger like equation for a particle of fictitious mass $p_+$
moving in the transverse plane to $x_3$. Correspondingly its free Green
function is given by
\begin{align}
G_s^{(0)}(x_+^1,\v{x}_t^1;x_+^0,\v{x}_t^0)&=\exp\left[+i\frac{m^2}{2p_+}(x_+^1-x_+^0)\right]\nonumber\\
&\left\{\frac{p_+}{2\pi(x_+^1-x_+^0)}\exp\left[+i\frac{p_+}{2(x_+^1-x_+^0)}(\v{x}_t^1-\v{x}_t^0)^2\right]\right\}.
\end{align}
This function propagates the free solutions between two space-time points in
light cone variables. In order to add an interaction, we simply use the
perturbative definition of the interacting Green function and take into
account that the propagation is not constrained, in general, to a plane, if
the interaction term is not, so
\begin{align}
&G_s^{(n)}(x_+^1,x_-^0,\v{x}_t^1;x_+^0,x_-^0,\v{x}_t^0)=G_s^{(0)}(x_+^1,\v{x}_t^1;x_+^0,\v{x}_t^0)+\nonumber\\
&-ig\int_{x_+^0}^{x_+^1}
dx_+\int d^{3}x
G_s^{(0)}(x_+^1,\v{x}_t^1;x_+,\v{x}_t)A_0^{(n)}(x)G_s^{(0)}(x_+,\v{x}_t;x_+^0,\v{x}_t^0)+(...)
\end{align}
and, since the free propagation does not depend on the $x_-$ variable, one
finds
\begin{align}
G_s^{(n)}(x_+^1,x_-^0,\v{x}_t^1;x_+^0,x_-^0,\v{x}_t^0)&=G_s^{(0)}(x_+^1,\v{x}_t^1;x_+^0,\v{x}_t^0)\nonumber\\
-ig\int_{x_+^0}^{x_+^1} dx_+ \int d^2\v{x}_t\medspace
G_s^{(0)}(&x_+^1,\v{x}_t^1;x_+,\v{x}_t) \nonumber\\
&\cdot\left\{\int dx_-
A_0^{(n)}(x)\right\}G_s^{(0)}(x_+,\v{x}_t;x_+^0,\v{x}_t^0)+(...)\nonumber\\
\equiv\int \mathcal{D}^2\v{r}_t \exp\bigg[i\int^{x_+^1}_{x_-^0} dx_+ &\left(\frac{p_+}{2}\dot{\v{r}}_t^2-g\int^{+\infty}_{-\infty} dx_-A_0^{(n)}(x)\right)\bigg]
\end{align}
which is the perturbative expansion of a path integral in the effective potential
\begin{align}
-ig\int dx_- A_0^{(n)}(\v{x}):=-ig\int dx_-
A_0^{(n)}\left(\v{x}_t,\frac{x_+-x_-}{\sqrt{2}}\right).
\end{align}
Since the interaction terms stops being a function of the $x_+$ variable once
integrated in $x_-$, consequently, the path integration affects only to the
free propagator yielding
\begin{align}
G_s^{(n)}(x_1,x_0)=G_s^{(0)}(x_1,x_0)\exp\left[-ig\int^{+\infty}_{-\infty}
    dx_-A_0^{(n)}(x)\right].
\end{align}
The above propagator has lost the local dependence of the field with
coordinate $x_3$ as a result of the time evaluation of $x_-$ at $x_0=\pm
\infty$.

%% file: chapter2/chapter2.tex
\chapter{High energy multiple scattering}
With the development of the wave function in the previous chapter we are now
in position to define the scattering amplitude, that is, the amplitude of
finding the particle with some momentum and spin due to the effect of the
medium
\cite{scott1949,gnedin1964,weinberg1966,glauber1970,kogut1970,bjorken1971}. We
will find that the obtained result preserves the unitarity of the emerging
wave by proving the optical theorem, and that the series expansion in the
coupling $g=Ze^2$ reproduces the perturbative description of the same
problem. As previously stated, the evaluation of the wave or the scattering
amplitudes for a particular configuration of the medium constituents is not
interesting and we will instead look for averaged squared amplitudes and
observables. In this way we will find that these averaged quantities for $n$
scattering sources are always expressible as exponentiations of the single
$n$=1 case.  In the process of averaging we will discover also the emergence
of interference phenomena related to multiple scattering events which suggests
a splitting of the cross sections into two contributions.  In a certain limit,
namely for macroscopic mediums, the resulting cross sections become totally
incoherent and therefore the scattering process will acquire an statistical
interpretation in terms of probability distributions, whereas for mediums of
microscopic size diffractive phenomena and thus, medium coherence, have a
prominent role over the incoherent limit. We will also demonstrate that the
incoherent contribution leads to the well-known Moliere's derivation
\cite{moliere1948} based on markovian arguments \cite{bethe1953} and we will
compute the averaged momentum transfer $\langle\v{q}^2\rangle$. Finally we
will introduce a beyond eikonal evaluation by considering a non vanishing
longitudinal momentum change. We will show that the beyond eikonal squared
amplitude, when averaged over mediums of macroscopic dimensions, reduces to
the pure eikonal scattering for on-shell particles. However, longitudinal
momentum changes at different energy states can produce interferences in the
squared amplitudes, ultimately leading to incoherence phenomena when creation
or annihilation of particles is also considered, as we will see in the next
chapter.
\section{The scattering amplitude}
\label{sec:section_2_1}
We depart from an initial asymptotic free fermion with momentum $p_i$ and spin
$s_i$, whose state will be denoted $\psi_i^{(0)}(x)$, 
\begin{align}
\psi_i^{(0)}(x)=\sqrt{\frac{m}{p_i^0}}u_{s_i}(p_i)e^{-ip_i\cdot x}.
\end{align}
Due to the effect of the external field \eqref{general_external_field}
produced by the medium, in the asymptotic final state one finds the
superposition of two states, the former wave itself and an infinite
superposition of states of deflected momentum $p_f$ and spin $s_f$. This reads
\begin{equation}
\psi_f^{(n)}(x)=\psi_i^{(0)}(x)+
\sum_{s_f=1,2}\int\frac{d^3\v{p}_f}{(2\pi)^3}M_{s_fs_i}^{(n)}(p_f,p_i)\sqrt{\frac{m}{p_f^0}}u_{s_f}(p_f)e^{-ip_f\cdot
  x}.
\label{outcomingstatescattering}
\end{equation}
The weights are referred to as the scattering amplitude or {\it M}-matrix, this
is, the amplitude to find the wave with deflected momentum $p_f$ and spin
$s_f$. The asymptotic initial state can be absorbed in the scattering
amplitude if we define 
\begin{equation}
\psi_f^{(n)}(x)=
\sum_{s_f=1,2}\int\frac{d^3\v{p}_f}{(2\pi)^3}S_{s_fs_i}^{(n)}(p_f,p_i)\sqrt{\frac{m}{p_f^0}}u_{s_f}(p_f)e^{-ip_f\cdot
  x},
\label{outcomingstatescattering_smatrix}
\end{equation}
where the {\it S}-matrix includes the no collision
distribution plus the collisional {\it M}-matrix distribution, and by direct
inspection of \eqref{outcomingstatescattering} and
\eqref{outcomingstatescattering_smatrix} verifies then
\begin{align}
S_{s_fs_i}^{(n)}(p_f,p_i) \equiv M_{s_fs_i}^{(n)}(p_f,p_i)+(2\pi)^3\delta^3(\v{p}_f-\v{p}_i)\delta_{s_fs_i}.
\end{align}
In order to find the scattering amplitude we rewrite 
the emerging wave in \eqref{outcomingstatescattering} by using the
Lippmann-Schwinger recursive equation,
\begin{align}
\psi_f^{(n)}(x)&=\psi_i^{(0)}(x)+\int d^4y
S_F(x-y)g\gamma^0A_0^{(n)}(\v{y})\psi^{(n)}(y)
=\psi_i^{(0)}(x)+\psi_{diff}^{(n)}(x).
\end{align}
Inserting the result \eqref{wave_integration}, valid for the wave in the high
energy limit, we find
\begin{align}
\psi_{diff}^{(n)}(x)=\int d^4y
S_F(x-y)&g\gamma^0A_0^{(n)}(\v{y})
\cdot\left\{\left(1+i\frac{\gamma_k\gamma_0}{2p_i^0}\partial_k\right)W^{(n)}(\v{y},p_i)\right\}\psi_i^{(0)}(y).\label{diffracted_part_wave1}
\end{align}
In order to express this diffracted wave as an infinite
superposition broadened around $\v{p}_i$, we integrate the Feynman-Stueckelberg
propagator \eqref{electron_propagator} for energy positive solutions getting
\begin{align}
S_F^{(+)}(x)=\frac{1}{i}\int
\frac{d^3\v{p}}{(2\pi)^3}\frac{\slashed{p}+m}{2p_0}e^{-ip\cdot
  x}=\frac{1}{i}\sum_{s}\int
\frac{d^3\v{p}}{(2\pi)^3}\frac{m}{p_0}u_{s}(p)\otimes\bar{u}_{s}(p)e^{-ip\cdot
    x},\label{propagator_energy_positive}
\end{align}
where the energy momentum $p_0(\v{p})=\sqrt{\v{p}^2+m^2}$ relation
arising in the $p_0$ integration over the positive pole has been left 
implicit and the completeness relation
\begin{align}
\sum_{s=1,2}u_{s}(p)\otimes\bar{u}_{s}(p)=\frac{\slashed{p}+m}{2m},\label{completeness_relation}
\end{align}
has been used. By inserting \eqref{propagator_energy_positive} in
\eqref{diffracted_part_wave1} one finds
\begin{align}
\psi_{diff}^{(n)}(x)&=\sum_{s_1=1,2}\int
\frac{d^3\v{p}}{(2\pi)^3}\sqrt{\frac{m}{p^0_f}}u^{s_f}(p_f)e^{-ip_f\cdot
  x}\int d^4ye^{i(p_f-p_i)y}\left(-igA_0^{(n)}(\v{y})\right)\nonumber\\
&\cdot
  \sqrt{\frac{m}{p^0_f}} \bar{u}_{s_f}(p_f)\gamma^0
\left\{\left(1+i\frac{\gamma_k\gamma_0}{2p_i^0}\partial_k\right)
  W^{(n)}(\v{y},p_i)\right\}\sqrt{\frac{m}{p_i^0}}u_{s_i}(p_i)\nonumber\\
&\equiv\sum_{s_f=1,2}\int
\frac{d^3\v{p_f}}{(2\pi)^3}\sqrt{\frac{m}{p_f^0}}u_{s_f}(p_f)e^{-ip_f\cdot
  x}M_{s_fs_i}^{(n)}(p_f,p_i),
\label{diffracted_part_wave2}
\end{align}
where the integration must be carried on-shell. The expression
\eqref{diffracted_part_wave2} is interpretable as a superposition with amplitude
$M_{s_fs_i}^{(n)}(p_f,p_i)$ of being at the state $(p_f,s_f)$ given by
\begin{align}
\psi_f^{(0)}(x)=\sqrt{\frac{m}{p_f^0}}u_{s_f}(p_f)e^{-ip_fx}.
\end{align}
Thus we define the amplitude, following \eqref{outcomingstatescattering}, as
\begin{align}
M_{s_fs_i}^{(n)}(p_f,p_i)&=\int d^4y e^{i(p_f-p_i)\cdot
  y}\left(-igA_0^{(n)}(\v{y})\right)\\
&\cdot\sqrt{\frac{m}{p_f^0}}\bar{u}_{s_f}(p_f)\gamma^0\left\{\left(1+i\frac{\gamma_k\gamma_0}{2p_i^0}\partial_k\right)
  W^{(n)}(\v{y},p_i)\right\}u_{s_i}(p_i)\sqrt{\frac{m}{p_i^0}}.\nonumber
\label{elasticamplitude}
\end{align}
The quantity in the bottom line of the above expression, which we call $J$,
contains the non perturbative information of the diffracted wave,
\begin{align}
J=\sqrt{\frac{m}{p_f^0}}\bar{u}_{s_f}(p_f)\gamma_0\left\{\left(1+i\frac{\gamma_k\gamma_0}{2p_i^0}\partial_k\right)
  W_0^{(n)}(\v{y},p_i)\right\}u_{s_i}(p_i)\sqrt{\frac{m}{p_i^0}}.
\end{align}
Had we expanded the phase in the coupling we would have found
\begin{align}
W^{(n)}(\v{y},p_i)= 1 +\mathcal{O}\left(\frac{g}{\beta_i}\right)\rightarrow J
= \sqrt{\frac{m}{p_f^0}}\bar{u}_{s_f}(p_f)\gamma_0
u_{s_i}(p_i)\sqrt{\frac{m}{p_i^0}}+\mathcal{O}\left(\frac{g}{\beta_i}\right)
\end{align}
which, together the Fourier transform of of $A_0^{(n)}(\v{y})$
would produce the first perturbative term of the scattering matrix,
representing all single kicks combinatory with the medium elements. However, 
as we stated before, we expect that the perturbative
processes of high order are the relevant ones, then we have to compute the term
non perturbatively. We have
\begin{align}
J&=\sqrt{\frac{m}{p_f^0}}\bar{u}_{s_f}(p_f)\left(\gamma_0+\frac{i}{2p_i^0}\gamma_k^\dag\frac{\partial}{\partial
  x_k}\right)
u_{s_i}(p_i)\sqrt{\frac{m}{p_i^0}} W^{(n)}(\v{y},p_i).
\end{align}
Notice that, although the derivative correction is suppressed with an
overall factor $1/p_i^0$, the spinorial terms may introduce $p_i^0$ corrections. 
It is easy to show that for elastic scattering this is not the
case, since the above expression produces in the high energy limit, $p_f\approx p_i$,
\begin{align}
J\simeq\delta^{s_f}_{s_i}\Bigg(1&-i\frac{\beta}{2p_i^0}\partial_3\Bigg)\medspace
W^{(n)}(\v{y},p_i)=\delta^{s_f}_{s_i}\left(1-\frac{gA_0^{(n)}(\v{y})}{2p_i^0}\right)W^{(n)}(\v{y},p_i).
\end{align}
In this limit spin keeps unchanged $s_f=s_i$ since the spin-flip
amplitudes are suppressed a factor $1/p_i^0$ with respect to the
non-flip case. The term arising in the phase derivative can be 
neglected so
\begin{align}
\frac{g\braket{A_0^{(n)}(\v{y})}}{2p_i^0}\ll 1 \to J\simeq\sqrt{\frac{m}{p_f^0}}\bar{u}_{s_f}(p_f)\gamma_0u_{s_i}(p_i)\sqrt{\frac{m}{p_i^0}} W^{(n)}(\v{y},p_i)
\end{align}
where the Glauber condition \eqref{glaubercondition1} has been used.
It can be globally neglected, then, provided that the external interaction is
sufficiently smooth and $p_i^0\to\infty$. By performing the time integral
we get the energy conservation so $\beta_i=\beta_f\equiv\beta$ and
\begin{align}
M_{s_fs_i}^{(n)}(p_f,p_i) =
2\pi\delta(p_f^0-p_i^0)\sqrt{\frac{m}{p_f^0}}\bar{u}_{s_f}(p_f)&\gamma_0u_{s_i}(p_i)\sqrt{\frac{m}{p_i^0}}\\
&\times\int d^3\v{y} e^{-i\v{q}\cdot
  \v{y}}\left(-igA_0^{(n)}(\v{y})\right)W^{(n)}(\v{y},p_0)\nonumber,
\end{align}
where the momentum transfer with the external field is denoted as
$\v{q}=\v{p}_f-\v{p}_i$. The remaining integral can be performed by placing
$\v{p}_i$ along the z axis and taking the high energy limit
$q_z\ll q_t$. We find the asymptotic value
\begin{align}
\int
d^3\v{y}\medspace &e^{-i\v{q}\cdot
    \v{y}}\left(-igA_0^{(n)}(\v{y})\right)W^{(n)}(\v{y},p)\nonumber\\&\approx\beta\int
d^2\v{y}_t\medspace e^{-i\v{q}_t\cdot\v{y}_t}\int_{-\infty}^{+\infty} dy_3
\left(-i\frac{g}{\beta}A_0^{(n)}(\v{y})\right)\exp\left[-i\frac{g}{\beta}\int_{-\infty}^{y_3}
  dy_3' \medspace A_0^{(n)}(\v{y})\right]\nonumber\\
&=\beta\int
d^2\v{y}_t\medspace e^{-i\v{q}_t\cdot\v{y}_t}\left(\exp\left[-i\frac{g}{\beta}\int_{-\infty}^{+\infty}
  dy_3 \medspace A_0^{(n)}(\v{y})\right]-1\right).
\end{align}
Finally, the amplitude of a change of 4-momentum $q=p_1-p_0$ under the field of
$n$ sources, at high energy and at all orders in the coupling is
given \eqref{outcomingstatescattering} by 
\begin{align}
M_{s_fs_i}^{(n)}(p_f,p_i)=2\pi\delta(p_f^0-p_i^0)&\beta \sqrt{\frac{m}{p_f^0}}\bar{u}_{s_f}(p_f)\gamma_0u_{s_i}(p_i)\sqrt{\frac{m}{p_i^0}}
\label{scatteringmatrix}\\&\times
\int
d^2\v{y}_t\medspace e^{-i\v{q}_t\cdot\v{y}_t}\left(\exp\left[-i\frac{g}{\beta}\int_{-\infty}^{+\infty}
  dy_3 \medspace A_0^{(n)}(\v{y})\right]-1\right).
\nonumber
\end{align}
Similarly, by including the no collision amplitude \eqref{outcomingstatescattering_smatrix}, we write
\begin{align}
S_{s_fs_i}^{(n)}(p_f,p_i)=2\pi\delta(p_f^0-p_i^0)&\beta \sqrt{\frac{m}{p_f^0}}\bar{u}_{s_f}(p_f)\gamma_0u_{s_i}(p_i)\sqrt{\frac{m}{p_i^0}}
\label{scatteringmatrix_smatrix}\\&\times
\int
d^2\v{y}_t\medspace e^{-i\v{q}_t\cdot\v{y}_t}\exp\left[-i\frac{g}{\beta}\int_{-\infty}^{+\infty}
  dy_3 \medspace A_0^{(n)}(\v{y})\right].
\nonumber
\end{align}
It results convenient to extract the energy and spin 
conservation delta out of the above amplitudes by defining the
quantities 
\begin{align}
M_{s_fs_i}^{(n)}(p_f,p_i)=2\pi\delta(p_f^0-p_i^0)\delta^{s_f}_{s_i}\beta F_{el}^{(n)}(\v{q})
,\medspace \medspace S_{s_fs_i}^{(n)}(p_f,p_i)=2\pi\delta(p_f^0-p_i^0)\delta^{s_f}_{s_i}\beta S_{el}^{(n)}(\v{q}),\label{scatteringmatrix_fscatteringmatrix_relation}
\end{align}
where $F_{el}^{(n)}(\v{q}_t)$ denotes the integral part of \eqref{scatteringmatrix}
\begin{align}
F_{el}^{(n)}(\v{q}_t)\equiv \int
d^2\v{y}_t\medspace e^{-i\v{q}_t\cdot\v{y}_t}\left(\exp\left[-i\frac{g}{\beta}\int_{-\infty}^{+\infty}
  dy_3 \medspace A_0^{(n)}(\v{y})\right]-1\right).
\label{fscatteringmatrix}
\end{align}
and $S_{el}^{(n)}(\v{q}_t)$ denotes the integral part of
\eqref{scatteringmatrix_smatrix}
\begin{align}
S_{el}^{(n)}(\v{q}_t)\equiv \int
d^2\v{y}_t\medspace e^{-i\v{q}_t\cdot\v{y}_t}\exp\left[-i\frac{g}{\beta}\int_{-\infty}^{+\infty}
  dy_3 \medspace A_0^{(n)}(\v{y})\right].
\label{sscatteringmatrix}
\end{align}
Since $\beta \to 1$ and energy and spin are
preserved $F^{(n)}(\v{q}_t)$ contains all the relevant physics. 
In this approximation, as a result of taking the $q_z\to 0$ limit, the 
scattering amplitude stops being able to resolve any source structure in the 
$z$ direction, since the eikonal phase in 
\eqref{fscatteringmatrix} appears 
integrated in $y_3 =\pm \infty$. This approximation then 
erased the locality in $y_3$ of the wave
\eqref{wave_integration}, which satisfied that at point $y_3$ electron is
only affected by sources verifying $z_i<y_3$, a sign of 
a gradual transformation of time causality to $y_3$ causality 
as a result of taking the $\beta\to 1$ limit. 
Finally, we can connect this result with Glauber's spinless amplitude
$f(\theta)$. For a Schrodinger wave the scattered state can be
written as 
\begin{align}
\psi_f(x)=\psi_i(x)+f(\theta)\frac{e^{-i|\v{p}_i||\v{x}|}}{|\v{x}|}
\label{outcomingstatescattering_glauber}
\end{align}
where $\theta=q/\beta p_i^0$ is the scattering angle.
The relation between $f(\theta)$ and the {\it M}-matrix is geometric.
 Since $|f(\theta)|^2$ is the intensity in the solid angle interval $\Omega$
 and $\Omega+d\Omega$, then dividing by time and the incoming flux, from
 \eqref{outcomingstatescattering} and  \eqref{outcomingstatescattering_glauber}
\begin{align}
\int \big|f(\theta)\big|^2d\Omega = \int \frac{d^3\v{p}_f}{(2\pi)^3}
\big|M^{(n)}_{s_fs_i}(p_f,p_i)\big|^2 \to f(\theta)=\frac{|\v{p}_i|}{2\pi i} F_{el}^{(n)}(\v{q}_t).
\end{align}
Glauber derived the above amplitude for a Schrodinger particle, thus in principle-
 omitting relativistic and spin effects, something which seems contradictory
with the high energy limit. His description is, however, completely accurate 
in the limit $\beta\to 1$ since the high energy limit
of the Dirac equation is well approximated by a Schrodinger like equation.
\section{Perturbative expansion and strong coupling}
\label{sec:section_2_2}
The high energy integration of the scattering amplitude can be 
expanded in the coupling $g=Ze^2$. Order to order it has to agree 
with the standard perturbative representation of the elastic scattering
with the medium. We note first that the integration of the field in
the phase produces for a general interaction \eqref{general_external_field}
\begin{align}
\chi_0^{(n)}(\v{y}_t)\equiv\int_{-\infty}^{+\infty} dy_3\medspace A_0^{(n)}(\v{y})&= \sum_{i=1}^n \int_{-\infty}^{+\infty} dy_3 \int
\frac{d^3\v{q}}{(2\pi)^3}e^{i\v{q}_t\cdot(\v{y}_t-\v{r}_t^i)+iq_3(y_3-r_3^i)}\hat{A}_0^{(1)}(\v{q})\nonumber\\&=
\sum_{i=1}^n \int \frac{d^2\v{q}_t}{(2\pi)^2}e^{i\v{q}_t\cdot(\v{y}_t-\v{r}_t^i)}\hat{A}_0^{(1)}(\v{q}_t,0).\label{chi_general}
\end{align}
The above result means that the $y_3$ coordinates of each center are lost, the
entire medium is seen as an infinitesimal sheet and any of the single
collisions are considered totally eikonal $q_z=0$. For the Debye screened
interaction \eqref{yukawa_external_field}, in particular, we find
\begin{align}
\chi_0^{(n)}(\v{y}_t)=\int^{+\infty}_{-\infty} dy_3\medspace
A_0^{(n)}(\v{y})=2\sum_{i=1}^n K_0(\mu_d |\v{y}_t-\v{r}_t^i|).
\label{chi_yukawa}
\end{align}
At leading order in $g$, using \eqref{chi_general} we obtain from
\eqref{fscatteringmatrix} 
\begin{align}
F_{el}^{(n)}(\v{q}_t)&=
\int d^2\v{y}_te^{-i\v{q}_t\cdot\v{
    y}_t}
\left(-i\frac{g}{\beta}\sum_{i=1}^n \int
\frac{d^2\v{k}_t}{(2\pi)^2}e^{i\v{k}_t\cdot(\v{y}_t-\v{r}_t^i)}A_0^{(1)}(\v{k}_t)\right)+\mathcal{O}\left(\frac{g^2}{\beta^2}\right)\nonumber\\&=-i\frac{g}{\beta}\hat{A}_0^{(1)}(\v{q}_t)\sum_{i=1}^ne^{-i\v{q}_t\cdot\v{r}_t^i}+\mathcal{O}\left(\frac{g^2}{\beta^2}\right)
\label{scatteringmatrix_f}
\end{align}
\begin{figure}[ht]
\centering
\includegraphics[scale=0.4]{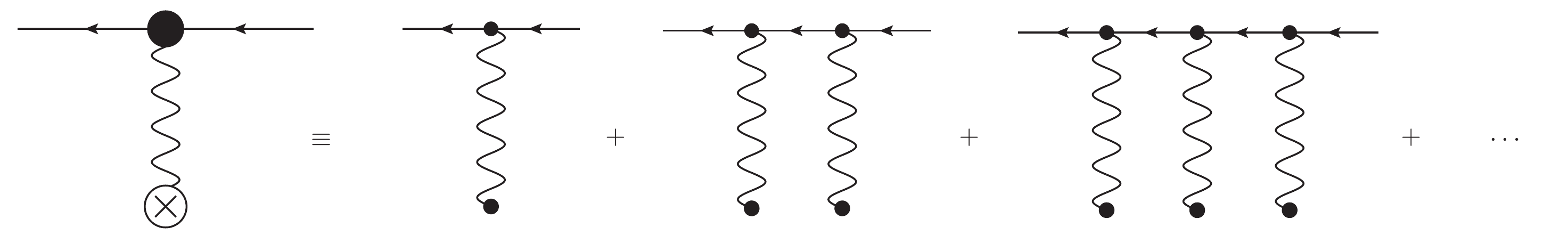}
\caption{Diagrammatic representation of $\big|F_{el}^{(1)}(\v{q})\big|^2$ for a
  single center up to 3rd order in the coupling $g=Ze^2$.}
\label{fig:figure_2_7}
\end{figure}
So amplitude \eqref{scatteringmatrix} at first order in the coupling
simplifies to
\begin{align}
M_{s_fs_i}^{(n)}(p_f,p_i)= 2\pi
\delta(p_f^0-p_i^0)\sqrt{\frac{m}{p_f^0}}&\bar{u}_{s_f}(p_f)\gamma_0u_{s_i}(p_i)\sqrt{\frac{m}{p_i^0}}\nonumber\\
&\times (-ig)\hat{A}_0^{(1)}(\v{q}_t)\left(\sum_{i=1}^{n}e^{-i\v{q}_t\cdot\v{r}_{t}^i}\right)
+\mathcal{O}\left(\frac{g^2}{\beta^2}\right)\label{scatteringmatrix_leadingorder}
\end{align}
For the particular case of the Debye screened interaction, using
\eqref{yukawa_external_field} or \eqref{chi_yukawa} we find a 
superposition of $n$ single Mott amplitudes since
\begin{align}
-ig\hat{A}_0^{(1)}(\v{q})=-\frac{i4\pi Ze^2}{\v{q}^2+\mu_d^2}.
\end{align}
\begin{figure}[ht]
\centering
\includegraphics[scale=0.7]{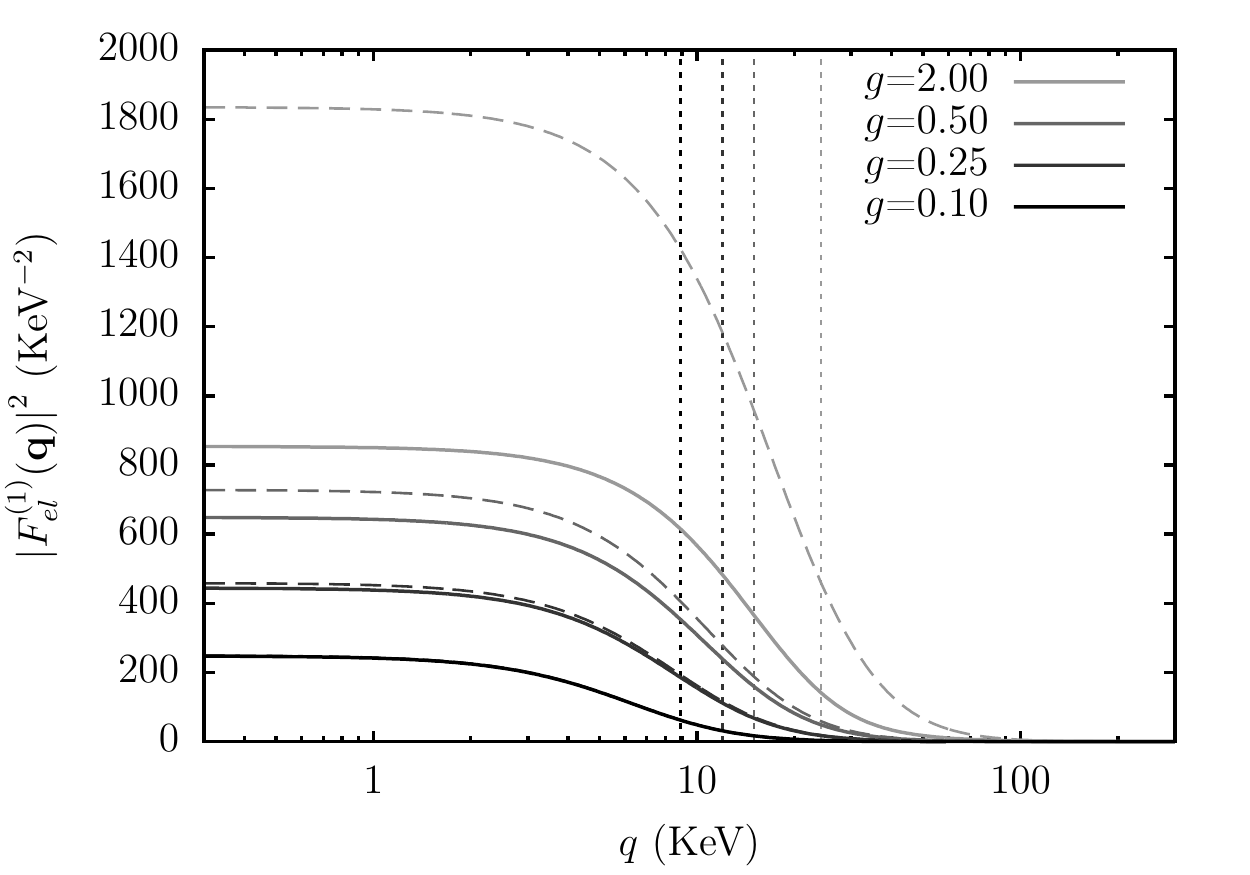}
\caption{Saturation effect of the squared elastic amplitude
  $\big|F_{el}^{(1)}(\v{q})\big|^2$ for a single center at increasing coupling
  $g=Ze^2$ (continuous lines) compared to the leading order 
  approximation (dashed lines) as a function of the transverse momentum change
$\v{q}$. Vertical small-dashed lines represent the respective $\mu_d$.}
\label{fig:figure_2_1}
\end{figure}
Only at leading order (l.o.) in $g$ 
the total amplitude of each center simply adds with a
phase related to its position,
\begin{equation}
\left(M_{s_fs_i}^{(n)}(p_f,p_i)\right)_{l.o.}= \left(\sum_{i=1}^{n}e^{-i\v{q}_t\cdot\v{r}_{t}^i}\right)\left(M_{s_fs_i}^{(1)}(p_f,p_i)\right)_{l.o.}\label{multiple_single_relation}
\end{equation}
For larger values of $g$ a numerical integration of \eqref{scatteringmatrix}
has to be carried. In Figure \ref{fig:figure_2_1} various cases of the
scattering amplitude \eqref{scatteringmatrix} are shown and compared with the
leading order approximation \eqref{scatteringmatrix_leadingorder}. We see that
for large enough $q$ the $1/q^4$ tail of the Rutherford scattering is found
and the arbitrary coupling evaluation matches the leading order approximation.
A closed form of the amplitude (\ref{scatteringmatrix}) for a single center
can be given if $\mu_d\to 0$.  In this limit the leading order of $F^{(1)}(q)$
is well defined but the next orders are not. In order to see it let us rescale
expression (\ref{scatteringmatrix_f}), for $n=1$ and the Debye interaction
\eqref{chi_yukawa}, as
\begin{align}
F_{el}^{(1)}(\v{q})=\frac{2\pi}{q^2}\int
 dy \medspace y J_0(y) \left\{\exp\left[-i\frac{2g}{\beta}K_0(\mu_d y /q)\right]-1\right\}
\end{align}
In order to take the limit $\mu_d\to 0$ we use the relations
\begin{align}
\lim_{\mu_d\to 0}K_0(\mu_d y/q)\approx \lim_{\mu_d\to 0} \log(2q/\mu
y)-\gamma,\medspace J_n(x)e^{1+2ia}=2^{2ia+1}\frac{\Gamma(1+ig/\beta)}{\Gamma(-ig/\beta)}.
\end{align}
As a consequence a running coupling is found, of the form
\begin{align}
\lim_{\mu_d\to
  0}F_{el}^{(1)}(\v{q})= \alpha_r(q) \frac{4\pi}{\v{q}^2},\medspace 
\alpha_r(q)\equiv \frac{1}{2}\frac{1}{q^{2ig\beta}}
\frac{\Gamma(1+ig/\beta)}{\Gamma(-ig/\beta)}\cdot \lim_{\mu_d\to
  0}\exp\left[\frac{2ig}{\beta}\log(\mu_d)\right]\label{coulombamplitude}
\end{align}
The above amplitude has a global divergent phase in the remaining limit which
does not affect the first order in $g$, but it does in the next orders. This
term was conjectured by Dalitz as a way of resuming the Coulomb divergencies,
and was later proved correct by Weinberg. Equation \eqref{coulombamplitude}
also preserves the quantization levels despite being a high energy
approximation, since the same arguments for deriving \eqref{scatteringmatrix}
still hold when the small angle limit is taken. Correspondingly, in the
forward zone of the amplitude \eqref{coulombamplitude} the poles of the gamma
function, accessible at very small $\theta$ are given by $ig/\beta=-n$, where
at low energies $\beta^2\simeq 2p_0/m-2$. We then write
\begin{align}
\beta^2=-\frac{g^2}{n^2}\medspace\medspace\medspace\medspace\to\medspace\medspace\medspace\medspace
p_0=m-\frac{m Z^2e^4}{2n^2}
\end{align}
which correspond to the quantization levels of a single external Coulomb field.
\section{Total cross section and optical theorem}
\label{sec:section_2_3}
The optical theorem relates the total cross section with the
forward part of the amplitude as a consequence of the unitarity constraint of $S_{s_fs_i}^{(n)}(p_f,p_i)$. We now check that the obtained scattering amplitude 
\eqref{scatteringmatrix} obeys this relation. 
The total cross section can be evaluated
by taking the square of the amplitude, summing over
final states and normalizing by the incoming flux $\beta$ and
time $T$,
\begin{equation}
\sigma_{tot}^{(n)}\equiv \frac{1}{T\beta} \sum_{s_f}\int
\frac{d^3\v{p}_f}{(2\pi)^3}\left|M_{s_fs_i}^{(n)}(p_f,p_i)\right|^2
\label{crosssectiondefinition}.
\end{equation}
At leading order in $\beta\to 1$ we find using \eqref{scatteringmatrix},
\begin{align}
&\int
\frac{d^3\v{p}_f}{(2\pi)^3}\left|M_{s_fs_i}^{(n)}(p_f,p_i)\right|^2=\int
\frac{d^3\v{p}_f}{(2\pi)^3}\bigg|2\pi\delta(q^0)\delta_{s_fs_i}\beta\bigg|^2\int
d^2\v{x}_t\int d^2\v{y}_t\medspace
e^{-i\v{q}_t\cdot(\v{x}_t-\v{y}_t)}\nonumber\\
&\cdot\bigg(1-\exp\left[-i\frac{g}{\beta}\chi_0^{(n)}(\v{x}_t)\right]-\exp\left[+i\frac{g}{\beta}\chi_0^{(n)}(\v{y}_t)\right]
+\exp\left[-i\frac{g}{\beta}\chi_0^{(n)}(\v{x}_t)+i\frac{g}{\beta}\chi_0^{(n)}(\v{y}_t)\right]\bigg).
\end{align}
The momentum integration takes advantage of the slow dependence of the 
spinorial term with $\v{p}_f$ with respect to the term inside the integral. In
order to be consistent with the approximation, we apply the same limit in the
integration variables at equation (\ref{crosssectiondefinition}).
In the high energy limit we have $\v{p}^2d\Omega\approx d^2\v{p}_t$ so
\begin{align}
\frac{d^3\v{p}}{(2\pi)^3}\approx
 \frac{1}{(2\pi)^3}dp \medspace d^2\v{p}_t=
\frac{1}{(2\pi)^3}\frac{p_0}{p}dp_0 \medspace d^2 \v{p}_t=
\frac{1}{(2\pi)^3}\frac{1}{\beta}dp_0\medspace d^2\v{q}_t,
\end{align}
Using this approximation, the transverse integration of (\ref{crosssectiondefinition}) is trivial and one easily finds
\begin{align}
\sum_{s_f}\int
\frac{d^3\v{p}_f}{(2\pi)^3}&\left|M_{s_fs_i}^{(n)}(p_f,p_i)\right|^2=2\pi\delta(0) \delta_{s_is_i}\beta \int d^2\v{x}_t \left|\exp\left[-i\frac{g}{\beta}\chi_0^{(n)}(\v{x}_t)\right]-1\right|^2\nonumber\\
&=2\pi\delta(p_i^0-p_i^0) \delta_{s_is_i} \beta \int d^2\v{x}_t \left(2-2\Re\left( \exp\left[-i\frac{g}{\beta}\chi_0^{(n)}(\v{x}_t)\right]\right)\right)\nonumber.
\end{align}
We notice the above result is just twice the negative real part of the
diagonal values of \eqref{scatteringmatrix}, so one can write the relation
\begin{align}
\sum_{s_f}\int
\frac{d^3\v{p}_f}{(2\pi)^3}&\left|M_{s_fs_i}^{(n)}(p_f,p_i)\right|^2=-2\Re M_{s_is_i}^{(n)}(p_i,p_i),
\end{align}
which constitutes the well known optical theorem. A more general form of the
optical theorem can be derived
\begin{align}
\sum_{s}\int
\frac{d^3\v{p}}{(2\pi)^3}M_{ss_i}^{(n),*}(p,p_f)M_{s_is}^{(n)}(p,p_i)=-
M_{s_fs_i}^{(n)}(p_f,p_i)-M_{s_is_f}^{(n),*}(p_i,p_f).
\end{align}
The proof is very simple, we have using \eqref{scatteringmatrix},
\begin{align}
&\sum_{s}\int
\frac{d^3\v{p}}{(2\pi)^3}\medspace M_{ss_f}^{(n),*}(p,p_f)M_{ss_i}^{(n)}(p,p_i)=
2\pi\delta(p_f^0-p_i^0)\delta_{s_fs_i}\beta\nonumber\\
&\cdot \int d^2\v{x}_t
e^{-i\v{q}_t\cdot\v{x}_t}
\cdot\left[\exp\left(-i\frac{g}{\beta}\chi_0^{(n)}(\v{x}_t)\right)-1\right]\left[\exp\left(+i\frac{g}{\beta}\chi_0^{(n)}(\v{x}_t)\right)-1\right]\nonumber\\
&=2\pi\delta(p_f^0-p_i^0)\delta_{s_fs_i}\beta \int d^2\v{x}_t
e^{-i\v{q}_t\cdot\v{x}_t}
\bigg[1-\exp\left(-i\frac{g}{\beta}\chi_0^{(n)}(\v{x}_t)\right)\nonumber\\
&+1-\exp\left(+i\frac{g}{\beta} \chi_0^{(n)}(\v{x}_t)\right)\bigg]
\equiv-M_{s_fs_i}^{(n)}(p_f,p_i)-M_{s_is_f}^{(n),*}(p_i,p_f).
\end{align}
The infinite arising in the conservation delta $2\pi\delta(0)\equiv T$, due to
an integration of a time-independent quantity, accounts for the uniform rate
of scattering in time. On the other hand the $\beta$ accompanying the
amplitude is just the incoming flux,
\begin{align}
\sqrt{\frac{m}{p_i^0}}\bar{u}_{s_i}(p_i)\gamma_3 u_{s_i}(p_i)\sqrt{\frac{m}{p_i^0}}=\frac{p_i^3}{p_i^0}=\beta.
\end{align}
Correspondingly, as read from
\eqref{crosssectiondefinition}, the rate of total scattered particles 
per unit of time and per unit of incoming flux is given by
\begin{align}
\sigma_{tot}^{(n)}=-2\Re F_{el}^{(n)}(\v{0})=\frac{4\pi}{|\v{p}_i|}\Im f^{(n)}(0)\label{opticaltheorem}.
\end{align}
Although scattering amplitude (\ref{scatteringmatrix}) only allows changes in 
perpendicular momentum it results exact if we replace $\v{q}_t$ with
$\v{q}$, since the longitudinal momentum change is implicitly fixed 
by the energy conservation delta. This can be made manifest for
interactions with azimuthal symmetry. We find
\begin{align}
\chi_0^{(n)}(\v{y}_t)=\frac{1}{\beta}\int_{-\infty}^{+\infty}dy_3 \medspace
A_0^{(n)}(|\v{y}_t|,y_3)\equiv\chi_0^{(n)}(y_t).
\end{align}
With this extra symmetry, we can integrate in the angular parameter in order to find
\begin{align}
f^{(n)}(\theta)=\frac{p_i^0}{i}
 \int_{-\infty}^{+\infty}
 dy_t y_t J_0(qy_t) \left\{\exp\left[-i\frac{g}{\beta}\chi_0^{(n)}(y_t)\right]-1\right\},
\end{align}
where now $\v{q}$ is the complete momentum change. Sum in the plane $\v{q}_t$ 
is equivalent to a sum in the sphere. To show that, we check the optical
theorem in the sphere. Using
\begin{align}
\v{q}^2 = \v{p}_f^2+\v{p}_i^2-2\v{p}_f\cdot\v{p}_i= 2p^2
\left(1-\cos^2\theta\right)=4p^2\sin^2\left(\frac{\theta}{2}\right),
\end{align}
we get
\begin{align}
d\Omega = \sin\theta d\theta d\varphi = \frac{q
  dqd\varphi}{p^2},
\end{align}
so the total cross section reads
\begin{align}
\sigma_{tot}^{(n)}=2\pi\bigg|\int_{0}^{+\infty} dx_t
  \medspace x_t \left(\exp\left[-i\frac{g}{\beta}
    \chi_0^{(n)}(x_t)\right]-1\right)\bigg|^2
\int_0^{2p} dq\medspace q J_0(qx_t)J_0(qy_t).
\end{align}
We can now use the orthogonality relation
\begin{align}
\lim_{p\to \infty}  \int_0^{2p} dq \medspace q J_0(qx_t)J_0(qy_t) =
\frac{1}{x_t}\delta(x_t-y_t),
\end{align}
in order to find
\begin{align}
\sigma_{tot}^{(n)}&=2\pi \int_{0}^{+\infty} dx_t \medspace
  x_t\left|\exp\left[-i\frac{g}{\beta} \chi_0^{(n)}(x_t)\right]-1\right|^2\nonumber\\
&=2 \Re  \int
  d^2\v{x}_t\left(1-\exp\left[-i\frac{g}{\beta} \chi_0^{(n)}(\v{x}_t)\right]\right) = \frac{4\pi}{|\v{p}_i|}\Im f^{(n)}(0),
\end{align}
which agrees with equation (\ref{opticaltheorem}). This implies that
only in the infinite momentum frame operations over full momentum change
$\v{q}^2d\Omega$ can be replaced with operations
over the transverse momentum change $d^2\v{q}_t$.

With the above tools we are in position to compute 
the total elastic cross section for the Debye interaction \eqref{yukawa_external_field} with \eqref{chi_yukawa}. The single case
for small values of the coupling $g$ is given, following \eqref{scatteringmatrix_f}, by
\begin{align}
F_{el}^{(1)}(\v{q})=&\int d^2\v{y}_t
e^{-i\v{q}_t\cdot\v{y}_t}\left\{\exp\bigg[-i\frac{g}{\beta}\chi_0^{(1)}(\v{y})\bigg]-1\right\}\nonumber\\=&\int d^2\v{y}_t
e^{-i\v{q}_t\cdot\v{y}_t}
\cdot\left\{-i\frac{g}{\beta}\chi_0^{(1)}(\v{y}_t)-\frac{g^2}{2\beta^2}\left(\chi_0^{(1)}(\v{y}_t)\right)^2+\mathcal{O}\left(\frac{g^3}{\beta^3}\right)\right\}\nonumber\\
=&-i\frac{g}{\beta}\left(\frac{4\pi}{\v{q}_t^2+\mu_d^2}\right)-\frac{g^2}{2\beta^2}\left(\frac{16\pi\arcsinh(q_t/2\mu_d)}{q_t\sqrt{\v{q}_t^2+4\mu_d^2}}\right)+\mathcal{O}\left(\frac{g^3}{\beta^3}\right).
\end{align}
Then we find from equation (\ref{opticaltheorem}) 
\begin{equation}
\sigma_{tot}^{(1)}=-2\Re F_{el}^{(1)}(\v{0})=\frac{4\pi g^2}{\beta^2\mu^2}+\cdots,\label{cross_section_t_1_leading}
\end{equation}
and we define for later uses the oscillatory part also
\begin{equation}
\sigma_{osc}^{(1)}=+2\Im F^{(1)}(\v{0})=-\frac{8\pi g}{\beta\mu^2}+\cdots.\label{cross_section_i_1_leading}
\end{equation}
\begin{figure}[ht]
\centering
\includegraphics[scale=0.7]{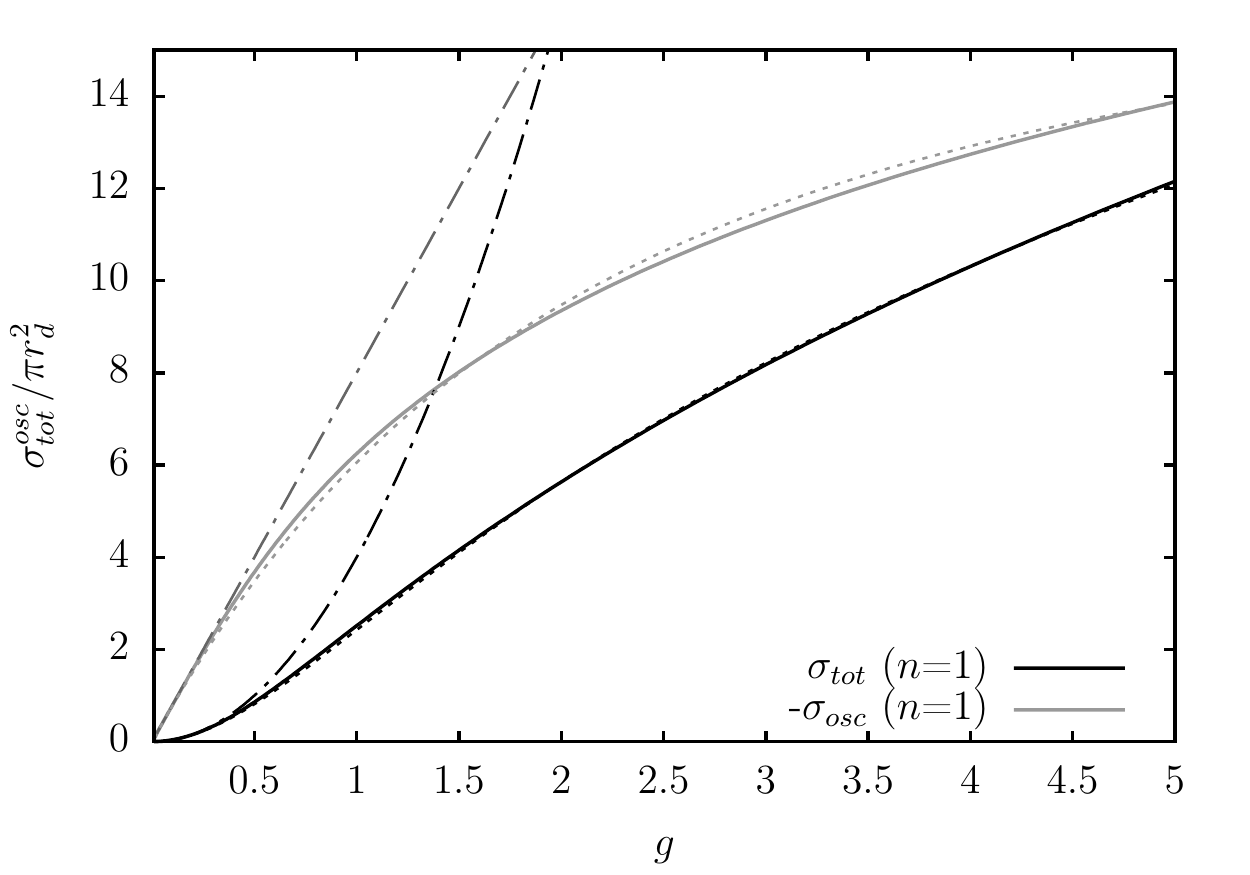}
\caption{Saturation effect of the elastic cross section
  for a single center at increasing coupling
  using expressions \eqref{cross_section_t_1} and \eqref{cross_section_i_1}
  (continuous lines), compared to the leading perturbative
  order at equations \eqref{cross_section_t_1_leading} and
  \eqref{cross_section_i_1_leading} (dot dashed lines), together with the
  approximants given at \eqref{cross_section_approximants} (dot lines), as a
  function of the coupling $g=Ze^2$ with running $\mu_d=\alpha^2m_eZ^{1/3}$.}
\label{fig:figure_2_2}
\end{figure}
For arbitrary larger values of the coupling $g$ the above values are
not applicable. In those cases the single scattering amplitude stops
being perturbative in $g$ and we write instead
\begin{align}
\sigma_{tot}^{(1)}&=4\pi
\int_{0}^{\infty}dy\medspace
y\left(1-\cos\left[\frac{2g}{\beta}K_0(\mu_d y)\right]\right)\nonumber\\
&=\frac{4\pi}{\mu_d^2}\int_0^{\infty} ds \medspace s \left(1-\cos\left[\frac{2g}{\beta}K_0(s)\right]\right)\equiv\frac{4\pi}{\mu_d^2}\Theta_1\left(\frac{g}{\beta}\right)\label{cross_section_t_1},
\end{align}
for the total cross section and
\begin{align}
\sigma_{osc}^{(1)}&=-4\pi
\int_{0}^{\infty}dy\medspace y\sin\left[\frac{2g}{\beta}K_0(\mu_dy)\right]\nonumber\\
&=-\frac{4\pi}{\mu_d^2}\int_0^{\infty} ds \medspace s \sin\left[\frac{2g}{\beta}K_0(s)\right]\equiv-\frac{4\pi}{\mu_d^2}\Theta_2\left(\frac{g}{\beta}\right)\label{cross_section_i_1},
\end{align}
for the oscillatory part of the cross section. The behavior of $\sigma_{tot}^{(n)}$
and $\sigma_{osc}^{(n)}$ is shown in Figure \ref{fig:figure_2_2}. Function $\Theta_1(g/\beta)$ grows as $g^2/\beta^2$ for
small values and then grows logarithmically at larger values of the parameter,
and $\Theta_2(g/\beta)$ grows linearly for small parameters and grows
logarithmically at larger values. A good approximation, valid in the range
$[0,10]$ of the parameter $y=g/\beta$ at the $5\%$ level, is given by the expressions
\begin{align}
\Theta_1(y)=\frac{y^2}{1+a_1y^{a_2}},\medspace\medspace\medspace\medspace\medspace\medspace\medspace\medspace \Theta_2(y)=2\frac{y}{1+b_1y^{b_2}},\label{cross_section_approximants}
\end{align}
where $a_1=0.65$ and $a_2=1.5$, and $b_1=0.32$ and $b_2=1.1$.
\section{Multiple scattering effects}
\label{sec:section_2_4}
We have up to now provided some basic properties of the amplitude for a
general interaction and presented the main results for the particular $n$=1
case, but without taking into account the effects of the multiple scattering.
In this section we will treat with detail the large $n$ limit of the squared
amplitudes and the cross sections. As it is customary we use the shorthands
\begin{align}
-i\frac{g}{\beta}\int^{+\infty}_{-\infty} dy_3\medspace
A_0^{(n)}(\v{y})=-i\frac{g}{\beta}\sum_{i=1}^n \chi_0^{(1)}(\v{y}_t-\v{r}_{t}^i)\equiv -i\frac{g}{\beta}\sum_{i=1}^n
\chi_0^i(\v{y}_t).
\end{align}
The average over medium configurations of the square of
(\ref{scatteringmatrix}) reads, at leading order in $\beta\to 1$,
\begin{align}
\left\langle
\left|M_{s_fs_i}^{(n)}(p_f,p_i)\right|^2\right\rangle=2\pi\delta(p_f^0-p_i^0)\delta_{s_fs_i}
\beta \left\langle
\left|F_{el}^{(n)}(\v{q}_t)\right|^2\right\rangle,
\end{align}
where we erase an overall factor $2\pi\delta(0)$ and $\beta$ when dividing by
time and incoming flux. The relevant quantity in the above expression is
$F_{el}^{(n)}(\v{q})$, which contains the information of the medium
configuration
\begin{align}
\left\langle\left|F_{el}^{(n)}(\v{q}_t)\right|^2\right\rangle&=\int d^2\v{x}_t \int d^2\v{y}_t\medspace
e^{-i\v{q}_t\cdot(\v{x}_t-\v{y}_t)}
\left\langle1
-\exp\left[-i\frac{g}{\beta}\sum_{i=1}^n  \chi_0^i(\v{x}_t)\right]\right.\\
&-\left.\exp\left[+i\frac{g}{\beta}\sum_{i=1}^n
  \chi_0^i(\v{y}_t)\right]
+\exp\left[-i\frac{g}{\beta}\sum_{i=1}^n\chi_0^{i}(\v{x}_t)+i\frac{g}{\beta}\sum_{i=1}^n\chi_0^{i}(\v{y}_t)\right]\right\rangle\nonumber.
\end{align}
By taking the average over medium configurations only the multiple scattering
effect of the medium geometry is contemplated. Particular configurations of
the sources are ignored. We will assume that the target is a cylindrical
medium of volume $V=l\Omega$, with $l$ the length and $\Omega=\pi R^2$ the
transverse area. Then the above average is understood as
\begin{align}
\left\langle\left|F_{el}^{(n)}(\v{q})\right|^2\right\rangle=\frac{1}{V^n}\int_V
d^3\v{r}_1...d^3\v{r}_n\left|F_{el}^{(n)}(\v{q})\right|^2
=\frac{1}{\Omega^n}\int_V d^2\v{r}_1^t...d^2\v{r}_n^t\left|F_{el}^{(n)}(\v{q})\right|^2,
\end{align}
so we find that it reduces to an average over a single center of the form
\begin{align}
&\left\langle\left|F_{el}^{(n)}(\v{q})\right|^2\right\rangle=\int
d^2\v{x}_t \medspace d^2\v{y}_t\medspace
e^{-i\v{q}_t\cdot(\v{x}_t-\v{y}_t)}\nonumber\\
&\left[1-\left(\frac{1}{\Omega}\int_\Omega
d^2\v{r}_t\exp\left[-i\frac{g}{\beta}\chi_0^{(1)}(\v{x}_t-\v{r}_t)\right]\right)^n
-\left(\frac{1}{\Omega}\int_\Omega
d^2\v{r}_t\exp\left[+i\frac{g}{\beta}\chi_0^{(1)}(\v{y}_t-\v{r}_t)\right]\right)^n\right.\nonumber\\
&+\left.\left(\frac{1}{\Omega}\int_\Omega
d^2\v{r}_t\exp\left[-i\frac{g}{\beta}\chi_0^{(1)}(\v{x}_t-\v{r}_t)\right]\exp\left[
+i\frac{g}{\beta}\chi_0^{(1)}(\v{y}_t-\v{r}_t)\right]\right)^n\right].
\end{align}
The above expression is, however, highly oscillatory and unsuitable for numerical
evaluation. One can make a standard approximation, valid for large number of
scattering centers $n\gg 1$. By defining the density of sources $n_0=n/V$ we observe
\begin{align}
\left(\frac{1}{\Omega}\int_\Omega
d^2\v{r}_t\right.&\left.\exp\left[-i\frac{g}{\beta}\chi_0^{(1)}(\v{x}_t-\v{r}_t)\right]\right)^n\nonumber\\
&=\left(1+\frac{1}{\Omega}\int_\Omega
d^2\v{r}_t\left(\exp\bigg[-i\frac{g}{\beta}\chi_0^{(1)}(\v{x}_t-\v{r}_t)\bigg]-1\right)\right)^n\nonumber\\
&=\left(1+\frac{n_0l}{n}\int_\Omega
d^2\v{r}_t\left(\exp\bigg[-i\frac{g}{\beta}\chi_0^{(1)}(\v{x}_t-\v{r}_t)\bigg]-1\right)\right)^n.
\end{align}
Since the interaction vanishes at transverse distances larger than
$\mu_d^{-1}$ then the integral is bounded so, for large enough $n$ we can do
\begin{align}
\lim_{n\to\infty}\left(\frac{1}{\Omega}\int_\Omega
d^2\v{r}_t\right.&\left.\exp\left[-i\frac{g}{\beta}\chi_0^{(1)}(\v{x}_t-\v{r}_t)\right]\right)^n\nonumber\\
=&\exp\left[n_0l\int_\Omega
d^2\v{r}_t\left(\exp\bigg[-i\frac{g}{\beta}\chi_0^{(1)}(\v{x}_t-\v{r}_t)\bigg]-1\right) \right].
\end{align}
And finally
\begin{align}
\left\langle\left|F_{el}^{(n)}(\v{q})\right|^2\right\rangle&=\int
d^2\v{x}_t \medspace d^2\v{y}_t\medspace
e^{-i\v{q}_t\cdot(\v{x}_t-\v{y}_t)}\label{squared_scattering_amplitude}\\
\times\Bigg(1&-\exp\left[n_0l\int_\Omega
d^2\v{r}_t\bigg(\exp\bigg[-i\frac{g}{\beta}\chi_0^{(1)}(\v{x}_t-\v{r}_t)\bigg]-1\bigg)\right]
\nonumber\\
&-\exp\left[n_0l\int_\Omega
d^2\v{r}_t\bigg(\exp\bigg[+i\frac{g}{\beta}\chi_0^{(1)}(\v{y}_t-\v{r}_t)\bigg]-1\bigg)\right]
\nonumber\\
&+\exp\left[n_0l\int_\Omega
d^2\v{r}_t\bigg(\exp\bigg[-i\frac{g}{\beta}\chi_0^{(1)}(\v{x}_t-\v{r}_t)+i\frac{g}{\beta}\chi_0^{(1)}(\v{y}_t-\v{r}_t)\bigg]-1\bigg)\right]\Bigg).\nonumber
\end{align}
The above expression is already suitable for numerical evaluation.  We notice,
however, that it can be split into two parts which admit a clear physical
interpretation. One part is related to the coherent scattering which can be
interpreted as the scattering in an averaged medium, and the other to an
incoherent contribution. To show this fact we use the relation $M=S-1$ leading
to
\begin{align}
&\left\langle
M_{s_fs_i}^{(n)}(p_f,p_i)\left(M_{s_fs_i}^{(n)}(p_f,p_i)\right)^*\right
\rangle=\bigg|\bigg\langle S_{s_fs_i}^{(n)}(p_f,p_i)-1\bigg\rangle\bigg|^2\nonumber\\
&\left(\left\langle
S_{s_fs_i}^{(n)}(p_f,p_i)\left(S_{s_fs_i}^{(n)}(p_f,p_i)\right)^*
\right\rangle-\left\langle
S_{s_fs_i}^{(n)}(p_f,p_i)\right\rangle\left\langle\left(S_{s_fs_i}^{(n)}(p_f,p_i)\right)^*
\right\rangle\right),
\end{align}
where we added and subtracted the term
\begin{align}
\pm \bigg\langle S_{s_fs_i}^{(n)}(p_f,p_i)\bigg\rangle\left\langle\left(S_{s_fs_i}^{(n)}(p_f,p_i)\right)^*
\right\rangle,
\end{align}
in order to find the well known result from statistics
$\braket{x^2}=\braket{x}^2+\sigma^2$. Two contributions appear with a clear
physical interpretation. We will call these contributions the coherent and incoherent average, denoted as
\begin{align}
\left<\left|M_{s_fs_i}^{(n)}(p_f,p_i)\right|^2\right>=\Pi_2^{(n)}(p_f,p_i)+\Sigma_{2}^{(n)}(p_f,p_i),
\end{align}
with the coherent contribution given by
\begin{align}
\Pi_2^{(n)}(p_f,p_i)\equiv\bigg|\bigg\langle S_{s_fs_i}^{(n)}(p_f,p_i)-1\bigg\rangle\bigg|^2=\bigg|\bigg\langle M_{s_fs_i}^{(n)}(p_f,p_i)\bigg\rangle\bigg|^2,
\end{align}
and the incoherent contribution being
\begin{align}
\Sigma_2^{(n)}(p_f,p_i)\equiv\left\langle
S_{s_fs_i}^{(n)}(p_f,p_i)\left(S_{s_fs_i}^{(n)}(p_f,p_i)\right)^*
\right\rangle
-\bigg\langle
S_{s_fs_i}^{(n)}(p_f,p_i)\bigg\rangle\left\langle\left(S_{s_fs_i}^{(n)}(p_f,p_i)\right)^*
\right\rangle.
\end{align}
The first contribution consists in the square of the averaged
amplitude. Correspondingly, as usual by dividing by incoming flux and time and
factorizing the conservation deltas we define
\begin{align}
\Pi_2^{(n)}(p_f,p_i)&=2\pi\delta(p_f^0-p_i^0)\delta_{s_fs_i}\beta \hat{\Pi}_2^{(n)}(\v{q},l),
\end{align}
where the relevant coherent average is given by
\begin{align}
\hat{\Pi}_2^{(n)}&(\v{q},l)=\left|\left\langle
F_{el}^{(n)}(\v{q})\right\rangle\right|^2\label{coherent_contribution_anysize}\\
&=\left|\int
  d^2\v{x}_te^{-i\v{q}_t\cdot\v{x}_t}\left(\exp\left[n_0l\int_\Omega
d^2\v{r}_t\left(\exp\bigg[-i\frac{g}{\beta}\chi_0^{(1)}(\v{x}_t-\v{r}_t)\bigg]-1\right)
\right]-1\right)\right|^2.\nonumber
\end{align}
This contribution consists in a coherent superposition of each
center at the level of the amplitude. We can integrate
(\ref{coherent_contribution_anysize}) for a cylinder whose transverse dimensions
greatly exceed the dimensions of a single scatterer, $R\gg 1/\mu_d$. In this case 
\begin{align}
&\int_\Omega d^2\v{r}_t
\left(\exp\bigg[-i\frac{g}{\beta}\chi_0^{(1)}(\v{x}_t-\v{r}_t)\bigg]-1\right) =F_{el}^{(1)}(\v{0}).
\end{align}
Inserting this result in (\ref{coherent_contribution_anysize}) we find
the larger size approximation of the coherent contribution
\begin{align}
\Pi_2^{(n)}(p_f,p_i)=2\pi\delta(p_f^0-p_i^0)\delta_{s_fs_i}\beta\left|(2\pi)^2\delta^2(\v{q}_t)\left(\exp\bigg[
  n_0l F_{el}^{(1)}(\v{0}) \bigg]-1\right)\right|^2.\label{coherent_contribution_infinitesize}
\end{align}
Due to symmetry arguments, the infiniteness of the medium transverse direction
transforms the coherent part into a pure forward contribution. 
The values of the single scattering 
amplitude $F_{el}^{(1)}(\v{0})$ in the forward direction appearing 
in the last equation must be related to the single elastic cross 
section 
\begin{align}
F_{el}^{(1)}(\v{0})= \frac{1}{2}\left(-\sigma_{tot}^{(1)}+i\sigma_{osc}^{(1)}\right),
\end{align}
and can be read from equations \eqref{cross_section_t_1} and
\eqref{cross_section_i_1} for arbitrary coupling or from
\eqref{cross_section_t_1_leading} and \eqref{cross_section_i_1_leading} at
leading order provided that the single collision case is perturbative in
$g$. In any case, we notice that this does not imply the single scattering
regime, since $n_0$ is arbitrary.  For mediums of finite size, however, the
coherent term represents the modification of the single scattering matrix due
to the border effects. In order to show this fact we write the relation
\begin{align}
\exp\bigg[-i\frac{g}{\beta}\chi_0^{(1)}(\v{x}_t-\v{r}_t)\bigg] = 1+ \frac{1}{(2\pi)^2}\int
d^2\v{k}_t \medspace e^{+i\v{k}_t\cdot(\v{x}_t-\v{r}_t)}F_{el}^{(1)}(\v{k}_t),
\label{wtrick}
\end{align}
which integrated in a finite section $\Omega$ produces an extra
window function
\begin{align}
& \int_\Omega d^2\v{r}_t
\left(\exp\bigg[-i\frac{g}{\beta}\chi_0^{(1)}(\v{x}_t-\v{r}_t)\bigg]-1\right) =  
 \int \frac{d^2\v{k}_t}{(2\pi)^2} e^{+i\v{k}_t\cdot\v{x}_t}F_{el}^{(1)}(\v{k}_t)\int_\Omega
 d^2\v{r}_t e^{-i\v{k}_t\cdot\v{r}_t}.
\end{align}
The window function for the case of the finite cylinder acquires the form
\begin{align}
W_{\Omega}(\v{k}_t,R) \equiv \int_\Omega
 d^2\v{r}_t e^{-i\v{k}_t\cdot\v{r}_t} \to W_{cyl}(\v{k}_t,R) = \frac{2\pi R}{|\v{k}_t|} J_1(|\v{k}_t|R).\label{window_function}
\end{align}
where $J_1(x)$ is the Bessel function of the first kind. For an arbitrary geometry, the window function momentum domain has an oscillatory
behavior in $1/R$ modulating the $1/\v{q}^4$
fall off. This can be directly observed at low density by expanding
\eqref{coherent_contribution_anysize} in $n_0$ yielding
\begin{align}
\Pi_2^{(n)}(p_1,p_0)&=2\pi\delta(p_f^0-p_i^0) \delta_{s_fs_i} \beta \bigg| n_0l W_\Omega(\v{q}_t,R)F_{el}^{(1)}(\v{q}_t)\bigg|^2.\label{coherent_contribution_anysize_lowdensity}
\end{align}
As expected for a coherent contribution this term is of order $n^2$, so at
low density the averaged amplitude squared is $n^2$ times the squared elastic
amplitude of a single center, with a window function containing information of
the medium geometry. This window can be accounted as a border diffractive
effect which factorizes from the single elastic amplitude only at low
densities. For a cylinder medium of radius $R$, in particular, we find
\begin{align}
\Pi_2^{(n)}(p_1,p_0)= 4n^2\frac{J_1^2(|\v{q}_t|R)}{{\v{q}_t^2}R^2}\left|M_{s_fs_i}^{(1)}(p_1,p_0)\right|^2+\mathcal{O}(n^4).\label{coherent_contribution_cylinder_lowdensity}
\end{align}
We proceed to compute now the incoherent contribution to the scattering
amplitude. Using \eqref{sscatteringmatrix}, the first term is given by
\begin{align}
\bigg\langle &S_{s_fs_i}^{(n)}(p_1,p_0)\bigg(S_{s_fs_i}^{(n)}(p_1,p_0)\bigg)^*\bigg\rangle
=2\pi\delta(p_1^0-p_0^0) \delta_{s_fs_i}  \beta \left\langle
S_{el}^{(n)}(\v{q})S_{el}^{(n),*}(\v{q})\right\rangle,
\end{align}
where the incoherent average restricts to the integral parts
\begin{align}
\left\langle
S_{el}^{(n)}(\v{q})\right.&\left.S_{el}^{(n),*}(\v{q})\right\rangle=
\int d^2\v{x}_t\int
d^2\v{y}_t\medspace e^{-i\v{q}_t\cdot(\v{x}_t-\v{y}_t)}\\
&\bigg\{\exp\bigg[n_0l\int_\Omega
d^2\v{r}_t\left(\exp\bigg[-i\frac{g}{\beta}\chi_0^{(1)}(\v{x}_t-\v{r}_t)+i\frac{g}{\beta}\chi_0^{(1)}(\v{y}_t-\v{r}_t)\bigg]-1\right)
\bigg].\nonumber
\end{align}
It results convenient to rewrite the terms in the exponent as squared single
amplitudes. To do that we notice
\begin{align}
\exp&\bigg[-i\frac{g}{\beta}\chi_0^{(1)}(\v{x}_t-\v{r}_t)+i\frac{g}{\beta}\chi_0^{(1)}(\v{y}_t-\v{r}_t)\bigg]-1
\nonumber\\=&\bigg(\exp\bigg[-i\frac{g}{\beta}\chi_0^{(1)}(\v{x}_t-\v{r}_t)\bigg]-1\bigg)+\bigg(\exp\bigg[
+i\frac{g}{\beta}\chi_0^{(1)}(\v{y}_t-\v{r}_t)\bigg]-1\bigg)\nonumber \\
&+\bigg(\exp\bigg[-i\frac{g}{\beta}\chi_0^{(1)}(\v{x}_t-\v{r}_t)\bigg]-1\bigg)\bigg(\exp\bigg[
+i\frac{g}{\beta}\chi_0^{(1)}(\v{y}_t-\v{r}_t)\bigg]-1\bigg).
\end{align}
By taking the infinite medium limit $R\gg 1/\mu_d$ 
the first two integrals are just forward single amplitudes,
\begin{align}
n_0 l \int_\Omega d^2\v{r}_t
\left(\exp\bigg[-i\frac{g}{\beta}\chi_0^{(1)}(\v{x}_t-\v{r}_t)\bigg]-1\right) &= n_0 l
F_{el}^{(1)}(\v{0})\nonumber\\
n_0 l \int_\Omega d^2\v{r}_t
\left(\exp\bigg[+i\frac{g}{\beta}\chi_0^{(1)}(\v{y}_t-\v{r}_t)\bigg]-1\right) &= n_0 l
F_{el}^{(1),*}(\v{0}),
\end{align}
whereas for the mixed term we find, using \eqref{wtrick}
\begin{align}
n_0 l &\int_\Omega d^2\v{r}_t \bigg(\exp\bigg[-i\frac{g}{\beta}\chi_0^{(1)}(\v{x}_t-\v{r}_t)\bigg]-1\bigg)\bigg(\exp\bigg[
+i\frac{g}{\beta}\chi_0^{(1)}(\v{y}_t-\v{r}_t)\bigg]-1\bigg)\nonumber\\
=\medspace &n_0l\int\frac{d^2\v{k}_t}{(2\pi)^2}e^{+i\v{k}_t\cdot(\v{x}_t-\v{y}_t)}\left|F_{el}^{(1)}(\v{k}_t)\right|^2.
\end{align}
The second term of the incoherent contribution is given by
\begin{align}
\bigg\langle S_{s_fs_i}^{(n)}(p_f,p_i)\bigg\rangle \bigg\langle S_{s_fs_i}^{(n),*}(p_f,p_i)\bigg\rangle
=2\pi\delta(p_f^0-p_i^0) \delta_{s_fs_i} \beta \left\langle
S_{el}^{(n)}(\v{q})\right\rangle\left\langle S_{el}^{(n),*}(\v{q})\right\rangle,
\end{align}
where similarly
\begin{align}
\left\langle
S_{el}^{(n)}(\v{q})\right\rangle\left\langle S_{el}^{(n),*}(\v{q})\right\rangle&=
\int d^2\v{x}_t\int
d^2\v{y}_t\medspace e^{-i\v{q}_t\cdot(\v{x}_t-\v{y}_t)}\nonumber\\
\cdot\exp\bigg[&n_0l\int_\Omega d^2\v{r}_t
  \bigg(\exp\bigg[-i\frac{g}{\beta}\chi_0^{(1)}(\v{x}_t-\v{r}_t)\bigg]-1\bigg)\nonumber\\
+&n_0l\int_\Omega
  d^2\v{r}_t
  \bigg(\exp\bigg[+i\frac{g}{\beta}\chi_0^{(1)}(\v{y}_t-\v{r}_t)\bigg]-1\bigg)\bigg].
\end{align}
Proceeding in the same way, for a medium satisfying $R\gg \mu_d^{-1}$ 
we simply have
\begin{align}
\bigg\langle S_{s_fs_i}^{(n)}(p_f,p_i)\bigg\rangle \bigg\langle S_{s_fs_i}^{(n),*}(p_f,p_i)\bigg\rangle
=&2\pi\delta(p_f^0-p_i^0) \delta_{s_fs_i} \beta\int d^2\v{x}_t\int
d^2\v{y}_t\medspace e^{-i\v{q}_t\cdot(\v{x}_t-\v{y}_t)}\nonumber\\
&\cdot\exp\bigg[n_0lF_{el}^{(1)}(\v{0})+n_0lF_{el}^{(1),*}(\v{0})\bigg].
\end{align}
Correspondingly after summing these two terms one finds
\begin{align}
\Sigma_2^{(n)}(p_f,p_i)=&2\pi\delta(q^0)\delta_{s_fs_i}\beta \hat{\Sigma}_2^{(n)}(\v{q},l),
\end{align}
where the relevant incoherent average is given by
\begin{align}
\hat{\Sigma}_2^{(n)}(\v{q},l)&=\exp\left[2n_0l\Re F_{el}^{(1)}(\v{0})\right]\int d^2\v{x}_t d^2\v{y}_te^{-i\v{q}_t\cdot(\v{x}_t-\v{y}_t)}\nonumber\\
&\times\left\{\exp\left[n_0l\int\frac{d^2\v{k}_t}{(2\pi)^2}e^{+i\v{k}_t\cdot(\v{x}_t-\v{y}_t)}\left|F_{el}^{(1)}(\v{k}_t)\right|^2\right]-1\right\}.
\end{align}
The above form of the incoherent contribution is related to Moliere's theory
of scattering \cite{moliere1948,bethe1953} since by using
\eqref{opticaltheorem},
\begin{align}
\hat{\Sigma}_2^{(n)}(\v{q},l)&=\Omega \exp\bigg[-n_0l\sigma_{tot}^{(1)}\bigg]\int
d^2\v{x}_t e^{-i\v{q}_t\cdot \v{x}_t}\nonumber\\
&\cdot\left\{\exp\left[n_0l\int\frac{d^2\v{k}_t}{(2\pi)^2}e^{+i\v{k}_t\cdot\v{x}_t}\left|F_{el}^{(1)}(\v{k}_t)\right|^2\right]-1\right\}\label{incoherent_contribution_infinitesize},
\end{align}
where the single cross section can be read from equation
\eqref{cross_section_t_1_leading} for small coupling or from equation
\eqref{cross_section_t_1} for general coupling, and the trivial 
integral in one of the impact parameters has produced and overall factor 
$\Omega=\pi R^2$ accounting for transverse homogeneity when $R\to\infty$. 
Except for the $-1$, accounting for boundary effects and allowing 
the integration for screened interactions,
\eqref{incoherent_contribution_infinitesize}
constitutes a solution of the Moliere equation. The $-1$ is strictly
necessary to integrate in impact parameter for any screened interaction.
The Fourier transform
of the single amplitude $F_{el}^{(1)}(\v{q})$ in the exponential at
\eqref{incoherent_contribution_infinitesize} will be denoted as 
\begin{equation}
\sigma_{el}^{(1)}(\v{x})=\int\frac{d^2\v{k}_t}{(2\pi)^2}e^{+i\v{k}_t\cdot\v{x}_t}\left|F_{el}^{(1)}(\v{k}_t)\right|^2
\label{sigmax}.
\end{equation}
and can be numerically computed for arbitrary $g$ using
\eqref{scatteringmatrix_f} representing the all orders interaction with each
single center. The extension to $n$ centers is just its
exponentiation. For low coupling we can use
\begin{equation}
F_{el}^{(1)}(\v{k}_t)=-\frac{i}{\beta}\frac{4\pi g}{\v{k}_t^2+\mu_d^2}+\mathcal{O}(g^2)\to \sigma_{el}^{(1)}(\v{x}_t)= \frac{4\pi g^2}{\beta^2\mu_d^2}
\mu_d |\v{x}_t|K_1(\mu_d |\v{x}_t|)+\mathcal{O}(g^3).
\end{equation}
which after inserted in \eqref{incoherent_contribution_infinitesize} produces
a suitable form for fast numerical calculations. Observe that we called this
function $\sigma^{(1)}(\v{x}_t)$ because it is related to the single cross
section as $\sigma^{(1)}(\v{0})\equiv\sigma_{tot}^{(1)}$. An expansion in
$n_0$ produces
\begin{align}
\Sigma_2^{(n)}(p_f,p_i) &=
  \Omega  2\pi\delta(p_f^0-p_i^0)\delta_{s_fs_i}\beta\label{incoherent_contribution_infinitesize_lowdensity}\\
&\cdot\int
d^2\v{x}_t \medspace e^{-i\v{q}_t\cdot \v{x}_t}
\left\{n_0l\int\frac{d^2\v{k}_t}{(2\pi)^2}e^{+i\v{k}_t\cdot\v{x}_t}\left|F_{el}^{(1)}(\v{k}_t)\right|^2\right\}+\mathcal{O}(n_0^2)\nonumber\\
=(n_0\pi R^2l) 2\pi&\delta(p_f^0-p_i^0)\delta_{s_fs_i}
 \beta\left|F_{el}^{(1)}(\v{q}_t)\right|^2+\mathcal{O}(n_0^2)\equiv
n \left|M_{s_fs_i}^{(1)}(p_f,p_i)\right|^2+\mathcal{O}(n^2)\nonumber.
\end{align}
Only at low densities the probability of changing
$\v{q}_t$ due to the joint effect of $n$ centers is just $n$ times the
probability of changing $\v{q}_t$ due to a single center. This relation does
not hold for higher densities and, in particular, for saturation densities
an asymptotic form is going to be later defined. At low $n$ and $R\gg 1/\mu_d$
we can join equations \eqref{coherent_contribution_anysize_lowdensity} and \eqref{incoherent_contribution_infinitesize_lowdensity} leading to
\begin{equation}
\left<\left|M_{s_fs_i}^{(n)}(p_f,p_i)\right|^2\right>\simeq \left\{n+4n^2\frac{J_1^2(|\v{q}_t|R)}{\v{q}_t^2R^2}\right\}\left|M_{s_fs_i}^{(1)}(p_f,p_i)\right|^2\label{squared_scattering_amplitude_lowdensity}.
\end{equation}
The above result may have been obtained from a
direct expansion in the number of centers. This approach will give us some
insight on the interference pattern and the statistical interpretation 
of the coherent and incoherent contributions. Let us denote
\begin{align}
\Gamma_i=\exp\bigg(-i\frac{g}{\beta}\chi(\v{x}_t-\v{r}_t^i)\bigg)-1,
\end{align}
in such a way that the elastic amplitude can be rewritten as
\begin{align}
&F_{el}^{(n)}(\v{q})=\int d^2\v{x}_t e^{-i\v{q}_t\cdot\v{x}_t}
\bigg(\prod_{i=1}^{n}(\Gamma_i+1)-1\bigg)\nonumber\\
&=\int d^2\v{x}_t
e^{-i\v{q}_t\cdot\v{x}_t}
\bigg(\sum_{i=1}^{n}\Gamma_i+\sum_{i=1}^n\sum_{j=i+1}^n\Gamma_i\Gamma_j+\cdots\bigg)\nonumber
=\sum_{i=1}^n I_i +\sum_{i=1}^{n}\sum_{j=i+1}^n I_{ij}+\cdots,
\end{align}
which is an expansion in the number of collisions. Consider first the terms
where only one $(\Gamma_i)$ collision appears in the amplitude. We are left with
\begin{align}
\sum_{i=1}^n I_i (\v{q})= \sum_{i=1}^n\int d^2\v{x}
e^{-i\v{q}\cdot\v{x}}\bigg(\exp\bigg(-i\frac{g}{\beta}\chi_0(\v{x}-\v{r}_i)\bigg)-1\bigg)=\bigg(\sum_{i=1}^{n}e^{-i\v{q}\cdot\v{r}_i}\bigg)F_{el}^{(1)}(\v{q}).
\end{align}
Any other term of higher order can be always written as a convolution with an
additional phase factor of the form
\begin{align}
I_{ij}(\v{q})=\int \frac{d^2\v{q}_i}{(2\pi)^2} \frac{d^2\v{q}_j}{(2\pi)^2} e^{-i\v{q}_i\cdot\v{r}_i-i\v{q}_j\cdot\v{r}_j}F_{el}^{(1)}(\v{q}_i)F_{el}^{(1)}(\v{q}_j)(2\pi)^2\delta^2(\v{q}-\v{q}_i-\v{q}_j),
\end{align}
and so on. Now consider the squared amplitude, in this form it reads
\begin{align}
\left|F_{el}^{(n)}(\v{q})\right|^2= \sum_{i=1}^{n}\sum_{j=1}^n
I_iI_j^{*}+\sum_{i=1}^{n}\sum_{j=1}^{n}\sum_{k=j+1}^n\bigg(I_iI_{jk}^*+I_{i}^*I_{jk}\bigg)+\cdots.
\end{align}
The first contribution is given by
\begin{align}
\sum_{i=1}^{n}\sum_{j=1}^nI_iI_j^*=\sum_{i=1}^{n}\sum_{j=1}^n
e^{-i\v{q}\cdot(\v{r}_i-\v{r}_j)}\left|F_{el}^{(1)}(\v{q})\right|^2=\bigg(n+\sum_{i=1}^{n}\sum_{j\neq
i}^ne^{-i\v{q}\cdot(\v{r}_i-\v{r}_j)}\bigg)\left|F_{el}^{(1)}(\v{q})\right|^2.
\end{align}
The interpretation of the two above terms is clear. The first term gives the
contribution from $n$ independent collisions, whereas the second term gives
the interference, at this order, between them. Notice that at $\v{q}=0$ we
have
\begin{align}
\left.\sum_{i=1}^{n}\sum_{j\neq
i}^ne^{-i\v{q}\cdot(\v{r}_i-\v{r}_j)}\right|_{\v{q}=0}=n(n-1)
\end{align}
which is the constructive interference between the $n$ centers. For $\v{q}\neq
0$ this factor is not positive defined, so that one can not interpret it as a
probability density. One can average, however, over center configurations
obtaining, for a cylinder of section $\Omega=\pi R^2$ 
\begin{align}
\left\langle \sum_{i=1}^{n}\sum_{j\neq
i}^ne^{-i\v{q}\cdot(\v{r}_i-\v{r}_j)} \right\rangle = \frac{n(n-1)}{\Omega^2}\int_{\Omega}
d^2\v{r}_i\int_\Omega d^2\v{r}_j e^{-i\v{q}\cdot(\v{r}_i-\v{r}_j)} =
4n(n-1)\frac{J_1^2(qR)}{(qR)^2}.
\end{align}
In the $R\to\infty$ limit we retrieve the forward constraint since the
contribution is given by $n(n-1)/R^2\delta(\v{q})$. Thereby, the contribution
to the scattering by decomposition in single scatterings is given by
\begin{align}
\left|F_{el}^{(n)}(\v{q})\right|^2\simeq
\left\{n+4n(n-1)\frac{J_1^2(qR)}{(qR)^2}\right\}\left|F_{el}^{(1)}(\v{q})\right|^2,
\end{align}
the first term being the independent (incoherent) sum of the $n$ scattering
centers and the second term the (coherent) interference between them, which results in the original
scattering amplitude squared modulated by the Fourier transform of the shape
of the medium, matching the result
\eqref{squared_scattering_amplitude_lowdensity}. The coherent term has interference peaks at $q=0$ and $q=5.14/R$. For
large $R$ the second term will be negligible for $q$ larger than about
$q=(4\pi nl/R)^{1/3}$. For the terms with two collisions in the amplitude we find
\begin{align}
\sum_{i=1}^{n}\sum_{j\neq i}^n\sum_{k=1}^{n}\sum_{l\neq k}^n I_{ij}I^*_{kl} &=
\int \frac{d^2\v{q}_i}{(2\pi)^2}\int \frac{d^2\v{q}_j}{(2\pi)^2}\int
\frac{d^2\v{q}_k}{(2\pi)^2}\int \frac{d^2\v{q}_l}{(2\pi)^2}
W_2(\v{q})\nonumber\\
&\cdot F_{el}^{(1)}(\v{q}_i)F_{el}^{(1)}(\v{q}_j)F_{el}^{(1),*}(\v{q}_k)F_{el}^{(1),*}(\v{q}_l),
\end{align}
where the diffractive window function reads now
\begin{align}
W_2(\v{q})&=\sum_{i=1}^n\sum_{j\neq i}^n\sum_{k=1}^n\sum_{l\neq k}^n
e^{-i\v{q}_i\cdot\v{r}_i-i\v{q}_j\cdot\v{r}_j+i\v{q}_k\cdot\v{r}_k+i\v{q}_l\cdot\v{r}_l}\nonumber\\
&=\sum_{k=i=1}^n\sum_{l=j\neq  i}^n+\sum_{i=1}^n\sum_{j\neq i}^n\sum_{k\neq i}^n\sum_{l\neq k \neq j}e^{-i\v{q}_i\cdot\v{r}_i-i\v{q}_j\cdot\v{r}_j+i\v{q}_k\cdot\v{r}_k+i\v{q}_l\cdot\v{r}_l},
\end{align}
where again we have split the sum into a diagonal and a non diagonal part. The
diagonal part is given by
\begin{align}
W_2^{dg}(\v{q})&= \frac{n(n-1)}{\Omega^2}\int_\Omega d^2\v{r}_i\int_\Omega
d^2\v{r}_j e^{-ir_i\cdot (\v{q}_i-\v{q}_k)-i\v{r}_j\cdot(\v{q}_j-\v{q}_l)}
\nonumber\\
&=
\frac{n(n-1)}{\Omega^2}\frac{2\pi R J_1(|q_i-q_k|R)}{|q_i-q_k|}\frac{2\pi R J_1(|q_j-q_l|R)}{|q_j-q_l|},
\end{align}
which simplifies to $\delta(\v{q}_i-\v{q}_k)\delta(\v{q}_j-\v{q}_l)$ in the
limit $R\to\infty$ producing
\begin{align}
\left(\sum_{i=1}^{n}\sum_{j\neq i}^n\sum_{k=1}^{n}\sum_{l\neq k}^n
I_{ij}I^*_{kl}\right)^{dg}=\int\frac{d^2\v{q}_i}{(2\pi)^2}\int\frac{d^2\v{q}_j}{(2\pi)^2}F_{el}^{(1)}(\v{q}_i)F_{el}^{(1)}(\v{q}_j)(2\pi)^2\delta^2(\v{q}-\v{q}_i-\v{q}_j),
\end{align}
that is, a convolution of two independent collisions with different centers
leading to a total momentum change $\v{q}$. The non diagonal part will
contribute through convolutions of the same type to the forward
amplitude. Therefore, the structure of the expansion in $n$ is always split
into a contribution which in the large medium limit $R\gg \mu_d^{-1}$ can be
interpreted in probabilistic terms, corresponding to the expansion in $n_0$ of
equation \eqref{incoherent_contribution_infinitesize}, and a contribution
accounting for diffractive effects.

\begin{figure}[ht]
\centering
\includegraphics[scale=0.7]{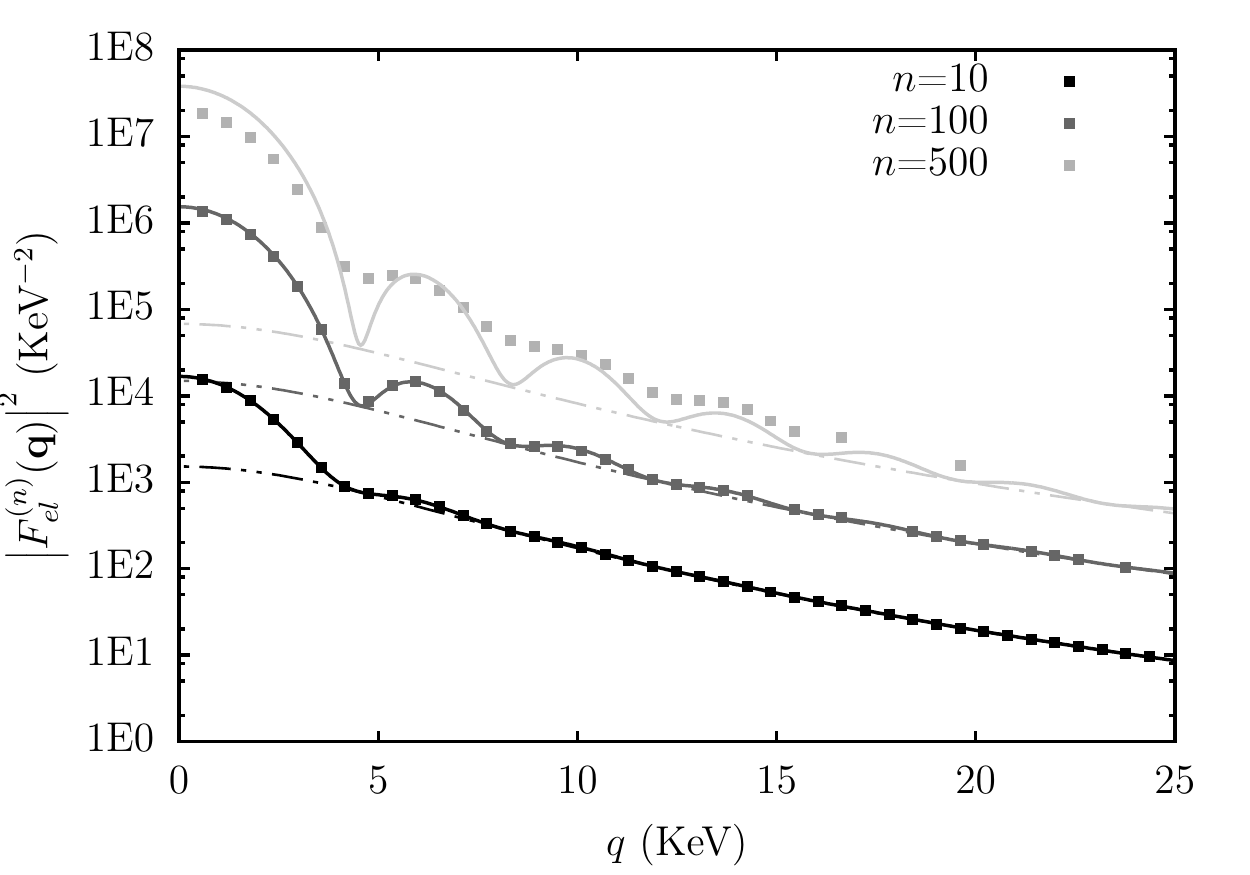}
\caption{Squared elastic amplitude
  $\big|F_{el}^{(n)}(\v{q})\big|^2$ after traversing a medium with $Z=7$ 
  corresponding to a coupling of $g=0.05$, Debye mass of $\mu_d$=71 KeV for
  increasing values of $n$ centers distributed in cylinder of radius 
  $R=6r_d$ and length $l=R/10$. Dots are direct evaluation of
  \eqref{squared_scattering_amplitude},
  whereas continuum lines are the low density result
  \eqref{squared_scattering_amplitude_lowdensity} and dot-dashed lines
  are the incoherent contribution \eqref{incoherent_contribution_infinitesize}.}
\label{fig:figure_2_3}
\end{figure}

For general $n$ and $g$ the probability of finding the 
fermion with final momentum $p_f$
is given necessarily by a numerical evaluation of the full expression
\eqref{squared_scattering_amplitude} or by using the coherent \eqref{coherent_contribution_anysize} and 
incoherent \eqref{incoherent_contribution_infinitesize} terms, easier
to compute than \eqref{squared_scattering_amplitude}. In Figure \ref{fig:figure_2_3} the
probability of finding an electron with transverse momentum $\v{q}$ is
shown and compared with the leading order in $n$ approximation together
with the incoherent contribution.
One can check that the
optical theorem is valid also for the medium averaged cross sections. Indeed,
\begin{align}
\sum_{s_f}\int \frac{d^3\v{p}_f}{(2\pi)^3}
\left\langle\left|M_{s_fs_i}^{(n)}(p_f,p_i)\right|^2\right\rangle =
2\pi\delta(p_i^0-p_i^0)\delta_{s_is_i}\beta \int\frac{d^2\v{q}}{(2\pi)^2}\left\langle\left|F_{el}^{(n)}(\v{q})\right|^2\right\rangle.
\end{align}
By using \eqref{squared_scattering_amplitude} we find
\begin{align}
\int\frac{d^2\v{q}}{(2\pi)^2}&\left\langle\left|F_{el}^{(n)}(\v{q})\right|^2\right\rangle
=\\
&2\Re\int
d^2\v{x}_t\medspace \left(1-\exp\left[n_0l\int_\Omega
d^2\v{r}_t\bigg(\exp\bigg[-i\frac{g}{\beta}\chi_0^{(1)}(\v{x}_t-\v{r}_t)\bigg]-1\bigg)\right]\right),\nonumber
\end{align}
so finally
\begin{align}
\int \frac{d^3\v{p}_f}{(2\pi)^3}
\left\langle\left|M_{s_fs_i}^{(n)}(p_f,p_i)\right|^2\right\rangle &=
2\pi\delta(p_i^0-p_i^0)\delta_{s_is_i}\beta 2\Re \left\langle
F_{el}^{(n)}(\v{0})\right\rangle \nonumber\\
&\equiv-2\Re \left\langle M_{s_is_i}^{(n)}(p_i,p_i)\right\rangle.
\end{align}
This is remarkable, since the averaging at the level of the amplitudes is
rather different than the averaging at the level of the cross
sections. Similar result can be obtained by direct integration of the
coherent and incoherent contributions 
\eqref{coherent_contribution_infinitesize} and \eqref{incoherent_contribution_infinitesize}
arriving to
\begin{align}
&\sigma_{tot}^{(n)}=\sum_{s_f}\int
\frac{d^3\v{p}_f}{(2\pi)^3}\left[\Pi_2^{(n)}(p_f,p_i)+\Sigma_2^{(n)}(p_f,p_i)\right]\label{cross_section_n}\\
&=2\pi
R^2 \Re\left[1-\exp\bigg(n_0lF_{el}^{(1)}(\v{0})\bigg)\right]=2\pi R^2\left[1-\exp\left(-\frac{n_0l\sigma_{tot}^{(1)}}{2}\right)\cos\left(\frac{n_0l\sigma_{osc}^{(1)}}{2}\right)\right].\nonumber
\end{align}
where we used the single cross section definitions \eqref{cross_section_t_1}
and \eqref{cross_section_i_1}. At small density we simply have
\begin{align}
\sigma_{tot}^{(n)}=2\pi R^2 \frac{n_0l\sigma_{tot}^1}{2}+\cdots = n\sigma_{tot}^{(1)},
\end{align}
as expected. In the limit of very large density, instead, we get
$\sigma_{tot}^{(n)}=2\pi R^2$, which is the correct high energy limit for the cross
section of a black disk of radius $R$. Oscillations of \eqref{cross_section_n}
are a diffractive effect due to the coherent scattering of the entire medium. Indeed, for the
split contributions to the total cross section we find in the incoherent part
\begin{align}
\sigma_{inc}^{(n)}\equiv \sum_{s_f}\int \frac{d^3\v{p}_f}{(2\pi)^3}
\Sigma_2^{(n)}(p_f,p_i)= \pi R^2\left[1-\exp\left(-\frac{n_0l\sigma_{tot}^{(1)}}{2}\right)\right]\label{cross_section_n_incoherent},
\end{align}
whereas for the coherent contribution
\begin{align}
\sigma_{coh}^{(n)}&\equiv \sum_{s_f}\int \frac{d^3\v{p}_f}{(2\pi)^3}
\Pi_2^{(n)}(p_f,p_i)\nonumber\\
&=\pi R^2\left[1+\exp\left(-n_0l\sigma_t^{(1)}\right)-2\exp\left(-\frac{n_0l\sigma_{tot}^{(1)}}{2}\right)\cos\left(\frac{n_0l\sigma_i^{(1)}}{2}\right)\right].\label{cross_section_n_coherent}
\end{align}
Oscillations in \eqref{cross_section_n} with surface density $n_0l$ due
to a coherently acting medium have a 
period $n_0l\sigma_{osc}^{(1)}/2$. Since $\sigma_{osc}^{(1)}$ is of order $\mathcal{O}(g)$ whereas 
$\sigma_{tot}^{(1)}$ is of order $\mathcal{O}(g^2)$ for small coupling the oscillations will be
clearly seen in the total cross section. We see that for 
$n_0l\approx 2/\sigma_{tot}^{(1)}$ the total cross section
saturates. This defines a saturation scale given by
\begin{align}
n_0l = \frac{2}{\sigma_t^{(1)}}\approx \frac{\mu_d^2}{2\pi g^2},
\end{align}
where the last equality holds only for small coupling, or, using $n_0=n/\pi
R^2l$,
\begin{align}
n_{sat} = \frac{R^2\mu_d^2}{2g^2}.
\end{align}
\begin{figure}[ht]
\includegraphics[scale=0.54]{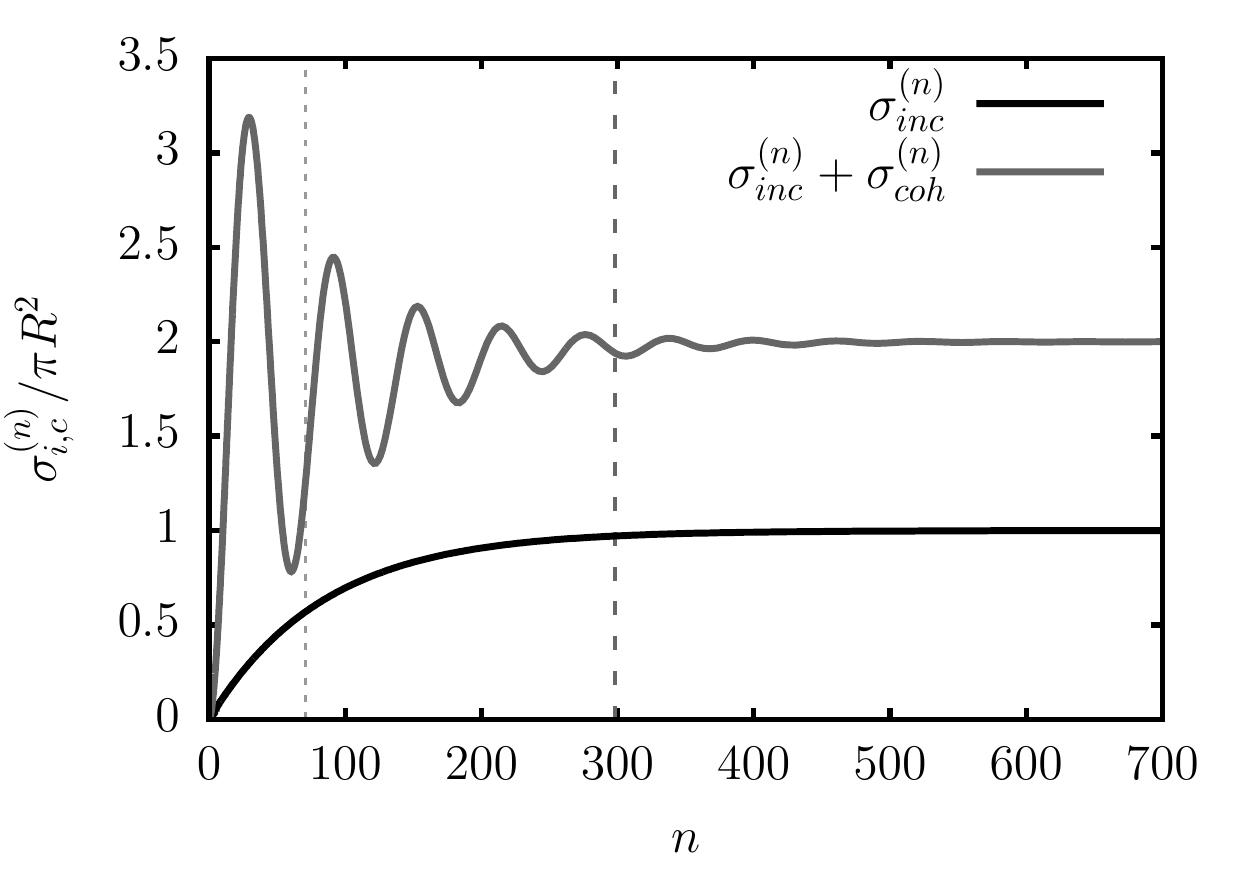}
\includegraphics[scale=0.54]{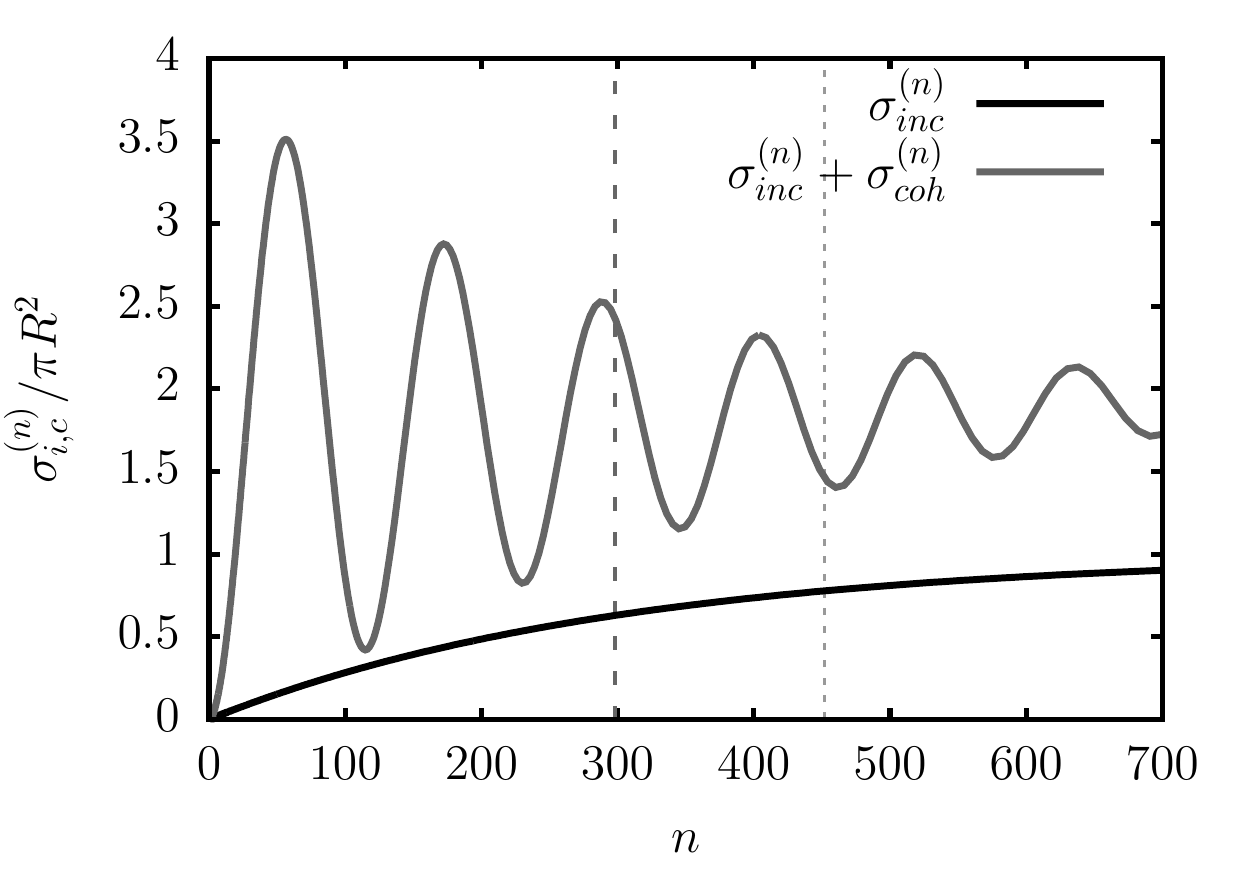}
\caption{Total cross section in terms of the incoherent and coherent
  contributions as a function of the number of centers $n$ for a medium with 
  $g=0.5$ (top) and $g=0.25$ (bottom). Radius and length are chosen as
  $R=6r_d$ and $l=3r_d$, $\mu_d=\alpha_e m Z^{1/3}$. Short-dashed vertical lines correspond to the
  saturation scale $n_{sat}$ whereas long-dashed vertical lines correspond to 
  the degenerated limit $n_0/r_d^3$=1 in which doubled sources should be
  absorbed into a coupling redefinition.}
\label{fig:figure_2_4}
\end{figure}
In Figure \ref{fig:figure_2_4} we show the fully integrated total cross section
as a function of the number of centers $n$, for a couple of values of the
coupling $g$. For smaller values the oscillation is larger and lasts
longer. The saturation value may be achieved independently of the degenerate
scale, that is, the density at which centers start to overlap. 
This occurs more easily in mediums of small thickness $l$ and coupling $g$, as expected.
\section{Relation to Moliere's theory and Fokker-Planck approximation.}
\label{sec:section_2_5}
When the medium transverse dimension extends far beyond the dimensions of a
single scatterer and, thus, can be considered infinite, and the coherent
diffractive part can be neglected, the emergence of
 a infinite transverse symmetry has to lead to 
a diffusive behavior of the momentum distributions.
Notice that since the oscillations are of 
width $1/R$, in
the infinite transverse limit $R\gg \mu_d^{-1}$ the coherent term stops being
observable and transforms into a pure forward component
\begin{align}
\lim_{R\to\infty} \left<\left|M_{s_fs_i}^{(n)}(p_f,p_i)\right|^2\right>
=2\pi\delta(p_f^0-p_i^0)\delta_{s_fs_i}\beta
\pi R^2 (2\pi)^2\delta(\v{q}_t^2)+ \Sigma_{2}^{(n)}(p_f,p_i),
\end{align}
and the process is dominated by incoherent scattering. 
\begin{figure}[ht]
\centering
\includegraphics[scale=0.8]{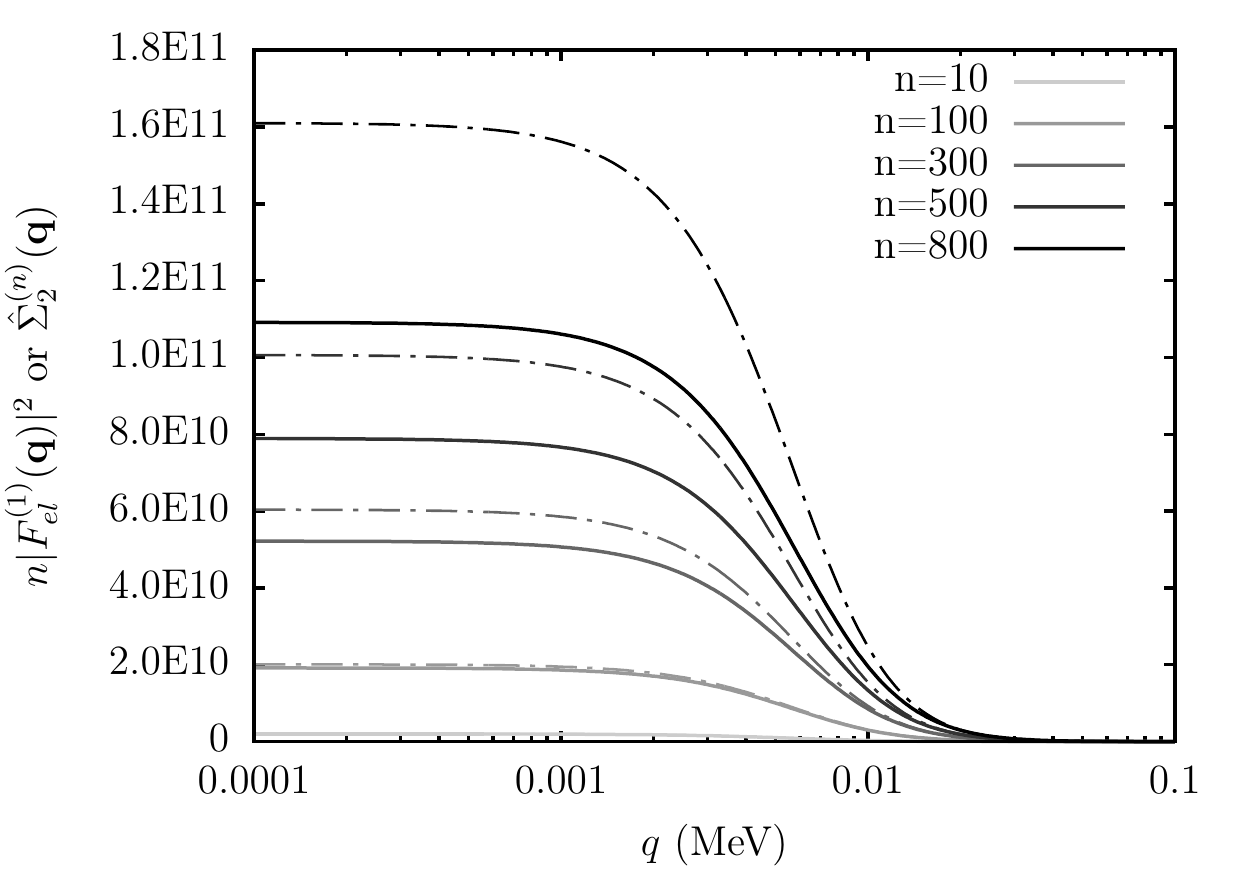}
\caption{Incoherent (solid lines) and incoherent $n_0$ expansions 
  (dot-dashed lines) contribution
  to the squared elastic amplitude as a function of $\v{q}$ for various
  number of centers $n$ for a medium with 
  $Z=10$. Radius and length are chosen as
  $R=6r_d$ and $l=3r_d$, $\mu_d=\alpha_e m Z^{1/3}$.}
\label{fig:figure_2_6}
\end{figure}
In order to find the corresponding transport equation we take the derivative
of \eqref{incoherent_contribution_infinitesize}
with respect to $l$ obtaining
\begin{align}
\frac{\partial}{\partial l} \Sigma_2^{(n)}(p_f,p_i) = 
2\pi\delta(p_f^0-p_i^0)\delta_{s_fs_i}\beta \left(\frac{\partial}{\partial l}\hat{\Sigma}_2^{(n)}(\v{q},l)\right),
\end{align}
By deriving we find
\begin{align}
\frac{\partial}{\partial l}&\hat{\Sigma}_2^{(n)}(\v{q},l)=\frac{\partial}{\partial l}\bigg\{\Omega\exp\bigg[-n_0l\sigma_{el}^{(1)}(\v{0})\bigg]\int
d^2\v{x}_t e^{-i\v{q}_t\cdot \v{x}_t}\bigg(
\exp\bigg[n_0l\sigma_{el}^{(1)}(\v{x}_t)\bigg]-1\bigg)\bigg\} \nonumber\\
=&- n_0\sigma_{el}^{(1)}(\v{0}) \hat{\Sigma}_2^{(n)}(\v{q},l)+n_0\Omega \exp\bigg[-n_0l\sigma_{el}^{(1)}(\v{0})\bigg]\int
d^2\v{x}_t e^{-i\v{q}_t\cdot \v{x}_t}\sigma_{el}^{(1)}(\v{x}_t)\nonumber\\
&+n_0\Omega\exp\bigg[-n_0l\sigma_{el}^{(1)}(\v{0})\bigg]\int
d^2\v{x}_t e^{-i\v{q}_t\cdot \v{x}_t}\sigma_{el}^{(1)}(\v{x}_t)\bigg(
\exp\bigg[n_0l\sigma_{el}^{(1)}(\v{x}_t)\bigg]-1\bigg),
\end{align}
where we subtracted and added a $1$ in order to express the
result in terms of already defined quantities. The first contribution
produces a term proportional to $\Sigma_2^{(n)}(\v{q}_t,l)$,
whereas the second contribution can be expressed as
\begin{align}
\exp\bigg[-n_0l\sigma_{el}^{(1)}(\v{0})\bigg]\int
d^2\v{x}_t \medspace e^{-i\v{q}_t\cdot \v{x}_t}n_0\sigma_{el}^{(1)}(\v{x}_t)
= n_0\exp\bigg[-n_0l\sigma_{el}^{(1)}(\v{0})\bigg]\left|F_{el}^{(1)}(\v{q})\right|^2.
\end{align}
The third contribution can be expressed, using the definition of
$\hat{\Sigma}_2^{(n)}(\v{q},l)$ and (\ref{sigmax}) and applying the
convolution theorem, as
\begin{align}
&n_0\Omega\exp\bigg[-n_0l\sigma_{el}^{(1)}(\v{0})\bigg]
\cdot \int
d^2\v{x}_t \medspace e^{-i\v{q}_t\cdot \v{x}_t}\sigma_{el}^{(1)}(\v{x}_t)\bigg\{
\exp\bigg[n_0l\sigma_{el}^{(1)}(\v{x}_t)\bigg]-1\bigg\}\nonumber\\
&=n_0 \int \frac{d^2\v{k}}{(2\pi)^2} \left|F_{el}^{(1)}(\v{k})\right|^2\hat{\Sigma}_2^{(n)}(\v{q}-\v{k},l).
\end{align}
Consequently, joining the three terms, one finds 
\begin{align}
\frac{\partial \hat{\Sigma}_2^{(n)}(\v{q},l)}{\partial l} &= n_0\int
\frac{d^2\v{k}}{(2\pi)^2} \left(
\left|F_{el}^{(1)}(\v{k})\right|^2\hat{\Sigma}_2^{(n)}(\v{q}-\v{k};l)-\left|F_{el}^{(1)}(\v{k})\right|^2\hat{\Sigma}_2^{(n)}(\v{q},l)\right)\nonumber\\
&+n_0\Omega\exp\bigg[-n_0l\sigma_t^{(1)}\bigg]\left|F_{el}^{(1)}(\v{q})\right|^2.\nonumber
\end{align}
The above equation tells us that the number of states with transverse
momentum $\v{q}$ at $l+\delta l$  is
(a) fed with the number of states which were with transverse momentum
$\v{q}-\v{k}$ at $l$ and  collided with amplitude $F_{el}^{(1)}(\v{k})$ 
with the layer $\delta l$ gaining $\v{k}$ ,
achieving total amount $\v{q}$ and (b) decreased by the number of
states which already were with momentum $\v{q}$ at $l$ and
experimented a collision of amplitude $F_{el}^{(1)}(\v{k})$ for any $\v{k}$ 
in the layer $\delta l$ and (c) the number of states which at $l+\delta l$ did
not experiment any collision yet with the medium, so they acquire momentum
$\v{q}_t$ into their first collision at $l$. 
As expected, at large distances $l$ the probability
of finding this kind of events is negligible, since the third term is
exponentially suppressed with $l$.
 For large deeps in the medium $l\gg 1/n_0\sigma_t^{(1)}$ then, one arrives to a Master equation
\begin{align}
\frac{\partial \hat{\Sigma}_2^{(n)}(\v{q},l)}{\partial l} = n_0\int
&\frac{d^2\v{k}_t}{(2\pi)^2}\left\{
\left|F_{el}^{(1)}(\v{k})\right|^2\hat{\Sigma}_2^{(n)}(\v{q}-\v{k},l)-\left|F_{el}^{(1)}(\v{k})\right|^2\hat{\Sigma}_2^{(n)}(\v{q},l)\right\}.\label{masterequation}
\end{align}
By performing one of the trivial integrals in the momentum $\v{k}$ one finds
the total cross section of $n=1$ center and correspondingly, the actual form
of the Moliere's equation
\begin{align}
\frac{\partial \hat{\Sigma}_2^{(n)}(\v{q},l)}{\partial l} = -n_0 \sigma_t^{(1)}\Sigma_2^{(n)}(\v{q},l)+ n_0\int
\frac{d^2\v{k}}{(2\pi)^2}
\left|F_{el}^{(1)}(\v{k})\right|^2\hat{\Sigma}_2^{(n)}(\v{q}-\v{k},l).\label{moliereequation}
\end{align}
which was derived following homogeneity arguments to the form of any transport
equation. Solution to this equation is given by
\eqref{incoherent_contribution_infinitesize} without the $-1$ term, which is
responsible of the discarded boundary term.  This term is important, however,
for mediums of any size to achieve a convergent form for
$\hat{\Sigma}_2^{(n)}(\v{q},l)$. In the high density regime we notice that
since
\begin{align}
\nabla^2_t \sigma_{el}^{(1)}(\v{x}_t) \bigg|_{\v{x}_t=\v{0}}< 0,
\end{align}
then \eqref{incoherent_contribution_infinitesize} can be
asymptotically solved. The exponential term 
$n_0\sigma_{el}^{(1)}(\v{x}_t)$ quickly decays in a short $\v{x}$ range with
respect to the momentum term $\v{q}_t\cdot\v{x}_t$, so we only
need to know the short $\v{x}_t$ behavior of $\sigma_{el}^{(1)}(\v{x}_t)$
given by,
\begin{align}
n_0l\sigma_{el}^{(1)}(\v{x}_t)=n_0l\sigma_{el}^{(1)}(\v{0})-\frac{1}{2}\hat{q}l\v{x}_t^2+\mathcal{O}(\v{x}_t^4).
\end{align}
In this regime of high saturation of centers one easily finds using last
expression and equation \eqref{incoherent_contribution_infinitesize}, a 
Fokker-Planck equation whose solution is a Gaussian distribution
\begin{align}
\Sigma_2^{(n)}(\v{q},l)&=
 \int d^2\v{x}\medspace
e^{-i\v{x}\cdot\v{q}}\left\{\exp\left[-\frac{1}{2}\hat{q}l\v{x}^2\right]-\exp\left[-n_0l\sigma_{el}^{(1)}(\v{0})\right]\right\}\nonumber\\
&=
(2\pi)^2\left\{\frac{1}{2\pi\hat{q}l}\exp\left[-\frac{\v{q}^2}{2\hat{q}l}\right]-\delta^{(2}(\v{q}_t)\exp\left[-n_0l\sigma_{el}^{(1)}(\v{0})\right]\right\}.
\end{align}
In this asymptotic region the $-1$ term is not needed anymore
except to take into account border effects. However, since we 
are assuming $n_0l\gg 1$ within this approximation we neglect the 
second term and write from here onwards
\begin{align}
\Sigma_{G}^{(n)}(\v{q}_t,l)=\frac{2\pi}{\hat{q}l}\exp\left[-\frac{\v{q}_t^2}{2\hat{q}l}\right].\label{gaussianapproximation}
\end{align}
The obtained distribution is now normalized meaning that the border term
does not contribute. Under this approximation the averaged squared momentum after
traversing a distance $l$ verifies $\langle\v{q}^2\rangle=2\hat{q}l$ but leads to a 
high suppression of the long $\v{q}$ tail of the original distribution 
\eqref{incoherent_contribution_infinitesize}.
\section{The $\v{p}_t^2$ average value}
\label{sec:section_2_6}
Let us compute the average value of the transverse momentum after a
single collision. If the coupling $g=Ze^2$ is low enough to
guarantee that the amplitude can be given at leading order we write
\begin{align}
\sigma_{tot}^{(1)}= \int\frac{d^2\v{p}_t}{(2\pi)^2}\frac{(4\pi
  g)^2}{(\v{p}_t^2+\mu_d^2)^2}=\frac{4\pi g^2}{\mu_d^2},
\end{align}
so the average value is just
\begin{align}
\langle \v{p}_t^2\rangle^{(1)} = \frac{1}{\sigma_{tot}^{(1)}} \int
  \frac{d^2\v{q}_t}{(2\pi)^2}\medspace\v{q}_t^2
  \left|F_{el}^{(1)}(\v{q}_t)\right|^2=4\pi\mu_d^2\int\frac{d^2\v{p}_t}{(2\pi)^2}\frac{\v{p}_t^2}{(\v{p}_t^2+\mu_d^2)^2}.
\end{align}
In this optical approximation the integral is logarithmically divergent, since 
omitting the energy conservation delta in $M_{s_fs_i}^{(1)}(p_f,p_i)$ 
leads to an infinite range of perpendicular momentum. Using the energy 
conservation delta we actually find
\begin{align}
 \v{q}^2 = (\v{p}_1-\v{p}_0)^2 = \beta^2E^2+\beta^2E^2-2\beta^2E^2\cos\theta=
 4E^2\sin^2\theta/2,
\end{align}
so we have
\begin{align}
\langle \v{p}_t^2\rangle^{(1)} &\equiv  \frac{1}{\sigma_{tot}^{(1)}} \int
  \frac{d^3\v{p}_f}{(2\pi)^3}\medspace(\v{p}_f-\v{p}_i)^2
  \left|M_{s_fs_i}^{(1)}(p_f,p_i)\right|^2 \nonumber\\
&= \mu_d^2 \int d\theta
\frac{\sin^3\theta/2\cos\theta/2}{\left(\sin^2\theta/2+\left(\frac{\mu_d}{2E}\right)^2\right)^2}=\mu_d^2\left(2\log\bigg(\frac{2E}{\mu_d}\bigg)-1\right)\label{momentum_transfer_1}.
\end{align}
In order to compute the momentum change after passing a length $l$ 
through a medium we will consider that $R\gg \mu_d^{-1}$ so the coherent
contribution reduces to a forward delta and it can be neglected. We have then
\begin{align}
\langle \v{p}_t^2 \rangle^{(n)} = \int \frac{d^2\v{q}}{(2\pi)^2}\medspace 
\left(\v{q}^2\hat{\Sigma}_2^{(n)}(\v{q},l)+\v{q}^2\hat{\Pi}_2^{(n)}(\v{q},l)\right)=\int \frac{d^2\v{q}}{(2\pi)^2}\medspace 
\v{q}^2\hat{\Sigma}_2^{(n)}(\v{q},l).
\end{align}
We will look for a momentum additivity rule by defining a diffusion equation
for $\langle\v{q}^2(l)\rangle$. One finds, deriving with respect to depth $l$,
\begin{align}
\frac{\partial}{\partial l} \langle \v{p}_t^2\rangle^{(n)} &= \int\frac{d^2\v{q}}{(2\pi)^2}\medspace 
\v{q}^2 \left(\frac{\partial}{\partial l} \hat{\Sigma}_2^{(n)}(\v{q}^2,l)\right)= n_0\exp\left(-n_0l\sigma_t^{(1)}\right)\langle \v{p}_t^2\rangle^{(1)}\\
\medspace &-n_0\sigma_t^{(1)}\langle \v{p}_t^2
\rangle^{(n)}+n_0 \int \frac{d^2\v{q}}{(2\pi)^2}\medspace\v{q}^2 \int 
\frac{d^2\v{k}}{(2\pi)^2}\medspace
\left|F_{el}^{(1)}(\v{k}^2)\right|^2\hat{\Sigma}_2^{(n)}(\v{q}-\v{k},l)\nonumber.
\end{align}
By defining $\v{q}'=\v{q}-\v{k}$ the above relation simplifies to
\begin{align}
\frac{\partial}{\partial l} \langle \v{p}_t^2\rangle^{(n)} &=
n_0\sigma_t^{(1)}\left(1-\exp\left[-n_0l\sigma_t^{(1)}\right]\right)\langle\v{p}_t^2\rangle^{(1)}+n_0\exp\left[-n_0l\sigma_t^{(1)}\right]\langle\nonumber
\v{p}_t^2\rangle^{(1)}\nonumber\\
&=n_0\sigma_t^{(1)}\langle\v{p}_t^2\rangle^{(1)}\equiv 2\hat{q},
\end{align}
which implies that the average squared momentum change is additive in the
traveled length
\begin{align}
\left\langle\v{p}_t^2(l)\right\rangle = n_0\sigma_t^{(1)}\langle\v{p}_t^2\rangle^{(1)}l=2\hat{q}l\label{momentum_transfer_n}.
\end{align}
Here, we can measure the length or depth into the medium $l$ in units of 
the quantity $\lambda=1/n_0\sigma_t^{(1)}$. The number of times $\eta=l/\lambda$ 
we walk into the medium the average transverse momentum is just
$\eta$ times the transverse momentum change in one collision
\eqref{momentum_transfer_1}. Correspondingly the quantity $\eta=l/\lambda$ is
interpretable as the average number of collisions and $\lambda$ can be
understood as the mean free path, that is, the average distance between 
two consecutive collisions. Notice that the attenuation exponents 
\begin{align}
P(l)=\exp\bigg(-n_0\sigma_t^{(1)}l\bigg),
\end{align}
which appear in the normalization of the incoherent contribution
\eqref{incoherent_contribution_infinitesize} acquire now the sense of a
probability distribution. Namely, the probability of penetrating into the
medium a length $l$ without undergoing any collision, so $1-P(l)$ is just the
normalization of the incoherent contribution, namely the probability of
penetrating $l$ and having, at least, one collision. The medium parameter
\begin{align}
\hat{q}=\frac{1}{2}n_0\sigma_t^{(1)}\langle \v{p}_t^2\rangle^{(1)},
\end{align}
is known as the transport coefficient. For a screened Coulomb interaction and
at lowest order in the coupling we find
\begin{align}
\hat{q}\simeq 4\pi g^2 n_0 \left(\log\left(\frac{2E}{\mu_d}\right)-\frac{1}{2}\right),
\end{align}
which means that, in principle, it has to be fixed in terms of the initial
energy. If inserted in the Gaussian approximation
\eqref{gaussianapproximation} the transport parameter $\hat{q}$ severely 
modifies the momentum distribution to compensate the exponentially suppressed
$\v{p}_t$ tail of the Gaussian distribution while matching the average squared
momentum transfer.
\section{Beyond eikonal scattering}
\label{sec:section_2_7}
We have given a picture of multiple scattering departing from a pure eikonal
limit, in which the longitudinal dimension of the medium is never resolved
since longitudinal momentum change verifies $q_3=0$.  The optical phase
\eqref{wave_integration} appearing in amplitude \eqref{elasticamplitude} can
be approximated at leading order in $q_3$, since the slow $q_3$ dependence of
a single interaction $A_0^{(1)}(\v{q})\simeq A_0^{(1)}(\v{q}_t,0)$ can be
neglected compared with the rapid phase oscillations $q_3y_3$ at large
distances.  This leads to a combination of step functions
\begin{align}
\int^{y_3}_{-\infty} ds \medspace A_0^{(n)}(\v{y}_t,s) &\simeq\sum_{i=1}^n
\int^{y_3}_{-\infty}ds \int \frac{dq_3}{(2\pi)} e^{-iq_3(s-r_3^i)}
\int \frac{d^2\v{q}_t}{(2\pi)^2}
e^{-i\v{q}\cdot (\v{y}_t-\v{r}_t^i)}\hat{A}_0^{(1)}(\v{q}_t,0)\nonumber\\
&=
\sum_{i=1}^n \Theta\big(y_3-r_3^i\big)\chi_0^{(1)}\big(\v{y}_t-\v{r}_t^i\big),
\end{align}
where $\Theta(y)$ is the Heaviside step function. The above result says that
the wave \eqref{wave_integration} at $y_3$ is affected by the set of centers
at the left of $y_3$, and changes abruptly at the passage of each center. In
this way it preserves the internal structure of the longitudinal organization
of the medium although it looses the internal structure of the longitudinal
dimensions of each center.  We will now consider a set of $n_1\equiv n(z_1)$ centers
laying in a sheet of vanishing thickness $\delta z\to 0$ at coordinate
$z$. The elastic amplitude of this sheet reads from \eqref{elasticamplitude}
\begin{align}
&M_{s_1s_0}^{(n_1)}(p_1,p_0)=2\pi\delta(q^0)
\delta^{s_1}_{s_0}\beta \int d^3\v{y} \medspace e^{-i\v{q}\cdot
  \v{y}}\frac{\partial }{\partial
  y_3}\exp\left(-i\frac{g}{\beta}\int_{-\infty}^{y_3}ds\medspace 
A_0^{(m)}(\v{y})\right)\nonumber \\
&=2\pi\delta(q^0)\delta^{s_1}_{s_0}\beta e^{-iq_3z}
\int d^2\v{x}_t \medspace e^{-i\v{q}_t\cdot
  \v{x}_t}\left(\exp\bigg[-i\frac{g}{\beta}\sum_{i=1}^{m}\chi(\v{y}_t-\v{r}_t^i)\bigg]-1\right),
\end{align}
where the integration has been carried by parts. This amplitude
reproduces the diffracted part of the wave, so it includes at least one
collision. We can write for the total wave the total amplitude $S=M+1$
\begin{align}
S_{s_1s_0}^{(n_1)}(p_1,p_0)=2\pi\delta(q^0)
\delta^{s_1}_{s_0}\beta e^{-iq_3z}\int d^2\v{y}_t \medspace e^{-i\v{q}_t\cdot
  \v{y}_t}\exp\bigg[-i\frac{g}{\beta}\sum_{i=1}^{m}\chi(\v{y}_t-\v{r}_t^i)\bigg],
\end{align}
which leaves opened the possibility of non interacting at all in $z$. We can
consider the medium as a set of $n$ layers at $z_1,z_2,\ldots,z_n$ with
$n(z_i)$ scattering centers. The total number of centers in the medium is then
\begin{align}
\sum_{i=1}^n n(z_i)\equiv N.
\end{align}
The amplitude of emerging with momentum $p_n$ and spin $s_n$ after traversing
the $n$ sheets from $z_1$ to $z_n$ is given at high energies by the convolution
\begin{align}
S_{s_ns_0}^{(N)}(p_n,p_0)\equiv\left(\prod_{i=1}^{n-1}\int\frac{d^3\v{p}_i}{(2\pi)^3}\right)\left(\prod_{i=0}^{n}S_{s_{i+1}s_i}^{n(z_i)}(p_{i+1},p_i)\right),\label{orderedsmatrix_discretized_a}
\end{align}
sum in repeated indices assumed. From this we simply get
\begin{align}
M_{s_ns_0}^{(N)}(p_n,p_0) = S_{s_ns_0}^{(N)}(p_n,p_0)-S_{s_ns_0}^{(0)}(p_n,p_0).\label{orderedmmatrix_discretized_a}
\end{align}
A path integral representation of the above amplitudes can be written. By
integrating in internal momenta and summing over intermediating spins we
obtain
\begin{align}
S_{s_ns_0}^{(N)}(p_n,p_0) &= 2\pi\delta(p_n^0-p_0^0)
\delta^{s_n}_{s_0}\beta 
\left(\prod_{i=i}^{n-1}\int\frac{d^2\v{p}^t_i}{(2\pi)^2}\right)\nonumber\\
&\left(\prod_{i=1}^{n}\int
d^2\v{x}^t_i\exp\left(-i\v{q}_i\cdot\v{x}_i-i\frac{g}{\beta}\sum_{j=1}^{n(z_i)}
\chi(\v{x}_{i}^t-\v{r}_j^t)\right)\right),\label{orderedsmatrix_discretized_b}
\end{align}
where $\v{q}_i=\v{p}_i-\v{p}_{i-1}$ is a complete 3-momentum change and the
energy conservation deltas have been used to fix the longitudinal components
as $p_z\simeq p_0^0-\v{p}_t^2/(2p_0^0)$.  We rewrite the terms according to
\begin{align}
-i\sum_{i=1}^n\v{q}_i\cdot\v{x}_i=-i\v{p}_n\cdot\v{x}_n+i\sum_{i=1}^{n-1}\v{p}_i\cdot\delta\v{x}_{i}+i\v{p}_0\cdot\v{x}_1,
\end{align}
with $\delta\v{x}_i\equiv \v{x}_{i+1}-\v{x}_i$. If we also define $\delta
z_i=z_{i+1}-z_i$ we then obtain
\begin{align}
&S_{s_ns_0}^{(N)}(p_n,p_{0})=2\pi\delta(p_n^0-p_0^0) \delta^{s_n}_{s_0}
  \beta \left(\prod_{i=1}^n\int d^2 \v{x}_{i}^t\right)
e^{-i\v{p}_{n}\cdot\v{x}_{n}+i\v{p}_{0}\cdot\v{x}_1}\nonumber\\
&\left(\prod_{i=1}^{n-1}\int\frac{d^2\v{p}^t_i}{(2\pi)^2}\exp\left(+i\v{p}_i^t\cdot\delta\v{x}_{i}^t-i\frac{\big(\v{p}_{i}^t\big)^2}{2p_0^0}\delta
z_i-i\frac{g}{\beta}\sum_{j=1}^{n(z_i)}\chi(\v{x}_i^t-\v{r}_j^t)\right)\right).
\end{align}
Upon performing the momentum integrals and taking the $\delta z\to 0$ limit we
find
\begin{align}
&\lim_{\delta z\to 0} S_{s_ns_0}^{(N)}(p_n,p_{0})=
  2\pi\delta(p_n^0-p_0^0) \delta^{s_n}_{s_0}\beta \int d^2\v{x}_n^t \int d^2\v{x}_1^t e^{-i\v{p}_{n}\cdot\v{x}_{n}+i\v{p}_{0}\cdot\v{x}_1}\nonumber\\ &\int_{\v{x}_t(z_1)}^{\v{x}_t(z_n)}
\mathcal{D}^2\v{x}_t(z)\medspace\exp\left(i\int_{z_1}^{z_n}dz\medspace \bigg(\frac{p_0^0}{2}\dot{\v{x}}_t^2(z)-\frac{g}{\beta}\sum_{k=1}^{n(z)}\chi_0^{(1)}(\v{x}_t(z)-\v{r}_t^i(z))\bigg)\right),\label{orderedsmatrix_continuum}
\end{align}
Notice that the evaluation of this amplitude for the most probable path is
given by the Euler-Lagrange equation
\begin{align}
\frac{d\v{p}_t}{dz}&=g
\v{\nabla}\left(\int^{+\infty}_{-\infty} dt \medspace
 A_0^{n(z)}(\v{x}+\v{\beta}t)\right),
\end{align}
which is a classical trajectory approximation assuming a straight
propagation. The average of the square of \eqref{orderedmmatrix_discretized_a}
over medium configurations can be written as before as the sum of an incoherent
and a coherent contribution
\begin{align}
\left\langle
M_{s_ns_0}^{(N)*}(p_n,p_0)M_{s_ns_0}^{(N)}(p_n,p_0)\right\rangle = \Sigma_2^{(N)}(p_n,p_0)+\Pi_s^{(n)}(p_n,p_0),
\end{align}
where the incoherent contribution is given by the quantity
\begin{align}
\Sigma_2^{(N)}(p_n,p_0)=\left\langle
\left(S_{s_ns_0}^{(N)}(p_n,p_0)\right)^*S_{s_ns_0}^{(N)}(p_n,p_0)\right\rangle-\left\langle
S_{s_ns_0}^{(N)}(p_n,p_0)\right\rangle^*\left\langle S_{s_ns_0}^{(N)}(p_n,p_0)\right\rangle,
\end{align}
and the coherent contribution is given by the averaged amplitude squared
\begin{align}
\Pi_2^{(N)}(p_n,p_0)=\left|\left\langle
S_{s_ns_0}^{(N)}(p_n,p_0)-S_{s_ns_0}^{(0)}(p_n,p_0)\right\rangle\right|^2.
\end{align}
The cancellation of the longitudinal phases in the macroscopic limit, since
the transverse momentum in the conjugated amplitude equals the transverse
momentum in the amplitude, leads to an incoherent contribution of the form
\begin{align}
\Sigma_2^{(N)}(p_n,p_0)=2\pi\delta(p_n^0-p_0^0)\beta_p \delta^{s_n}_{s_0}\pi
R^2\prod_{i=1}^{n-1}\left(\int \frac{d^2\v{p}_i^t}{(2\pi)^2}\right)\prod_{i=1}^{n}\Bigg(\int d^2\v{x}_i^te^{-i\delta\v{p}_i^t\cdot\v{x}_i^t}\Bigg)\nonumber\\
\times\left\{\prod_{i=1}^n\exp\bigg(\delta
z_in_0(z_i)\left(\sigma_{el}^{(1)}(\v{x}_i^t)-\sigma_{el}^{(1)}(\v{0})\right)\bigg)-\prod_{i=1}^n\exp\bigg(-\delta
z_in_0(z_i)\sigma_{el}^{(1)}(\v{0})\bigg)\right\},
\end{align}
where we divided by time $2\pi\delta(0)\equiv T$ and incoming flux $\beta$.
Correspondingly the integration in internal momenta is trivial and the
internal structure of the scattering distribution is lost. By taking the
$\delta z_i\to 0$ limit we obtain
\begin{align}
\Sigma_2^{(N)}(p_n,p_0)=2\pi\delta(p_n^0-p_0^0)\beta \delta^{s_n}_{s_0}\pi
R^2\exp\left(-\sigma_{el}^{(1)}(\v{0})\int_{z_1}^{z_n} dz \medspace
n_0(z)\right)\nonumber\\
\times\int d^2\v{x}_te^{-i(\v{p}_n^t-\v{p}_0^t)\cdot\v{x}_t}\left(\exp\left(\sigma_{el}^{(1)}(\v{x}_t)\int_{z_1}^{z_n} dz \medspace n_0(z)\right)-1\right).
\end{align}
The evaluation of the coherent contribution follows the same steps. The
average of the amplitude produces
\begin{align}
&\left\langle S_{s_ns_0}^{(N)}(p_n,p_0)\right\rangle =
2\pi\delta(p_n^0-p_0^0)\beta\delta^{s_n}_{s_0}\prod^{n-1}_{i=1}\left(\frac{i\delta
z_i}{2\pi p_0^0}\right)\prod_{i=1}^n\left(\int d^2\v{x}_t^i\right)\\
&\times\exp\left(-i\v{p}_n^t\cdot\v{x}_n^t+i\sum_{i=1}^{n-1}\frac{p_0^0}{2}\left(\frac{\delta \v{x}_i^t}{\delta
z_i}\right)^2\delta z_i+\sum_{i=1}^{n}\delta z_i n_0(z_i)\pi_{el}^{(1)}(\v{x}_i^t)+i\v{p}_0^t\v{x}_1^t\right)\nonumber,
\end{align}
where in analogy with the function $\sigma_{el}^{(1)}(\v{x})$ for the
incoherent contribution we defined the Fourier transform of the single elastic
amplitude convoluted with the window function of the medium as
\begin{align}
\pi_{el}^{(1)}(\v{x})\equiv \int \frac{d^2\v{q}_t}{(2\pi)^2}e^{i\v{q}\cdot\v{x}}F_{el}^{(1)}(\v{q})W_\Omega(\v{q},R).
\end{align}
By taking the $\delta z\to 0$ limit the above average transforms into a path
integral in the transverse plane where the time variable is the $z$ position
as
\begin{align}
\left\langle S_{s_ns_0}^{(N)}(p_n,p_0)\right\rangle=
2\pi\delta(p_n^0-p_0^0)\beta_p\delta^{s_n}_{s_0}\int d^2\v{x}_n^te^{-i\v{p}_n^t\cdot\v{x}_n^t}\int d^2\v{x}_1^t
e^{+i\v{p}_0^t\v{x}_1^t}\nonumber\\
\int \mathcal{D}^2\v{x}_t(z)\exp\left(i\int^{z_n}_{z_1}
dz\left(\frac{p_0^0}{2}\dot{\v{x}}_t^2(z)-in_0(z)\pi_{el}^{(1)}\Big(\v{x}_t(z)\Big)\right)\right).
\end{align}
Then the beyond eikonal evaluation of the coherent contribution is given by
\begin{align}
&\Pi_2^{(N)}(p_{n},p_0)=2\pi\delta(p_n^0-p_0^0)\beta\delta^{s_n}_{s_0}\Bigg|\int d^2\v{x}_n^te^{-i\v{p}_n^t\cdot\v{x}_n^t}\int d^2\v{x}_1^t
e^{+i\v{p}_0^t\v{x}_1^t}\\&\times\int \mathcal{D}^2\v{x}_t(z)\exp\left(i\int^{z_n}_{z_1}
dz\medspace\frac{p_0^0}{2}\dot{\v{x}}_t^2(z)\right)\left(\exp\left(\int_{z_1}^{z_n}
dz \medspace n_0(z)\pi_{el}^{(1)}\Big(\v{x}_t(z)\Big)\right)-1\right)\Bigg|^2\nonumber,
\end{align}
where we divided by time and incoming flux. The above beyond eikonal results
will become useful when evaluating emission processes occurring in a multiple
scattering scenario. As we will see in the next chapter, when an energy gap is
considered between the state of the traveling particle and its conjugate,
corresponding to an energy carried by the emitted particle, a non vanishing
longitudinal phase will lead to coherence effects in the intensity spectrum.

%% file: chapter3/chapter3.tex
\chapter{High energy emission}

In the previous chapters we worked out the basic concepts and tools towards an
evaluation of the amplitude and the intensity of a high energy fermion
transiting from $(p_\alpha,s_\alpha)$ to $(p_\beta,s_\beta)$ due to the effect
of the multiple scattering sources in a medium.  We will consider now a single
photon bremsstrahlung occurring while this multiple scattering develops. For
this purpose we take advantage of the fact that, for sufficiently low photon
energies, the elastic amplitude has to factorize from the emission amplitude
\cite{weinberg1965}. In consequence, as we will show, our previous results
dealing with the pure elastic problem have to be recovered and be still valid
for the evaluation of the radiation of soft quanta off high energy fermions.

The emission intensity in a multiple scattering scenario has been predicted by
Ter-Mikaelian \cite{termikaelian1953} and Landau and Pomeranchuk
\cite{landau1953a} to be suppressed with respect to a naive picture consisting
in an incoherent sum of single Bethe-Heitler \cite{bethe1934}
intensities. Longitudinal phases in the scattering amplitudes regulate the
amount of matter which can be considered a single and independent
emitter. These phases control the coherence in the sum of all the involved
Feynman diagrams and grow with the distance, with the photon energy and with
the accumulated angle of the electron-photon pair. When the coherence length
extends beyond one scattering length, comprising several collisions on
average, the total intensity is not anymore the sum of the single
Bethe-Heitler intensities at each collision and, thus, the intensity is
substantially reduced. This effect, known as the Landau-Pomeranchuk-Migdal
(LPM) suppression, leads to a stopping/energy-loss by bremsstrahlung in
condensed media significatively smaller than the one expected for a set of
unrelated collisions. The implications of this suppression spanned several
fields of high energy physics
\cite{landau1953b,stanev1982,sorensen1992a,wang1994a}
and are still open nowadays. Although the first indirect measurements of this
suppression were early given through cosmic ray showers \cite{fowler1959} soon
after its prediction, it was not until recently that various experiments,
first at SLAC \cite{anthony1995,anthony1996,anthony1997} and then at CERN
\cite{hansen2003,hansen2004}, measured in detail the phenomenon. Recent and
renewed interest has also suscited the LPM suppression of gluons in the
presence of the QCD media produced in high energy collisions at RHIC and
LHC. The formation of new states of the matter, namely the quark-gluon plasma
(QGP), are studied, among other probes, through the energy loss pattern of
partons while traveling by this colored QCD medium\footnote{A more detailed
  study of the LPM effect and its phenomenology in QCD will be given in the
  next chapters.}.

Owing to the amount of work behind this subject it becomes necessary to
introduce a brief historical remark. The first evaluation of this effect was
given by Landau \cite{landau1953b} with a classical calculation for a medium
of semi-infinite length. In his work, the external field interaction is
replaced with the Fokker-Planck approximation, leading to a total squared
momentum transfer satisfying a Gaussian distribution. The computation of
Landau produces a differential energy intensity which vanishes as
$\sqrt{\omega}$ in the soft limit $\omega\to 0$ while approaches the
incoherent sum of Bethe-Heitler intensities for larger $\omega$. Landau's
evaluation was later extended by Migdal \cite{migdal1956} to the quantum case
by means of a Boltzmann transport equation for an averaged target. As it was,
maybe, better and later explained in the rederivation by Bell \cite{bell1958},
Migdal's result shown that, except for spinorial corrections in the hard part
of the photon spectrum, the LPM suppression is still a classical effect, in
correspondence with the classical behavior of the infrared divergence
\cite{bethe1934,weinberg1965}.

Following these two seminal works, a finite target evaluation was soon
introduced first by Ternovskii \cite{ternovskii1961} by using Migdal's
transport approach, and later by Shul'ga and Fomin \cite{shulga1978} through
the classical limit. In both works radiation is computed in the softest regime
$\omega\to 0$ and, unlike to the infinite suppression predicted by Landau and
Migdal, the resulting intensity is found to saturate into a Bethe-Heitler
power-like law in which the medium coherently acts as a single radiation
entity. Both results, however, are only given for this coherent limit and in
the saturation regime, in which the electron momentum transfer at the end of
the process greatly exceeds its mass in magnitude. As we will later see, this
coherent regime, which was not considered neither by Landau nor by Migdal, can
be treated as part of a more general picture, corresponding to the Weinberg's
soft photon theorem \cite{weinberg1965}.

Very diverse and more recent calculations have appeared since then. We cite
the work of Blankenbecler and Drell
\cite{blankenbecler1987,blankenbecler1996,blankenbecler1997a,blankenbecler1997b},
who established a formalism in which the finite target case can be generally
computed in the Fokker-Planck approximation. They adopted an approach in which
a double integral in longitudinal position accounts for the photon emission
point, both in the amplitude and its conjugate, and then the relevant quantity
modulating the interference behavior is just the accumulated longitudinal
momentum change between these two points. Their result is given for arbitrary
lengths, in such a way that, asymptotically for very low $\omega$ and in the
large length limit, they recover Landau's and Migdal's predictions except for
constant factors. Simultaneously to their work, and with the scope of
developing a bremsstrahlung formalism for QCD matter, Zakharov
\cite{zakharov1996a,zakharov1996b,zakharov1998} also treated the semi-infinite
case in the Fokker-Planck approximation by reformulating Migdal's transport
approach through path integrals in the transverse plane, leading to the same
result. In his approach, however, the finite target implementation
\cite{zakharov1996b} is not easy and requires a different treatment. In the
same line the BDMPS group \cite{baier1996} following a transport approach very
similar to the one by Migdal and, omitting the collinear divergence introduced
by the neglection of the electron mass (leading to a fail in reproducing the
Bethe-Heitler limit) found a result which was shown \cite{baier1998a} to be
equivalent to the one by Zakharov in the soft limit. We have to cite also the
works of Baier and Katkov \cite{vbaier1998,vbaier2004}, who put some focus
into accounting the Coulomb corrections to the Fokker-Planck approximation,
and also in computing the finite size case, among other interesting
situations. Their finite size result is comparable to the one found by
Ternovskii, but frequently the various cases of study lack a general
formulation and require different treatments. Similar predictions were later
given by Wiedemann and Gyulassy \cite{wiedemann1999}, extending Zakharov path
integral result to account for the angular dependence and the finite size case
in a very similar way to the finite size evaluation by Blankenbecler and
Drell. For further details on the history behind these works extensive and
interesting reviews \cite{klein1998,vbaier2005} have been written.

A comprehensive non-perturbative QFT description of the emission of quanta for
the general case of a finite/structured target, which takes into account not
only the LPM effect, but dielectric and transition radiation effects as a
particular cases of the same phenomena, admitting an evaluation for general
interactions beyond the Fokker-Planck approximation and which takes care of the
angular distributions is still missing. Reasons for this situation are
diverse and probably related to the complexity and particularities of the vast
part of the calculations \cite{klein1998} and the usual approach of
considering \textit{ab initio} the intensity for an averaged target of
infinite transverse size, instead of an adequate quantum mechanical definition
at the level of the amplitude. As a result the diagrammatic structure is
hidden, the transverse coherence effects are never considered and the soft
photon limit and length constraints are often misunderstood.

Taking these considerations into account, in Section \ref{sec:section_3_1} we
find the emission amplitude and we briefly explain the LPM effect by taking
the simpler classical limit. The main purpose of this section is to relate the
LPM effect and the Weinberg's soft photon theorem as part of the same
coherence phenomena. With these ideas in mind we define then in Section
\ref{sec:section_3_2} the quantum amplitude \cite{feal2018b} in terms of the
scattering amplitudes found in the previous chapter, with the aim of taking
the square of this amplitude in Section \ref{sec:section_3_3} in order to
evaluate the photon intensity.  We show that the intensity can be split
into a coherent and an incoherent contribution, related to transverse
interference effects, just as we did with the pure elastic case. The coherent
contribution will be found to be a pure quantum mechanical contribution which
does not admit a statistical interpretation, whereas the incoherent
contribution results into a mixed contribution which, in the infinite
transverse size limit for the medium, reproduces the macroscopic/classical
limit and thus admits an statistical interpretation. The resulting expression
in the discrete limit, which is the central result of the present work, admits
a numerical evaluation by Monte Carlo methods for general interactions,
arbitrary medium lengths, angular distributions and medium effects in the
photon dispersion relation. In Section \ref{sec:section_3_4} we take the
continuous limit and we find that our result is equivalent to a path integral
formulation which can be solved in the Fokker-Planck approximation. This
result will be used as a check for the numerical evaluation of our discretized
approach. Migdal's/Zakharov and BDMPS results are recovered as two particular
cases.
\section{Amplitude and the classical LPM effect}
\label{sec:section_3_1}
\begin{figure}[ht]
\centering
\includegraphics[scale=0.7]{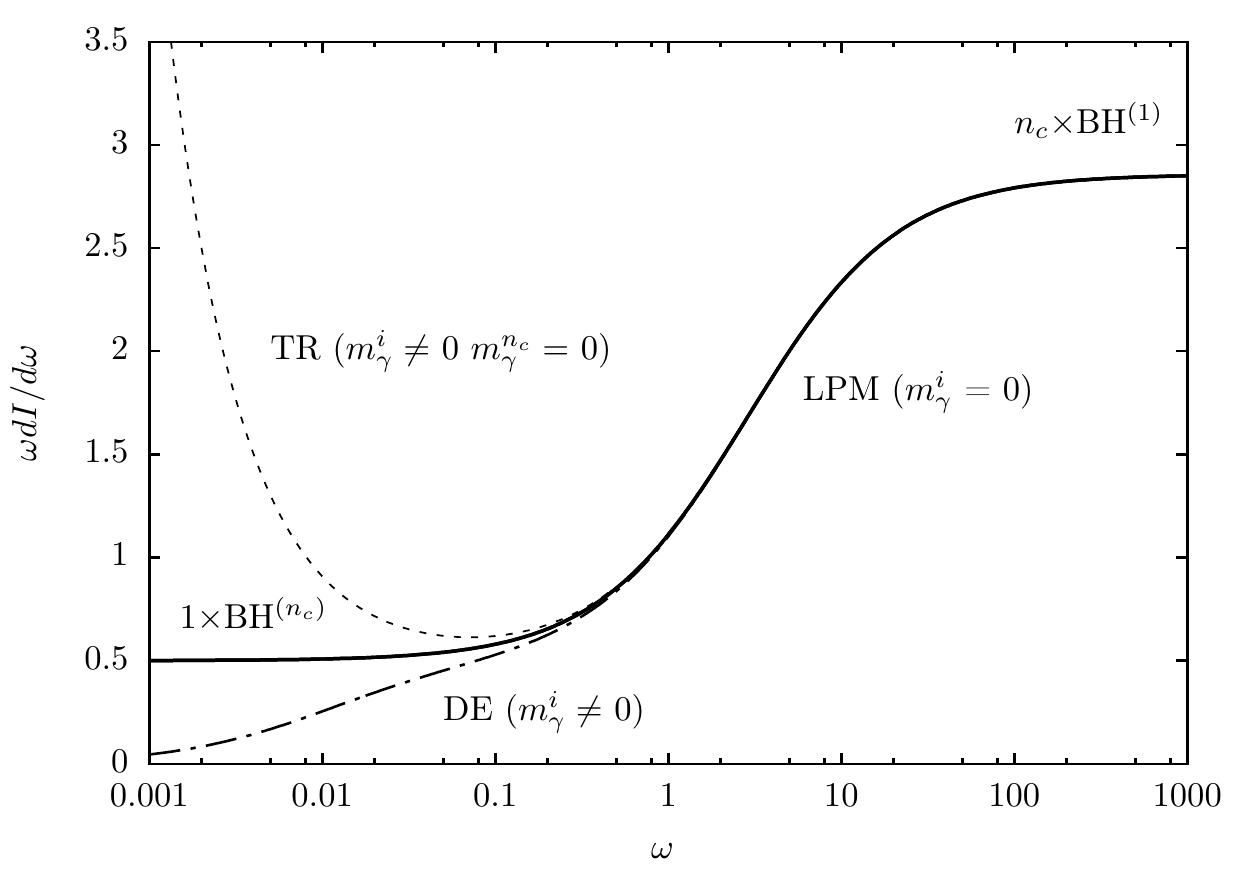}
\caption{Schematic representation of the regimes of the radiation and the LPM
  effect (solid line), the lower plateau is given by
  \eqref{classicalintensity_coherentplateau} and the upper plateau is
  given by \eqref{classicalintensity_incoherentplateau}. Also shown are the
  dielectric effect (dot dashed line) and the transition radiation effect
  (dotted line).}
\label{fig:figure_3_0}
\end{figure}

We consider the amplitude of emission of a photon due to the effect of the
multiple scattering sources in a medium. Let the photon be considered free
after the emission and let the 4-momentum be denoted by
$k\equiv(\omega,\v{k})$. Medium effects in the energy momentum relation or
refractive index can be introduced by defining an effective photon mass
$m_\gamma$. Then the photon's velocity reads
\begin{align}
\beta_k=\sqrt{1-m_\gamma^2/\omega^2},
\end{align}
and the momentum can be written as $k=\omega(1,\beta_k\hat{\v{k}})$, where
$\hat{\v{k}}$ is the unitary vector in the direction of $\v{k}$. A photon
with this momentum $k$ and polarization $\lambda$ is given by
\begin{align}
A_{\mu}^\lambda(x)=\mathcal{N}(k) \epsilon_\mu^\lambda e^{ik\cdot x},
\end{align}
where $\mathcal{N}(k)=\sqrt{4\pi/2\omega}$ is the normalization. The new
electron state under the effect of this emitted photon
$\psi^{(n)}_{\gamma}(x)$ can be always written
as the superposition
\begin{align}
\psi^{(n)}_{\gamma}(x)=\psi_{i}^{(n)}+e\int d^4y S_F^{(n)}(x-y)\gamma^\mu A_\mu^\lambda(y)\psi_{\gamma}^{(n)}(y)\label{psi_gamma_solution},
\end{align}
where $\psi_{i}^{(n)}(y)$ is a solution to the electron state in absence of
interaction with the emitted photon $A_\mu^\lambda(y)$, but subject to the
interaction with the photons from the external field
\eqref{general_external_field} of the medium. It verifies then
\begin{align}
\left(i\gamma^\mu\partial_\mu-gA_0^{(n)}(x)-m\right)\psi_{i}^{(n)}(x)=0,
\label{psi_dirac_equation}
\end{align}
where the external field $A_0^{(n)}(x)$ is given by
\eqref{yukawa_external_field} and $S_F^{(n)}(x)$ is the adequate propagator,
given in particular by \eqref{propagationrelation} at high energies. From the
equation \eqref{psi_gamma_solution} we easily find the required relation
\begin{align}
\left(i\gamma^\mu\partial_\mu-gA_0^{(n)}(x)-m\right)S_F^{(n)}(x-y)=\delta^4(x-y).
\end{align}
The wave \eqref{psi_gamma_solution} can be always expanded in the original
basis of scattered states with the $(n)$ sources. Correspondingly, the
amplitude of finding the wave $\psi_\gamma^{(n)}(x)$ in the final scattered
state $\psi^{(n)}_{f}(x)$ is given, using \eqref{psi_gamma_solution}, by
\begin{align}
\mathcal{S}_{em}^{(n)}=\int d^3\v{x} \medspace\psi^{(n),\dag}_{f}(x)
\psi^{(n)}_{i}(x)+e\int d^3\v{x}\medspace \psi^{(n),\dag}_{f}(x)S_F^{(n)}(x-y)A_\mu^\lambda(y)\psi_{\gamma}^{(n)}(y).
\end{align}
The first term is just the pure elastic scattering contribution. Indeed by
expanding the scattered states as in \eqref{outcomingstatescattering_smatrix}
and taking into account the unitarity of the $S$-matrix we just find
\begin{align}
\int d^3\v{x}\medspace \psi^{(n),\dag}_{f}(x)
\psi^{(n)}_{i}(x)= S^{(n)}_{s_fs_i}(p_f,p_i).
\end{align}
We are instead interested in the second contribution, which constitutes the
radiative correction to the elastic scattering. By using
\eqref{propagationrelation}, we find the conjugate relation
\begin{align}
 \bar{\psi}_f^{(n)}(y)\equiv i\int d^3\v{x}\medspace \psi_f^{(n),\dag}(x) S_F^{(n)}(x-y),
\end{align}
and since $\gamma_0^2=1$ then 
\begin{align}
\mathcal{S}_{em}^{(n)}=S^{(n)}_{s_fs_i}(p_f,p_i)-ie\int d^4y  \bar{\psi}_f^{(n)}(y) \gamma^\mu A_\mu^\lambda(y)\psi_\gamma^{(n)}(y)\label{amplitude_emission_s}.
\end{align}
We then define the amplitude of going from $(p_i,s_i)$ to $(p_f,s_f)$
\textit{while} emitting a photon, neglecting the pure elastic contribution, as
the contribution
\begin{align}
\mathcal{M}_{em}^{(n)} \equiv -ie \int d^4y \medspace  \bar{\psi}_f^{(n)}(y) \gamma^\mu A_\mu^\lambda(y)\psi_i^{(n)}(y)+\mathcal{O}(e^2)\label{amplitude_emission_m},
\end{align}
where we expanded in the last step
$\psi_\gamma^{(n)}(x)=\psi_i^{(n)}(x)+\mathcal{O}(e)$, thus the above relation
is only valid at leading order in $e=\sqrt{\alpha}$ for the interaction
between the emitted photon and the electron. Evaluating
\eqref{amplitude_emission_m} for a single source ($n$ = 1) leads to the
well-known Bethe-Heitler bremsstrahlung amplitude \cite{bethe1934}. We are
interested, instead, in the phenomenology of the multiple scattering case,
which leads to interference effects in the squared amplitude
\eqref{amplitude_emission_m}, as predicted by Ter-Mikaelian
\cite{termikaelian1953} and Landau and Pomeranchuk
\cite{landau1953a,landau1953b}. In order to understand these interferences we
make use of the classical behavior of the infrared divergence
\cite{weinberg1965}, i.e. since \eqref{amplitude_emission_m} has a pole in
$\omega\to 0$ the number of photons diverges and a classical evaluation holds
\cite{weinberg1995}. In that case we can replace the electron current by the
classical path
\begin{align}
J_{k}(x)\equiv\bar{\psi}_{f}^{(n)}(x)\gamma_k\psi_{i}^{(n)}(x)\to J_{k}(x)=v_k(t)\medspace\delta^3(\v{x}-\v{x}(t)),
\end{align}
where $\v{v}(t)\equiv \dot{\v{x}}(t)$ is the electron velocity. Then \eqref{amplitude_emission_m} acquires the classical form
\begin{align}
\mathcal{M}_{em}^{(n)}=-ie\mathcal{N}(k)\int^{+\infty}_{-\infty} dt \medspace
\bigg(\frac{\v{k}}{\omega}\times \v{v}(t)\bigg)\exp\bigg(i\omega
t-i\v{k}\cdot\v{x}(t)\bigg), \label{amplitude_emission_m_classical}
\end{align}
where the polarization vector is assumed to be normalized and orthogonal to
the photon momentum $\v{\epsilon}^\lambda(k)\cdot\v{k}=0$. Equation
\eqref{amplitude_emission_m_classical} can be understood as a sum over all the
instants $t$ at which the photon can be emitted. To see this fact we
discretize the electron trajectory \cite{bell1958} as $\v{v}_j$ for
$j=1,\ldots,n_c+1$ and piecewise path
$\v{x}_j=\v{x}_{j-1}+\v{v}_{j-1}(t_j-t_{j-1})$, where $n_c$ is the number of
collisions. From \eqref{amplitude_emission_m_classical} we get
\begin{align}
\mathcal{M}_{em}^{(n)}=e\mathcal{N}(k)\frac{1}{\omega}\sum_{j=1}^{n_c}\v{\delta}_j \medspace
e^{i\varphi_j},\label{amplitude_emission_m_classical_discretized}
\end{align}
which is a sum of $n_c$ single Bethe-Heitler amplitudes \cite{bethe1934} of
the form
\begin{align}
\v{\delta}_j \equiv \v{k} \times \bigg(\frac{\v{v}_{j+1}}{\omega-\v{k}\cdot\v{v}_{j+1}}-\frac{\v{v}_{j}}{\omega-\v{k}\cdot\v{v}_{j}}\bigg),\label{delta_definition}
\end{align}
interfering with a phase $i\varphi_j\equiv i\omega
t_j-i\v{k}\cdot\v{x}_j$. The evaluation of the square of
\eqref{amplitude_emission_m_classical_discretized} leads
to a total emission intensity between the photon's solid angle $\Omega_k$ and
$\Omega_k+d\Omega_k$ given by
\begin{align}
\omega \frac{dI}{d\omega d\Omega_k} = \frac{e^2}{(2\pi)^2} \left(\sum_{i=1}^{n_c}
\v{\delta}_i^2+2\Re\sum_{i=1}^{n_c}\sum_{j=1}^{i-1} \v{\delta}_i\cdot \v{\delta}_j
e^{i\varphi_{ij}}\right),\label{classicalintensity}
\end{align}
where we have split the sum in a diagonal and non diagonal contribution. It is
clear that the phase between two emission points controls the sum in
\eqref{classicalintensity}. By now we assume for simplicity that
$\beta_k$ = 1, then the phase difference can be written as
\begin{align}
i\varphi_j^i \equiv ik_\mu (x^\mu_j-x^\mu_i) = i\int_{z_i}^{z_j} dz \frac{k_\mu p^\mu(z)}{p_0}
= i\omega(1-\beta_p)\int^{z_j}_{z_i} dz +i\omega \int_{z_i}^{z_j} dz
\frac{\delta\v{p}^2(z)}{2\beta_p p_0^2}, \label{phase_difference}
\end{align}
where $p\equiv (p_0,\v{p})$ is the electron 4-momentum, its velocity is then
given by
\begin{align}
\beta_p=\sqrt{1-m^2_e/p_0^2},
\end{align}
and $\delta\v{p}^2(z)$ the accumulated momentum change at $z$ as measured with
respect to the photon direction $\hat{\v{k}}$. This phase grows with $\omega$,
with the emission angle and of course with accumulated distance in general.
When the phase between each two consecutive collisions satisfies
$\varphi_j^{j+1}\gg 1$ then the high oscillatory behavior cancels the
non-diagonal term in \eqref{classicalintensity} and
\begin{align}
\omega \frac{dI_{sup}}{d\omega d\Omega_k} = \frac{e^2}{(2\pi)^2} \left(\sum_{i=1}^{n_c}
\v{\delta}_i^2\right),\label{classicalintensity_incoherentplateau}
\end{align}
which is a totally incoherent sum of $n_c$ single Bethe-Heitler
\cite{bethe1934} intensities. In this regime all the available deflections of
the electron emit as independent sources of bremsstrahlung. This occurs at
frequencies larger than $\omega_s$ which by using the condition
$\langle\varphi_j^{j+1}\rangle\simeq 1$ and Eqs. \eqref{phase_difference} and
\eqref{momentum_transfer_n} produces 
\begin{align}
\omega_s\simeq \frac{1}{\lambda}\frac{p_0^2}{m_e^2+\hat{q}l},\label{saturation_frequency}
\end{align}
where $\lambda$ is the average distance between two consecutive collisions or
the mean free path, $l$ the medium length and $l\gg \lambda$ was assumed. In
this limit the radiation behaves as if the $n_c$ scattering centers were
separated an infinite distance and the total bremsstrahlung cross section is
just the sum of the individual cross sections.  On the contrary, if the phase
is minimal so it satisfies $\varphi_0^{n_c}\ll 1$ we find from
\eqref{amplitude_emission_m_classical_discretized}
\begin{align}
\omega \frac{dI_{inf}}{d\omega d\Omega_k} = \frac{e^2}{(2\pi)^2} \left(\sum_{i=1}^{n_c}
\v{\delta}_i\right)^2,\label{classicalintensity_coherentplateau}
\end{align}
which is a Bethe-Heitler intensity with initial velocity $\v{v}_1$ and final
velocity $\v{v}_{n_c+1}$, thus corresponding to the coherent deflection with
all the medium. Indeed for negligible phases one finds
\begin{align}
\sum_{i=1}^{n_c} \v{\delta}_i = \v{k} \times \bigg(\frac{\v{v}_{n_c+1}}{\omega-\v{k}\cdot\v{v}_{n_c+1}}-\frac{\v{v}_{1}}{\omega-\v{k}\cdot\v{v}_{1}}\bigg),\label{last_first_classical}
\end{align}
and the internal structure of scattering becomes irrelevant for the
process. This occurs for frequencies lower than $\omega_c$, where this
frequency is given by $\langle\varphi_0^{n_c}\rangle\simeq 1$ thus by using
\eqref{momentum_transfer_n} and \eqref{phase_difference} we find
\begin{align}
\omega_c\simeq \frac{1}{l}\frac{p_0^2}{m_e^2+\hat{q}l}.\label{coherent_frequency}
\end{align}
In this regime the only independent source of bremsstrahlung is the medium as
a whole, and the emitted photons are the first and the last ones, which is
nothing but the soft photon theorem \cite{weinberg1965}\footnote{In some LPM
  literature this is known as the Ternovskii-Shul'ga-Fomin
  \cite{ternovskii1961,shulga1998} term, however their result restricts to the
  saturation regime $n_c\gg 1$ under the Fokker-Planck approximation for the
  distribution of $\v{v}_i$.}. The LPM effect refers, then, to the suppression
from the incoherent plateau \eqref{classicalintensity_incoherentplateau} of
maximal intensity to the coherent plateau
\eqref{classicalintensity_coherentplateau} of minimal intensity, and both
regimes are given by a Bethe-Heitler like law. Observe that by using
\eqref{saturation_frequency} and \eqref{coherent_frequency} the ratio of the
characteristic frequencies is just the average number of collisions,
$\omega_s/\omega_c=n_c$. A schematic representation of this behavior is
depicted in Figure \ref{fig:figure_3_0}.
It results interesting to relate this phase interference with the uncertainty
relation between longitudinal distance and longitudinal momentum
change. Phases and denominators at
\eqref{amplitude_emission_m_classical_discretized} agree with the infrared
pole of the off-shell electron propagator,
\begin{align}
(p+k)^2-m^2=p^2+k^2+2p\cdot k -m^2= 2p_\mu k^\mu,
\end{align}
By rewriting the above relation in terms of the elastic momentum change of an
off-shell electron $q=(0,\v{q})$ with $q=p_f+k-p_i$ we find
\begin{align}
(p+q)-m^2= p^2+q^2+2p\cdot q-m^2 \approx 2p\cdot q = -2\v{p}\cdot\v{q},
\end{align}
where we used the fact that at high energies $\v{p}\gg \v{q}$.  Following
this, phase \eqref{phase_difference} agrees, at high energies, with the
accumulation of momentum change in the dominant direction of $\v{p}$, which
can be taken along the $z$ direction and we can write
\begin{align}
i\varphi_{j}^{i} = -i \int^{t_j}_{t_i} dt \hat{\v{p}}\cdot \v{q}(t) = -i
\int^{t_j}_{t_i} dt \medspace q_z(t).\label{accumulated_momentum_change}
\end{align}
We assume for simplicity $q_z$ constant in the following interpretation. Then
when $1/q_z$ becomes larger than the medium length $l$ the medium is seen as a
infinitely thin sheet, being impossible for the traveling electron and photon
pair to resolve its interior structure. On the other hand, when $1/q_z$ is of
the order or smaller than $l$, the internal medium structure in the distance
$1/q_z$ can not be resolved and the amplitude is the incoherent sum of its
coherent $lq_z$ parts. This sum saturates when $1/q_z$ is of the order of the
average distance between scatterings.

On the other hand medium effects in the photon dispersion relation severely
change this picture in the soft limit. In that case the emitted photon field
can be considered to satisfy a Helmholtz equation with a source term
\begin{align}
\left(\nabla^2-\frac{\partial^2}{\partial^2 x_0}\right)\v{E}(x)=
4\pi\frac{\partial \v{J}(x)}{\partial x_0},
\end{align}
where the medium current is given by $\v{J}=n_0Ze\dot{\v{r}}$. Then by using
$m_e\ddot{\v{r}}=e\v{E}$ we find
\begin{align}
\omega^2-\v{k}^2 = \frac{4\pi n_0Ze^2}{m_e} \equiv m_\gamma^2.
\end{align}
Since $k=\omega(1,\beta_k\hat{\v{k}})$ and $p=p_0(1,\beta_p\hat{\v{p}})$ we
can write for denominators at \eqref{delta_definition} and for the phase
\eqref{phase_difference}
\begin{align}
\frac{k_\mu p^\mu}{p_0}=\omega(1-\beta_k\beta_p\hat{\v{k}}\cdot \hat{\v{p}})=\omega(1-\beta_k)+\omega\beta_k(1-\beta_p\hat{\v{k}}\cdot \hat{\v{p}}).
\end{align}
For the energies considered here we will assume always $\omega\gg
m_\gamma$. Then, with the required accuracy, we notice that $\beta_k$ can be
made unity in the second term of the right hand side but $(1-\beta_k)\simeq
m^2_\gamma/2\omega^2$ so
\begin{align}
\left.\frac{k_\mu p^\mu}{p_0}\right|_{m_\gamma}\simeq
\frac{m^2_\gamma}{2\omega}+\omega(1-\beta_p\hat{\v{k}}\cdot \hat{\v{p}}) = \frac{m^2_\gamma}{2\omega}+\left.\frac{k_\mu p^\mu}{p_0}\right|_{m_\gamma=0}.
\end{align}
This phase difference further suppresses the coherent plateau at frequencies
lower than $\omega_{de}$, when the extra term $m_\gamma^2/2\omega$ becomes of
the order of the massless term, i.e. $\omega_{de}\simeq m_\gamma^2l\omega_c$,
since for frequencies lower than $\omega_{de}$ \eqref{delta_definition}
vanishes. This effect is known as the dielectric suppression. However, if this
photon mass becomes local in $z$ due to an structured or a finite target, then
the correction $m_\gamma^2/2\omega$ changes for each
\eqref{delta_definition}. In particular, for finite and homogeneous mediums,
one can consider that the first and all the interior photons have
$m_\gamma\neq$ 0 whereas the last photon propagates in the vacuum and thus
$m_\gamma$ = 0. By using \eqref{classicalintensity_coherentplateau} this leads
to a big resonance difference in the coherent plateau for frequencies lower
than $\omega_{de}$ and the intensity is dramatically enhanced. This effect is
called transition radiation. Both the dielectric suppression and the transition
radiation effect are qualitatively shown in Figure \ref{fig:figure_3_0}.
\section{Amplitude in the quantum approach}
\label{sec:section_3_2}
Since the number of photons diverges at both regimes,
Eqs. \eqref{classicalintensity_coherentplateau} and
\eqref{classicalintensity_incoherentplateau}, it is clear that, except for
hard photon corrections, the previous classical behavior has to be recovered
in a quantum evaluation. The electron states entering
\eqref{amplitude_emission_m} are given by the superposition
\eqref{outcomingstatescattering_smatrix}
\begin{align}
\psi_i^{(n)}(x)=\psi_i^{(0)}(x)+\sum_{s=1,2}\int
\frac{d^3\v{p}}{(2\pi)^3}u_{s}(p)\sqrt{\frac{m_e}{p^0}}e^{-ip\cdot
  x}M_{ss_0}^{(n)}\big(p,p_0;z,z_0\big),\label{incoming_electron}
\end{align}
for the incoming wave and
\begin{align}
\psi_f^{(n)}(x)=\psi_f^{(0)}(x)+\sum_{s=1,2}\int
\frac{d^3\v{p}}{(2\pi)^3}u_{s}(p)\sqrt{\frac{m_e}{p^0}}e^{-ip\cdot
  x}M_{ss_n}^{(n)}\big(p,p_n;z,z_n\big),\label{outcoming_electron}
\end{align}
for the outcoming wave, where the elastic amplitudes
$M_{s_as_b}^{(n)}\big(p_a,p_b;z_a,z_b\big)$ are beyond eikonal, and thus ordered,
evaluations of the scattering amplitudes \eqref{orderedmmatrix_discretized_a}
to keep track of the accumulated longitudinal momentum change, as suggested by
\eqref{accumulated_momentum_change}. The existence of the unscattered states
$\psi_i^{(0)}(x)$ and $\psi_f^{(0)}(x)$ in all the process produces in
\eqref{amplitude_emission_m} a term of the form
\begin{align}
\mathcal{M}_{em}^{(0)}&\equiv-ie \int d^4y \medspace  \bar{\psi}_f^{(0)}(y)
\gamma^\mu A_\mu^\lambda(y)\psi_i^{(0)}(y)\nonumber\\
&=-ie\mathcal{N}(k)(2\pi)^4\delta^{4}(p_n+k-p_0)\sqrt{\frac{m}{p_n^0}}\bar{u}_{s_n}(p_n)\epsilon_\mu^\lambda\gamma^\mu
u_{s_0}(p_0)\sqrt{\frac{m}{p_0^0}}=0,\label{amplitude_emission_m_vacuum}
\end{align}
which corresponds to the vacuum emission diagram and vanishes due to energy
momentum conservation. Then it has to be explicitely removed, as we would do
in a perturbative evaluation, in order to avoid the divergences inserted by
its four Dirac deltas. We define, then, the amplitude of emission of a photon
\textit{while} interacting at least once with the medium. By inserting
\eqref{incoming_electron} and \eqref{outcoming_electron} in
\eqref{amplitude_emission_m} and subtracting
\eqref{amplitude_emission_m_vacuum} we get
\begin{align}
&\mathcal{M}_{em}^{(n)}-\mathcal{M}_{em}^{(0)}=-ie\mathcal{N}(k)\sum_{s,s'}\int \frac{d^3\v{p}}{(2\pi)^3}
\left\{\sqrt{\frac{m}{p_0^0-\omega}}\bar{u}_{s}(p)\epsilon_\mu^\lambda\gamma^\mu
u_{s'}(p+k)\sqrt{\frac{m}{p_0^0}}\right\}\nonumber\\
\times&\int^{+\infty}_{-\infty} dz \medspace
S_{s_ns}^{(n)}\big(p_n,p;z_n,z\big)\exp \bigg(-i q(p,p+k)z\bigg) S_{s's_0}^{(n)}\big(p+k,p_0;z,z_0\big) -\mathcal{M}_{em}^{(0)}\label{amplitude_emission_m_momentum}.
\end{align}
The quantity appearing in the phase at $z$ is the
longitudinal momentum change at the emission vertex, which reads
\begin{align}
q(p,p+k)&\equiv
p_{p_0^0-\omega}^z(\v{p}_t)+k^z_\omega(\v{k}_t)-p_{p_0^0}^z(\v{p}_t+\v{k}_t),\nonumber\\
q(p-k,p)&\equiv p_{p_0^0-\omega}^z(\v{p}_t-\v{k}_t)+k^z_\omega(\v{k}_t)-p_{p_0^0}^z(\v{p}_t),\label{longitudinal_momentum_change_vertex}
\end{align}
where we explicitely subscripted the energy of $\v{p}$ and $\v{k}$ in the $z$
components of the momenta. This produces in the high energy limit
\begin{align}
q_z(p,p+k)\simeq 
-\frac{\omega}{2p_0^0(p_0^0-\omega)}\left(m_e^2+\left(\v{p}_t-\frac{p_0^0-\omega}{\omega}\v{k}_t\right)^2\right)=-\frac{k_\mu
p^\mu}{p_0^0},\label{pole_after}
\end{align}
if we carry the integration in $\v{p}$ at
\eqref{amplitude_emission_m_momentum} in the leg after the emission, of
modulus $\beta p^0=\beta(p_0^0-\omega)$, or
\begin{align}
q_z(p-k,p)\simeq 
-\frac{\omega}{2p_0^0(p_0^0-\omega)}\left(m_e^2+\left(\v{p}_t-\frac{p_0^0}{\omega}\v{k}_t\right)^2\right)=-\frac{k_\mu
p^\mu}{p_0^0-\omega},\label{pole_before}
\end{align}
had we chosen $\v{p}$ in the leg just before the emission, of modulus $\beta
p^0=\beta p_0^0$. Relations \eqref{pole_after} and \eqref{pole_before} are
just the pole of the off-shell fermionic propagator, after and before the
emission point, respectively. As seen from
\eqref{longitudinal_momentum_change_vertex}, in the high energy limit these
two cases are related by a shift in the transverse plane
$\v{p}_t\to\v{p}_t\pm\v{k}_t$, which equals to cross the emission vertex in
one or another direction.  We observe at equation
\eqref{amplitude_emission_m_momentum} that the electron can penetrate till any
depth $z$ into the medium with amplitude $S_{s's_0}^{(n)}(p+k,p_0;z,z_0)$
given by \eqref{orderedsmatrix_discretized_a} and energy $p_0^0$. This
amplitude leaves open the possibility that the electron can interact or not
with any center between $z_0$ to $z$. At that point it emits a photon, loosing
a momentum $k$ and thus energy $\omega$, and changing spin from $s'$ to $s$,
the amplitude of this vertex factorizing as
\begin{align}
f_{ss'}^\lambda(p,p+k)\equiv \sqrt{\frac{m_e}{p_0^0-\omega}}\bar{u}_{s}(p)\epsilon_\mu^\lambda\gamma^\mu
u_{s'}(p+k)\sqrt{\frac{m_e}{p_0^0}},\label{emission_spinorial_vertex}
\end{align}
and continues propagating 
elastically till $z_n$ with energy $p_0^0-\omega$, 
interacting or not, with amplitude $S_{s_ns}^{(n)}(p_n,p;z_n,z)$. Finally, we sum
over emission points $dz$, which automatically inserts the adequate 
poles of the fermion propagators $(k_\mu p^\mu)^{-1}$ at the phase. This is
clearly seen by integrating by parts, in which case one gets from
\eqref{amplitude_emission_m_momentum} the alternative expression
\begin{align}
\mathcal{M}_{em}^{(n)}-&\mathcal{M}_{em}^{(0)}=-e\mathcal{N}(k)\sum_{s,s'}\int
\frac{d^3\v{p}}{(2\pi)^3}\int^{+\infty}_{-\infty} dz\medspace \exp \bigg(-iq(p,p+k)z\bigg)\label{emissionamplitude_fixedphase}\\
&\times\frac{d}{dz}
\left\{S_{s_ns}^{(n)}(p_n,p;z_n,z)\frac{f_{ss'}^\lambda(p,p+k)}{q(p,p+k)}S_{s's_0}^{(n)}(p+k,p_0;z,z_1) \right\}-(n=0).
\nonumber
\end{align}
As an illustrative example we compute the simplest case, a set of $n$ centers
distributed at equal $z_1$ coordinate. A perturbative expansion of the beyond
eikonal elastic amplitudes as
\begin{align}
S_{s_as_b}^{(n)}(p_a,p_b;z_a,z_b)=\delta_{s_b}^{s_a}(2\pi)^3\delta^3(\v{p}_a-\v{p}_b)+2\pi\delta(p_a^0-p_b^0)\nonumber\\
\times\sqrt{\frac{m_e}{p_a^0}}\bar{u}_{s_a}(p_a)\gamma_0
u_{s_b}(p_b)\sqrt{\frac{m_e}{p_b^0}}\frac{-4\pi i
  Ze^2}{\v{q}^2+\mu_d^2}\sum_{i=1}^{n}e^{-i\v{q}\cdot\v{r}_i},
\end{align}
produces for \eqref{emissionamplitude_fixedphase} with
$\v{q}=\v{p}_1+\v{k}-\v{p}_0$, at leading order in $g=Ze^2$ 
\begin{align}
\mathcal{M}_{em}^{(n)}&-\mathcal{M}_{em}^{(0)}=e\mathcal{N}(k)2\pi\delta(p_1^0+\omega-p_0^0)\frac{4\pi i
  Ze^2}{\v{q}^2+\mu_d^2}\sum_{i=1}^{n}e^{-i\v{q}\cdot\v{r}_i}\\
\times&\left\{\frac{f_{s_1s'}^\lambda(p_1,p_1+k)}{q(p_1,p_1+k)}e^{-iq(p_1,p_1+k)z_1}\sqrt{\frac{m_e}{p_0^0}}\bar{u}_{s'}(p_1+k)\gamma_0
u_{s_0}(p_0)\sqrt{\frac{m_e}{p_0^0}}\right.\nonumber\\
-&\left.\sqrt{\frac{m_e}{p_0^0-\omega}}\bar{u}_{s_1}(p_1)\gamma_0
u_{s}(p_0-k)\sqrt{\frac{m_e}{p_0^0-\omega}}\frac{f_{ss_0}^\lambda(p_0-k,p_0)}{q(p_0-k,p_0)}e^{-iq(p_0-k,p_0)z_1}\right\}.\nonumber
\end{align}
And with the help of the completeness relation \eqref{completeness_relation} and 
\eqref{pole_after}, \eqref{pole_before} and \eqref{emission_spinorial_vertex}
\begin{align}
\frac{p_0^0-\omega}{k_\mu p_0^\mu}&\sum_{s}\sqrt{\frac{m_e}{p_0^0-\omega}}u_{s}(p_0-k)\bar{u}_{s}(p_0-k)\sqrt{\frac{m_e}{p_0^0-\omega}}=\frac{\slashed{p}_0-\slashed{k}+m_e}{2k_\mu
    p_0^\mu},\nonumber\\
\frac{p_0^0}{k_\mu p_1^\mu}&\sum_{s'}\sqrt{\frac{m_e}{p_0^0}}u_{s'}(p_1+k)\bar{u}_{s'}(p_1+k)\sqrt{\frac{m_e}{p_0^0}}=\frac{\slashed{p}_1+\slashed{k}+m_e}{2k_\mu
    p_1^\mu},
\end{align}
we recover the more familiar form of the Bethe-Heitler amplitude. For mediums
of non negligible thickness the above amplitude stops being applicable, since
each elastic amplitude at $z_i$ carries its own local phase. If we discretize
the medium in $n$ sheets and thus $p_i$ from $i=1,\ldots,n$ the beyond
eikonal elastic amplitudes can be read from
\eqref{orderedsmatrix_discretized_a} or
\eqref{orderedsmatrix_discretized_b}. By inserting these amplitudes in
\eqref{amplitude_emission_m_momentum} one gets three terms, so that we write
\begin{align}
\mathcal{M}_{em}^{(n)}=e\mathcal{N}(k)\left(\prod_{i=1}^{n-1}\int\frac{d^3\v{p}_i}{(2\pi)^3}\right) \left(\mathcal{M}_f+\mathcal{M}_{int}+\mathcal{M}_i\right).
\end{align}
The first of them represents photons emitted in the last leg
$(+\infty,z_n)$. Indeed using \eqref{pole_after} we get
\begin{align}
\mathcal{M}_f\equiv \frac{f_{s_ns}^\lambda(p_n,p_n+k)}{k_\mu
  p^\mu_{n}/p_0^0}&\exp\left(+i\frac{k_\mu p^\mu_n}{p_0^0}z_n\right)S_{ss_{n-1}}^{n(z_n)}(p_n+k,p_{n-1})\left(\prod_{i=1}^{n-1}S_{s_{i+1}s_i}^{n(z_i)}(q_i)\right),\label{amplitude_emission_m_final}
\end{align}
with $q_i= p_i-p_{i-1}$ and $S_{s_{i+1}s_i}^{n(z_i)}(q_i)\equiv
S_{s_{i+1}s_i}^{n(z_i)}(p_i,p_i)$. The inner term represents photons emitted
at any of the $(z_1,z_{n})$ internal legs,
\begin{align}
\mathcal{M}_{int}&\equiv-\sum_{k=1}^{n-1}\left(\prod_{i=k+1}^n
S_{s_is_{i-1}}^{n(z_i)}(q_i)\right)\frac{f_{s_ks'_k}^\lambda(p_k,p_k+k)}{k_\mu p^\mu_k/p_0^0}\label{amplitude_emission_m_interior}\\
\times &
\left(\exp\left(+i\frac{k_\mu p^\mu_k}{p_0^0}z_{k+1}\right)-\exp\left(+i\frac{k_\mu p^\mu_k}{p_0^0}z_k\right)\right)S_{s'_ks_{k-1}}^{n(z_k)}(p_k+k,p_{k-1})\left(\prod_{i=1}^{k-1}
S_{s_is_{i-1}}^{n(z_i)}(q_i)\right).\nonumber
\end{align}
Finally the term consisting in photons emitted in the first leg ($z_1,-\infty)$, given by
\begin{align}
\mathcal{M}_{i}&=-\left(\prod_{i=2}^{n}S_{s_is_{i-1}}^{n(z_i)}(q_i)\right)S_{s_1s}^{n(z_1)}(p_1,p_0-k)\frac{f_{ss_0}^\lambda(p_0-k,p_0)}{k_\mu
p^\mu_{0}/(p_0^0-\omega)}\exp\left(+i\frac{k_\mu p^\mu_0}{p_0^0-\omega}z_1\right).\label{amplitude_emission_m_initial}
\end{align}
\begin{figure}[ht]
\centering
\includegraphics[scale=0.35]{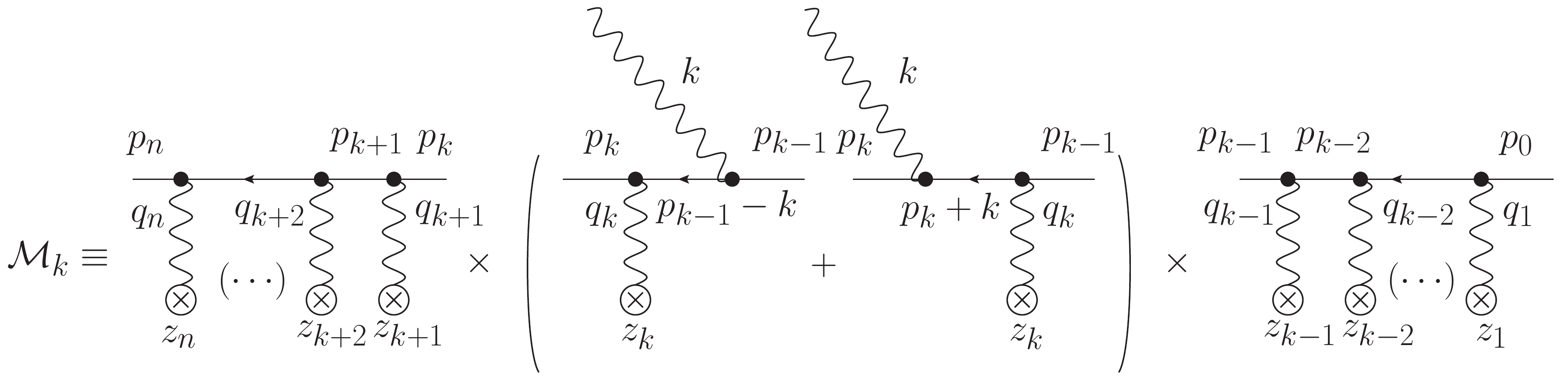}
\caption{Diagrammatic representation of the terms in the sum \eqref{amplitude_emission_m_allterms}.}
\label{fig:figure_3_1}
\end{figure}
Terms \eqref{amplitude_emission_m_final} and
\eqref{amplitude_emission_m_initial} are the only ones surviving when
$\omega\to 0$, since the phase in \eqref{amplitude_emission_m_interior}
vanish, leading to the soft photon theorem. They correspond to the classical
equation \eqref{last_first_classical} leading to the coherent plateau in the
squared amplitude. A joint expression of these three terms and its vacuum
subtraction can be given by reorganizing the sum, which in turns corresponds
to the integration by parts at \eqref{emissionamplitude_fixedphase}. This
produces
\begin{align}
\mathcal{M}_{em}^{(n)}=e\mathcal{N}(k)\left(\prod_{i=1}^{n-1}\int\frac{d^3\v{p}_i}{(2\pi)^3}\right)\times\sum_{k=1}^n
\mathcal{M}_k,
\end{align}
where
\begin{align}
&\mathcal{M}_k = \left(\prod_{i=k+1}^n
S_{s_is_{i-1}}^{n(z_i)}(q_i)\right)\times\left\{\frac{f_{s_ks}^\lambda(p_k,p_k+k)}{k_\mu
  p^\mu_k/p_0^0}\exp\left(+i\frac{k_\mu
  p^\mu_k}{p_0^0}z_k\right)S^{n(z_k)}_{ss_{k-1}}(p_k+k,p_{k-1})\right.\nonumber\\
&-\left.S^{n(z_k)}_{s_ks}(p_k,p_{k-1}-k)\frac{f_{ss_{k-1}}^{\lambda}(p_{k-1}-k,p_{k-1})}{k_\mu
  p^\mu_{k-1}/(p_0^0-\omega)}\exp\left(+i\frac{k_\mu
  p^\mu_{k-1}}{p_0^0-\omega}z_k\right)\right\}\times\left(\prod_{i=1}^{k-1}
S_{s_is_{i-1}}^{n(z_i)}(q_i)\right).\label{mk_definition}
\end{align}
The Feynman diagram structure of this joint expression is simple and can be
seen in Figure \ref{fig:figure_3_1}. We notice that with this notation the
vacuum subtracted term has been also discretized as
\begin{align}
\mathcal{M}_k^{(0)} = \left(\prod_{i=k+1}^n
S_{s_is_{i-1}}^{(0)}(q_i)\right) \times\left\{\frac{f_{s_ks}^\lambda(p_k,p_k+k)}{k_\mu
  p^\mu_k/p_0^0}\exp\left(+i\frac{k_\mu
  p^\mu_k}{p_0^0}z_k\right)S^{(0)}_{ss_{k-1}}(p_k+k,p_{k-1})\right.\nonumber\\
-\left.S^{(0)}_{s_ks}(p_k,p_{k-1}-k)\frac{f_{ss_{k-1}}^{\lambda}(p_{k-1}-k,p_{k-1})}{k_\mu
  p^\mu_{k-1}/(p_0^0-\omega)}\exp\left(+i\frac{k_\mu
  p^\mu_{k-1}}{p_0^0-\omega}z_k\right)\right\}\times\left(\prod_{i=1}^{k-1}
S_{s_is_{i-1}}^{(0)}(q_i)\right).\label{mk0_definition}
\end{align}
Then we write 
\begin{align}
\mathcal{M}_{em}^{(n)}-\mathcal{M}_{em}^{(0)}=e\mathcal{N}(k)\left(\prod_{i=1}^{n-1}\int\frac{d^3\v{p}_i}{(2\pi)^3}\right)\times\sum_{k=1}^n
\Bigg(\mathcal{M}_k-\mathcal{M}_k^{(0)}\Bigg).\label{amplitude_emission_m_allterms}
\end{align}
The classical correspondence of this quantum amplitude has been given in
\eqref{amplitude_emission_m_classical_discretized} and is going to be
recovered in the averaged square of \eqref{amplitude_emission_m_allterms}
for a medium of macroscopic transverse size.
\section{Intensity}
\label{sec:section_3_3}
Relation \eqref{amplitude_emission_m_allterms} can be squared and then
averaged over medium configurations and initial spins, and summed over final
states and photon polarizations in order to evaluate the intensity of photons
in a particular direction. For that purpose we choose, as with the elastic
scattering, a cylinder of transverse area $\pi R^2$ and length $l$. The
unpolarized intensity of photons in the energy interval $\omega$ and
$\omega+d\omega$ and in the solid angle $\Omega_k$ and $\Omega_k+d\Omega_k$,
per unit of medium transverse area and time, is given by
\begin{align}
 dI \equiv \frac{ \omega^2 d\omega d\Omega_k}{(2\pi)^3} \frac{1}{\beta_p T\pi R^2}
 \frac{1}{2}\sum_{s_ns_0}\sum_\lambda \int \frac{d^3\v{p}_n}{(2\pi)^3}
 \left\langle\left|\mathcal{M}_{em}^{(n)}-\mathcal{M}^{(0)}_{em}\right|^2\right\rangle,
\end{align}
where we divided by $\beta_p$, the incoming electron flux, time $T$,
accounting for time translation invariance, and medium transverse area
$\Omega=\pi R^2$. In the process of averaging over medium configurations we notice that
\begin{align}
\left\langle\left|\mathcal{M}_{em}^{(n)}-\mathcal{M}^{(0)}_{em}\right|^2\right\rangle = \left\langle\mathcal{M}^{(n)}_{em}
\left(\mathcal{M}^{(n)}_{em}\right)^*\right\rangle-\left\langle
\mathcal{M}^{(n)}_{em}\right\rangle\left\langle\left(
\mathcal{M}^{(n)}_{em}\right)^*\right\rangle+\left| \left\langle
\mathcal{M}_{em}^{(n)}\right\rangle-\mathcal{M}_{em}^{(0)}\right|^2.
\end{align}
Here we added and subtracted the term $\left\langle
\mathcal{M}^{(n)}_{em}\right\rangle\left\langle\left(
\mathcal{M}^{(n)}_{em}\right)^*\right\rangle$ as before. Then we define the
incoherent contribution to the emission as
\begin{align}
\Sigma_{em}^{(n)} \equiv \left\langle\mathcal{M}^{(n)}_{em}
\left(\mathcal{M}^{(n)}_{em}\right)^*\right\rangle-\left\langle
\mathcal{M}^{(n)}_{em}\right\rangle\left\langle\left(
\mathcal{M}^{(n)}_{em}\right)^*\right\rangle,\label{incoherent_average_emission_definition}
\end{align}
which as we will shown, in the limit $R\gg \mu_d^{-1}$ can be interpreted in
probabilistic terms and, except for quantum corrections in the hard part of
the spectrum, leads to the classical behavior of the infrared divergence and
the LPM effect. We also find a coherent contribution
\begin{align}
\Pi_{em}^{(n)} \equiv \left| \left\langle
\mathcal{M}_{em}^{(n)}-\mathcal{M}_{em}^{(0)}\right\rangle\right|^2,\label{coherent_average_emission_definition}
\end{align}
which is just the averaged emission amplitude squared, and encodes the
quantum diffractive behavior of the LPM effect. As we will later show, this
contribution in the $R\gg \mu_d^{-1}$ limit can be omitted since it represents
the negligible high energy transition radiation of the electron.
\subsubsection{Transverse-coherent contribution}

The transverse coherent contribution to the intensity consists in the averaged
emission amplitude squared. It is given by
\begin{align}
 dI_{coh} \equiv \frac{ \omega^2 d\omega d\Omega_k}{(2\pi)^3} \frac{1}{\beta_p T\pi R^2}
 \frac{1}{2}\sum_{s_ns_0}\sum_\lambda \int \frac{d^3\v{p}_n}{(2\pi)^3}\Pi_{em}^{(n)}.
\end{align}
For microscopic mediums it contains the quantum diffractive behavior of the
medium boundaries in the transverse plane, thus it can be interpreted as the
contribution related to medium transverse coherence. For macroscopic mediums
this diffractive behavior, as we will show, constraints the electron
propagation to the forward direction, thus this contribution reduces in that
limit to the electron transition radiation related to its energy gap and the
longitudinal boundaries. Using \eqref{amplitude_emission_m_allterms} and
\eqref{coherent_average_emission_definition} we get
\begin{align}
\Pi_{em}^{(n)}(k) = e^2\mathcal{N}^2(k)\left|\left(\prod_{i=1}^{n-1}\int\frac{d^3\v{p}_i}{(2\pi)^3}\right)\left(\sum_{k=1}^n
\langle\mathcal{M}_k\rangle-\mathcal{M}_{k}^{(0)}\right)\right|^2,
\end{align}
where the required averages on the right hand side of the above equation
affect only to the amplitudes containing centers $\mathcal{M}_k$ and are of
the form
\begin{align}
\langle\mathcal{M}_k\rangle = &\left(\prod_{i=k+1}^{n}\left\langle
S_{s_is_{i-1}}^{n(z_i)}(p_i,p_{i-1})\right\rangle\right)\left(\prod_{i=1}^{k-1}\left\langle
S_{s_is_{i-1}}^{n(z_i)}(p_i,p_{i-1})\right\rangle\right)\label{coherent_k_1}\\
\times&\left(\frac{f_{s_ks}^\lambda(p_k,p_k+k)}{k_\mu
  p^\mu_{k-1}/p_0^0}\exp\left(+i\frac{k_\mu p_k^\mu}{p_0^0}z_k\right)\left\langle
S_{ss_{k-1}}^{n(z_k)}(p_k+k,p_{k-1})\right\rangle\right.\nonumber\\
&-\left.\left\langle
S_{s_ks}^{n(z_k)}(p_k,p_{k-1}-k)\right\rangle \frac{f_{ss_{k-1}}^\lambda(p_{k-1}-k,p_{k-1})}{k_\mu
  p^\mu_{k-1}/(p_0^0-\omega)}\exp\left(+i\frac{k_\mu p_{k-1}^\mu}{p_0^0-\omega}z_k\right)\right).\nonumber
\end{align}
At this point we note that the elastic amplitudes in \eqref{coherent_k_1} for
the layers between $n$ and $k+1$ verify $p_i^0=p_0^0-\omega$, the advanced and
retarded elastic amplitudes at the layer $k$ verifies $p_k^0=p_0^0$ and
$p_k^0=p_0^0-\omega$, respectively, and the elastic amplitudes between $k$ and
$1$ verify $p_i^0=p_0^0$. We briefly review the results of the averaging
process, \textit{c.f.} Section \ref{sec:section_2_4}. A single coherent average at
the layer $i$ of $n(z_i)$ scattering sources over a cylinder of transverse
area $\pi R^2$ and length $\delta z$ produces
\begin{align}
\left\langle S_{s_is_{i-1}}^{n(z_i)}(p_i,p_{i-1})\right\rangle
=(2\pi)^3\delta(\v{q}_i)\delta_{s_{i-1}}^{s_{i}}
+
2\pi\beta_p\delta(q_i^0)
\delta_{s_{i-1}}^{s_i}\exp\left(-iq_i^zz_i\right)\nonumber\\
\times\int
d^2\v{x}_te^{-i\v{q}_i^t\cdot \v{x}_t} \left(\left\langle
\exp\left[-i\frac{g}{\beta_p}\sum_{k=1}^{n(z_i)}\chi_0^{k}(\v{x}_t)\right]\right\rangle-1\right),
\end{align}
where we used the relation
$S_{s_is_{i-1}}^{n(z_i)}(p_i,p_{i-1})=M_{s_is_{i-1}}^{n(z_i)}(p_i,p_{i-1})+(2\pi)^3\delta^3(\v{p}_i-\v{p}_{i-1})\delta_{s_{i-1}}^{s_{i}}$. In
the large but finite $R\gg \mu_d$ limit this average can be well approximated
as
\begin{align}
\int d^2\v{x}e^{-i\v{q}\cdot \v{x}} \left\langle
\exp\left[-i\frac{g}{\beta}\sum_{k=1}^{n(z_i)}\chi_0^{k}(\v{x})\right]\right\rangle\simeq\int d^2\v{x} e^{-i\v{q}\cdot\v{x}}\exp\Bigg(n_0(z_i)\delta z \pi_{el}^{(1)}(\v{x}) \Bigg).
\end{align}
The function $\pi_{el}^{(1)}(\v{x})$ is the Fourier transform of the single
elastic amplitude for a collision with a single scattering source coherently
distributed at the amplitude level over a cylinder of radius $R$, and given by
\begin{align}
\pi^{(1)}_{el}(\v{x})=\int \frac{d^2\v{q}}{(2\pi)^2}
  e^{+i\v{q}\cdot\v{x}}F_{el}^{(1)}(\v{q})W_{cyl}(\v{q},R),\medspace W_{cyl}(\v{q},R)=\frac{2\pi R}{|\v{q}|}J_1(|\v{q}|R),\label{pi_el_single_definition}
\end{align}
where $W_{cyl}(\v{q},R)$ is the window function of the cylinder and $J_1(x)$
the Bessel function of the first kind. The function $\pi^{(1)}_{el}(\v{x})$
has a typical width $R$, in contrast to dimensions of a single scattering
source $r_d=1/\mu_d$. It results convenient to separate the momentum
distribution from the spin content and the longitudinal phases. Then we write
\begin{align}
\left\langle S_{s_is_{i-1}}^{n(z_i)}(p_i,p_{i-1})\right\rangle
=\delta_{s_{i-1}}^{s_{i}}\left(\phi_{coh}^{(0)}(\delta
p_i,\delta z)+\phi_{coh}^{(n)}(\delta
p_i,\delta z)\right)\exp\left(-i\delta p_i^zz_i\right),\label{single_coherent_average}
\end{align}
where we defined the no collision $(0)$ and the collision $(n)$ 
coherently averaged amplitudes for the matter in $\delta z$ as
\begin{align}
\phi_{coh}^{(0)}(\delta
p_i,\delta z)&\equiv (2\pi)^3\delta(\delta\v{p}_i),\nonumber\\
 \phi_{coh}^{(n)}(\delta
p_i,\delta z)&\equiv 2\pi\beta\delta(\delta p_i^0)\int d^2\v{x} e^{-i\delta\v{p}_i^t\cdot\v{x}}\left(\exp\bigg(n_0(z_i)\delta z \pi_{el}^{(1)}(\v{x}) \bigg)-1\right).
\end{align}
These distributions act over the electron wave function and can only be
interpreted in probabilistic terms in the squared amplitude. They preserve
spin, lead to a momentum distribution of typical width $1/R$ and add a local
longitudinal phase of the form $\delta p_i^zz_i$ at each step, responsible of
modulating the quantum LPM effect. The first section of coherent scatterings
produces, by reiterative use of the relation \eqref{single_coherent_average},
\begin{align}
\prod_{i=1}^{k-1}&\left\langle
S_{s_is_{i-1}}^{n(z_i)}(p_i,p_{i-1})\right\rangle=\delta_{s_0}^{s_{k-1}}\left(\prod_{i=1}^{k-1}\left(\phi_{coh}^{(n)}(\delta
p_i,\delta z)+\phi_{coh}^{(0)}(\delta p_i,\delta z)\right)\right)
\nonumber\\
&\times\exp\left(-iz_{k-1}p_{p_0^0}^z(\v{p}_{k-1}^t)+i\left(\sum_{i=1}^{k-2}\delta z
p_{p_0^0}^z(\v{p}_i^t)\right)+iz_1p_{p_0^0}^z(\v{p}_0^t)\right),\label{c_k-1_1}
\end{align}
where $\delta z = z_{i+1}-z_i$ and we rearranged the phase by summing by
parts. Subindices in the longitudinal momenta denote the energy, read from the
energy conservation deltas either at $\phi_{coh}^{(n)}(\delta p_i)$ or
$\phi_{coh}^{(0)}(\delta p_i)$. They fix the longitudinal momentum in terms of
the transverse momentum. Similarly the advanced average at the emission layer
$k$ produces, using \eqref{single_coherent_average},
\begin{align}
\left\langle
S_{ss_{k-1}}^{n(z_k)}(p_k+k,p_{k-1})\right\rangle&=\delta_{s_{k-1}}^{s}\Bigg(\phi_{coh}^{(n)}(\delta
p_k+k,\delta z)+\phi_{coh}^{(0)}(\delta p_k+k,\delta z)\Bigg)\nonumber\\
&\times\exp\Bigg(-iz_kp_{p_0^0}(\v{p}_k^t+\v{k}_t)+iz_kp_{p_0^0}(\v{p}_{k-1}^t)\Bigg),\label{c_k+}
\end{align}
whereas the average just after the emission, where the electron energy is now
$p_0^0-\omega$, produces
\begin{align}
\left\langle
S_{s_ks}^{n(z_k)}(p_k,p_{k-1}-k)\right\rangle=&\delta^{s_k}_s\Bigg(\phi_{coh}^{(n)}(\delta
p_k+k,\delta z)+\phi_{coh}^{(0)}(\delta p_k+k,\delta z)\Bigg)\nonumber\\
&\times\exp\Bigg(-iz_kp_{p_0^0-\omega}(\v{p}_k^t)+iz_kp_{p_0^0-\omega}^z(\v{p}_{k-1}^t-\v{k}^t)\Bigg).\label{c_k-}
\end{align}
For the last set of coherent scatterings with energy $p_0^0-\omega$ we find,
using reiteratively \eqref{single_coherent_average},
\begin{align}
\prod_{i=k+1}^{n}&\left\langle
S_{s_is_{i-1}}^{n(z_i)}(p_i,p_{i-1})\right\rangle=\delta_{s_{k}}^{s_n}\left(\prod_{i=k+1}^{n}\left(\phi_{coh}^{(n)}(\delta
p_i,\delta z)+\phi_{coh}^{(0)}(\delta p_i,\delta z)\right)\right)\nonumber\\
&\times\exp\left(-iz_{n}p_{p_0^0-\omega}^z(\v{p}_{n}^t)+i\left(\sum_{i=k+1}^{n-1}\delta
z p_{p_0^0-\omega}^z(\v{p}_i^t)\right)+iz_{k+1}p_{p_0^0-\omega}^z(\v{p}_k^t)\right).\label{c_k+1_n}
\end{align}
Since $\omega\ll p_0^0$ we assume that $\beta_p$ can be taken unaltered in all
the process and the photon effect in the elastic amplitude neglected. The
leading contribution of the photon to the elastic propagation is enclosed
instead in the energy gap of the longitudinal phases and in the loss of
$\v{k}_t$ at the emission layer $z_k$.  The insertion of
  \eqref{c_k-1_1},\eqref{c_k+}, \eqref{c_k-} and \eqref{c_k+1_n} in
  \eqref{coherent_k_1} produces then a soft photon factorization of the form
\begin{align}
\left\langle\mathcal{M}_k\right\rangle &=\left(\prod_{i=1}^{n}\left(\phi_{coh}^{(n)}(\delta
p_i,\delta z_i)+\phi_{coh}^{(0)}(\delta p_i,\delta z_i)\right)\right)J_k,
\end{align}
where we shifted the electron momentum variable from the emission point $k$
onwards, as $\v{p}_i^t+\v{k}_t\to \v{p}_i^t$. The emission current is given
by
\begin{align}
J_k=e^{i\varphi_k}\Bigg\{f_k^{(+)}
-f_k^{(-)}\Bigg\},
\end{align}
where the phase of each element is given by
\begin{align}
i\varphi_k\equiv -ip_{p_0^0-\omega}^z(\v{p}_n^t-\v{k}_t)z_n&+i\sum_{i=k}^{n-1}\delta z_i
p_{p_0^0-\omega}^z(\v{p}_i^t-\v{k}_t) -ik_zz_k\nonumber\\
&+i\sum_{i=1}^{k-1}\delta
z_ip_{p_0^0}^z(\v{p}_i^t)+ip_{p_0^0}^z(\v{p}_0^t)z_1,
\end{align}
and we defined the following shorthands for the emission vertex together with
the corresponding propagator
\begin{align}
f_k^{(+)}&\equiv \frac{f_{s_ns_0}(p_k,p_k+k)}{k_\mu
p_k^\mu/p_0^0}, \medspace\medspace\medspace\medspace\medspace\medspace\medspace\medspace\medspace\medspace\medspace\medspace\medspace\medspace\medspace\medspace\medspace f_k^{(-)}\equiv \frac{f_{s_ns_0}(p_{k-1}-k,p_{k-1})}{k_\mu
p_{k-1}^\mu/(p_0^0-\omega)}.\label{vertex_propagator_shorthands}
\end{align}
The photon longitudinal momentum can be written in the high energy limit as
$k_z\simeq\omega-\v{k}_t^2/2\omega$. Then the phase can be rearranged as
\begin{align}
i\varphi_k=&+i\left(\omega\frac{m_e^2}{2p_0^0(p_0^0-\omega)}+i\frac{\v{k}_t^2}{2\omega}\right)z_k\nonumber\\
&+i\frac{\left(\v{p}_n^t-\v{k}_t\right)^2}{2(p_0^0-\omega)}z_n-i\sum_{i=k}^{n-1}\delta z_i
\frac{\left(\v{p}_i^t-\v{k}_t\right)^2}{2(p_0^0-\omega)}-i\sum_{i=1}^{k-1}\delta
z_i \frac{(\v{p}_i^t)^2}{2p_0^0}-i\frac{(\v{p}_0^t)^2}{2p_0^0}z_1.\label{phase_definition_coherentaverage}
\end{align}
The boundary terms $z_n$ and $z_1$ can be omitted if desired since they will
cancel in the squared amplitude. Observe that if a term of the form $J_kJ_l^*$
is evaluated for the same electron trajectory in the amplitude and its
conjugate we recover the classical phase,
\begin{align}
J_kJ_l^*\propto \exp\left(+i\sum_{i=l}^{k-1}\delta
z_i\left(m_e^2+\left(\v{p}_i^t-\frac{p_0^0}{\omega}\v{k}_t\right)^2\right)\right)\simeq\exp\left(+i\int^{z_k}_{z_l}dz\medspace
k_\mu x^\mu(z)\right),
\end{align}
where we used \eqref{pole_before} and assumed $k>l$. It is straightforward to
prove that the term containing no interaction with any of the layers can be
written in a similar manner to this one. With the replacement
$\phi_{coh}^{(n)}+\phi_{coh}^{(0)}\to \phi_{coh}^{(0)}$ we find
\begin{align}
\mathcal{M}_{k}^{(0)} &=\left(\prod_{i=1}^{n}\left(\phi_{coh}^{(0)}(\delta p_i,\delta z_i)\right)\right)J_k,
\end{align}
and we just find the overall vacuum subtraction of the contribution
$\langle\mathcal{M}_k\rangle$. Since the sum of
$\langle\mathcal{M}_k\rangle-\mathcal{M}_{k}^{(0)}$ ought to be squared
further simplifications can be done. We notice that, \textit{c.f.} Appendix
\ref{appendix1},
\begin{align}
\frac{\omega^2}{2}\sum_\lambda\sum_{s_ns_0}&\left(f_k^{(+)}-f_k^{(-)}\right)\left(f_l^{(+)}-f_l^{(-)}\right)^*=h_n(y)\v{\delta}_k\cdot\v{\delta}_l+h_s(y)\delta_k^s\delta_l^s,
\end{align}
where the spin flip and non flip currents are given, respectively, by
\begin{align}
\v{\delta}_k^n\equiv \frac{\v{k}\times \v{p}_k}{k_\mu
  p^\mu_k}-\frac{\v{k}\times \v{p}_{k-1}}{k_\mu
  p^\mu_{k-1}}, \medspace\medspace\medspace\medspace\medspace\medspace \delta_k^s \equiv \frac{\omega
  p_k^0}{k_\mu p^\mu_k}-\frac{\omega p_{k-1}^0}{k_\mu p^\mu_{k-1}}.
\end{align}
Here $y=\omega/p_0^0$ is the fraction of energy carried by the photon,
$p_k^0=p_0^0-\omega$ and $p_{k-1}^0=p_0^0$ and the functions $h_n(y)$ and
$h_s(y)$ are given by
\begin{align}
h^n(y)=\frac{1}{2}(1+(1-y)^2), \medspace\medspace\medspace\medspace\medspace\medspace h^s(y)=\frac{1}{2}y^2.
\end{align}
This indicates that we can split the emission current into two independent
contributions, $\v{J}_k^n$ and $J_k^s$ which can be separately squared and then
multiplied with $h^n(y)$ and $h^s(y)$. These currents are given by
\begin{align}
\v{J}_k^n\equiv=\frac{1}{\omega^2}\v{\delta}_k^ne^{i\varphi_k},\medspace\medspace\medspace\medspace\medspace J_k^s\equiv\frac{1}{\omega^2}\delta_k^s e^{i\varphi_k},
\end{align}
where the phase $\varphi_k$ is given by
\eqref{phase_definition_coherentaverage}. Then we have found a transverse
coherent average of the emission intensity given by
\begin{align}
\left(\Pi_{em}^{(n)}\right)_{sn}=\frac{e^2\mathcal{N}^2}{\omega^2}\left|\left(\prod_{i=1}^{n-1}\frac{d^3\v{p}_i}{(2\pi)^3}\right)\left(\prod_{i=1}^{n}\left(\phi_{coh}^{(n)}(\delta
p_i,\delta z_i)+\phi_{coh}^{(0)}(\delta p_i,\delta
z_i)\right)\right.\right.\nonumber\\
-\left.\left.\prod_{i=1}^{n}\left(\phi_{coh}^{(0)}(\delta p_i,\delta
z_i)\right)\right)\left(\sum_{k=1}^{n}\v{\delta}_k^ne^{i\varphi_k}\right)\right|^2,\label{coherent_average_spinnoflip}
\end{align}
for the spin no-flip contribution and 
\begin{align}
\left(\Pi_{em}^{(n)}\right)_{sf}=\frac{e^2\mathcal{N}^2}{\omega^2}\left|\left(\prod_{i=1}^{n-1}\frac{d^3\v{p}_i}{(2\pi)^3}\right)\left(\prod_{i=1}^{n}\left(\phi_{coh}^{(n)}(\delta
p_i,\delta z_i)+\phi_{coh}^{(0)}(\delta p_i,\delta
z_i)\right)\right.\right.\nonumber\\
-\left.\left.\prod_{i=1}^{n}\left(\phi_{coh}^{(0)}(\delta p_i,\delta
z_i)\right)\right)\left(\sum_{k=1}^{n}\delta_k^se^{i\varphi_k}\right)\right|^2,\label{coherent_average_spinflip}
\end{align}
for the spin flip contribution. The coherent average contribution to the
photon intensity is then given by the sum of these two terms once integrated
in the final electron momentum
\begin{align}
\omega\frac{dI_{coh}}{d\omega d\Omega_k} &= \left(\frac{\omega}{2\pi}\right)^3
 \frac{1}{\pi R^2T}\int\frac{d^3\v{p}_n}{(2\pi)^3}\left(h^n(y)\left(\Pi_{em}^{(n)}\right)_{sn}+h^s(y)\left(\Pi_{em}^{(n)}\right)_{sf}\right).\label{coherent_average_emission}
\end{align}
A particular case of the above result consists in taking the macroscopic $R\to
\infty$ limit. Then $W_{cyl}(\v{q},R)=(2\pi)^2\delta^2(\v{q})$,
$\pi_{el}^{(1)}(\v{x})=F_{el}^{(1)}(\v{0})$ and thus 
\begin{align}
\phi_{coh}^{(n)}(\delta p_i,\delta z_i)=
(2\pi)^3\delta^3(\delta\v{p}_i)\left(\exp\left(n_0(z_i)\delta z_i F_{el}^{(1)}(\v{0})\right)-1\right),
\end{align}
which leads to a pure forward propagation of the electron at the level of the
amplitude. Then the only difference of states in the functions
$\v{\delta}_i^n$ and $\delta_i^s$ corresponds to the energy gap 
of the electron momentum and thus
\begin{align}
\sum_{k=1}^{n}\delta_k^se^{i\varphi_k}=\left\{\frac{\beta_f\sin\theta}{1-\beta_f\cos\theta}-\frac{\beta_i\sin\theta}{1-\beta_i\cos\theta}\right\}\sum_{k=1}^ne^{i\varphi_k},
\end{align}
where $\theta$ is the angle between the photon and the initial electron
direction $\hat{\v{p}}_0$ and $\beta_f$ and $\beta_i$ are the final and
initial electron velocities. Similarly the spin flip contribution produces
\begin{align}
\sum_{k=1}^{n}\delta_k^s=e^{i\varphi_k} \left\{\frac{\beta_f}{1-\beta_f\cos\theta}-\frac{\beta_i}{1-\beta_i\cos\theta}\right\}\sum_{k=1}^ne^{i\varphi_k},
\end{align}
where the phases are also constrained to the forward direction and given in
this case by
\begin{align}
i\varphi_k=i\omega\frac{m_e^2}{2p_0^0(p_0^0-\omega)}z_k+i\frac{\v{k}_t^2}{2\omega}z_k\equiv i\frac{k_\mu p_0^\mu}{p_0^0}z_k,
\end{align}
as expected. At high energies the negligible energy gap in the velocities
$\beta_i$ and $\beta_f$ of the current can be neglected, as we already did in
the elastic weights. Then $\v{\delta}^n_k=0$ and $\delta_k^s=0$, which means
that at high energies in the macroscopic $R\to\infty$ limit, the coherent
average contribution cancels, $\left(\Pi_{em}^{(n)}\right)_{sn}=0$ and
$\left(\Pi_{em}^{(n)}\right)_{sf}=0$, as expected.
\subsubsection{Transverse-incoherent contribution}
The transverse-incoherent contribution to the emission intensity corresponds
to the averaged squared amplitude.  Using
\eqref{amplitude_emission_m_allterms} and
\eqref{incoherent_average_emission_definition} we get
\begin{align}
\Sigma_{em}^{(n)} =e^2\mathcal{N}^2(k)\left(\prod_{i=1}^{n-1}\int\frac{d^3\v{p}_i}{(2\pi)^3}\frac{d^3\v{u}_i}{(2\pi)^3}\right)\times\sum_{j,k=1}^n
\left(\langle\mathcal{M}_k\mathcal{M}_j^*\rangle-\langle\mathcal{M}_k\rangle\langle \mathcal{M}_j^*\rangle\right).
\end{align}
The averages on the right hand side of the above equation are of the form
\begin{align}
\langle\mathcal{M}_k\mathcal{M}_j^*\rangle = A_{j+1}^n
\left((f_j^{+})^*A_j^{+}-A_j^{-}(f_j^{-})^*\right)A^{j-1}_{k+1}\left(f_k^+A_k^+-A_k^-f_k^-\right)A_1^{k-1},\label{incoherent_kj_1}\\
\langle\mathcal{M}_k\rangle\langle\mathcal{M}_j^*\rangle = B_{j+1}^n
\left((f_j^{+})^*B_j^{+}-B_j^{-}(f_j^{-})^*\right)B^{j-1}_{k+1}\left(f_k^+B_k^+-B_k^-f_k^-\right)B_1^{k-1},\label{incoherent_kj_2}
\end{align}
where $f_j^{+}$ and $f_j^{-}$ are the emission vertex and propagator
shorthands given at \eqref{vertex_propagator_shorthands} but with the local
phases, thus given by
\begin{align}
f_k^{(+)}&\equiv \frac{f_{s_ks}^\lambda(p_k,p_k+k)}{k_\mu
  p^\mu_k/p_0^0}\exp\left(+i\frac{k_\mu p^\mu_k}{p_0^0}\right),\nonumber\\
f_k^{(-)}&\equiv \frac{f_{ss_{k-1}}^\lambda(p_{k-1}-k,p_{k-1})}{k_\mu p^\mu_{k-1}/(p_0^0-\omega)}\exp\left(+i\frac{k_\mu p^\mu_{k-1}}{p_0^0-\omega}\right),
\end{align}
and the terms $A^j_k$
and $B^j_k$ are shorthands for the incoherent averages of the squared
amplitudes corresponding to the layers of scatterers from $z_k$ to $z_j$. Let
$\delta p_i = p_i-p_{i-1}$ and $\delta u_i = u_i-u_{i-1}$. If the medium
verifies that the transverse dimensions greatly exceed the dimensions of a
scatterer, that is, $R\gg \mu_d^{-1}$, we get from
\eqref{amplitude_emission_m_allterms}, \textit{c.f.} Section \ref{sec:section_2_4},
elements of the form
\begin{align}
\left\langle
S_{s_is_{i-1}}^{n(z_i)}(\delta p_i)S_{r_ir_{i-1}}^{n(z_i),*}(\delta u_i)\right\rangle
=\delta_{s_i}^{s_{i-1}}\delta_{r_i}^{r_{i-1}}(2\pi)^2\delta^2(\delta\v{p}_i^t-\delta\v{u}_i^t)\nonumber\\
\times 2\pi\beta_u\delta(\delta u_i^0)
\exp\bigg(-i\left(\delta p_i^z-\delta u_i^z\right) z_i\bigg)
 \phi_{inc}(\delta p_i).\label{single_average_element_a}
\end{align}
In \eqref{single_average_element_a} if we assume that the energy gap due to
the photon is negligible in the elastic part $\beta_u\simeq \beta_p \simeq
\beta$, then $\phi(q)$ acquires a probabilistic interpretation in all the
range and can be split into two contributions
\begin{align}
\phi_{inc}(q) &= e^{-n_0(z_i)\delta z
  \sigma_t^{(1)}}
(2\pi)^3\delta^3(\v{q})+2\pi\delta (
q_0)\beta_p \hat{\Sigma}_2(\v{q},\delta z)\equiv\phi_{inc}^{(0)}(q)+\phi_{inc}^{(n)}(q),\label{phi_inc_definition}
\end{align}
where $n_0(z_i)$ is the density of scattering centers at the layer at $z_i$ of
thickness $\delta z$, the single elastic cross section $\sigma_t^{(1)}$ is
given by \eqref{cross_section_t_1} at arbitrary coupling or by
\eqref{cross_section_t_1_leading} at small coupling and the collisional
distribution after an incoherent scattering with the layer
$\hat{\Sigma}_2(\v{q},\delta z)$ is given by
\eqref{incoherent_contribution_infinitesize} without the $\Omega=\pi R^2$
factor. We have found, then, the probability of no colliding with the matter
in $\delta z$, given by $\exp\left(-n_0(z_i)\delta z \sigma_t^{(1)}\right)$,
times the corresponding forward distribution $\delta^3(\v{q}_i)$, or the
incoherent contribution $\hat{\Sigma}_2(\v{q},\delta z)$ in case of scattering
with the matter in $\delta z$. In this transport spin is preserved and the
transverse momentum change in the conjugated amplitude $\delta\v{u}_i^t$
equals the one in the amplitude $\delta\v{p}_i^t$, as a result of the
macroscopic limit $R\gg \mu_d^{-1}$. By using the energy conservation deltas
the longitudinal phases are of the form
\begin{align}
\delta p_i^z-\delta u_i^z \equiv \left(p^z(\v{p}_i^t)-p^z(\v{p}_{i-1}^t)\right)-\left(u^z(\v{u}_i^t)-u^z(\v{u}_{i-1}^t)\right),
\end{align}
Similarly, the coherent averages of the kind $B$ are found to be
\begin{align}
\left\langle
S_{s_is_{i-1}}^{n(z_i)}(p_i,p_{i-1})\right\rangle\left\langle S_{r_ir_{i-1}}^{n(z_i),*}(u_i,u_{i-1})\right\rangle
=\delta_{s_i}^{s_{i-1}}\delta_{r_i}^{r_{i-1}}(2\pi)^2\delta^2(\delta\v{p}_i^t-\delta\v{u}_i^t)\nonumber\\
\times 2\pi\beta\delta(\delta u_i^0)
\exp\bigg(-i\left(\delta p_i^z-\delta u_i^z\right) z_i\bigg)
 \phi_{inc}^{(0)}(\delta p_i).\label{single_average_element_b}
\end{align}
With these tools we are in position of evaluating all the required terms. The
first path of elastic scatterings at \eqref{incoherent_kj_1} corresponds to
the passage from the beginning of the medium $z_1$ to previous layer where the
photon is emitted in one of the amplitudes, i.e. $z_{k-1}$
\begin{align}
A^{k-1}_1\equiv \left(\prod_{i=1}^{k-1}
\left\langle S_{s_is_{i-1}}^{n(z_i)}(p_i,p_{i-1})S_{r_ir_{i-1}}^{n(z_i),*}(u_i,u_{i-1})\right\rangle\right).
\end{align}
Using \eqref{single_average_element_a} we find $p_i^0=u_i^0\equiv p_0^0$ for
$i=1,\ldots,k-1$ and, since $\v{u}_0^t=\v{p}_0^t$ then from
\eqref{single_average_element_a} we get $\v{u}_i^t=\v{p}_i^t$ for
$i=1,\ldots,k-1$. The longitudinal phase with these constraints then vanishes
\begin{align}
\delta p_i^z-\delta u_i^z= \left(p_{p_0^0}^z(\v{p}_i^t)-p_{p_0^0}^z(\v{p}_{i-1}^t)\right)-\left(p_{p_0^0}^z(\v{u}_i^t)-p_{p_0^0}^z(\v{u}_{i-1}^t)\right)=0.
\end{align}
At \eqref{single_average_element_a} we use
$(2\pi)^2\delta^2(\v{u}_i^t-\v{p}_i^t)2\pi\beta_u\delta(u_i^0-p_0^0)=(2\pi)^3\delta^3(\v{u}_i-\v{p}_i)$
so for the set of scatterings from $z_1$ to $z_{k-1}$ we find a simple
convolution of incoherent elastic scatterings without longitudinal phases,
\begin{align}
A^{k-1}_{1}= \delta^{s_{k-1}}_{s_0}\delta^{r_{k-1}}_{s_0}
\left(\prod_{i=1}^{k-1}\phi_{inc}(\delta p_i)\times (2\pi)^3\delta^3(\v{u}_i-\v{p}_i)\right).\label{a_k-1_1}
\end{align}
The averages at $z_k$ are given by two different terms. The term corresponding
to a retarded emission after the collision is given by
\begin{align}
A^+_k=\left\langle
S_{ss_{k-1}}^{n(z_k)}(p_k+k,p_{k-1})\right.&\left.S_{r_kr_{k-1}}^{n(z_k),*}(u_k,u_{k-1})\right\rangle,
\end{align}
where both amplitudes are on-shell with $p_{k-1}^0=u_{k-1}^0=p_0^0$. Since we
inherit $\v{u}_{k-1}=\v{p}_{k-1}$ from \eqref{a_k-1_1} we find, using
\eqref{single_average_element_a}, that $\v{p}_k^t+\v{k}^t=\v{u}_k^t$. The
longitudinal phase in \eqref{single_average_element_a} then vanishes again
since
\begin{align}
\delta p_k^z-\delta u_k^z= \left(p_{p_0^0}^z(\v{p}_k^t+\v{k}^t)-p_{p_0^0}^z(\v{p}_{k-1}^t)\right)-\left(p_{p_0^0}^z(\v{u}_k^t)-p_{p_0^0}^z(\v{u}_{k-1}^t)\right)=0.
\end{align}
Then using $(2\pi)^2\delta^2(\v{p}_k^t+\v{k}^t-\v{u}_k^t)
(2\pi)\beta\delta(p_k^0+\omega-u_k^0)=(2\pi)^3\delta^3(\v{p}_k+\v{k}-\v{u}_k)$
we find
\begin{align}
A^+_k=\delta_{s_{k-1}}^s\delta_{r_{k-1}}^{r_k}\phi_{inc}(\delta p_k+k)\times
(2\pi)^3\delta^3(\v{p}_k+\v{k}-\v{u}_k).\label{a_k_+}
\end{align}
The term corresponding to an advanced emission to the collision at $z_k$ is
instead given by
\begin{align}
A^-_k=\left\langle
S_{s_ks}^{n(z_k)}(p_k,p_{k-1}-k)\right.&\left.S_{r_kr_{k-1}}^{n(z_k),*}(u_k,u_{k-1})\right\rangle,
\end{align}
where one of the amplitudes is on-shell with $p_{k}^0=p_0^0-\omega$ whereas
the other with $u_{k}^0=p_0^0$. As before from \eqref{a_k-1_1} we use
$\v{u}_{k-1}=\v{p}_{k-1}$ and thus from \eqref{single_average_element_a} we
get $\v{p}_k^t+\v{k}^t=\v{u}_k^t$. In this case the longitudinal phase can be
rearranged as
\begin{align}
\delta p_k^z-\delta u_k^z=
\left(p_{p_0^0-\omega}^z(\v{p}_k^t)-p_{p_0^0-\omega}^z(\v{p}_{k-1}^t-\v{k}_t)\right)-\left(p_{p_0^0}^z(\v{u}_k^t)-p_{p_0^0}^z(\v{u}_{k-1}^t)\right)\nonumber\\
=\left(p_{p_0^0-\omega}^z(\v{p}_k^t)-p_{p_0^0-\omega}^z(\v{p}_{k-1}^t-\v{k}_t)\right)-\left(p_{p_0^0}^z(\v{p}_k^t+\v{k}^t)-p_{p_0^0}^z(\v{p}_{k-1}^t)\right).
\end{align}
By adding and subtracting $\pm p_\omega^z(\v{k}_t)$ and using
\eqref{pole_after} and \eqref{pole_before} this can be written as
\begin{align}
\delta p_k^z-\delta
u_k^z&=\left(p_{p_0^0-\omega}^z(\v{p}_k^t)+p_\omega^z(\v{k}_t)-p_{p_0^0}^z(\v{p}_k^t+\v{k}^t)\right)\\
&-\left(p_{p_0^0-\omega}^z(\v{p}_{k-1}^t-\v{k}_t)+p_\omega^z(\v{k}_t)-p_{p_0^0}^z(\v{p}_{k-1}^t)\right)=-\frac{k_\mu
p_k^\mu}{p_0^0}+\frac{k_\mu p_{k-1}^{\mu}}{p_0^0-\omega},\nonumber
\end{align}
so that the advanced emission term at $z_k$ reduces to an incoherent elastic
scattering carrying an extra longitudinal phase, 
\begin{align}
A^-_k=\delta_{s}^{s_k}\delta_{r_{k-1}}^{r_k}\phi_{inc}(\delta p_k+k)\times
(2\pi)^3\delta^3(\v{p}_k+\v{k}-\v{u}_k)\exp\left(+i\frac{k_\mu
p_k^\mu}{p_0^0}z_k-i\frac{k_\mu p_{k-1}^{\mu}}{p_0^0-\omega}z_k \right).\label{a_k_-}
\end{align}
The intermediate step of the scatterings corresponds to the passage of the
electron from the layer immediately after the emission $z_{k+1}$ in one of the
amplitudes, to the layer before the emission $z_{j-1}$ the other amplitude. It
is given by
\begin{align}
A^{j-1}_{k+1}=\left(\prod_{i=k+1}^{j-1}
\left\langle S_{s_is_{i-1}}^{n(z_i)}(p_i,p_{i-1})S_{r_ir_{i-1}}^{n(z_i),*}(u_i,u_{i-1})\right\rangle\right).
\end{align}
In this section of scatterings one of the amplitudes is on-shell with
$p_i^0=p_0^0-\omega$ but the other with $u_i^0=p_0^0$. We inherit the boundary
condition $\v{p}_k+\v{k}=\v{u}_k$ either from \eqref{a_k_+} or \eqref{a_k_-}
and then we find from \eqref{single_average_element_a}
$\v{p}_i^t+\v{k}^t=\v{u}_i^t$ for $i=k+1,\ldots,j-1$. The longitudinal phase
produces
\begin{align}
\delta p_i^z-\delta u_i^z=
\left(p_{p_0^0-\omega}^z(\v{p}_i^t)-p_{p_0^0-\omega}^z(\v{p}_{i-1}^t)\right)&-\left(p_{p_0^0}^z(\v{u}_i^t)-p_{p_0^0}^z(\v{u}_{i-1}^t)\right)\\
=\left(p_{p_0^0-\omega}^z(\v{p}_i^t)-p_{p_0^0-\omega}^z(\v{p}_{i-1}^t)\right)&-\left(p_{p_0^0}^z(\v{p}_i^t+\v{k}^t)-p_{p_0^0}^z(\v{p}_{i-1}^t+\v{k}^t)\right).\nonumber
\end{align}
As before we can add and subtract $p_\omega^z(\v{k}^t)$ and 
reorganize the terms as in
\eqref{longitudinal_momentum_change_vertex} in order to obtain, using \eqref{pole_after},
\begin{align}
\delta p_i^z-\delta
u_i^z&=\left(p_{p_0^0-\omega}^z(\v{p}_i^t)+p_\omega^z(\v{k}^t)-p_{p_0^0}^z(\v{p}_i^t+\v{k}^t)\right)\\
&-\left(p_{p_0^0-\omega}^z(\v{p}_{i-1}^t)+p_\omega^z(\v{k}^t)-p_{p_0^0}^z(\v{p}_{i-1}^t+\v{k}^t)\right)=
-\frac{k_\mu p_i^\mu}{p_0^0}+\frac{k_\mu p_{i-1}^\mu}{p_0^0}\nonumber.
\end{align}
Correspondingly the phase can be rearranged in the well known classical form
\begin{align}
-i\sum_{i=k+1}^{j-1}(\delta p_i^z-\delta
u_i^z)z_i&=-i\sum_{i=k+1}^{j-1}\frac{k_\mu
  p_{i-1}^\mu}{p_0^0}z_i+i\sum_{i=k+1}^{j-1}\frac{k_\mu
  p_i^\mu}{p_0^0}z_i\\
&=+i\frac{k_\mu p_{j-1}^\mu}{p_0^0}z_{j-1}-i\sum_{i=k+1}^{j-2} \frac{k_\mu
  p^\mu_i}{p_0^0}(z_{i+1}-z_i)-i\frac{k_\mu p^\mu_{k}}{p_0^0}z_{k+1}.\nonumber
\end{align}
This passage from $z_{k+1}$ to $z_{j-1}$ produces then the largest
contribution to the longitudinal phase accumulation,
\begin{align}
A^{j-1}_{k+1}= \delta^{s_{j-1}}_{s_{k}}\delta_{r_k}^{r_{j-1}}&\left(\prod_{i=k+1}^{j-1}(2\pi)^3\delta^3(\v{p}_i+\v{k}-\v{u}_i)\times\phi_{inc}(\delta p_i)\right)\label{a_j-1_k+1}\\
&\times\exp\left(+i\frac{k_\mu p_{j-1}^\mu}{p_0^0}z_{j-1}-i\sum_{i=k+1}^{j-2} \frac{k_\mu
  p^\mu_i}{p_0^0}(z_{i+1}-z_i)-i\frac{k_\mu p^\mu_{k}}{p_0^0}z_{k+1}\right).\nonumber
\end{align}
The term corresponding to a retarded emission after the collision at $z_j$ is
given by the incoherent average
\begin{align}
A_j^+=\left\langle
S_{s_js_{j-1}}^{n(z_j)}(p_j,p_{j-1})\right.&\left.S_{rr_{j-1}}^{n(z_j),*}(u_j+k,u_{j-1})\right\rangle.
\end{align}
Here one amplitude is on-shell with $p_{j-1}^0=p_0^0-\omega$ and the other is
instead with the initial energy $u_{j-1}^0=p_0^0$. Since from
\eqref{a_j-1_k+1} we can set $\v{p}_{j-1}+\v{k}=\v{u}_{j-1}$ then using
\eqref{single_average_element_a} we find $\v{p}_j^t=\v{u}_j^t$ and thus
\begin{align}
\delta p_j^z-\delta u_j^z=
\left(p_{p_0^0-\omega}^z(\v{p}_j^t)-p_{p_0^0-\omega}^z(\v{p}_{j-1}^t)\right)&-\left(p_{p_0^0}^z(\v{u}_j^t+\v{k}^t)-p_{p_0^0}^z(\v{u}_{j-1}^t)\right)\\
=\left(p_{p_0^0-\omega}^z(\v{p}_j^t)-p_{p_0^0-\omega}^z(\v{p}_{j-1}^t)\right)&-\left(p_{p_0^0}^z(\v{p}_j^t+\v{k}^t)-p_{p_0^0}^z(\v{p}_{j-1}^t+\v{k}^t)\right).\nonumber
\end{align}
By adding and subtracting $p_\omega^z(\v{k}^t)$ and using \eqref{pole_after} we
find again a longitudinal phase
\begin{align}
(\delta p_j^z-\delta
u_j^z)&=\left(p_{p_0^0-\omega}^z(\v{p}_j^t)+p_\omega^z(\v{k}^t)-p_{p_0^0}^z(\v{p}_j^t+\v{k}^t)
\right)\\
&-\left(p_{p_0^0-\omega}^z(\v{p}_{j-1}^t)+p_\omega^z(\v{k}^t)-p_{p_0^0}(\v{p}_{j-1}^t+\v{k}^t)\right)=-\frac{k_\mu
  p^\mu_j}{p_0^0}+\frac{k_\mu p_{j-1}^\mu}{p_0^0},\nonumber
\end{align}
so this term produces
\begin{align}
A_j^+=\delta_{s_{j-1}}^{s_j}\delta_{r_{j-1}}^{r}(2\pi)^3\delta^3(\v{p}_j-\v{u}_j)\times\phi(\delta p_j)\exp\left(+i\frac{k_\mu
  p^\mu_j}{p_0^0}z_j-i\frac{k_\mu p_{j-1}^\mu}{p_0^0}z_j\right).\label{a_j_+}
\end{align}
Similarly the term corresponding to an advanced emission previous to the
collision at $z_j$ is given by
\begin{align}
A_j^-=\left\langle
S_{s_js_{j-1}}^{n(z_j)}(p_j,p_{j-1})S_{s'_js'}^{n(z_j),*}(u_j,u_{j-1}-k)\right\rangle
\end{align}
where now both amplitudes are already on-shell with the final energy
$p_j^0=u_j^0=p_0^0-\omega$. As before from \eqref{a_j-1_k+1} we set
$\v{p}_{j-1}+\v{k}=\v{u}_{j-1}$ and then $\v{p}_j^t=\v{u}_j^t$ so the phase
vanishes
\begin{align}
\delta p_j^z-\delta u_j^z=
\left(p_{p_0^0-\omega}^z(\v{p}_j^t)-p_{p_0^0-\omega}^z(\v{p}_{j-1}^t)\right)&-\left(p_{p_0^0-\omega}^z(\v{u}_j^t)-p_{p_0^0-\omega}^z(\v{u}_{j-1}^t-\v{k})\right)=0,
\end{align}
and we find for this term
\begin{align}
A_j^-=\delta_{s_{j-1}}^{s_j}\delta_{r}^{r_{j}}(2\pi)^3\delta^3(\v{p}_j-\v{u}_j)\times\phi(\delta p_j)\label{a_j_-}
\end{align}
Finally, the last step of scatterings from $z_{j+1}$ to $z_n$ is given by the
incoherent averages
\begin{align}
A^n_{j+1}=\left(\prod_{i=j+1}^n
\left\langle S_{s_is_{i-1}}^{n(z_i)}(p_i,p_{i-1})S_{r_ir_{i-1}}^{n(z_i),*}(u_i,u_{i-1})\right\rangle\right).
\end{align}
In this case using either \eqref{a_j_+} or \eqref{a_j_-} the relation
$\v{p}_i+\v{k}=\v{u}_i+\v{k}$ holds for $i=j,\ldots,n$ and the longitudinal
phase vanishes in all the range since
\begin{align}
\delta p_i^z-\delta u_i^z= \left(p_{p_0^0-\omega}^z(\v{p}_i^t)-p_{p_0^0-\omega}^z(\v{p}_{i-1}^t)\right)-\left(p_{p_0^0-\omega}^z(\v{u}_i^t)-p_{p_0^0-\omega}^z(\v{u}_{i-1}^t)\right)=0,
\end{align}
so we get a simple convolution of incoherent elastic averages without
longitudinal phases, as it was the case of the electron path from $z_1$ to
$z_{k-1}$,
\begin{align}
A^n_{j+1}=\delta_{s_j}^{s_n}\delta_{r_j}^{s_n}
\times
\left(\prod_{i=j+1}^n(2\pi)^3\delta^3(\v{u}_i-\v{p}_i)\times\phi(\delta
p_i)\right).\label{a_n_j+1}
\end{align}
Upon inserting \eqref{a_k-1_1} \eqref{a_k_+} \eqref{a_k_-} \eqref{a_j-1_k+1}
\eqref{a_j_+} \eqref{a_j_-} and \eqref{a_n_j+1} in \eqref{incoherent_kj_1} we
find 
\begin{align}
&\left\langle\mathcal{M}_k^a\mathcal{M}_j^{a,*}\right\rangle = \left(\frac{f_{s_ns_0}^{\lambda,*}(p_j,p_j+k)}{k_\mu
  p^\mu_j/p_0^0}-\frac{f_{s_ns_0}^{\lambda,*}(p_{j-1}-k,p_{j-1})}{k_\mu
  p^\mu_{j-1}/(p_0^0-\omega)}\right)\label{incoherent_kj_1_result}\\
&\times\exp\left(-i\sum_{i=k}^{j-1} \frac{k_\mu
  p^\mu_i}{p_0^0}(z_{i+1}-z_i)\right)
\left(\frac{f_{s_ns_0}^\lambda(p_k,p_k+k)}{k_\mu
  p^\mu_k/p_0^0}-\frac{f_{s_ns_0}^{\lambda}(p_{k-1}-k,p_{k-1})}{k_\mu
  p^\mu_{k-1}/(p_0^0-\omega)}\right)\nonumber\\
&\times\left(\prod_{i=j}^n\phi_{inc}(\delta
p_i)\times(2\pi)^3\delta^3(\v{p}_i-\v{u}_i)\right)\left(\prod_{i=k+1}^{j-1}\phi_{inc}(\delta
p_i)\times(2\pi)^3\delta^3(\v{p}_i+\v{k}-\v{u}_i)\right)\nonumber\\
&\times\left(\phi_{inc}(\delta p_k+k)\times(2\pi)^3\delta^3(\v{p}_k+\v{k}-\v{u}_k)\right)
\left(\prod_{i=1}^{k-1}\phi_{inc}(\delta p_i)\times(2\pi)^3\delta^3(\v{p}_i-\v{u}_i)\right)\nonumber
\end{align}
The evaluation of term
$\langle\mathcal{M}_k^a\rangle\langle\mathcal{M}_j^{a,*}\rangle$ at
\eqref{incoherent_kj_2} follows the same steps as
\eqref{incoherent_kj_1}. Since the kinematical conditions in
\eqref{single_average_element_a} equally hold for for the averages
\eqref{single_average_element_b}, we only have to replace $\phi_{inc}(\delta
p_i)$ with the no collision distribution $\phi_{inc}^{(0)}(\delta p_i)$ in the
above expression. We get
\begin{align}
&\left\langle\mathcal{M}_k^a\right\rangle\left\langle\mathcal{M}_j^{a,*}\right\rangle = \left(\frac{f_{s_ns_0}^{\lambda,*}(p_j,p_j+k)}{k_\mu
  p^\mu_j/p_0^0}-\frac{f_{s_ns_0}^{\lambda,*}(p_{j-1}-k,p_{j-1})}{k_\mu
  p^\mu_{j-1}/(p_0^0-\omega)}\right)\label{incoherent_kj_2_result}\\
&\times\exp\left(-i\sum_{i=k}^{j-1} \frac{k_\mu
  p^\mu_i}{p_0^0}(z_{i+1}-z_i)\right)
\left(\frac{f_{s_ns_0}^\lambda(p_k,p_k+k)}{k_\mu
  p^\mu_k/p_0^0}-\frac{f_{s_ns_0}^{\lambda}(p_{k-1}-k,p_{k-1})}{k_\mu
  p^\mu_{k-1}/(p_0^0-\omega)}\right)\nonumber\\
&\times\left(\prod_{i=j}^n\phi_{inc}^{(0)}(\delta
p_i)\times(2\pi)^3\delta^3(\v{p}_i-\v{u}_i)\right)\left(\prod_{i=k+1}^{j-1}\phi_{inc}^{(0)}(\delta
p_i)\times(2\pi)^3\delta^3(\v{p}_i+\v{k}-\v{u}_i)\right)\nonumber\\
&\times\left(\phi_{inc}^{(0)}(\delta p_k+k)\times(2\pi)^3\delta^3(\v{p}_k+\v{k}-\v{u}_k)\right)
\left(\prod_{i=1}^{k-1}\phi_{inc}^{(0)}(\delta p_i)\times(2\pi)^3\delta^3(\v{p}_i-\v{u}_i)\right).\nonumber
\end{align}
With this result we can integrate in the auxiliary momentum $\v{u}_i$. Since
$\v{u}_n\equiv \v{p}_n$ there is an extra 3-Dirac delta accounting for time
translation invariance and transverse homogeneity. In addition if we assume
that $\omega\ll p_0^0$ then we can neglect the momentum carried by the photon
in the elastic distributions and a current is found which factorizes as
\begin{align}
\Sigma_{em}^{(n)}&=e^2\mathcal{N}^2(k) (2\pi)^2\delta^{(2}(\v{0}) 2\pi\beta_p \delta(0)\\
&\times\left(\prod_{i=1}^{n-1}\int\frac{d^3\v{p}_i}{(2\pi)^3}\right)\left\{\left(\prod_{i=1}^n\phi_{inc}^{(n)}(\delta p_i)+\phi_{inc}^{(0)}(\delta p_i)\right)-\left(\prod_{i=1}^n\phi_{inc}^{(0)}(\delta p_i)\right)\right\}
\left|\sum_{k=1}^{n} J_k\right|^2,\nonumber
\end{align}
where we defined the emission current as
\begin{align}
J_k\equiv \left(\frac{f_{s_ns_0}^\lambda(p_k-k,p_k)}{k_\mu
  p^\mu_k/(p_0^0-\omega)}-\frac{f_{s_ns_0}^{\lambda}(p_{k-1}-k,p_{k-1})}{k_\mu
  p^\mu_{k-1}/(p_0^0-\omega)}\right)
\exp\left(-i\sum_{i=1}^{k-1} \frac{k_\mu
  p^\mu_i}{(p_0^0-\omega)}\delta z\right).
\end{align}
By identifying $2\pi\delta(0)\equiv T$ and $(2\pi)^2\delta(\v{0})\equiv \pi
R^2$ we can finally integrate in the electron final momentum $\v{p}_n$ in
order to obtain
\begin{align}
\omega \frac{dI_{inc}}{d\omega d\Omega} &=
\frac{e^2}{(2\pi)^2}\left(\prod_{i=1}^{n}\int\frac{d^3\v{p}_i}{(2\pi)^3}\right)\left\{\left(\prod_{i=1}^n\phi_{inc}^{(n)}(\delta
p_i)+\phi_{inc}^{(0)}(\delta
p_i)\right)-\left(\prod_{i=1}^n\phi_{inc}^{(0)}(\delta
p_i)\right)\right\}\nonumber\\&\times
\frac{\omega^2}{2}\sum_{\lambda,s}\left|\sum_{k=1}^{n} J_k\right|^2.
\end{align}
The resulting expression is just the evaluation over the elastic distribution
at each layer, given either by $\phi_{inc}^{(n)}(\delta p_i,\delta z)$ in case
of collision or by $\phi_{inc}^{(0)}(\delta p_i,\delta z)$ in case of no
collision, of the current $\sum J_k$, which except for spin corrections in the
hard part of the spectrum agrees with the classical current. Finally, we
remove the squared no collision emission diagram, which corresponds to no
colliding at any of the layers. Further simplifications can be done for the
unpolarized and spin averaged intensity as we did in the coherent
contribution. In that case one can define two effective currents,
\textit{c.f.}  Appendix \ref{appendix1}, one corresponding to spin preserving
vertices and agreeing with the classical current, the other corresponding to
spin flipping vertices, correcting the classical contribution for hard photons
$y=\omega/p_0^0\sim 1$,
\begin{align}
&\omega \frac{dI_{inc}}{d\omega d\Omega} = \frac{e^2}{(2\pi)^2}\left(\prod_{i=1}^{n}\int\frac{d^3\v{p}_i}{(2\pi)^3}\right)\left\{\left(\prod_{i=1}^n\phi_{inc}^{(n)}(\delta p_i)+\phi_{inc}^{(0)}(\delta p_i)\right)-\left(\prod_{i=1}^n\phi_{inc}^{(0)}(\delta p_i)\right)\right\}\nonumber\\&\times
\left(h^n(y)\left|\sum_{k=1}^{n} \v{\delta}_k^n
\exp\left(-i\sum_{i=1}^{k-1} \frac{k_\mu
  p^\mu_i}{p_0^0-\omega}\delta z\right)\right|^2+h^s(y)\left|\sum_{k=1}^{n} \delta_k^s
\exp\left(-i\sum_{i=1}^{k-1} \frac{k_\mu
  p^\mu_i}{p_0^0-\omega}\delta z\right)\right|^2\right),\label{central_equation_qed}
\end{align}
where the spin non flip currents agree with the classical result
\eqref{delta_definition} 
\begin{align}
\v{\delta}_k \equiv \v{k} \times
\bigg(\frac{\v{p}_{k}}{k_\mu p^\mu_{k}}-\frac{\v{p}_{k-1}}{k_\mu p_{k-1}^\mu}\bigg),\label{spin_nonflip_current}
\end{align}
and the spin flip currents are given by
\begin{align}
\delta_k \equiv
\frac{p_0^0\omega}{\omega p_0^0-\v{k}\cdot\v{p}_{k}}-\frac{p_0^0\omega}{\omega p_0^0-\v{k}\cdot\v{p}_{k-1}}.\label{spin_flip_current}
\end{align}
The weighting functions of the two contributions are given by
$h^n(y)=(1+(1-y)^2)/2$ and $h^s(y)=y^2/2$. The phases  the single
Bethe-Heitler amplitudes \eqref{spin_nonflip_current} and
\eqref{spin_flip_current}, agreeing in this incoherent average with the
classical phases, are responsible of causing the interferences in their
squared sum, leading to the LPM, the dielectric and the transition radiation
effects. The interference pattern is essentially the same as the one discussed
in the classical correspondence at Section \ref{sec:section_3_2}. The
composition of the elastic propagations at each layer produces a squared
momentum transfer additive in the traveled length. If $\delta l$ verifies
$\delta l\leq \lambda$ where $\lambda=1/n_0\sigma_t^{(1)}$ is the average mean
free path, using \eqref{phi_inc_definition} and
\eqref{incoherent_contribution_infinitesize} we arrive at
\eqref{momentum_transfer_n}
\begin{align}
\frac{\partial}{\partial l} \langle \delta \v{p}^2(l)\rangle =
n_0\sigma_t^{(1)}\langle\delta\v{p}^2(\delta l)\rangle\equiv 2\hat{q},\label{momentum_transfer_equation}
\end{align}
where we define the transport coefficient $\hat{q}$. This allows to relate the
momentum transfer in a length $l$ with the momentum transfer in a single
collision $\delta l$, which for the Debye screened interaction produces \eqref{momentum_transfer_1}
\begin{align}
\langle\delta\v{p}^2(\delta l)\rangle
=\left(2\log\left(\frac{2p_0^0}{\mu_d}\right)-1\right)\mu_d^2\equiv \eta \mu_d^2,
\end{align}
and the long tail introduced a substantial correction $\eta$ to the naive
expected value $\mu_d^2$ after a single collision and a maximum momentum
transfer of $2p_0^0$ is allowed due to the energy conservation delta in
\eqref{phi_inc_definition}. However, since electron momenta appears convoluted
in \eqref{central_equation_qed} together with the Bethe-Heitler functions
$\v{\delta}_k$ at \eqref{spin_nonflip_current} or $\delta_k$ at \eqref{spin_flip_current}, the momentum transfer can be approximated as of the order $|\delta
\v{p}|\simeq 2.5m_e$ and by using $\mu_d=\alpha m_e Z^{1/3}$ we obtain instead
\begin{align}
\eta=2\left(\log\left(\frac{2.5m_e}{\mu_d}\right)-\frac{1}{2}\right)=2\left(\log\left(\frac{2.5}{\alpha  Z^{1/3}}\right)-\frac{1}{2}\right).\label{eta_definition}
\end{align}
This can be numerically checked and matches Bethe's \cite{bethe1934,tsai1963}
$\eta=2\log(183/Z^{1/3})$ prediction within less than a $3\%$ of deviation in
the range $Z=(1,100)$. The choice of $\eta$ defines $\hat{q}$ and thus the
Fokker-Planck approximation \eqref{gaussianapproximation} for the scattering
distribution $\hat{\Sigma}_2(\delta\v{p},z)$ in
\eqref{phi_inc_definition}. However, since this fix is only valid for $\delta
l\leq \lambda$ this approximation is valid only in the incoherent plateau,
where the single scattering regime holds. A local with $\omega$ definition of
$\eta$ and $\hat{q}$ has to be employed in order to match the Debye screened
intensity in all the spectrum. In this way, a single Fokker-Planck
approximation can not be used unless the medium length is assumed infinite, in
which case the lower plateau can be neglected and the fail of the
Fokker-Planck approximation becomes irrelevant. 

Intensity \eqref{central_equation_qed} is suitable for a numerical evaluation
for finite size targets under a general interaction. A Monte Carlo code has
been developed in which the electron is assumed to describe a piece-wise
zig-zag path \cite{bell1958} where the step size is taken as $\delta z =
0.1\lambda$. This leads to paths going from $\sim$ 1500 steps for the shortest
mediums to $\sim$ 250000 steps for the largest. The integration in the
electron momenta transforms into an average over the paths under the elastic
weight \eqref{phi_inc_definition}. In a typical run around $10^4$ paths had to
be computed in order to obtain reasonable precision, spanning $50$ photon
frequencies and $100$ photon angles. Several medium materials and lengths were
chosen in order to compare with SLAC \cite{anthony1997} and CERN
\cite{hansen2004} data.  
\begin{figure}[ht]
\centering
\includegraphics[scale=0.6]{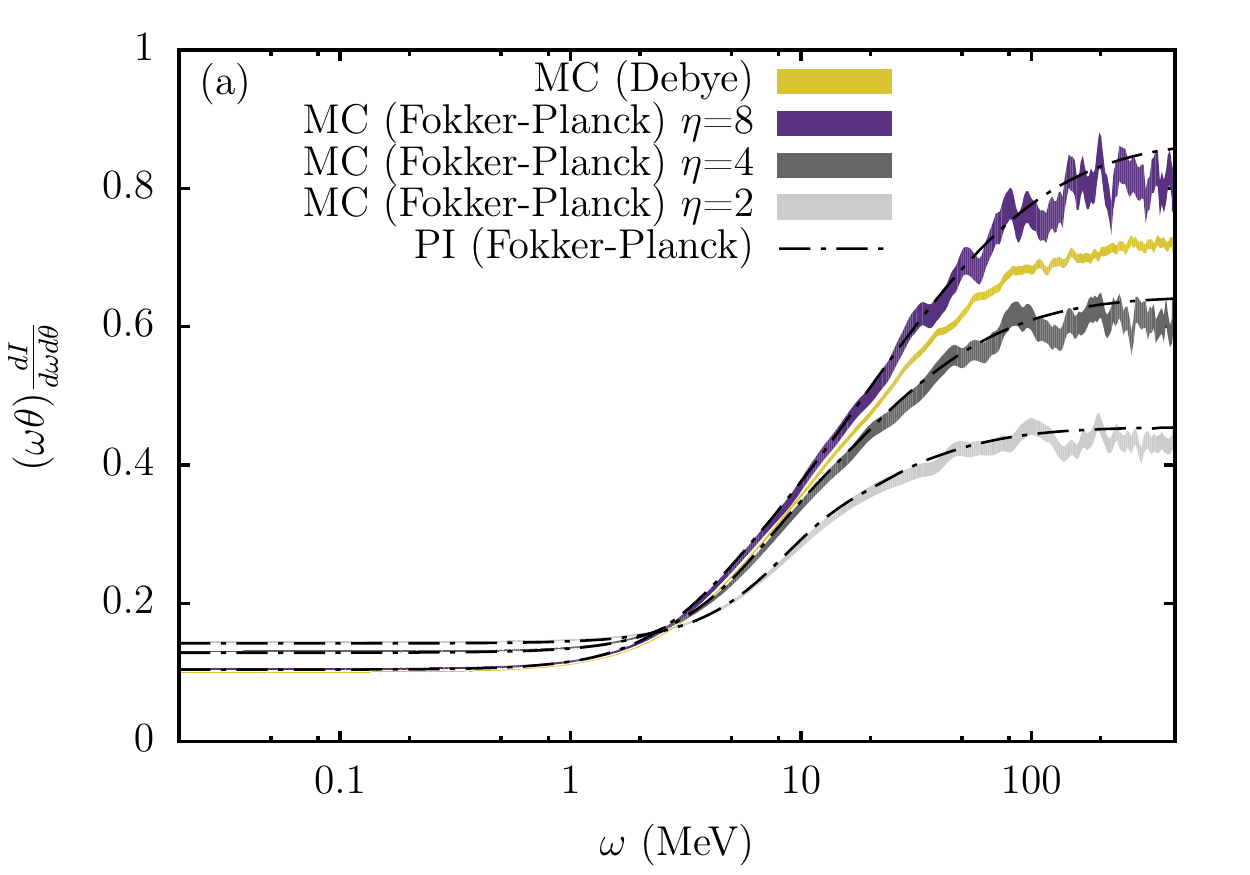}
\includegraphics[scale=0.6]{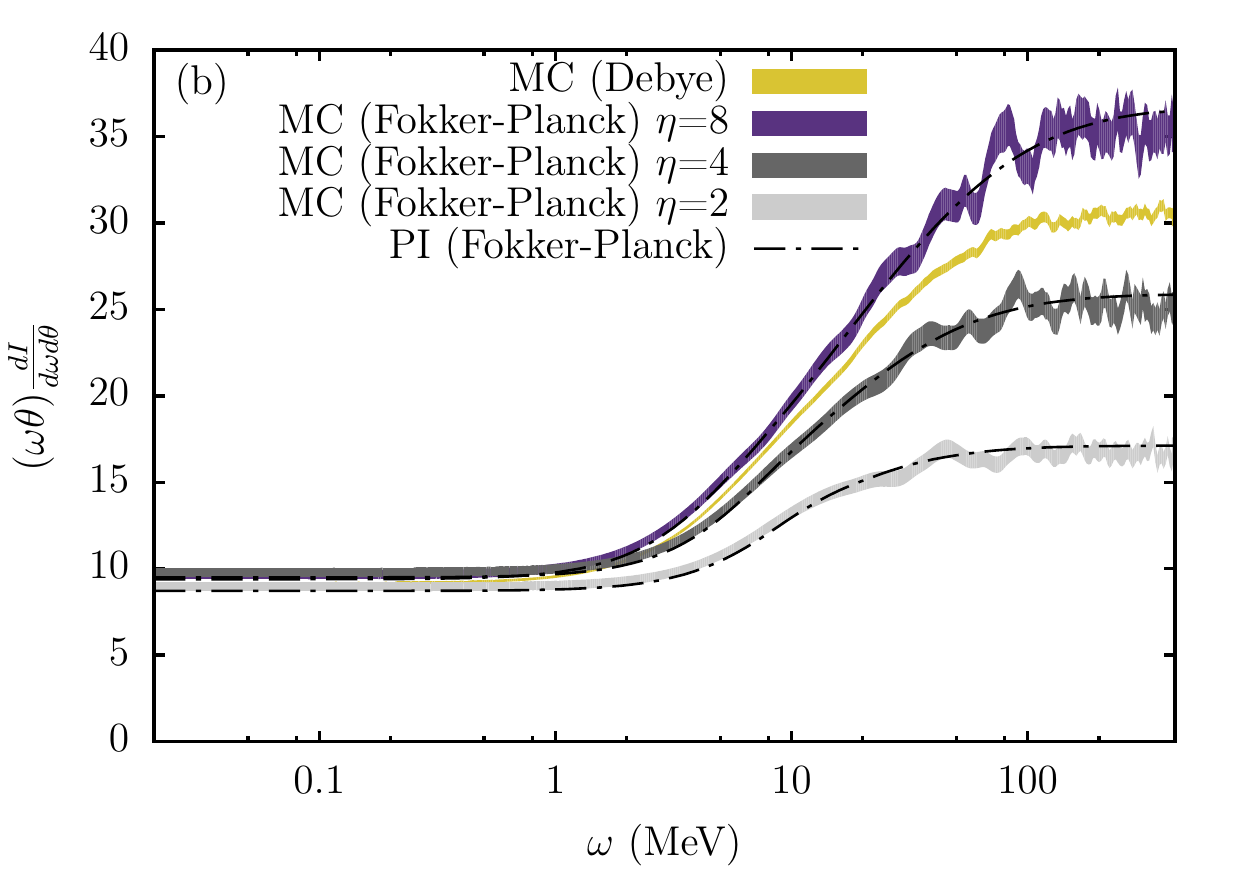}
\caption{Differential intensity of photons in the angle $\theta=0.01/\gamma_e$
  (a) and $\theta=0.5/\gamma_e$ (b) radiated from electrons of $p_0^0$ = 8
  GeV, $\gamma_e=p_0^0/m$, after traversing an Au sheet of $l$ = 0.0023 cm, as
  a function of the photon energy. Monte Carlo evaluation of
  \eqref{central_equation_qed} is shown in solid yellow line for the Debye
  interaction and for the Fokker-Planck approximation with $\eta$ = 8
  (purple), $\eta$ = 4 (dark grey) and $\eta$ = 2 (light grey). Also shown
  with dot-dashed lines the respective continuous limits of
  \eqref{central_equation_qed}, leading to the path integral in the
  Fokker-Planck approximation \eqref{central_equation_pathintegral_qed}.}
\label{fig:figure_3_2}
\end{figure}

We can follow an alternative approach which helps to qualitatively understand
the behavior of the found intensity if we observe that
\eqref{central_equation_qed} is just a sum of single Bethe-Heitler amplitudes
$\v{\delta}_k$ carrying a phase. These amplitudes are summed, squared and then
weighted by the incoherent averages \eqref{phi_inc_definition} of the electron
elastic intensity once the photon energy gap effect is neglected in its
velocity. This agrees with the classical intensity \eqref{classicalintensity}
and thus produces crossed terms $\sim\v{\delta}_k\v{\delta}_je^{i\varphi^j_k}$
interfering with a phase $\varphi^j_k$. We define the distance or coherence
length in which $\delta l=z_j-z_k$ in which $\varphi^j_k$ becomes larger than
unity. Using \eqref{phase_difference} and \eqref{momentum_transfer_equation}
we get
\begin{align}
\varphi^j_k=\frac{1}{p_0^0-\omega}\int^{z_j}_{z_k} dz\medspace k_\mu p^\mu(z) \simeq
\frac{\omega}{2p_0^0(p_0^0-\omega)}\left(m_e^2\delta l+\hat{q}(\delta l)^2\right)=1,
\end{align}
which by inversion produces a coherence length modulated by the photon frequency
\begin{align}
\delta l (\omega)\equiv \frac{m_e^2}{2\hat{q}}\left(\sqrt{1+\frac{8\hat{q}p_0^0(p_0^0-\omega)}{m_e^4\omega}}-1\right).\label{coherence_lenght}
\end{align}
This defines two characteristic frequencies of the LPM interference: the
frequency $\omega_c$ at which the coherence length becomes of the order of the
medium length $\delta l(\omega_c)=l$ thus $\omega_c\simeq
p_0^0(p_0^0-\omega)/(m_e^2l+\hat{q}l^2)$ and the frequency $\omega_s$ at which
the coherence length becomes of the order of a mean free path $\delta
l(\omega_s)=\lambda$ thus $\omega_s\simeq
p_0^0(p_0^0-\omega)/(m_e^2\lambda+\hat{q}\lambda^2)$. Since for $\delta l\geq
l$ there are no sources of scattering we further impose to
\eqref{coherence_lenght} $\delta l(\omega)=l$ for $\omega\geq \omega_c$. In a
coherence length the phase can be neglected so that the internal structure of
scattering becomes irrelevant, \textit{c.f.} Section 2.1, and the centers in
$\delta l(\omega)$ act like a single scattering source with a charge
equivalent to the total matter in $\delta l(\omega)$. Since there are
$l/\delta l(\omega)$ of these coherence lengths for a given $\omega$ we simply
write
\begin{align}
\omega \frac{dI_{inc}}{d\omega}(l)= \frac{l}{\delta l
  (\omega)}e^2\frac{d\Omega_k}{(2\pi)^2}\int
\frac{d^3\delta\v{p}}{(2\pi)^3}\left(h^n(y)|\v{\delta}_1^n|^2+h^s(y)|\delta_1^s|^2\right)\phi_{inc}^{(n)}(\delta\v{p},\delta l(\omega)).\label{heuristic_formula}
\end{align}
By inserting \eqref{phi_inc_definition} and integrating in the photon solid
angle $\Omega_k$ one gets
\begin{align}
\omega \frac{dI_{inc}}{d\omega}(l)=\frac{l}{\delta
  l(\omega)}\frac{e^2}{\pi^2}\int^{\pi}_0 d\theta
\sin(\theta)F(\theta)\hat{\Sigma}_2(\delta\v{p},\delta l(\omega)),
\end{align}
where $|\delta\v{p}|=2p_0^0\beta\sin(\theta)$ is the electron momentum change
and the $F(\theta)$ function is given by
\begin{align}
F(\theta)=\bigg[&\frac{1-\beta^2\cos\theta}{2\beta\sin(\theta/2)
\sqrt{1-\beta^2\cos^2(\theta/2)}}\nonumber\\
&\times\log\bigg[\frac{\sqrt{1-\beta^2
\cos^2(\theta/2)}+\beta\sin(\theta/2)}{\sqrt{1-\beta^2
\cos^2(\theta/2)}-\beta\sin(\theta/2)}\bigg]-1\bigg].
\end{align}
This last integral can be numerically evaluated both for the Debye screened
interaction for $\Sigma_2(\delta\v{p},\delta z)$ and for its Fokker-Planck
approximation \eqref{gaussianapproximation}, and the resulting values for
$\omega\gg \omega_s$ and $\omega \ll \omega_c$, i.e. for the incoherent and
coherent plateaus, respectively, are exact. In the Fokker-Planck approximation
one can further write an useful interpolating function for these two asymptotic
values,
\begin{align}
\omega \frac{dI_{inc}}{d\omega}(l)=\frac{l}{\delta l(\omega)}
\frac{2e^2}{\pi}\frac{1+n_m(\omega)}{3A+n_m(\omega)}
\log\bigg(1+An_m(\omega)\bigg),
\end{align}
where $n_m(\omega)\simeq 2\hat{q}\delta l(\omega)/m_e^2$ is a measure of the
number of transverse masses acquired in a coherence length and
$A=e^{-(1+\gamma)}$ where $\gamma$ is Euler's constant. Then the qualitative
behavior of the LPM effect is as follows. For frequencies lower than
$\omega_c$ the coherence length extends beyond the medium length and then the
radiation intensity consists in the difference of the first and last photon
diagrams squared. In this coherent plateau the vanishing phase causes trivial
convolutions of the elastic distributions at each layer and it can be shown
that the total distribution is then given by $\Sigma_2^{(n)}(\delta\v{p},l)$,
as stated in \eqref{heuristic_formula}. For frequencies larger than $\omega_c$
using \eqref{coherence_lenght} the number of independent emitters $l/\delta
l(\omega)$ grows with $\sqrt{\omega}$ whereas the charge of each emitter
logarithmically decreases with $\log(1/\sqrt{\omega})$. This enhancement from
the coherence plateau stops around $\omega_s$, when the coherence length
acquires the minimum thickness of matter $\lambda=1/n_0\sigma_t^{(1)}$ to
produce radiation, since in absence of collision both
\eqref{spin_nonflip_current} and \eqref{spin_flip_current} vanish. In this
incoherence plateau the radiation consists in the incoherent superposition of
the $n_c=l/\lambda$ single Bethe-Heitler intensities, where $n_c$ is the
average number of collisions. This maximal decoupling of the intensity can be
diagrammatically understood as if the phase difference introduced by the
internal fermion lines is so big that the electron emerging of an emission
after a collision is real and incoherent with the next emission diagram.
\begin{figure}[ht]
\centering
\includegraphics[scale=0.6]{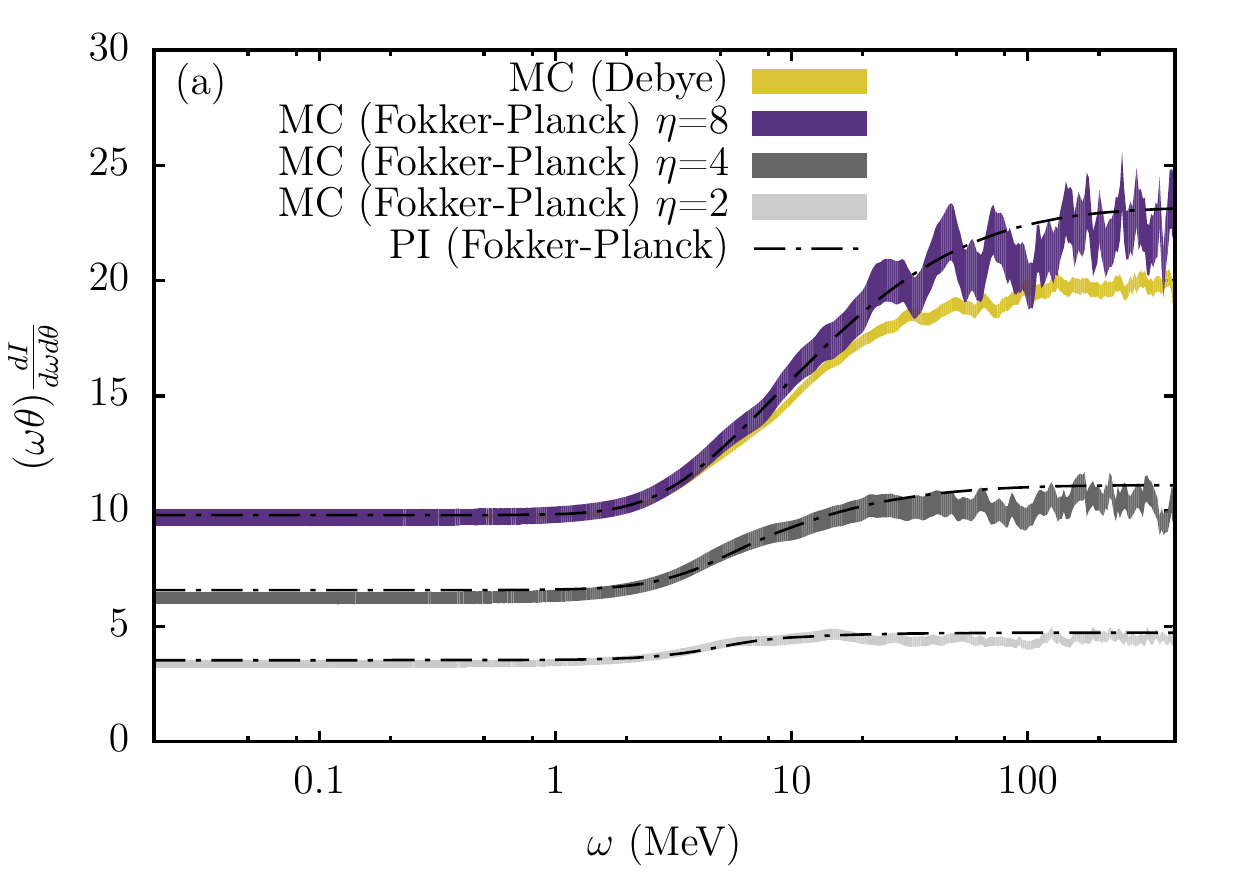}
\includegraphics[scale=0.6]{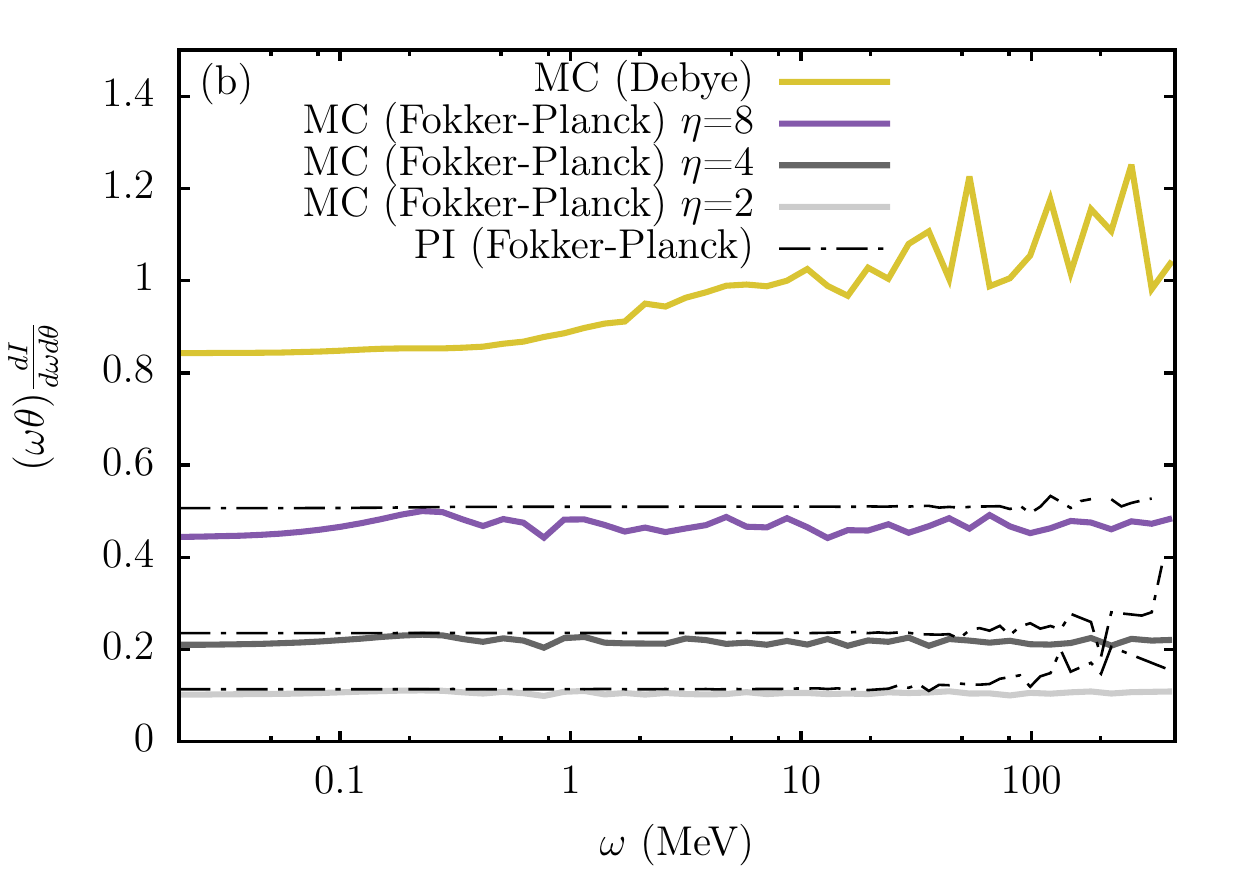}
\caption{Differential intensity of photons in the angle $\theta=2/\gamma_e$
  (a) and $\theta=10/\gamma_e$ (b) radiated from electrons of $p_0^0$ = 8 GeV,
  $\gamma_e=p_0^0/m$, after traversing a Gold sheet of $l$ = 0.0023 cm, as a
  function of the photon energy. Monte Carlo evaluation of
  \eqref{central_equation_qed} is shown in solid yellow line for the Debye
  interaction and for the Fokker-Planck approximation with $\eta$ = 8
  (purple), $\eta$ = 4 (dark grey) and $\eta$ = 2 (light grey). Also shown
  with dot-dashed lines the respective continuous limits of
  \eqref{central_equation_qed}, leading to the path integral in the
  Fokker-Planck approximation \eqref{central_equation_pathintegral_qed}.}
\label{fig:figure_3_3}
\end{figure}

The radiation intensity coming from electrons of $p_0^0 = $ 8 GeV for several
photon angles is depicted in Fig. \ref{fig:figure_3_2} and Fig.
\ref{fig:figure_3_3} both for the Debye screened interaction and the
Fokker-Planck evaluations of \eqref{central_equation_qed}. A target of Gold of
$l =$ 0.0023 cm is chosen which corresponds to an average of $n_c$ = 862
collisions. For this element we obtain a screening mass estimate of $\mu_d$ =
16 KeV and a transport parameter of $\hat{q}=(\eta/2)\times$1.89 KeV$^3$ with
$\eta\simeq$ 8 in order to match the angle-integrated incoherent plateau and
comparable to our estimate from \eqref{eta_definition} $\eta\simeq 7.76$. As
it can be clearly seen the Fokker-Planck approximation which matches the
incoherent (single) emission integrated spectrum, mismatches the unintegrated
spectrum. At the lower angles the Fokker-Planck approximation overestimates
the intensity by a $\sim$ 20$\%$ while at the larger angles the Fokker-Planck
approximation underestimates the intensity. For the larger angle $\theta =
10\gamma^{-1}_e$ in particular we find that only half of the real emission
is taken into account.
\begin{figure}[ht]
\centering
\includegraphics[scale=0.62]{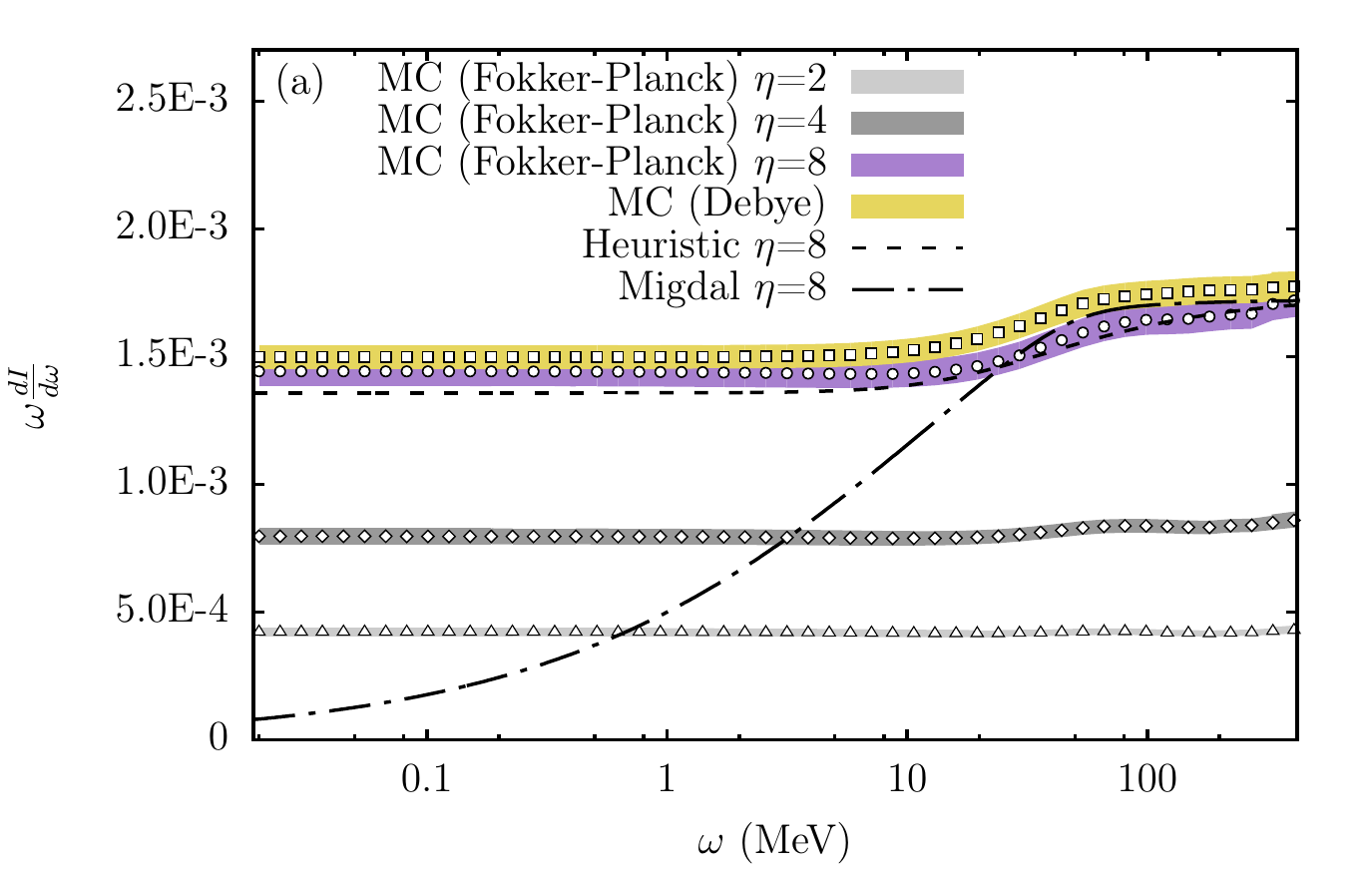}
\includegraphics[scale=0.62]{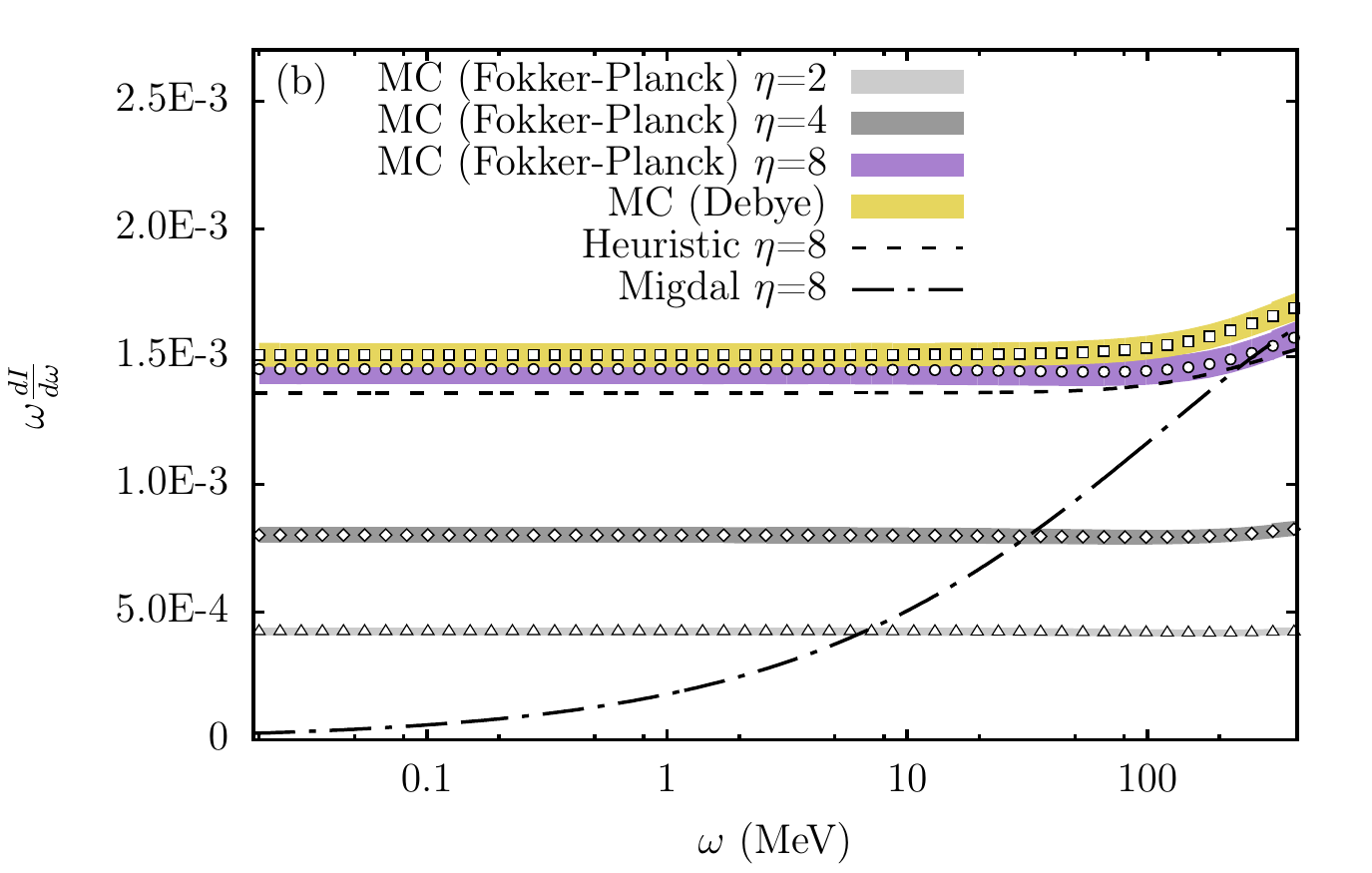}
\caption{Differential intensity of photons radiated from electrons of $p_0^0$
  = 8 GeV (a) and $p_0^0$ = 25 GeV (b) after traversing a Gold sheet of $l$ =
  0.00038 cm, as a function of the photon energy. Monte Carlo evaluation of
  \eqref{central_equation_qed} is shown for the Debye interaction (yellow and
  squares) and for the Fokker-Planck approximation with $\eta$ = 8 (purple and
  circles), $\eta$ = 4 (dark grey and diamonds) and $\eta$ = 2 (light grey and
  triangles). Also shown Migdal prediction \eqref{central_equation_qed}
  (dot-dashed line) with $\eta$ = 8 and our heuristic formula for finite size
  targets in the Fokker-Planck approximation (dashed line).}
\label{fig:figure_3_4}
\end{figure}
\begin{figure}[ht]
\centering
\includegraphics[scale=0.62]{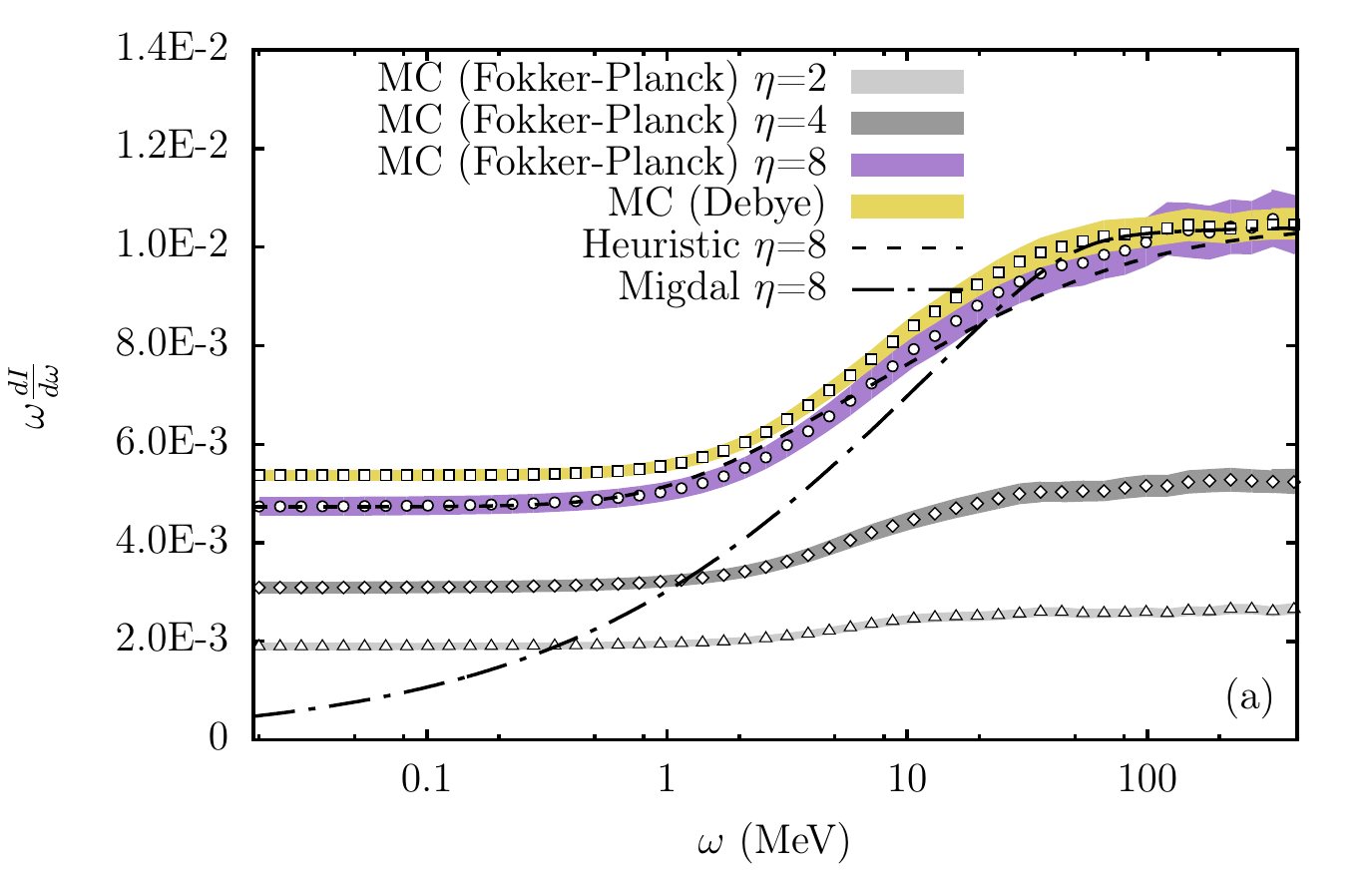}
\includegraphics[scale=0.62]{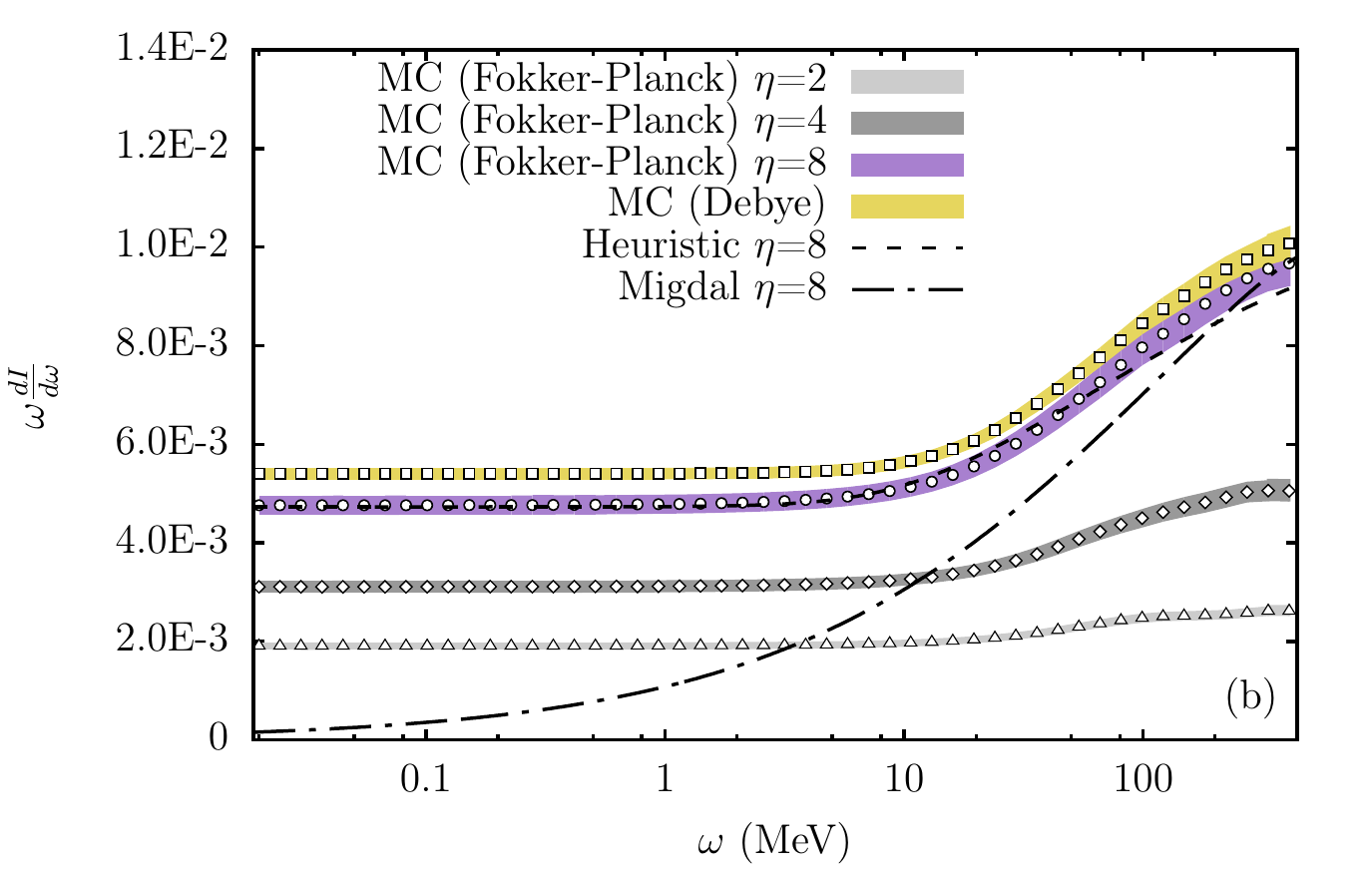}
\caption{Differential intensity of photons radiated from electrons of $p_0^0$
  = 8 GeV (a) and $p_0^0$ = 25 GeV (b) after traversing a Gold sheet of $l$ =
  0.0023 cm, as a function of the photon energy. Monte Carlo evaluation of
  \eqref{central_equation_qed} is shown for the Debye interaction (yellow and
  squares) and for the Fokker-Planck approximation with $\eta$ = 8 (purple and
  circles), $\eta$ = 4 (dark grey and diamonds) and $\eta$ = 2 (light grey and
  triangles). Also shown Migdal prediction \eqref{central_equation_qed} with
  $\eta$ = 8 (dot-dashed line) and our heuristic formula for finite size
  targets in the Fokker-Planck approximation (dashed line).}
\label{fig:figure_3_5}
\end{figure}
\begin{figure}[ht]
\centering
\includegraphics[scale=0.62]{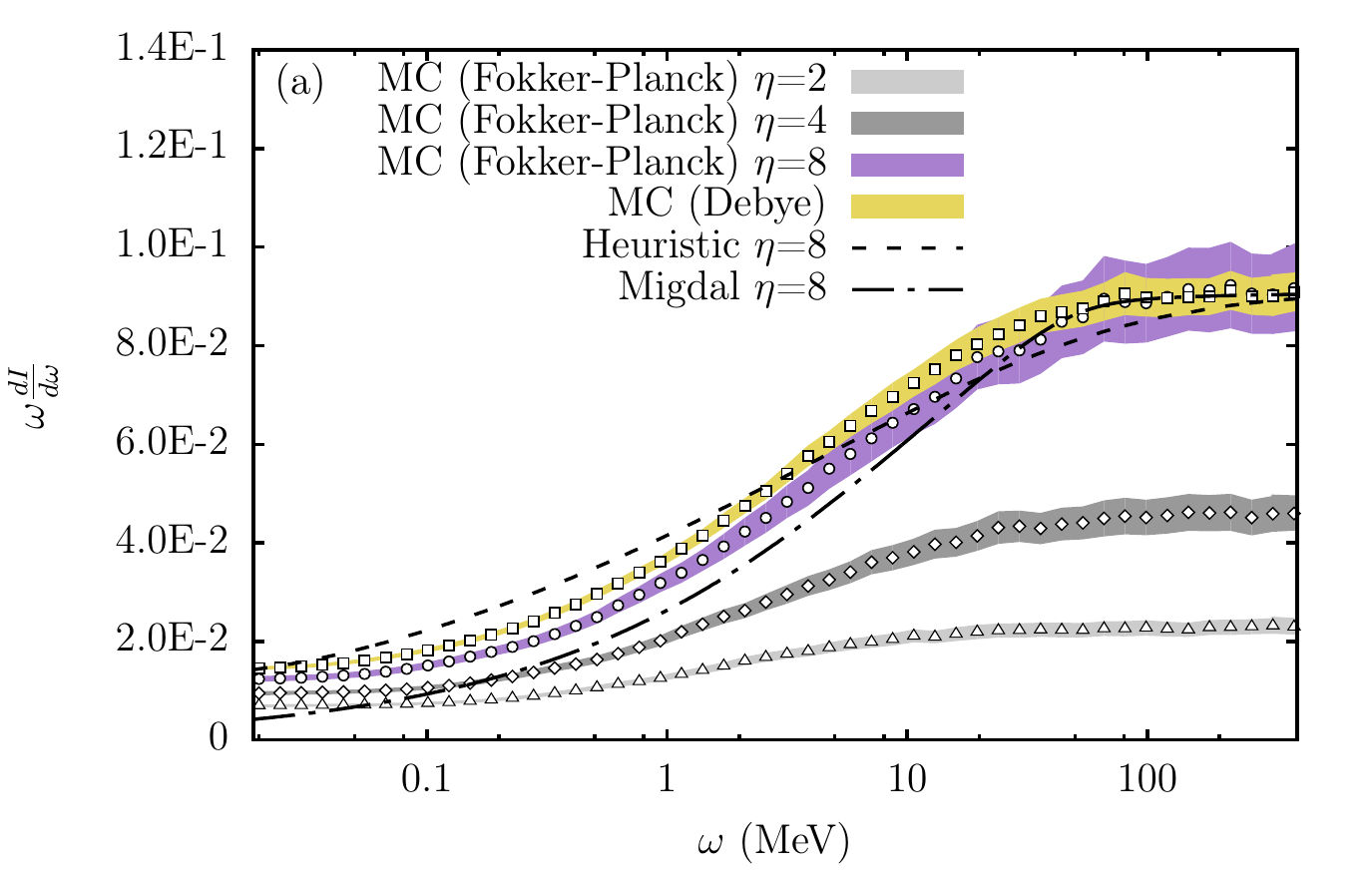}
\includegraphics[scale=0.62]{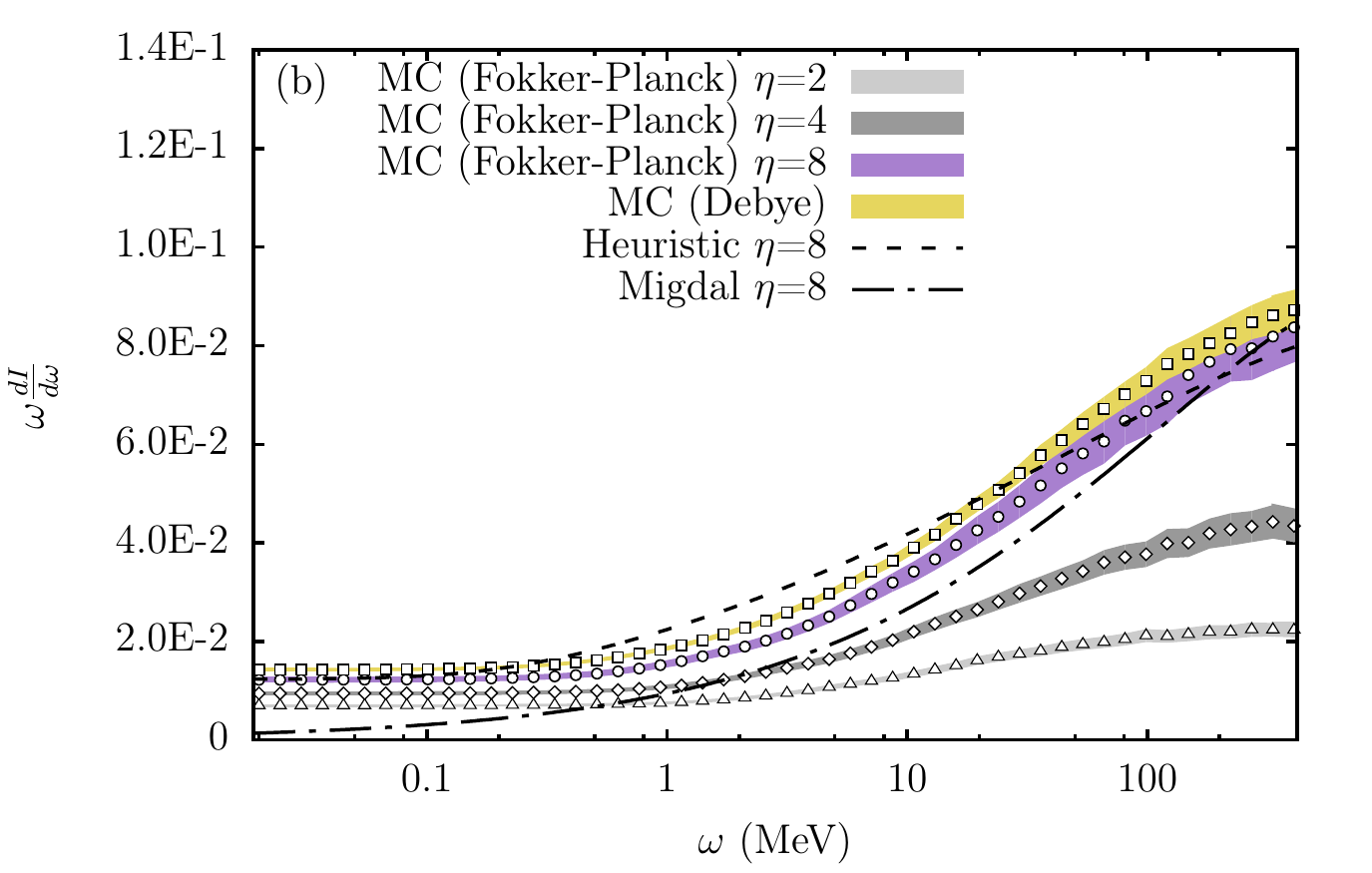}
\caption{Differential intensity of photons radiated from electrons of $p_0^0$
  = 8 GeV (a) and $p_0^0$ = 25 GeV (b) after traversing a Gold sheet of $l$ =
  0.0023 cm, as a function of the photon energy. Monte Carlo evaluation of
  \eqref{central_equation_qed} is shown for the Debye interaction (yellow and
  squares) and for the Fokker-Planck approximation with $\eta$ = 8 (purple and
  circles), $\eta$ = 4 (dark grey and diamonds) and $\eta$ = 2 (light grey and
  triangles). Also shown Migdal prediction \eqref{central_equation_qed} with
  $\eta$ = 8 (dot-dashed line) and our heuristic formula for finite size
  targets in the Fokker-Planck approximation (dashed line).}
\label{fig:figure_3_6}
\end{figure}
In Fig. \ref{fig:figure_3_4}, Fig. \ref{fig:figure_3_5} and
Fig. \ref{fig:figure_3_6} we show the angle-integrated spectrum for Gold
targets of $l =$ 0.00038 cm, $l =$ 0.0023 cm and $l =$ 0.02 cm, which
correspond to an average number of $n_c$ = 142, $n_c$ = 862 and $n_c$ = 7502
collisions, respectively. Two electron energies are shown, $p_0^0$ = 8 Gev and
$p_0^0$ = 25 GeV, and we present the Debye screened interaction and the
Fokker-Planck evaluations of \eqref{central_equation_qed}. For $l =$ 0.00038
cm the predicted characteristic frequencies are $\omega_c$ = 8 MeV and
$\omega_s =$ 1.1 GeV for electrons of $p_0$ = 8GeV, and $\omega_c$ = 80 MeV
and $\omega_s$ = 11 GeV for electrons of $p_0$ = 25 GeV. Since a small number
of collisions is occurring and the medium finiteness is taken into account,
the difference between the coherent plateau and the incoherent plateau is
small. For $l =$ 0.0023 cm the characteristic frequencies
\eqref{saturation_frequency} and \eqref{coherent_frequency} are given by
$\omega_c$ = 0.48 MeV and $\omega_s$ = 418 MeV for electrons of $p_0$ = 8 GeV,
and $\omega_c$ = 4.7 MeV and $\omega_s$ = 4 GeV for electrons of $p_0$ = 25
GeV. For $l =$ 0.02 cm we obtain $\omega_c$ = 8 KeV and $\omega_s$ = 60 MeV
for electrons of $p_0$ = 8 GeV, and $\omega_c$ = 80 keV and $\omega_s$ = 588
MeV for electrons of $p_0$ = 25 GeV. We also show our heuristic formula
\eqref{heuristic_formula} in the Fokker-Planck approximation, producing a
reasonable agreement with \eqref{central_equation_qed}, in particular in the
coherence and incoherence plateaus, where it becomes exact.
\begin{figure}[ht]
\centering
\includegraphics[scale=0.62]{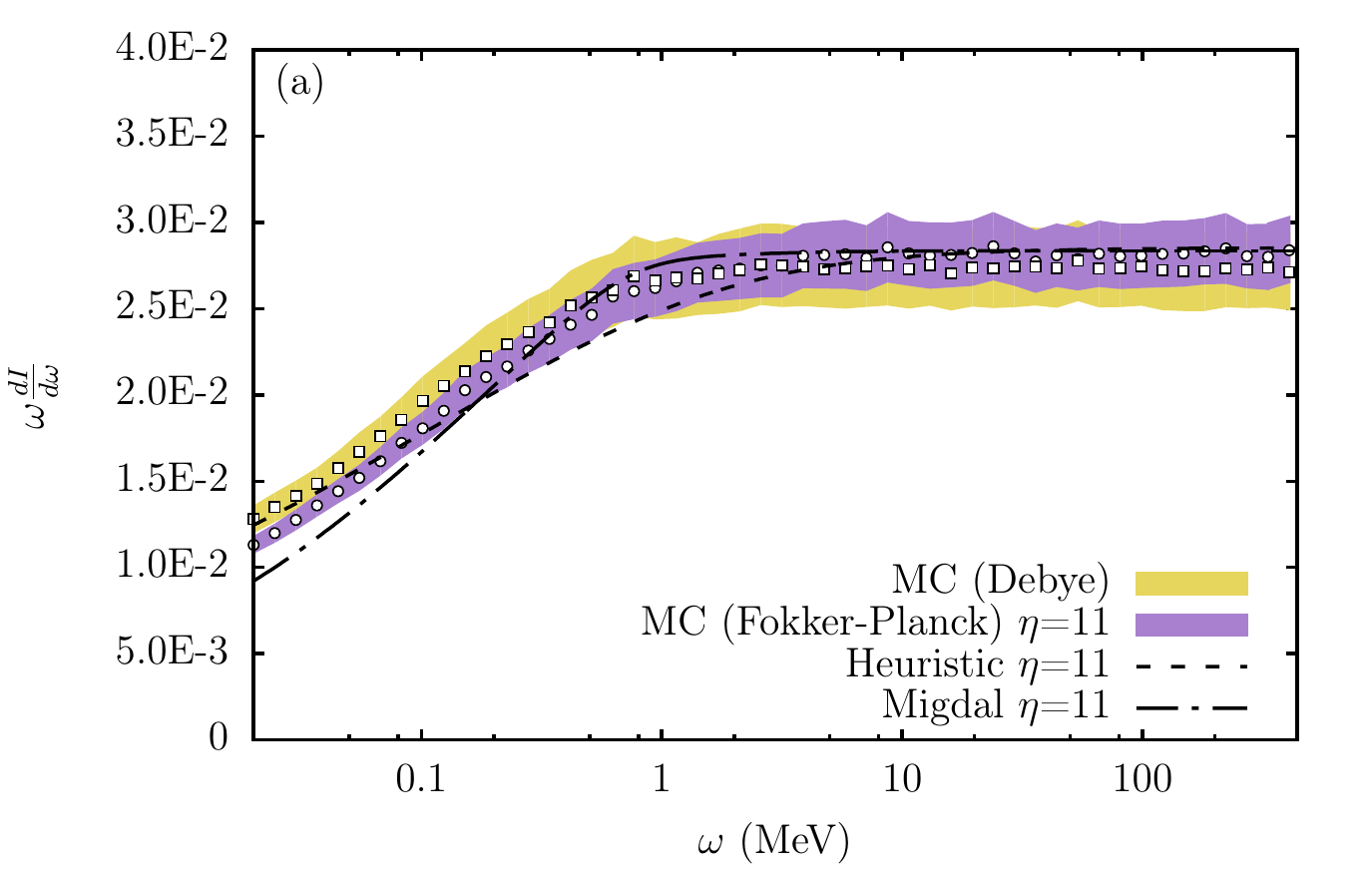}
\includegraphics[scale=0.62]{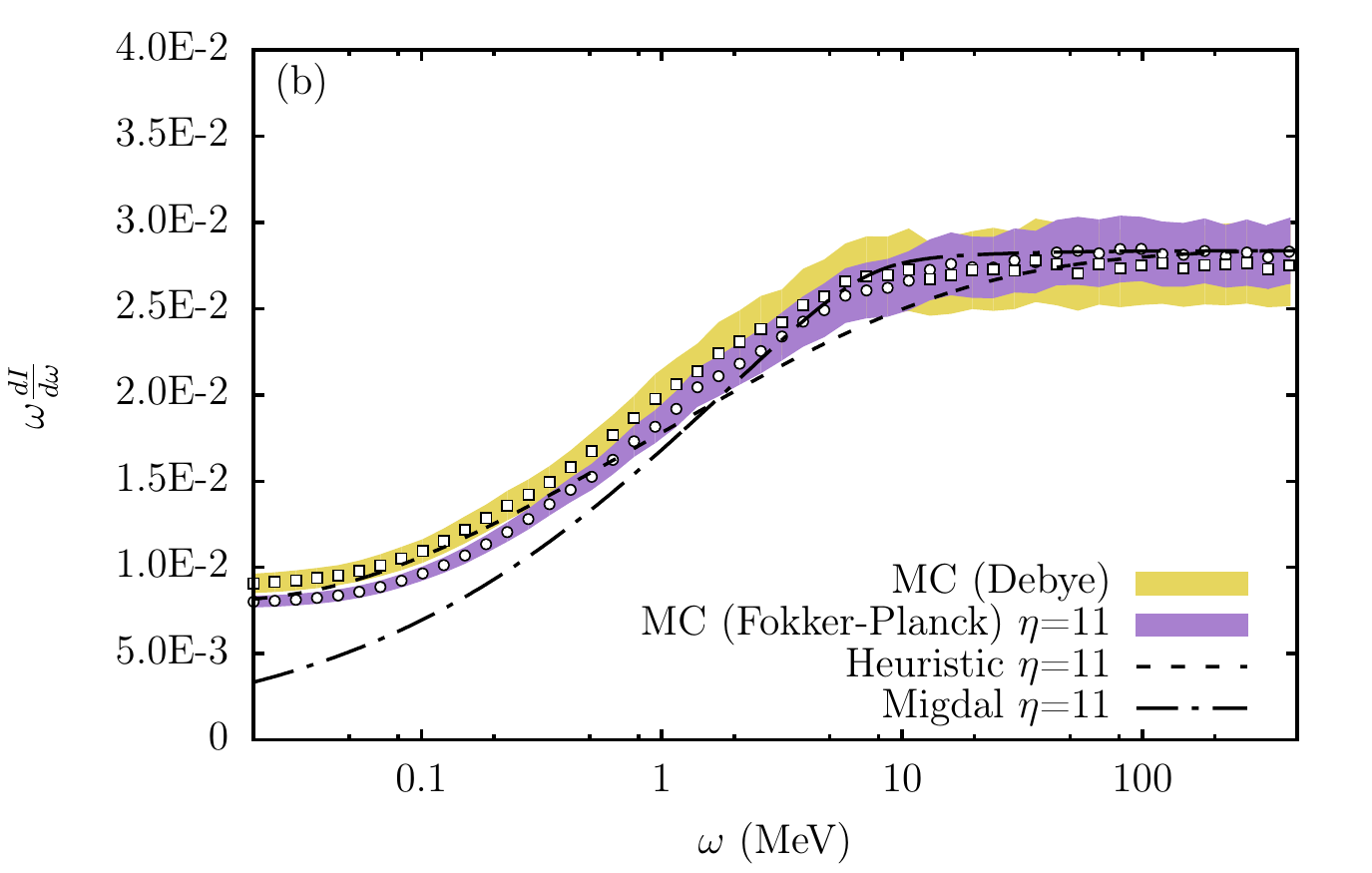}
\caption{Differential intensity of photons radiated from electrons of $p_0^0$
  = 8 GeV (a) and $p_0^0$ = 25 GeV (b) after traversing a Carbon sheet of $l$ =
  0.41 cm, as a function of the photon energy. Monte Carlo evaluation of
  \eqref{central_equation_qed} is shown for the Debye interaction (yellow and
  squares) and for the Fokker-Planck approximation with $\eta$ = 11 (purple
  and circles). Also shown Migdal prediction \eqref{central_equation_qed} with
  $\eta$ = 11 (dot-dashed line) and our heuristic formula for finite size
  targets in the Fokker-Planck approximation (dashed line).}
\label{fig:figure_3_7}
\end{figure}
In Fig. \ref{fig:figure_3_7} we show the angle-integrated spectrum for a
Carbon target of $l =$ 0.41 cm which produces an average number of $n_c =$
9521 collisions, for electron energies of $p_0^0$ = 8 GeV and $p_0^0$ = 25
GeV. Estimates of $\mu_d$ = 6.8 KeV, $\hat{q}=(\eta/2)\times 2.10\cdot
10^{-2}$ KeV$^3$ and $\eta$ = 11 are obtained. The evaluation of
\eqref{central_equation_qed} in the Debye screened interaction is shown
together with its Fokker-Planck approximation. The characteristic frequencies
are given by $\omega_c$ = 1 KeV and $\omega_s$ = 11 MeV for electrons of
$p_0^0$ = 8 GeV, and $\omega_c$ = 11 KeV and $\omega_s$ = 108 MeV for
electrons of $p_0^0$ = 25 GeV.
\begin{figure}[ht]
\centering
\includegraphics[scale=0.62]{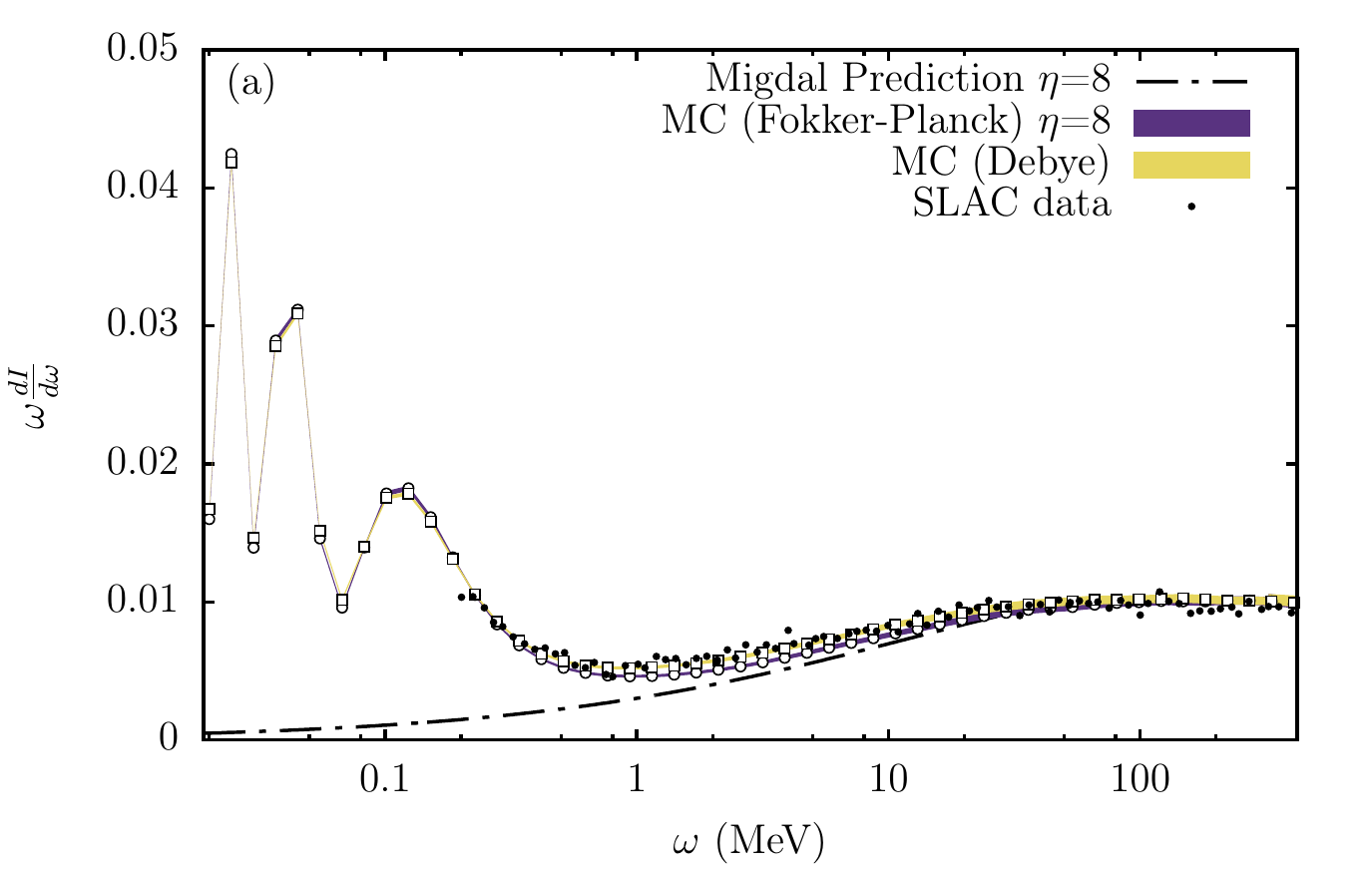}
\includegraphics[scale=0.62]{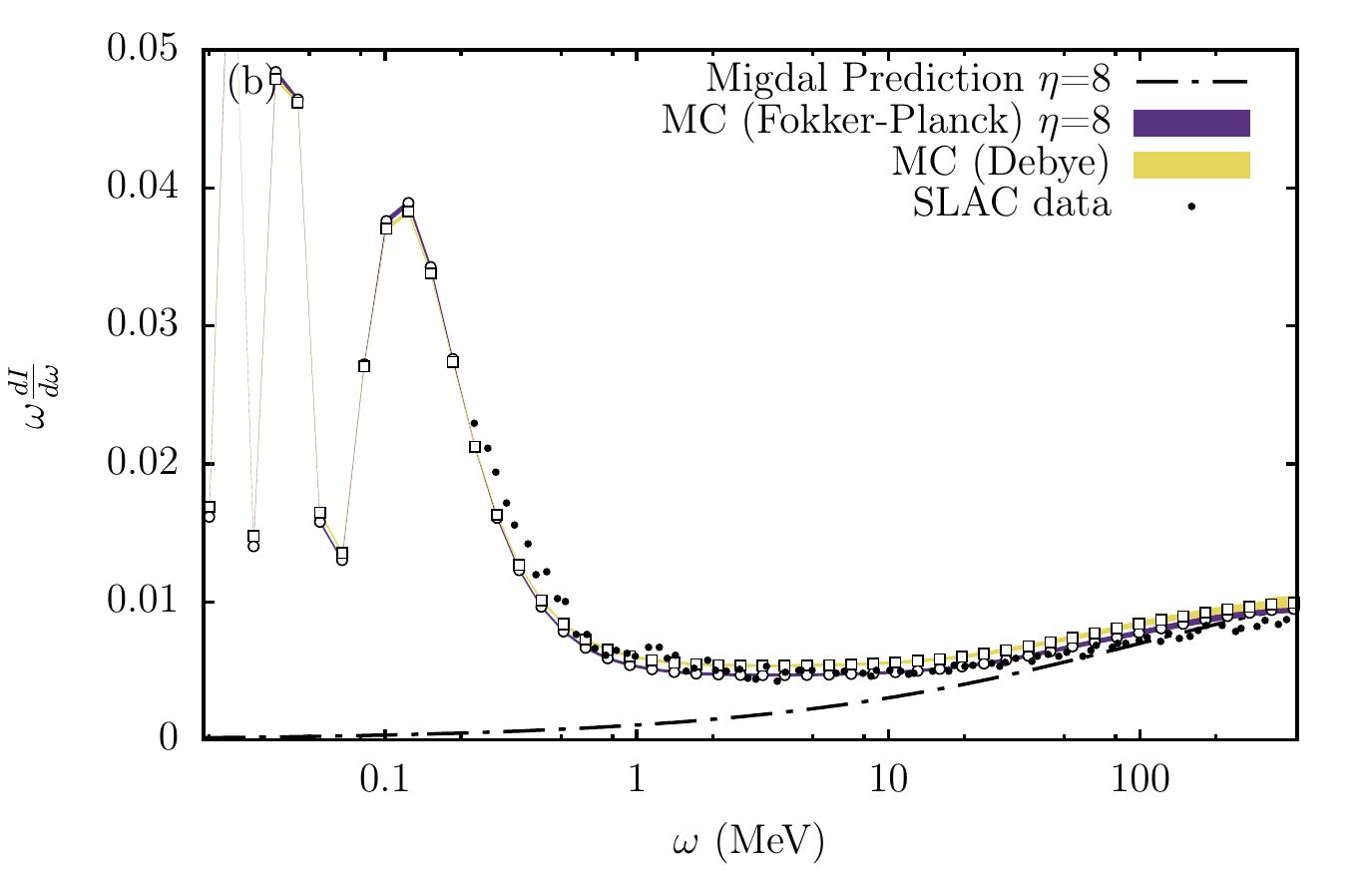}
\caption{Differential intensity of photons radiated from electrons of $p_0^0$
  = 8 GeV (a) and $p_0^0$ = 25 GeV (b) after traversing a Gold sheet of $l$ =
  0.0023 cm, as a function of the photon energy. Monte Carlo evaluation of
  \eqref{central_equation_qed} is shown for the Debye interaction (yellow and
  squares) and for the Fokker-Planck approximation with $\eta$ = 8 (purple
  and circles). Also shown Migdal prediction \eqref{central_equation_qed} with
  $\eta$ = 8 (dot-dashed line) and SLAC data.}
\label{fig:figure_3_8}
\end{figure}
The introduction of medium effects in the photon dispersion relation leads to
the dielectric and the transition radiation effects, \textit{c.f.} Section
3.2. The photon plasma frequency $\omega_p$ can be thought as an effective
photon mass $m_\gamma$ which introduces an extra term $m_\gamma^2/2\omega$ in
the resonances $k_\mu p^\mu(z)$ at the phases and propagators. Then, a strong
suppression occurs for frequencies lower than $\omega_{de}$, which is defined
as the frequency at which the extra term becomes of the same order than the
phase evaluated at $m_\gamma$ = 0, i.e. $\omega_{de}^2 =
m_\gamma^2l\omega_c$. In the particular case that the medium is finite and the
photon mass cannot be assumed global for all the emission diagrams, the last
photon verifies $m_\gamma$ = 0 and thus the intensity is instead dramatically
enhanced for frequencies $\omega \le \omega_{de}$. In Fig.\ref{fig:figure_3_8}
we show this dielectric and transition radiation effects of
\eqref{central_equation_qed} for a Gold target of $l $ = 0.0023 cm, which
introduces an effective photon mass of $m_\gamma$ = 0.08 KeV for all photons
except for the one in the last leg $m_\gamma$ = 0. The predicted
characteristic frequencies are $\omega_{de}$ = 0.6 MeV for $p_0^0$ = 8 GeV and
$\omega_{de}$ = 1.9 MeV for $p_0^0$ = 25 GeV. As it can be seen, for
frequencies lower than $\omega_{de}$ a dramatic enhancement of the coherent
plateau occurs, since the last leg diagram stops to be compensated due to the
dielectric suppression of the first leg diagram and the phase interference
between them grows as $m_\gamma^2/2\omega$. Also shown is the SLAC
experimental data \cite{anthony1997} for the same target, being in very good
agreement with our evaluation.
\begin{figure}[ht]
\centering
\includegraphics[scale=0.62]{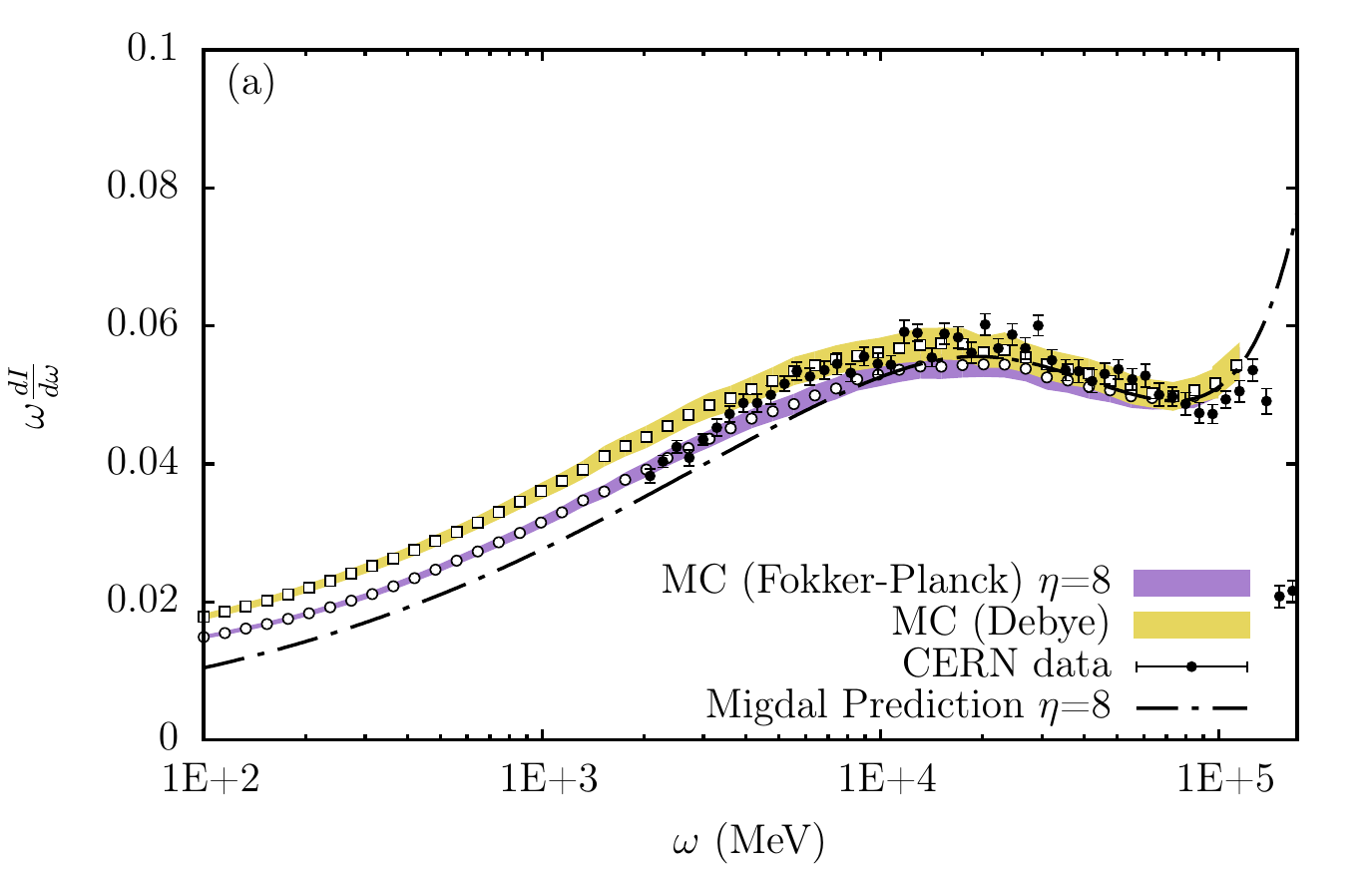}
\includegraphics[scale=0.62]{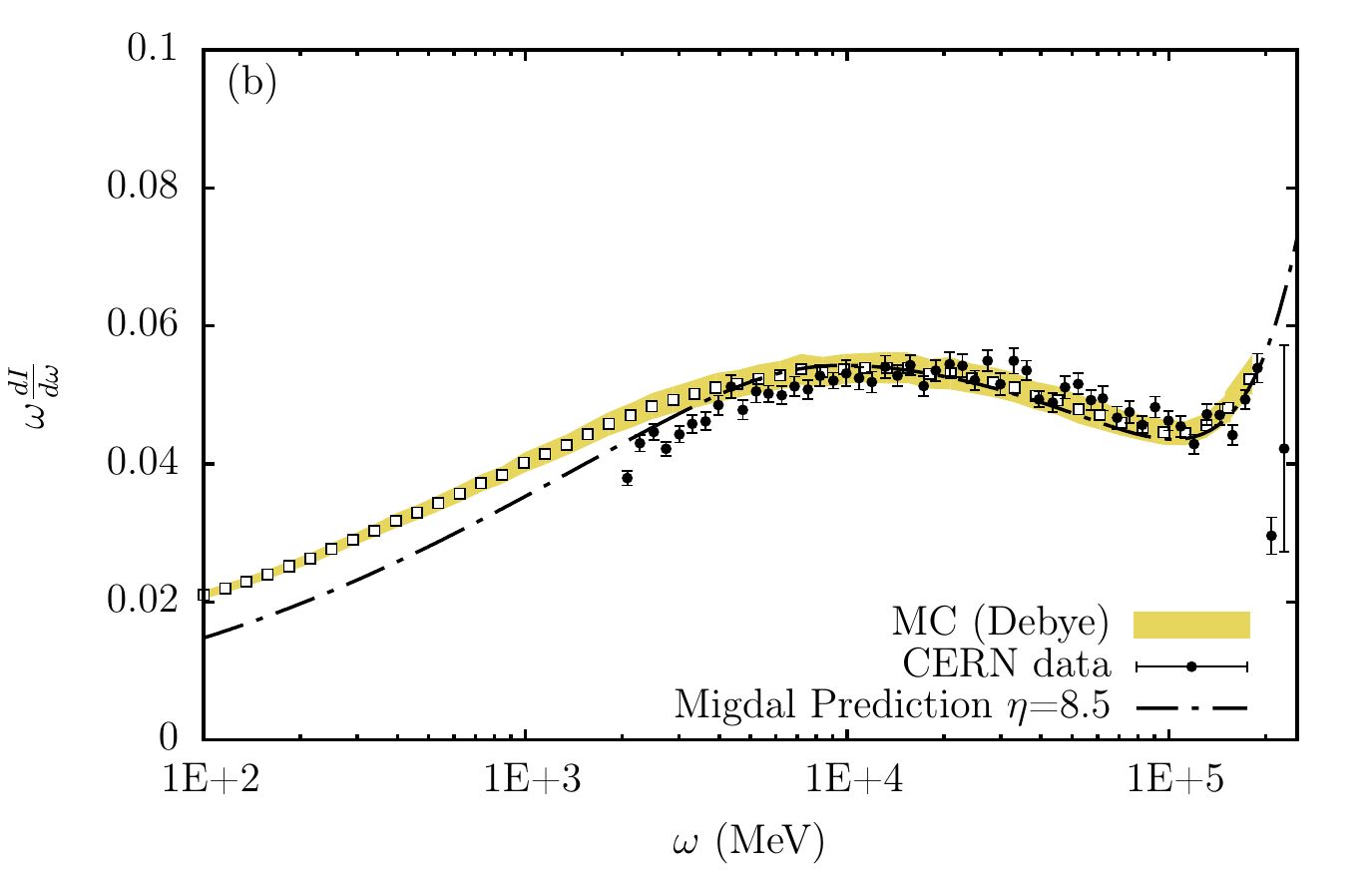}
\caption{Differential intensity of photons radiated from electrons of $p_0^0$
  = 149 GeV traversing a sheet of Iridium of $l$ = 0.0128 cm (a) and electrons
  of $p_0^0$ = 207 GeV traversing a sheet of Copper of $l$ = 0.063 cm (b), as
  a function of the photon energy. Monte Carlo evaluation of
  \eqref{central_equation_qed} is shown for the Debye interaction (yellow and
  squares) and for the Fokker-Planck approximation with $\eta$ = 8 (purple and
  circles). Also shown Migdal prediction \eqref{central_equation_qed} with
  $\eta$ = 8 (dot-dashed line) and SLAC data.}
\label{fig:figure_3_9}
\end{figure}
In Fig. \ref{fig:figure_3_9} we show our results for an Iridium target of $l$
= 0.0128 cm and electrons of $p_0^0$ = 149 GeV and a Copper target of $l$ =
0.063 and electrons of $p_0^0 =$ 207 GeV. Also shown is the CERN data
\cite{hansen2004} for the same scenario, which covers only a small part of the
suppression zone. Slight differences are found in the LPM  between
theoretical predictions and the experimental data.

\section{Continuous limit: a path integral}
\label{sec:section_3_4}
Several formalisms have been developed whose approach consists in a path
integral redefinition of the Boltzmann method used by Migdal. These works rely
on the integrability of the resulting path integrals in the
Fokker-Planck/Gaussian approximation for the interaction. While this provides
a powerful tool to evaluate the intensity we have shown that for finite size
targets and for the angular distributions of the final particles the
differences between the Fokker-Planck approximation and a direct evaluation of
\eqref{central_equation_qed} under a Debye screened interaction cannot be
reconciled into a single definition of the medium transport properties,
given by $\hat{q}$. In this section we will show that our result
\eqref{central_equation_qed} in the $\delta z \to 0$ continuous limit leads to
a path integral formulation which agrees with the finite size predictions of
\cite{wiedemann1999} except for the vacuum regularization, which is strictly
required to reproduce the right opacity/perturbative expansion but otherwise
irrelevant in the large $n_c$ limit and thus for the Fokker-Planck
approximation. It also recuperates the result of Migdal \cite{migdal1956},
Zakharov \cite{zakharov1998} and the BDMPS group \cite{baier1996} in the $l\to
\infty $ limit and under the right kinematical relaxations in the photon
angular integration.
\subsubsection{Transverse incoherent average}
We will restrict our evaluation to the spin non flip contribution. The spin
flip case follows the same steps. Intensity \eqref{central_equation_qed} is
split into two contributions, the first with the incoherent elastic weights
including collisions $\phi^{(n)}_{inc}(\delta p_i)+\phi^{(0)}_{inc}(\delta
p_i)$ and the second with the elastic weights omitting collisions
$\phi^{(0)}_{inc}(\delta p_i)$. We write
\begin{align}
\omega \frac{dI}{d\omega d\Omega}=\omega \frac{dI^{(n)}}{d\omega d\Omega}-\omega \frac{dI^{(0)}}{d\omega d\Omega}.
\end{align}
In order to take the continuous limit we notice that a single Bethe-Heitler
amplitude transforms into the infinitesimal change
\begin{align}
\v{\delta}_k   =\v{k} \times
\bigg(\frac{\v{p}_{k}}{k_\mu p^\mu_{k}}-\frac{\v{p}_{k-1}}{k_\mu
  p_{k-1}^\mu}\bigg) \to  \frac{\partial}{\partial z} \left(\frac{\v{k}\times \v{p}(z_k)}{k_\mu p^\mu(z_k)}\right).
\end{align}
This suggests rearranging the sum (integrating by parts) in order to find for
$\delta z \ll 1$
\begin{align}
\sum_{k=1}^{n} \v{\delta}_k^n
\exp\left(-i\sum_{i=1}^{k-1} \frac{k_\mu
  p^\mu_i}{p_0^0}\delta z\right) = \frac{\v{k}\times\v{p}_n}{k_\mu
  p_n^\mu}\exp\left(-i\sum_{i=1}^{n-1}\frac{k_\mu p_i^\mu}{p_0^0}\right)
\nonumber\\
+i \frac{\delta z}{p_0^0} \sum_{k=1}^{n-1}\v{k}\times \v{p}_k \exp\left(-i\sum_{i=1}^{k-1} \frac{k_\mu
  p^\mu_i}{p_0^0}\delta z\right)- \frac{\v{k}\times\v{p}_0}{k_\mu
  p_0^\mu}.
\end{align}
The first term corresponds to the photon emitted in the final leg, and in the
continuous limit corresponds to the integration between $[l,\infty)$. The
  middle term are just the internal photons and correspond to the integration
  in $(0,l)$. The third term is the photon emitted in the first leg and
  corresponds to the integration in $(-\infty,0]$. The boundary photons can be
taken into account by extending the integration of the interior term to
$(-\infty,+\infty)$ and neglecting the $+\infty$ and the $-\infty$
contribution.  The square of the interior sum can be written as
\begin{align}
&\left|\sum_{k=1}^{n} \v{\delta}_j^n
\exp\left(-i\sum_{i=1}^{k-1} \frac{k_\mu
  p^\mu_i}{p_0^0}\delta z\right)\right|^2 \\
&= \frac{1}{(p_0^0)^2}2\Re \sum_{j=1}^{n-1}\delta z \sum_{k=1}^{j-1}
\delta z \medspace \v{k}\times \v{p}_j \exp\left(-i\sum_{i=k}^{j-1} \frac{k_\mu
  p^\mu_i}{p_0^0}\delta z\right)\v{k}\times \v{p}_k.\nonumber
\end{align}
The presence of the phase from $z_k$ to $z_{j-1}$ suggests a separation of
\eqref{central_equation_qed} into three zones. We also separately consider
the contribution arising with the averages $\phi_{inc}^{(n)}(\delta
p_i)+\phi_{inc}^{(0)}(\delta p_i)$ from the contribution arising with the
averages $\phi_{inc}^{(0)}(\delta p_i)$. For this purpose we write
\begin{align}
&\omega \frac{dI^{(n)}_{inc}}{d\omega d\Omega} = \frac{e^2}{(2\pi)^2}h^n(y)
\frac{1}{(p_0^0)^2}2\Re\sum_{j=1}^{n}\delta z\sum_{k=1}^{j-1}\delta z\int \frac{d^3\v{p}_n}{(2\pi)^3}\label{path_integral_discrete}\\
&\int
\frac{d^3\v{p}_j}{(2\pi)^3}\int \frac{d^3\v{p}_k}{(2\pi)^3}\medspace P_{inc}^{(n)}(\v{p}_n,\v{p}_j)\medspace\v{k}\times \v{p}_j\medspace P^{(n)}_{\gamma}(\v{p}_j,\v{p}_k)
\medspace\v{k}\times \v{p}_k \medspace P_{inc}^{(n)}(\v{p}_k,\v{p}_0),\nonumber
\end{align}
and
\begin{align}
&\omega \frac{dI^{(0)}_{inc}}{d\omega d\Omega} = \frac{e^2}{(2\pi)^2}h^n(y)
\frac{1}{(p_0^0)^2}2\Re\sum_{j=1}^{n}\delta z\sum_{k=1}^{j-1}\delta z\int \frac{d^3\v{p}_n}{(2\pi)^3}\label{path_integral_discrete_regularizator}\\
&\int
\frac{d^3\v{p}_j}{(2\pi)^3}\int \frac{d^3\v{p}_k}{(2\pi)^3}\medspace P_{inc}^{(0)}(\v{p}_n,\v{p}_j)\medspace\v{k}\times \v{p}_j\medspace P^{(0)}_{\gamma}(\v{p}_j,\v{p}_k)
\medspace\v{k}\times \v{p}_k \medspace P_{inc}^{(0)}(\v{p}_k,\v{p}_0).\nonumber
\end{align}
The functions $P_{inc}^{(n)}(\v{p}_a,\v{p}_b)$ are the electron probability for
going from $\v{p}_b$ to $\v{p}_a$ including the no collision events. At high
energies then the convolution
\begin{align}
P_{inc}^{(n)}(\v{p}_a,\v{p}_b)&\equiv \left(\prod_{i=b+1}^{a-1}\int\frac{d^3\v{p}_i}{(2\pi)^3}\right)\left(\prod_{i=b+1}^a\left(\phi_{inc}^{(n)}(\delta
p_i)+\phi_{inc}^{(0)}(\delta
p_i)\right)\right),\label{pn_nogamma_definition}
\end{align}
and the functions $P_{inc}^{(0)}(\v{p}_a,\v{p}_b)$ are the probability of not
colliding thus producing a distribution $\v{p}_a=\v{p}_b$, given by
\begin{align}
P_{inc}^{(0)}(\v{p}_a,\v{p}_b)&\equiv\left(\prod_{i=b+1}^{a-1}\int\frac{d^3\v{p}_i}{(2\pi)^3}\right)\left(\prod_{i=b+1}^a\left(\phi_{inc}^{(0)}(\delta
p_i)\right)\right).\label{p0_nogamma_definition}
\end{align}
The functions $P_\gamma^{(n)}(\v{p}_a,\v{p}_b)$ and
$P_\gamma^{(0)}(\v{p}_a,\v{p}_b)$ convolute the former electron probabilities
for going from $\v{p}_b$ to $\v{p}_a$ with the internal phase
\eqref{longitudinal_momentum_change_vertex} introduced by the photon emission
at $z_k$ and $z_j$. These terms are instead given by
\begin{align}
P_{\gamma}^{(n)}(\v{p}_j,\v{p}_k)\equiv\left(\prod_{i=k+1}^{j-1}\int\frac{d^3\v{p}_i}{(2\pi)^3}\right)&\left(\prod_{i=k+1}^{j}\left(\phi_{inc}^{(n)}(\delta
p_i)+\phi_{inc}^{(0)}(\delta
p_i)\right)\right)\exp\left(-i\sum_{i=k+1}^{j-1} \frac{k_\mu
  p^\mu_i}{p_0^0}\delta z\right),
\end{align}
and
\begin{align}
P_{\gamma}^{(0)}(\v{p}_j,\v{p}_k)\equiv\left(\prod_{i=k+1}^{j-1}\int\frac{d^3\v{p}_i}{(2\pi)^3}\right)&\left(\prod_{i=k+1}^{j}\left(\phi_{inc}^{(0)}(\delta
p_i)\right)\right)\exp\left(-i\sum_{i=k+1}^{j-1} \frac{k_\mu
  p^\mu_i}{p_0^0}\delta z\right).
\end{align}
Here we join $\phi_{inc}^{(0)}(\delta p_i)+\phi_{inc}^{(n)}(\delta
p_i)$ into a single expression. Using \eqref{phi_inc_definition} we find
\begin{align}
\phi_{inc}^{(0)}(\delta p_i)+\phi_{inc}^{(n)}(\delta
p_i)&=2\pi\beta_p\delta(\delta p_i^0)\label{phi_inc_definition_joint}\\
&\times\int d^2\v{x}_i^t
e^{-i\delta\v{p}_i^t\cdot\v{x}_i^t}\exp\left(-n_0(z_i)\delta
z \medspace\bigg(\sigma_{el}^{(1)}(\v{0})-\sigma_{el}^{(1)}(\v{x}_i^t)\bigg)\right)\nonumber. 
\end{align}
The momentum integrations at $P_{inc}^{(n)}(\v{p}_n,\v{p}_j)$ and
$P_{inc}^{(n)}(\v{p}_k,\v{p}_0)$ are trivial, as expected, since in the
absence of longitudinal phases the convolution of infinitesimal eikonal
transports has to respect additivity. One finds for the first set of
scatterings using \eqref{phi_inc_definition_joint} and
\eqref{pn_nogamma_definition}
\begin{align}
P_{inc}^{(n)}(\v{p}_n,\v{p}_j)&=2\pi\beta_p(p_n^0-p_j^0)\\
&\times\int d^2\v{x}_n^t
e^{-i(\v{p}_n^t-\v{p}_j^t)\cdot\v{x}_n^t}\exp\left(-\bigg(\sigma_{el}^{(1)}(\v{0})-\sigma_{el}^{(1)}(\v{x}_n^t)\bigg)\sum_{i=j}^{n}\delta
  z \medspace n_0(z_i)\right),\nonumber
\end{align}
and for the last set of scatterings
\begin{align}
P_{inc}^{(n)}(\v{p}_k,\v{p}_0)&=2\pi\beta_p(p_k^0-p_0^0)\\
&\times\int d^2\v{x}_k^t
e^{-i(\v{p}_k^t-\v{p}_0^t)\cdot\v{x}_k^t}\exp\left(-\bigg(\sigma_{el}^{(1)}(\v{0})-\sigma_{el}^{(1)}(\v{x}_k^t)\bigg)\sum_{i=1}^{k}\delta
  z \medspace n_0(z_i)\right).\nonumber
\end{align}
The momentum integration in the function $P_{\gamma}^{(n)}(\v{p}_j,\v{p}_k)$
has to be performed, however, taking into account the internal longitudinal
phase. By using \eqref{longitudinal_momentum_change_vertex} we find
\begin{align}
\sum_{i=k+1}^{j-1}\frac{k_\mu p_i^\mu}{p_0^0}\delta z=&\frac{\omega
  m_e^2}{2p_0^0(p_0^0-\omega)}(z_j-z_{k+1})+\frac{\omega}{2p_0^0(p_0^0-\omega)}\sum_{i=k+1}^{j-1}\delta z\medspace\bigg(\v{p}_i^t-\frac{p_0^0-\omega}{\omega}\v{k}^t\bigg)^2.
\end{align}
This suggest reorganizing the phase in the elastic weights in the same way
\begin{align}
-i\v{p}_j^t\cdot\v{x}_j^t+i\sum_{i=k+1}^{j-1}\bigg(\v{p}_i^t\cdot\delta\v{x}_i^t\bigg)+i\v{p}_k^t\cdot\v{x}_{k+1}^t
-\sum_{i=k+1}^{j-1}\delta z n_0(z_k)\medspace\bigg(\sigma_{el}^{(1)}(\v{0})-\sigma_{el}^{(1)}(\v{x}_i^t)\bigg),
\end{align}
where $\delta \v{x}_i^t = \v{x}_{i+1}^t-\v{x}_{i}^t$. This organization
produces for $P_\gamma(\v{p}_j,\v{p}_k)$ then
\begin{align}
P_\gamma^{(n)}(\v{p}_j,\v{p}_k)= 2\pi\beta_e\delta(p_j^0-p_k^0)\exp\left(-i\frac{\omega
  m_e^2}{2p_0^0(p_0^0-\omega)}(z_j-z_{k+1})\right)\left(\prod_{i=k+1}^{j}\int
d^2\v{x}_i^t\right)\nonumber\\ \left(\prod_{i=k+1}^{j-1}
\int\frac{d^2\v{p}_i^t}{(2\pi)^2}\right)\exp\left(-i\v{p}_j^t\cdot\v{x}_j^t-i\sum_{i=k+1}^{j-1}\delta
z\medspace\left(\frac{\omega}{2p_0^0(p_0^0-\omega)}\left(\v{p}_i^t-\frac{p_0^0-\omega}{\omega}\v{k}^t\right)^2\right.\right.\nonumber\\
\left.\left.-\v{p}_i^t\cdot\frac{\delta\v{x}_i}{\delta
  z}\right)+\sum_{i=k+1}^{j-1}\delta
z \medspace
n_0(z_k)\bigg(-\sigma_{el}^{(1)}(\v{0})+\sigma_{el}^{(1)}(\v{x}_k^t)\bigg)+i\v{p}_k^t\cdot\v{x}_{k+1}\right).
\end{align}
By integrating in the momentum variables we obtain
\begin{align}
&\left(\prod_{i=k+1}^{j-1}\int
\frac{d^2\v{p}_i}{(2\pi)^2}\right)\exp\left(-i\sum_{i=k+1}^{j-1}\delta
z\medspace\left(\frac{\omega}{2p_0^0(p_0^0-\omega)}\left(\v{p}_i^t-\frac{p_0^0-\omega}{\omega}\v{k}^t\right)^2-\v{p}_i^t\cdot\frac{\delta\v{x}_i}{\delta
  z}\right)\right)\nonumber\\
&=\left(\prod_{i=k+1}^{j-1}\frac{p_0^0(p_0^0-\omega)}{2\pi i \omega \delta z}\right)
\exp\left(i\sum_{i=k+1}^{j-1}\delta z\bigg(\frac{p_0^0(p_0^0-\omega)}{2\omega}\left(\frac{\delta\v{x}_i}{\delta
    z}\right)^2+\frac{p_0^0-\omega}{\omega}\v{k}\cdot\frac{\delta\v{x}_i}{\delta z}\bigg)\right).
\end{align}
The derivative term $\delta \v{x}_k/\delta z$ can be directly integrated since
it does not depend on the path. By taking the $\delta z\to 0$ limit we find
\begin{align}
P_\gamma^{(n)}(\v{p}_j,\v{p}_k)=
2\pi\beta_e\delta(p_j^0-p_k^0)\exp\left(-i\frac{\omega
  m_e^2}{2p_0^0(p_0^0-\omega)}(z_j-z_k)\right) \nonumber \int
d^2\v{x}_j^t\int
d^2\v{x}_{k}^t\\
\exp\left(-i\v{p}_j^t\cdot\v{x}_j^t+i\frac{p_0^0-\omega}{\omega}\v{k}^t\cdot\v{x}_j^t\right) \hat{P}_\gamma(\v{x}_j^t,\v{x}_j^t)
\exp\left(+i\v{p}_k^t\cdot\v{x}_{k}^t-i\frac{p_0^0-\omega}{\omega}\v{k}^t\cdot\v{x}_{k}^t\right),
\end{align}
where we defined the function $\hat{P}_\gamma^{(n)}(\v{x}_j^t,\v{x}_k^t)$, the
Fourier transform of $P_\gamma^{(n)}(\v{p}_j,\v{p}_k)$, as the path integral
\begin{align}
\hat{P}_\gamma(\v{x}_j^t,\v{x}_k^t)\equiv \int_{\v{x}_i^t}^{\v{x}_j^t}
\mathcal{D}\v{x}_t^2(z) \exp\left(i\int_{z_k}^{z_j}dz\medspace
  \left(\frac{p_0^0(p_0^0-\omega)}{2\omega}\dot{\v{x}_t}^2(z)\right.\right.\\
\left.\left.+in_0(z)\bigg(\sigma_{el}^{(1)}(\v{0})-\sigma_{el}^{(1)}\big(\v{x}_t(z)\big)\bigg)\right)\right)\nonumber.
\end{align}
Further simplifications can be done still at \eqref{path_integral_discrete}.
We notice that since at high energies $\v{k}\simeq (\v{k}_t,\omega-\v{k}_t^2/2\omega)$ and
$\v{p}\simeq(\v{p}_t,p_0^0-\omega-\v{p}_t^2/2(p_0^0-\omega))$ the polarizations can be written as 
\begin{align}
(\v{k}\times \v{p}_j)(\v{k}\times \v{p}_i)\simeq \omega^2\bigg(\v{p}_j^t-\frac{p_0^0-\omega}{\omega}\v{k}_t\bigg)\bigg(\v{p}_i^t-\frac{p_0^0-\omega}{\omega}\v{k}_t\bigg).
\end{align}
Since these terms appear also in the exponentials of
$P_\gamma(\v{p}_j,\v{p}_k)$ they can be introduced
as in \eqref{path_integral_discrete} with derivatives,
\begin{align}
(\v{k}\times \v{p}_j)P_\gamma(\v{p}_j,\v{p}_k)(\v{k}\times \v{p}_i)
= 2\pi\beta_e\delta(p_j^0-p_k^0)\exp\left(-i\frac{\omega
  m_e^2}{2p_0^0(p_0^0-\omega)}(z_j-z_k)\right) \nonumber \\\times\omega^2\int
d^2\v{x}_j^t\int
d^2\v{x}_{k}^t
\exp\left(-i\v{p}_j^t\cdot\v{x}_j^t+i\frac{p_0^0-\omega}{\omega}\v{k}^t\cdot\v{x}_j^t\right) \left(\frac{\partial}{\partial \v{x}_j^t}\cdot\frac{\partial}{\partial \v{x}_k^t}\hat{P}_\gamma(\v{x}_j^t,\v{x}_j^t)\right)\nonumber\\\times
\exp\left(+i\v{p}_k^t\cdot\v{x}_{k}^t-i\frac{p_0^0-\omega}{\omega}\v{k}^t\cdot\v{x}_{k}^t\right).
\end{align}
The remaining integrations in intermediate momenta $\v{p}_j$ and $\v{p}_k$ and
final momentum $\v{p}_n$ integrations are now trivial. If we define
the Fourier transform of $P_{inc}^{(n)}(\v{p}_k,\v{p}_0)$ as
\begin{align}
\hat{P}_{inc}^{(n)}(\v{x}_k^t)=\exp\left(-\bigg(\sigma_{el}^{(1)}(\v{0})-\sigma_{el}^{(1)}(\v{x}_k^t)\bigg)\int_0^{z_k}
dz \medspace n_0(z)\right),\nonumber
\end{align}
we finally obtain
\begin{align}
&\omega \frac{dI^{(n)}}{d\omega d\Omega_k} = \left(\frac{\omega e}{2\pi}\right)^2
\frac{h^n(y)}{(p_0^0)^2}2\Re\int_{-\infty}^{+\infty} dz_j \int_{-\infty}^{z_j} dz_k\exp\left(-i\frac{\omega
  m_e^2}{2p_0^0(p_0^0-\omega)}(z_j-z_k)\right)\nonumber\\&
\int d^2\v{x}_k^t
\hat{P}_{inc}^{(n)}(\v{x}_k^t)\exp\left(-i\left(\frac{p_0^0-\omega}{\omega}\v{k}^t-\v{p}_0^t\right)\cdot\v{x}_{k}^t\right)\left(\frac{\partial}{\partial \v{x}_j^t}\cdot\frac{\partial}{\partial \v{x}_k^t}\hat{P}^{(n)}_\gamma(\v{x}_j^t,\v{x}_k^t)\right)\Bigg|_{\v{x}_j=0}.\label{path_integral_continuous}
\end{align}
The vacuum overall subtraction at \eqref{central_equation_qed} has to be
still evaluated in this continuous limit. However, it is easy to see that we
only have to make $\sigma_{el}^{(1)}(\v{x})=0$ in the above expression,
i.e. replace the functions $P_{inc}^{(n)}(\v{x})$ by $P_{inc}^{(0)}(\v{x})$
and $P_{\gamma}^{(n)}(\v{x})$ by $P_{\gamma}^{(0)}(\v{x})$ so we arrive at
\begin{align}
&\omega \frac{dI^{(0)}}{d\omega d\Omega_k} = \left(\frac{\omega e}{2\pi}\right)^2
\frac{h^n(y)}{(p_0^0)^2}2\Re\int_{-\infty}^{+\infty} dz_j \int_{-\infty}^{z_j} dz_k\exp\left(-i\frac{\omega
  m_e^2}{2p_0^0(p_0^0-\omega)}(z_j-z_k)\right)\nonumber\\&
\int d^2\v{x}_k^t
\hat{P}^{(0)}(\v{x}_k^t)\exp\left(-i\left(\frac{p_0^0-\omega}{\omega}\v{k}^t-\v{p}_0^t\right)\cdot\v{x}_{k}^t\right)\left(\frac{\partial}{\partial \v{x}_j^t}\cdot\frac{\partial}{\partial \v{x}_k^t}\hat{P}^{(0)}_\gamma(\v{x}_j^t,\v{x}_k^t)\right)\Bigg|_{\v{x}_j=0}.
\end{align}
This term provides the right perturbative/opacity expansion of the intensity
and becomes necessary when a small number of collisions is expected. It also
becomes necessary for an evaluation beyond the Fokker-Planck/Gaussian limit of
the path integrals, if it were possible, since it guarantees the convergency
at large impact parameters. The Fokker-Planck approximation, however,
does not require such a regularizator, since in the large number of collisions
the elastic incoherent averages $\Sigma_2^{(n)}(\v{q},\delta z)$ rapidly
converge at large $\v{x}$. We present now the Fokker-Planck approximation of
this continuous limit. Since the amplitude has been split in three
zones in the $z$ integration, the intensity leads to nine zones. Six of them,
however, are related by conjugation with the interchange of $z_j$ and $z_k$. So
we find,
\begin{align}
\int_l^\infty \int_l^\infty+\int_{-\infty}^0
\int_{-\infty}^0+\left(\int_l^\infty
\int_{-\infty}^0+\int_{-\infty}^0\int_l^\infty\right)+\int_0^l
\int_0^l\nonumber\\
+\left(\int_0^l
\int_{-\infty}^0+\int_{-\infty}^0\int_0^l\right)+\left(\int_0^l
\int_l^\infty+\int_l^\infty\int_0^l\right).
\end{align}
From here onwards we take the soft photon limit thus $p_0^0-\omega \simeq
p_0^0$ and $h^n(y)\simeq 1$ for simplicity. We also set the $z$ axis in the
initial electron direction so $\v{p}_0^t=\v{0}$. Using
\eqref{path_integral_continuous} the first term $(a)$, representing the square
of the photon emitted at the last leg, is given by
\begin{align}
&\omega \frac{dI^{(n)}_a}{d\omega d\Omega_k} = \left(\frac{e}{2\pi}\right)^2
\left(\frac{\omega}{p_0^0}\right)^2\int_{l}^{\infty} dz_j \int_{l}^{\infty} dz_k\exp\left(-i\frac{\omega
  m_e^2}{2(p_0^0)^2}(z_j-z_k)\right)\nonumber\\&
\times\int d^2\v{x}_k^t
\hat{P}_{inc}^{(n)}(\v{x}_k^t)\exp\left(-i\frac{p_0^0}{\omega}\v{k}^t\cdot\v{x}_{k}^t\right)\left(\frac{\partial}{\partial \v{x}_j^t}\cdot\frac{\partial}{\partial \v{x}_k^t}\hat{P}^{(n)}_\gamma(\v{x}_j^t,\v{x}_k^t)\right)\Bigg|_{\v{x}_j=0}.\label{path_integral_continuous_terma_definition}
\end{align}
We will assume constant density $n_0(z)\equiv n_0$. In the Fokker-Planck
approximation for $\hat{P}_{inc}^{(n)}(\v{x}_k^t)$ and
$\hat{P}^{(n)}_\gamma(\v{x}_j^t,\v{x}_k^t)$ we truncate at leading order
the interactions in $\v{x}$ 
\begin{align}
\exp\bigg(-n_0\delta z\sigma_{el}^{(1)}(\v{0})+n_0\delta
z\sigma_{el}^{(1)}(\v{x})\bigg)\approx\exp\left(-\frac{1}{2}\hat{q}\v{x}^2\right),
\end{align}
and the fall off of the interaction at large $\v{x}$ is now guaranteed due to
the screening neglection. The resulting momentum distribution is
Gaussian. Similarly in the propagation with phase
$\hat{P}^{(n)}_\gamma(\v{x}_j,\v{x}_k)$ we find a Gaussian path integral
\begin{align}
\exp\left(i\int_{z_k}^{z_j} dz\medspace 
  \left(\frac{(p_0^0)^2}{2\omega}\dot{\v{x}_t}^2(z)+in_0\left(\sigma_{el}^{(1)}(\v{0})-\sigma_{el}^{(1)}(\v{x}_t(z))\right)\right)\right)\nonumber\\
\approx\exp\left(i\int_0^l dz\medspace 
  \left(\frac{1}{2}m_{ef}\dot{\v{x}}^2(z)-\frac{1}{2}m_{ef}\Omega^2\v{x}^2\right)\right),\nonumber
\end{align}
whose effective mass can be read in the kinetic term $m_{ef}=(p_0^0)^2/\omega$
and then the harmonic oscillator frequency $\Omega$ has to be defined as
\begin{align}
m_{ef}\Omega^2 = -i\hat{q} \to \Omega =
\frac{1-i}{\sqrt{2}}\sqrt{\frac{\hat{q}\omega}{(p_0^0)^2}}.
\end{align}
With these definitions we simply outline the result of integrating
\eqref{path_integral_continuous_terma_definition},
\begin{align}
\omega \frac{dI^{(n)}_a}{d\omega d\Omega_k} =
\left(\frac{e}{2\pi}\right)^2\int^{\infty}_0dz\medspace z
\exp\left(-\frac{\omega}{2}\left(\frac{m_e}{p_0^0}\right)^2\eta_a^{(1)}-\frac{\v{k}_t^2}{2\omega}\eta_a^{(2)}\right)
\left(\v{k}_t^2\eta_a^{(3)}+\eta_a^{(4)}\right),
\end{align}
where the functions $\eta_a^{(i)}$ are defined as
\begin{align}
\eta_a^{(1)}\equiv z, \medspace
\eta_a^{(2)}\equiv\frac{z}{1+iz\Omega^2l},\medspace
\eta_a^{(3)}\equiv\frac{1}{(1+iz\Omega^2l)^3},\medspace \eta_a^{(4)}\equiv
\frac{2i\omega\Omega}{(1+iz\Omega^2l)^2}.
\end{align}
Similarly to this term we find the contribution $(b)$ corresponding to the
photon emerging from the first leg squared. It is given by
\begin{align}
&\omega \frac{dI^{(n)}_b}{d\omega d\Omega_k} = \left(\frac{e}{2\pi}\right)^2
\left(\frac{\omega}{p_0^0}\right)^2\int_{-\infty}^{0} dz_j \int_{-\infty}^{0} dz_k\exp\left(-i\frac{\omega
  m_e^2}{2(p_0^0)^2}(z_j-z_k)\right)\nonumber\\&\times
\int d^2\v{x}_k^t
\hat{P}_{inc}^{(n)}(\v{x}_k^t)\exp\left(-i\frac{p_0^0}{\omega}\v{k}^t\cdot\v{x}_{k}^t\right)\left(\frac{\partial}{\partial \v{x}_j^t}\cdot\frac{\partial}{\partial \v{x}_k^t}\hat{P}^{(n)}_\gamma(\v{x}_j^t,\v{x}_k^t)\right)\Bigg|_{\v{x}_j=0}.\label{path_integral_continuous_termb_definition}
\end{align}
It produces
\begin{align}
\omega \frac{dI^{(n)}_b}{d\omega d\Omega_k} &= \left(\frac{e}{2\pi}\right)^2
\int_0^\infty dz \medspace z
\exp\left(-\frac{\omega}{2}\left(\frac{m_e}{p_0^0}\right)^2\eta_b^{(1)}-\frac{\v{k}_t^2}{2\omega}\eta_b^{(2)}\right)\left(\v{k}_t^2\eta_b^{(3)}+\eta_b^{(4)}\right),
\end{align}
where as a consistency check the functions $\eta_b^{(i)}$ have to be given by
the $\hat{q}=0$ evaluation of the previous functions $\eta_a^{(i)}$. Indeed
\begin{align}
\eta_b^{(1)}=z,\medspace \eta_b^{(2)}=z, \medspace \eta_b^{(3)}=1, \medspace
\eta_b^{(4)}=0.
\end{align}
Correspondingly we can integrate in the longitudinal space to obtain
\begin{align}
\omega \frac{dI^{(n)}_b}{d\omega d\Omega_k} &= \left(\frac{e}{2\pi}\right)^2\v{k}_t^2
\int_0^\infty dz \medspace z
\exp\left(-\left(\frac{\omega}{2}\left(\frac{m_e}{p_0^0}\right)^2+\frac{\v{k}_t^2}{2\omega}\right)z\right)\nonumber=
\frac{\left(\frac{e}{2\pi}\right)^2\v{k}_t^2}{\left(\frac{\omega}{2} \left(\frac{m_e}{p_0^0}\right)^2+\frac{\v{k}^2}{2\omega}\right)^2}.
\end{align}
Observe that this is just $(\v{k}\times\v{p}_0)^2/(k_\mu p^\mu_0)^2$, as
expected. The next term $(c)$ corresponds to the interference between the
photon emitted in the last leg and the photon emitted in the first leg. It is
given by
\begin{align}
\omega \frac{dI^{(n)}_c}{d\omega d\Omega_k} &= \left(\frac{e}{2\pi}\right)^2
\left(\frac{\omega}{p_0^0}\right)^2\int_{l}^{\infty} dz_j \int_{-\infty}^{0} dz_k\exp\left(-i\frac{\omega
  m_e^2}{2(p_0^0)^2}(z_j-z_k)\right)\nonumber\\
\times\int &d^2\v{x}_k^t
\hat{P}_{inc}^{(n)}(\v{x}_k^t)\exp\left(-i\frac{p_0^0}{\omega}\v{k}^t\cdot\v{x}_{k}^t\right)\left(\frac{\partial}{\partial \v{x}_j^t}\cdot\frac{\partial}{\partial \v{x}_k^t}\hat{P}^{(n)}_\gamma(\v{x}_j^t,\v{x}_k^t)\right)\Bigg|_{\v{x}_j=0}+c.c.,\label{path_integral_continuous_termc_definition}
\end{align}
and produces
\begin{align}
\omega \frac{dI^{(n)}_c}{d\omega d\Omega_k} &= \left(\frac{e}{2\pi}\right)^22\Re \int_0^\infty dz_j \int_0^\infty dz_k
\exp\left(-\frac{\omega}{2}\left(\frac{m_e}{p_0^0}\right)^2\eta_c^{(1)}-\frac{\v{k}_t^2}{2\omega}\eta_c^{(2)}\right)\nonumber\\
&\times\left(\v{k}_t^2\eta_c^{(3)}+\eta_c^{(4)}\right),
\end{align}
where the functions $\eta_c^{(i)}$ are given by
\begin{align}
\eta_c^{(1)}&= il+z_j+z_k,\medspace \eta_c^{(2)}=\frac{i\sin(\Omega l)+\Omega\cos(\Omega l)(z_j+z_k)+iz_jz_k\Omega^2\sin(\Omega l)}{\Omega\cos(\Omega l)+iz_j\Omega^2\sin(\Omega l)}\nonumber\\
\eta_c^{(3)}&=\frac{\Omega^2}{\left(\Omega\cos(\Omega
  l)+iz_j\Omega^2\sin(\Omega l)\right)^2}, \medspace \eta_c^{(4)}=0.
\end{align}
The term $(d)$ contains all the photons emitted from the internal legs and
their respective interferences. It is given by
\begin{align}
\omega \frac{dI^{(n)}_d}{d\omega d\Omega_k} &= \left(\frac{e}{2\pi}\right)^2
\left(\frac{\omega}{p_0^0}\right)^2\int_{0}^{l} dz_j \int_{0}^{l} dz_k\exp\left(-i\frac{\omega
  m_e^2}{2(p_0^0)^2}(z_j-z_k)\right)\nonumber\\
\times\int &d^2\v{x}_k^t
\hat{P}_{inc}^{(n)}(\v{x}_k^t)\exp\left(-i\frac{p_0^0}{\omega}\v{k}^t\cdot\v{x}_{k}^t\right)\left(\frac{\partial}{\partial \v{x}_j^t}\cdot\frac{\partial}{\partial \v{x}_k^t}\hat{P}^{(n)}_\gamma(\v{x}_j^t,\v{x}_k^t)\right)\Bigg|_{\v{x}_j=0}+c.c.\medspace\medspace.\label{path_integral_continuous_termd_definition}
\end{align}
The integration produces, using $\delta z \equiv z_j-z_k$,
\begin{align}
\omega \frac{dI^{(n)}_d}{d\omega d\Omega_k} &= \left(\frac{e}{2\pi}\right)^22\Re \int_0^l dz_j \int_0^{z_j} dz_k
\exp\left(-\frac{\omega}{2}\left(\frac{m_e}{p_0^0}\right)^2\eta_d^{(1)}-\frac{\v{k}_t^2}{2\omega}\eta_d^{(2)}\right)\nonumber\\
&\times\left(\v{k}_t^2\eta_d^{(3)}+\eta_d^{(4)}\right),
\end{align}
where the functions $\eta_d^{(i)}$ are given by $\eta_d^{(1)}=i(z_j-z_k)$ and
\begin{align}
\eta_d^{(2)}&=\frac{i\sin\left(\Omega(z_j-z_k)\right)}{\Omega\cos\left(\Omega(z_j-z_k)\right)-z_k\Omega^2\sin\left(\Omega(z_j-z_k)\right)},\nonumber\\
\eta_d^{(3)}&=\frac{\cos\left(\Omega(z_j-z_k)\right)}{\left(\cos\left(\Omega(z_j-z_k)\right)-z_k\Omega\sin\left(\Omega(z_j-z_k)\right)\right)^3},\nonumber\\
 \eta_d^{(4)}&=\frac{2i\omega\Omega^2z_k}{\left(\cos\left(\Omega(z_j-z_k)\right)-z_k\Omega\sin\left(\Omega(z_j-z_k)\right)\right)^2}.
\end{align}
The term $(e)$ containing the interference between the internal photons and
the photon emitted in the first leg is given by
\begin{align}
\omega \frac{dI^{(n)}_e}{d\omega d\Omega_k} &= \left(\frac{e}{2\pi}\right)^2
\left(\frac{\omega}{p_0^0}\right)^2\int_{0}^{l} dz_j \int_{-\infty}^{0} dz_k\exp\left(-i\frac{\omega
  m_e^2}{2(p_0^0)^2}(z_j-z_k)\right)\label{path_integral_continuous_terme_definition}\\
\times\int &d^2\v{x}_k^t
\hat{P}_{inc}^{(n)}(\v{x}_k^t)\exp\left(-i\frac{p_0^0}{\omega}\v{k}^t\cdot\v{x}_{k}^t\right)\left(\frac{\partial}{\partial
  \v{x}_j^t}\cdot\frac{\partial}{\partial
  \v{x}_k^t}\hat{P}^{(n)}_\gamma(\v{x}_j^t,\v{x}_k^t)\right)\Bigg|_{\v{x}_j=0}+c.c.\medspace\medspace\medspace .\nonumber
\end{align}
This term produces
\begin{align}
\omega \frac{dI^{(n)}_e}{d\omega d\Omega_k} &=
-\left(\frac{e}{2\pi}\right)^22\Re \int_0^l dz_j \int_0^\infty dz_k\exp\left(-\frac{\omega}{2}\left(\frac{m_e}{p_0^0}\right)^2\eta_e^{(1)}-\frac{\v{k}_t^2}{2\omega}\eta_e^{(2)}\right)\nonumber\\
&\times\left(\v{k}_t^2\eta_e^{(3)}+\eta_e^{(4)}\right),
\end{align}
where the functions $\eta_e^{(i)}$ are given by
\begin{align}
\eta_e^{(1)}=i(z_j-z_k),\medspace \eta_e^{(2)}=-iz_k+i\frac{\sin(\Omega
  z_j)}{\Omega\cos(\Omega z_j)},\medspace \eta_e^{(3)}=\frac{1}{\cos^2(\Omega
  z_j)},\medspace \eta_e^{(4)}=0.
\end{align}
And finally the term $(f)$ containing the interferences between the photon
emitted in the last leg with the internal photons. It is given by
\begin{align}
\omega \frac{dI^{(n)}_f}{d\omega d\Omega_k} &= \left(\frac{e}{2\pi}\right)^2
\left(\frac{\omega}{p_0^0}\right)^2\int_{0}^{l} dz_j \int_{l}^{\infty} dz_k\exp\left(-i\frac{\omega
  m_e^2}{2(p_0^0)^2}(z_j-z_k)\right)\label{path_integral_continuous_termf_definition}\\
\times\int &d^2\v{x}_k^t
\hat{P}_{inc}^{(n)}(\v{x}_k^t)\exp\left(-i\frac{p_0^0}{\omega}\v{k}^t\cdot\v{x}_{k}^t\right)\left(\frac{\partial}{\partial
  \v{x}_j^t}\cdot\frac{\partial}{\partial
  \v{x}_k^t}\hat{P}^{(n)}_\gamma(\v{x}_j^t,\v{x}_k^t)\right)\Bigg|_{\v{x}_j=0}+c.c.,\nonumber
\end{align}
and produces using $t=(l-z_k)$
\begin{align}
\omega \frac{dI^{(n)}_f}{d\omega d\Omega_k} &=
\left(\frac{e}{2\pi}\right)^22\Im \int_0^l dz_j \int_0^\infty
dz_k\exp\left(-\frac{\omega}{2}\left(\frac{m_e}{p_0^0}\right)^2\eta_f^{(1)}-\frac{\v{k}_t^2}{2\omega}\eta_f^{(2)}\right)\nonumber\\
&\times\left(\v{k}_t^2\eta_f^{(3)}+\eta_f^{(4)}\right),
\end{align}
where the functions $\eta_f^{(i)}$ are given by $\eta_f^{(1)}=i\delta +z_k$
where $\delta = l-z_j$ and
\begin{align}
 \eta_f^{(2)}=\frac{i\sin\left(\Omega\delta\right)+z_k\Omega\cos\left(\Omega\delta\right)}{\Omega\cos\left(\Omega\delta\right)-z_j\Omega^2\sin\left(\Omega
   \delta\right)+iz_k\Omega^2\left(\sin\left(\Omega\delta\right)+z_j\Omega\cos\left(\Omega
   \delta\right)\right)}\nonumber\\
\eta_f^{(3)}=\frac{\cos(\Omega\delta)+iz_k\sin(\Omega\delta)}{\left(\cos\left(\Omega\delta\right)-z_j\Omega\sin\left(\Omega
   \delta\right)+iz_k\Omega\left(\sin\left(\Omega\delta\right)+z_j\Omega\cos\left(\Omega
   \delta\right)\right)\right)^3}\nonumber\\
\eta_f^{(3)}=\frac{2i\omega \Omega^2z_j}{\left(\cos\left(\Omega\delta\right)-z_j\Omega\sin\left(\Omega
   \delta\right)+iz_k\Omega\left(\sin\left(\Omega\delta\right)+z_j\Omega\cos\left(\Omega
   \delta\right)\right)\right)^2}.
\end{align}
The sum of the above six contributions constitute the path integral version of
the intensity in the Fokker-Planck approximation for mediums of arbitrary length
\begin{align}
\omega \frac{dI^{(n)}_{inc}}{d\omega d\Omega_k}-\omega\frac{dI_{inc}^{(0)}}{d\omega
  d\Omega_k}=\sum_{i=a}^f\omega \frac{dI^{(n)}_i}{d\omega d\Omega_k}.\label{central_equation_pathintegral_qed}
\end{align}
In Fig. \ref{fig:figure_3_2} and Fig. \ref{fig:figure_3_3} we show the
evaluation of these six path integrals
\eqref{central_equation_pathintegral_qed} for several photon angles and a Gold
target of $l$ = 0.0023 cm and an electron energy of $p_0^0$ = 8 GeV. We find a
complete agreement with the direct numerical evaluation of the
discretized approach \eqref{central_equation_qed}, as expected since we have
chosen for that purpose $\delta z\ll \lambda=1/n_0\sigma_t^{(1)}$.

If the semi-infinite medium length is to be considered the separation of the
space in three zones stops having sense. We can define in that case the
quantity 
\begin{align}
\omega \frac{dI^{(n)}_{inc}}{d\omega d\Omega_k}\equiv \int^{l}_0\int^{l}_0+\int_{-\infty}^0\int_{-\infty}^0=\omega \frac{dI^{(n)}_d}{d\omega d\Omega_k}+\omega \frac{dI^{(n)}_b}{d\omega d\Omega_k}.\label{migdal_qed_definition}
\end{align}
The integration in $\v{k}_t$ without kinematical restrictions can be easily be
done using $\omega^2d\Omega_k=d^2\v{k}_t$, producing
\begin{align}
\frac{1}{\omega^2}\int\frac{d^2\v{k}_t}{(2\pi)^2}
\exp\left(-\frac{\v{k}_t^2}{2\omega}\eta_d^{(2)}\right)\left(\v{k}_t^2\eta_d^{(3)}+\eta_d^{(4)}\right)=-\frac{1}{\pi}\frac{\Omega^2}{\sin^2\left(\Omega(z_j-z_k)\right)},
\end{align}
and
\begin{align}
\frac{1}{\omega^2}\int\frac{d^2\v{k}_t}{(2\pi)^2}
\exp\left(-\frac{\v{k}_t^2}{2\omega}\eta_b^{(2)}\right)\left(\v{k}_t^2\eta_b^{(3)}+\eta_b^{(4)}\right)=\frac{1}{\pi}\frac{1}{(z_j-z_k)^2}.
\end{align}
Correspondingly for the particular case $l\to\infty$ we find
\begin{align}
\omega \frac{dI^{(n)}_{inc}}{d\omega}=\frac{2e^2}{\pi}\Re\int_0^\infty &dz_j
\int_{0}^{z_j} dz_k
\exp\left(-i\frac{\omega}{2}\left(\frac{m_e}{p_0^0}\right)^2(z_j-z_k)\right)\nonumber\\
&\times\left(\frac{1}{(z_j-z_k)^2}-\frac{\Omega^2}{\sin^2\left(\Omega(z_j-z_k)\right)}\right)\label{migdal_prediction},
\end{align}
which is Migdal's result \cite{migdal1956,bell1958} for the $\phi(s)$
function. By performing one of the trivial integrals, in which a factor
proportional to $l$ arises, and by rotating the contour of integration to
avoid oscillations, we find a suitable expression for the intensity for a
medium of length $l\to\infty$, of the form
\begin{align}
\omega \frac{dI_{inc}^{(n)}}{d\omega} =l
\frac{2e^2}{\pi}\frac{|\Omega|}{\sqrt{2}}\int_0^\infty dz&
\exp\left(-\frac{z}{\sqrt{2}s}\right)\nonumber\\
&\times\left(\sin\left(\frac{z}{\sqrt{2}s}\right)+\cos\left(\frac{z}{\sqrt{2}{s}}\right)\right)\left(\frac{1}{z^2}-\frac{1}{\sinh^2(z)}\right),\label{migdal_solution}
\end{align}
where the parameter $s$ is given by
\begin{align}
s\equiv\frac{2}{\omega}\left(\frac{p_0^0}{m_e}\right)^2\left|\Omega\right|=\frac{2p_0^0}{m_e^2}\sqrt{\frac{\hat{q}}{\omega}}.
\end{align}
An useful approximant to the above integral within less than a 1$\%$ of
deviation in all the range is given by
\begin{align}
\omega \frac{dI^{(n)}_{inc}}{d\omega} \simeq
l\frac{2e^2}{3\pi}\sqrt{\frac{\hat{q}\omega}{(p_0^0)^2}}s\frac{1-1.52s^4+5.8s^5}{1+2.44s^5+2.73s^6}\label{migdal_approximant}.
\end{align}
The other relevant prescription for the infinite length approximation is
alternatively given by the quantity
\begin{align}
\left|\int_{-\infty}^l\right|^2-\left|\int_{-\infty}^0\right|^2=\int^l_0\int^l_0+2\Re\int_{-\infty}^0\int^l_0
\equiv \omega \frac{dI^{(n)}_d}{d\omega d\Omega_k}+\omega \frac{dI^{(n)}_e}{d\omega d\Omega_k},\label{bdmps_qed_definition}
\end{align}
As with the prescription leading to Migdal solution, since the final leg
photon disappears when $l\to\infty$ and the initial photon keeps impaired, it
has been removed in order to avoid a $\log(\omega)$ divergence in the angular
integration. The angle integration of the second term $(e)$ on the right hand
side produces
\begin{align}
\omega \frac{dI^{(n)}_e}{d\omega}=\frac{2e^2}{\pi}\Re\int_0^l &dz_j \int_0^\infty dz_k
\exp\left(-i\frac{\omega}{2}\left(\frac{m_e}{p_0^0}\right)^2(z_j-z_k)\right)\nonumber\\
&\times\frac{\Omega^2}{\left(\sin\left(\Omega z_j\right)-\Omega
  z_k \cos\left(\Omega z_j\right)\right)^2}.
\end{align}
Then by taking $l\to\infty$ we find
\begin{align}
\omega \frac{dI^{(n)}}{d\omega}=\frac{2e^2}{\pi}\Re\int_0^\infty &dz_j \int_0^\infty dz_k
\exp\left(-i\frac{\omega}{2}\left(\frac{m_e}{p_0^0}\right)^2(z_j-z_k)\right)\nonumber\\
&\times\left(\frac{\Omega^2}{\left(\sin\left(\Omega z_j\right)-\Omega
  z_k \cos\left(\Omega z_j\right)\right)^2}-\frac{\Omega^2}{\sin^2\left(\Omega(z_j-z_k)\right)}\right).
\end{align}
For the particular albeit unrealistic case in which the fermion mass can be
neglected the integral in $z_k$ can be performed and we find
\begin{align}
\omega \frac{dI^{(n)}_d}{d\omega}=\frac{2e^2}{\pi}\Omega\Re\int_0^l
dz \medspace \frac{\cos(\Omega z)}{\sin(\Omega z)},
\end{align}
for the first term and
\begin{align}
\omega \frac{dI^{(n)}_e}{d\omega}=-\frac{2e^2}{\pi}\Re\Omega\int_0^l &dz 
\frac{1}{\cos\left(\Omega z\right)\sin\left(\Omega z\right)}.
\end{align} 
Correspondingly
\begin{align}
\omega \frac{dI^{(n)}_d}{d\omega}+\omega
\frac{dI^{(n)}_e}{d\omega}&=\frac{2e^2}{\pi}\Re\Omega \int_0^l dz\left(
\frac{\cos(\Omega z)}{\sin(\Omega z)}-\frac{1}{\cos\left(\Omega
  z\right)\sin\left(\Omega z\right)}\right)\nonumber\\
&=\frac{2e^2}{\pi}\Re\left[\log\left(\cos\bigg(\Omega
l\bigg)\right)\right],\label{bdmps_qed_result}
\end{align}
which is the BDMPS result \cite{baier1996}. Unfortunately, the neglection of
the projectile mass leads to a continuous enhancement of the intensity in the
regime of maximal interference, i.e. $\omega \gg 1$, instead of the expected
Bethe-Heitler incoherent plateau \eqref{classicalintensity_incoherentplateau}.
Then the BDMPS result does not reproduce neither Weinberg's soft photon
theorem nor the Bethe-Heitler cross section.

In Figures \ref{fig:figure_3_4}, \ref{fig:figure_3_5}, \ref{fig:figure_3_6},
\ref{fig:figure_3_7}, \ref{fig:figure_3_8} and \ref{fig:figure_3_9} the
approximant \eqref{migdal_approximant} to the Migdal prediction
\eqref{migdal_solution} is shown together our evaluations in the Debye
screened interaction and the Fokker-Planck approximation for
\eqref{central_equation_qed}. For small size targets, compromising less than
$n_c$ = 10$^4$ collisions on average the Migdal prediction is not adequate,
since Weinberg's soft photon theorem is not considered.
\subsubsection{Transverse coherent average}
A reorganization of the coherent average currents in \eqref{coherent_average_spinnoflip} and \eqref{coherent_average_spinflip} can be done. We
will restrict to the spin non flip contribution, since the spin flipping
contribution follows the same steps. We also consider the limit $\omega\to 0$
in all the quantities except in the phases. Then we notice that
\begin{align}
\sum_{k=1}^{n}\v{\delta}_k^ne^{i\varphi_k}=\frac{\v{k}\times
  \v{p}_n}{k_\mu p_n^\mu}e^{i\varphi_n} +\sum_{k=1}^{n-1}\frac{\v{k}\times
  \v{p}_k}{k_\mu p_k^\mu}\left(e^{i\varphi_k}-e^{i\varphi_{k+1}}\right)-\frac{\v{k}\times
  \v{p}_0}{k_\mu p_0^\mu}e^{i\varphi_1}.
\end{align}
In the continuous limit the above relation is just an integration by parts,
where the two boundary terms correspond to an extension of the integration to
$(-\infty,z_1]$ and $[z_n,\infty)$ of the interior term. This interior term is
    given by
\begin{align}
\sum_{k=1}^{n-1}\frac{\v{k}\times
  \v{p}_k}{k_\mu p_k^\mu}\left(e^{i\varphi_k}-e^{i\varphi_{k+1}}\right)=\sum_{k=1}^{n-1}\frac{\v{k}\times
  \v{p}_k}{k_\mu p_k^\mu}e^{i\varphi_k}\left(1-e^{i\varphi_{k+1}-i\varphi_{k}}\right).
\end{align}
Following \eqref{phase_definition_coherentaverage} and \eqref{pole_before} the
phase difference is found to be
\begin{align}
i\varphi_{k+1}-i\varphi_k=\frac{i\omega(z_{k+1}-z_k)}{2p_0^0(p_0^0-\omega)}\left(m_e^2+\left(\v{p}_k^t-\frac{p_0^0}{\omega}\v{k}_t\right)^2\right)=i\frac{k_\mu
  p^\mu_k}{p_0^0-\omega}\delta z.
\end{align}
By taking the $\delta z \to 0 $ limit then we find
\begin{align}
\sum_{k=1}^{n-1}\frac{\v{k}\times
  \v{p}_k}{k_\mu
  p_k^\mu}e^{i\varphi}\left(1-\exp\left(i\frac{k_\mu p^\mu_k}{p_0^0-\omega}\delta
z\right)\right)\simeq -\frac{i}{p_0^0}\sum_{k=1}^{n-1}
\delta z\medspace \v{k}\times
  \v{p}_k e^{i\varphi_k}.
\end{align}
The quantity to evaluate at \eqref{coherent_average_spinnoflip} is
just the convolution of the above current with the elastic weights. For the
elements containing interaction this quantity can be always reorganized as
\begin{align}
&\left(\prod_{i=1}^{n-1}\frac{d^3\v{p}_i}{(2\pi)^3}\right)\left(\prod_{i=1}^{n}\left(\phi_{coh}^{(n)}(\delta
p_i,\delta z_i)+\phi_{coh}^{(0)}(\delta p_i,\delta
z_i)\right)\right)\left(\sum_{k=1}^{n}\v{\delta}_k^ne^{i\varphi_k}\right)\nonumber\\
&=-\frac{i}{p_0^0}\exp\left(i\frac{\omega m_e^2z_k}{2p_0^0(p_0^0-\omega)}+i\frac{\v{k}_t^2z_k}{2\omega}\right)\int\frac{d^3\v{p}_k}{(2\pi)^3} P_{coh}^{(n)}(\v{p}_n,\v{p}_k)\v{k}\times\v{p}_k P_{coh}^{(n)}(\v{p}_k,\v{p}_0),\label{coherent_average_intermediatestep1_continuous}
\end{align}
where the functions $P_{coh}^{(n)}(\v{p}_a,\v{p}_b)$ act at the level of the
amplitude and thus cannot be interpreted in probabilistic terms. 
If we define 
\begin{align}
p_n^z=-\frac{(\v{p}_n^t-\v{k}_t)^2}{2(p_0^0-\omega)}, \medspace\medspace\medspace\medspace\medspace
p_k^z=-\frac{(\v{p}_k^t-\v{k}_t)^2}{2(p_0^0-\omega)},\medspace\medspace\medspace\medspace\medspace p_0^z=-\frac{(\v{p}_0^t)^2}{2p_0^0},
\end{align}
and complete the boundary terms to total three momenta as
$\v{p}_t\cdot\v{x}_t+p_zz\equiv \v{p}\cdot\v{x}$ the propagators are given by
Fourier transforms as follows
\begin{align}
P_{coh}^{(n)}(\v{p}_k,\v{p}_0)\equiv& 2\pi\beta_p\delta(p_k^0-p_0^0)\left(\prod_{i=1}^{k-1}\frac{d^2\v{p}_i^t}{(2\pi)^2}\right)\left(\prod_{i=1}^{k}\frac{d^2\v{x}_i^t}{(2\pi)^2}\right)e^{-i\v{p}_k\cdot\v{x}_k+i\v{p}_{0}\cdot\v{x}_{1}}\nonumber\\
&\times\exp\left(-i\sum_{i=1}^{k-1}\delta z\left(
\frac{(\v{p}_i^t)^2}{2p_0^0}-\v{p}_i^t\cdot
\frac{\delta\v{x}_{i}^t}{\delta z}\right)+\sum_{i=1}^{k}\delta z n_0(z_i) \pi_{el}^{(1)}(\v{x}_i^t)\right),
\end{align}
and
\begin{align}
P_{coh}^{(n)}(\v{p}_n,\v{p}_k)&\equiv 2\pi\beta_p\delta(p_n^0-p_k^0)\left(\prod_{i=k+1}^{n-1}\frac{d^2\v{p}_i^t}{(2\pi)^2}\right)\left(\prod_{i=k+1}^{n}\frac{d^2\v{x}_i^t}{(2\pi)^2}\right)e^{-i\v{p}_n\cdot\v{x}_n+i\v{p}_{k}\cdot\v{x}_{k+1}}\nonumber\\
&\times\exp\left(-i\sum_{i=k+1}^{n-1}\delta z\left(
\frac{(\v{p}_i^t-\v{k}_t)^2}{2(p_0^0-\omega)}-\v{p}_i^t\cdot
\frac{\delta\v{x}_{i}^t}{\delta z}\right)+\sum_{i=k+1}^{n}\delta z n_0(z_i) \pi_{el}^{(1)}(\v{x}_i^t)\right),
\end{align}
where $\delta \v{x}_i^t \equiv \v{x}_{i+1}^t-\v{x}_i^t$. The integration in
internal momenta of the first passage produces 
\begin{align}
\left(\prod_{i=1}^{k-1}\frac{ip_0^0}{2\pi\delta
  z}\right)&\left(\prod_{i=2}^{k-1}\frac{d^2\v{x}_i^t}{(2\pi)^2}\right)\exp\left(+i\sum_{i=1}^{k-1}\delta
z_i \frac{p_0^0}{2}\left(\frac{\delta\v{x}_i^t}{\delta z_i}\right)^2+\sum_{i=1}^{k}\delta z_i
n_0(z_i)\pi_{el}^{(1)}(\v{x}_i^t)\right)\nonumber\\
\to &\int\mathcal{D}^2\v{x}_t(z)\exp\left(i\int^{z_k}_{z_1}dz\left(\frac{p_0^0}{2}\dot{\v{x}}_t^2(z)-in_0(z)\pi_{el}^{(1)}(\v{x}_t(z))\right)\right)
\end{align}
whereas the second passage produces
\begin{align}
&\left(\prod_{i=k+1}^{n-1}\frac{i(p_0^0-\omega)}{2\pi \delta
    z}\right)\left(\prod_{i=k+2}^{n-1}\frac{d^3\v{x}_i^t}{(2\pi)^2}\right)\exp\left(+i\sum_{i=k+1}^{n-1}\delta
  z_i
  \left(\frac{p_0^0-\omega}{2}\left(\frac{\delta\v{x}_i^t}{\delta z_i}\right)^2+\v{k}_t\cdot\frac{\delta\v{x}_i^t}{\delta
    z_i}\right)\right.\nonumber\\ &+\left.\sum_{i=k+1}^{n}\delta z_i
  n_0(z_i)\pi_{el}^{(1)}(\v{x}_i^t)\right)\to
  \exp\left(+i\v{k}_t\cdot(\v{x}_n^t-\v{x}_{k+1}^t)\right)\nonumber\\ &\times
  \int\mathcal{D}^2\v{x}_t(z)\exp\left(i\int^{z_n}_{z_{k+1}}dz\left(\frac{p_0^0-\omega}{2}\dot{\v{x}}_t^2(z)-in_0(z)\pi_{el}^{(1)}(\v{x}_t(z))\right)\right).
\end{align}
The remaining integration in $\v{p}_k$ can be done with some care. We notice
that
\begin{align}
\int\frac{d^2\v{p}_k^t}{(2\pi)^2}\left(\v{p}_t-\frac{p_0^0}{\omega}\v{k}_t\right)\exp\left(-i\v{k}_t\cdot\v{x}_{k+1}^t+i\v{p}_k^t\cdot(\v{x}_{k+1}^t-\v{x}_{k}^t)-i\delta
z
\frac{(\v{p}_k^t)^2}{2p_0^0}\right)\nonumber\\ =\left(i\frac{\partial}{\partial
\v{x}_{k}^t}-i\frac{p_0^0}{\omega}\left(\frac{\partial}{\partial
\v{x}_{k+1}^t}+\frac{\partial}{\partial
\v{x}_{k}^t}\right)\right)e^{-i\v{k}_t\cdot\v{x}_{k+1}^t}\frac{ip_0^0}{2\pi\delta
  z}\exp\left(+i\frac{(\v{x}_{k+1}^t-\v{x}_k^t)^2}{2p_0^0\delta
  z}\right).
\end{align}
In the limit $\delta z\to 0$ we find the representation of the Dirac delta
function and then
\begin{align}
\int\frac{d^3\v{p}_k}{(2\pi)^3}
P_{coh}^{(n)}(\v{p}_n,\v{p}_k)\v{k}\times\v{p}_k
P_{coh}^{(n)}(&\v{p}_k,\v{p}_0)=i\omega 2\pi\beta_p\delta(p_n^0-p_0^0)\int d^2\v{x}_n^t
e^{i(\v{k}_t-\v{p}_n^t)\cdot\v{x}_n^t}\nonumber\\ \times\int
d^2\v{x}_1^te^{+i\v{p}_0^t\cdot\v{x}_1^t}\int
d^2\v{x}_k^te^{-i\v{k}_t\cdot\v{x}_k^t}&\left\{\hat{P}_{coh}^{(n)}(\v{x}_n^t,\v{x}_k^t)\left(\left(\frac{p_0^0}{\omega}-1\right)\frac{\partial}{\partial
    \v{x}_k^t}\hat{P}_{coh}^{(n)}(\v{x}_k^t,\v{x}_1^t)\right)\right.\nonumber\\
+\left.\left(\frac{p_0^0}{\omega}\right.\right.&\left.\left.\frac{\partial}{\partial
  \v{x}_k^t}\hat{P}_{coh}^{(n)}(\v{x}_n^t,\v{x}_k^t)\right)\hat{P}_{coh}^{(n)}(\v{x}_k^t,\v{x}_1^t)\right\},\label{coherent_average_intermediatestep2_continuous}
\end{align}
or alternatively
\begin{align}
i\omega 2\pi\beta_p\delta(p_n^0-p_0^0)&\int d^2\v{x}_n^t e^{-i(\v{p}_n^t-\v{k}_t)\cdot\v{x}_n^t}\int
d^2\v{x}_1^te^{+i\v{p}_0^t\cdot\v{x}_1^t}\\
&\times\int
d^2\v{x}_k^te^{-i\v{k}_t\cdot\v{x}_k^t}\hat{P}_{coh}^{(n)}(\v{x}_n^t,\v{x}_k^t)\left\{-\frac{\v{k}_tp_0^0}{\omega}-i\frac{\partial}{\partial
  \v{x}_k^t}\right\}\hat{P}_{coh}^{(n)}(\v{x}_k^t,\v{x}_1^t),\nonumber
\end{align}
where the functions $\hat{P}_{coh}^{(n)}(\v{x}_k^t,\v{x}_1^t)$ an
$\hat{P}_{coh}^{(n)}(\v{x}_n^t,\v{x}_k^t)$ are the path integrals
\begin{align}
\hat{P}_{coh}^{(n)}(\v{x}_n^t,\v{x}_k^t)\equiv \int\mathcal{D}^2\v{x}_t(z)\exp\left(i\int^{z_n}_{z_{k}}dz\left(\frac{p_0^0-\omega}{2}\dot{\v{x}}_t^2(z)-in_0(z)\pi_{el}^{(1)}(\v{x}_t(z))\right)\right),
\end{align}
and
\begin{align}
\hat{P}_{coh}^{(n)}(\v{x}_k^t,\v{x}_1^t)\equiv\int\mathcal{D}^2\v{x}_t(z)\exp\left(i\int^{z_k}_{z_1}dz\left(\frac{p_0^0}{2}\dot{\v{x}}_t^2(z)-in_0(z)\pi_{el}^{(1)}(\v{x}_t(z))\right)\right).
\end{align}
It is easy to note that the term involving no collisions corresponds to making
$n=0$ in the above expressions. Then using
\eqref{coherent_average_spinnoflip}, \eqref{coherent_average_emission},
\eqref{coherent_average_intermediatestep1_continuous} and
\eqref{coherent_average_intermediatestep2_continuous}, we find the non flip
transverse coherent contribution in the continuous limit
\begin{align}
\omega& \frac{dI^{coh}}{d\omega d\Omega_k}=
\left(\frac{e}{2\pi}\right)^2\left(\frac{\omega}{p_0^0}\right)^2\frac{1}{\pi
R^2}\int\frac{d^2\v{p}_n^t}{(2\pi)^2}\nonumber\\
&\times\left|\int dz_k\exp\left(i\frac{\omega
  m_e^2z_k}{2p_0^0(p_0^0-\omega)}+i\frac{\v{k}_t^2z_k}{2\omega}\right)\int d^2\v{x}_n^t e^{-i(\v{p}_n^t-\v{k}_t)\cdot\v{x}_n^t}\int
d^2\v{x}_1^te^{+i\v{p}_0^t\cdot\v{x}_1^t}\right.\nonumber\\
&\times\left.\int d^2\v{x}_k^te^{-i\v{k}_t\cdot\v{x}_k^t}\hat{P}_{coh}^{(n)}(\v{x}_n^t,\v{x}_k^t)\left\{-\frac{\v{k}_tp_0^0}{\omega}-i\frac{\partial}{\partial
  \v{x}_k^t}\right\}\hat{P}_{coh}^{(n)}(\v{x}_k^t,\v{x}_1^t)-(n=0)\right|^2.
\end{align}
Since the Fourier transform of $\pi_{el}^{(1)}(\v{x}_t)$ is an even function
of $\v{q}_t$, the function has a saddle point in $\v{x}_t=0$. By assuming a
constant density $n_0(z)=n_0$ a series truncation at high number of collisions
holds and 
\begin{align}
-in_0\pi_{el}^{(1)}(\v{x}_t)\simeq-\frac{1}{2}\hat{q}\v{x}_t^2,
\end{align}
where by using \eqref{pi_el_single_definition} we find
\begin{align}
\hat{q}=-\frac{\partial^2}{\partial^2\v{x}_t}\pi_{el}^{(1)}(\v{x}_t)\Bigg|_{\v{x}_t=0}=4\pi
gR\int_0^\infty dq\frac{q^2}{q^2+\mu_d^2}J_1(qR) =4\pi g \mu_dRK_1(\mu_dR),
\end{align}
where $K_1(x)$ is the modified Bessel function. We observe that the transverse
coherent average leads to an equivalent charge given by the average
interaction potential. In this Fokker-Planck approximation the path integrals
have a Gaussian form and thus they are solvable,
\begin{align}
\hat{P}_{coh}^{(n)}(\v{x}_n^t,\v{x}_k^t)\equiv \int\mathcal{D}^2\v{x}_t(z)\exp\left(i\int^{z_n}_{z_{k}}dz\left(\frac{p_0^0-\omega}{2}\dot{\v{x}}_t^2(z)-\frac{1}{2}(p_0^0-\omega)\Omega_f^2\v{x}_t^2(z)\right)\right),
\end{align}
and 
\begin{align}
\hat{P}_{coh}^{(n)}(\v{x}_k^t,\v{x}_1^t)\equiv\int\mathcal{D}^2\v{x}_t(z)\exp\left(i\int^{z_k}_{z_1}dz\left(\frac{p_0^0}{2}\dot{\v{x}}_t^2(z)-\frac{1}{2}p_0^0\Omega_i^2\v{x}_t^2(z)\right)\right),
\end{align}
with real oscillator frequencies given by
\begin{align}
\Omega_f=\sqrt{\frac{4\pi g\mu_dRK_1(\mu_dR)}{p_0^0-\omega}},\medspace\medspace\medspace\medspace\Omega_i=\sqrt{\frac{4\pi g\mu_dRK_1(\mu_dR)}{p_0^0}}.
\end{align}
In the limit $R\to\infty$ both $\hat{q}$ and the oscillator frequencies vanish
and the coherent contribution vanishes when the $z$ integration in all the
space $(-\infty,\infty)$ is performed.

%% file: chapter4/chapter4.tex
\chapter{High energy multiple scattering in QCD}

In the following we consider a multiple scattering process for a high energy,
asymptotically free parton, traveling through colored condensed media. The
purpose of this chapter is to summarize a QCD equivalent of the formalism
already developed in the preceding chapters for QED. The main change with
respect to a QED scenario is, of course, the introduction of the color
structure. We only briefly review the SU(3) algebra relations required for our
purpose. In Section \ref{sec:section_4_1} we shortly look for the form of the
color field representing the medium, which will be considered static and
classical. As expected the resulting interaction will have matrix structure in
the color space of the target partons as the main difference with respect to
the scalar field of QED. In Section \ref{sec:section_4_2} we derive the high
energy integration of the scattering amplitude and its main properties, in the
same approach we took for the QED case. The amplitude is shown to inherit the
matrix structure of the target parton vertices and incorporate the matrix
structure of the traveling parton vertex. We show that even the squared single
elastic amplitude, after averaged over color target space, does not allow a
non-perturbative functional form in the high energy limit. In Section
\ref{sec:section_4_3} we obtain the probability of finding the quark in some
state after a multiple scattering process with a color and space averaged
medium consisting in $n$ partons, as a function of the single $n$ = 1 case. We
show that an splitting into an incoherent and a coherent contribution,
containing the probabilistic and the interference behavior, respectively, is
required in the microscopic limit, as we have already found in the QED
case. The incoherent contribution at low $n$ leads to the independent
superposition of the $n$ single squared scattering amplitudes, is positive
defined and then admits a statistical interpretation, whereas the coherent
contribution leads in the same low density regime to the $\sim n^2$ transverse
interferences between the different partons. We show that both contributions
admit a functional form for arbitrary $n$, in such a way that the low
collision limit is recovered. These forms are suitable for a numerical
evaluation under the realistic interaction, and particularly simple in their
Fokker-Planck approximations. The existence of the non-negligible coherent
contribution leads to an enhancement of the differential elastic cross
sections at low momentum changes for mediums of finite transverse size. This
enhancement depends on the geometry of the medium and contains the diffractive
behavior of its boundaries. The enhancement grows for large $n$ and tightens
up for large medium transverse sizes $R$, in such a way that for mediums of
infinite transverse area with respect to the parton direction it constraints
to a pure forward contribution, which thus does not contribute to the averaged
squared momentum change. In this limit, then, the incoherent behavior of the
scattering is found and the macroscopical limit recovered. Both the incoherent
and the coherent terms are shown to satisfy a transport equation for the
probability and for the amplitude, respectively. Finally in Section
\ref{sec:section_4_4} we extend the same results to a beyond eikonal
evaluation of the amplitudes and their squares, with the scope of using them
later in the study of some inelastic processes like gluon bremsstrahlung.
\section{Fields of color}
\label{sec:section_4_1}
Let us consider a single quark in the target medium making a transition from
the state $\psi_i(x)$, characterized by momentum $p_i$, spin $s_i$ and color
$a_i$, to the state $\psi_j(x)$, with $p_j$, $s_j$ and $a_j$. The transition
is mediated by a gluon, a field with structure of unitary, hermitian and
traceless matrix $t_{\alpha}$ \cite{gellmann1962} of the SU(N$_c$) algebra
\begin{align}
[t_\alpha,t_\beta]=if_{\alpha\beta}^\gamma t_\gamma,
\end{align}
where the number of colors $N_c$ = 3. Let also Latin letters run in the quark
color dimension $a=1,...,N_c$ and Greek letters in the gluon color dimension
$\alpha=1,...,N_c^2-1$. The structure constants can be expressed, using the
Jacobi identity, as the matrix elements of an adjoint representation of
SU(N$_c$),
\begin{align}
\left(T_\alpha\right)^{\gamma}_{\beta}=-if_{\alpha\beta}^\gamma,
\end{align}
satisfying the same algebra. The gluon carries an unit of color $c_j$ and one
of anticolor $\bar{c}_i$, an unit of spin $s_i-s_j$ and a fraction of momentum
$p_i-p_j$ off the quark. The matrix elements of the current of this process
are given by
\begin{equation}
\left(J_\alpha^\mu(x)\right)^{a_j}_{a_i} = g_s \bar{\psi}_j(x)\gamma^\mu t_\alpha \psi_i(x) = g_s(t_\alpha)^{a_j}_{a_i}
\sqrt{\frac{m}{p_j^0}}\bar{u}_{s_j}(p_j)\gamma^\mu
u_{s_i}(p_i)\sqrt{\frac{m}{p_i^0}}e^{+i(p_j-p_i)x},\label{quantumcurrent}
\end{equation}
where the quantity $g_s(t_\alpha)^{a_j}_{a_i}
\equiv\sqrt{\alpha_s}(t_\alpha)^{a_j}_{a_i}$ is to be interpreted as a colored
charge and $\alpha_s$ is the strong coupling constant. In particular for a
static current we would find $\left(J_\alpha^0(x)\right)^{a_k}_{a_i}=
g_s(t_\alpha)^{a_j}_{a_i}$ and $\v{J}_\alpha(x)=\v{0}$. The charge or strength
of the process depends then on the particular color transition choice of the
target quark and the color of the intermediating gluon. From this colored
current, as in ordinary electrodynamics, emanates a colored field or gluon,
which is just the propagation of the source current to any other point
\begin{equation}
A^\mu_\alpha(x) = \int d^4 y D^{\mu\nu}_{\alpha\beta}(x-y) J^\nu_\beta(y).
\label{gluonfielddefinition}
\end{equation}
sum over repeated indices assumed. From here onwards the gluon will be
considered massive, with effective mass $\mu_d$, in order to account for
screening effects at large distances in the QCD medium. The propagator in the
Feynman gauge for this gluon is given by
\begin{equation}
 \hat{D}^{\mu\nu}_{\alpha\beta}(q) = \delta_{\alpha\beta} \medspace g^{\mu\nu}
 \frac{4\pi}{-q^2+\mu_d^2}.
\label{gluonpropagator}
\end{equation}
From equations (\ref{quantumcurrent}), (\ref{gluonfielddefinition}) and
(\ref{gluonpropagator}) then we easily obtain 
\begin{equation}
A^\mu_\alpha(x) = \frac{4\pi g_s }{-(p_j-p_i)^2+\mu_d^2}\sqrt{\frac{m}{p_j^0}}\bar{u}_{s_j}(p_j)t_\alpha\gamma^\mu
u_{s_i}(p_i)\sqrt{\frac{m}{p_i^0}}e^{+i(p_j-p_i)x}.
\label{gluonfield}
\end{equation}
This gluon is ought to be absorbed by another traveling quark while
performing the transition from the state $\psi_k(x)$ to $\psi_l(x)$. The
amplitude of this process is given by
\begin{align}
M=-ig_s\int d^4x\medspace \bar{\psi}_l(x)t_\alpha\gamma_\mu\psi_k(x)A_\alpha^\mu(x),
\end{align}
Then omitting the momentum conservation delta and defining $q=p_j-p_i=p_l-p_k$
we find
\begin{align}
M\propto
\sqrt{\frac{m}{p_l^0}}\bar{u}_{s_l}(p_l)t_\alpha \gamma_\mu
u_{s_k}(p_k)\sqrt{\frac{m}{p_k^0}} \frac{4\pi ig_s^2
  }{q^2-\mu_d^2}\sqrt{\frac{m}{p_j^0}}\bar{u}_{s_j}(p_j)t_\alpha\gamma^\mu
u_{s_l}(p_l)\sqrt{\frac{m}{p_l^0}}.
\end{align}  
Compared to its QED analogous, this amplitude incorporates the matrix
structure of both quarks. Their elements can be written in general as
\begin{align}
\Big(M\Big)^{a_jb_j}_{a_ib_i} \propto  g_s^2(t_\alpha)^{a_j}_{a_i}(t_\alpha)^{a_l}_{a_k}=
\frac{1}{2}\delta_{a_j}^{a_k}\delta_{a_i}^{a_l}-\frac{1}{6}\delta_{a_j}^{a_i}\delta_{a_l}^{a_k}.
\end{align}
The particular color transition of the target and traveling parton is
normally not observed and we should average over initial configurations and
sum over final states in the square of $M$. We introduce the notation $d=N_c$
for the color dimension of quarks and and $d_A=N_c^2-1$ for the color
dimension of gluons. The average squared charge $C_R$ for a \textit{qq}
scattering at leading order in $g_s^2$ is then
\begin{figure}
\begin{minipage}{0.3\textwidth}
\begin{center}
\includegraphics[width=0.85\textwidth]{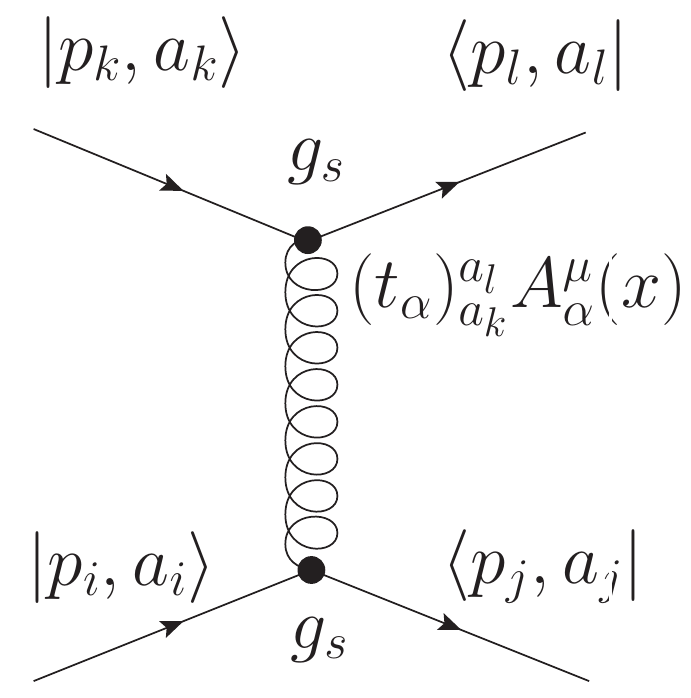}
\end{center}
\end{minipage}
\hfill
\begin{minipage}{0.3\textwidth}
\begin{center}
\includegraphics[width=0.85\textwidth]{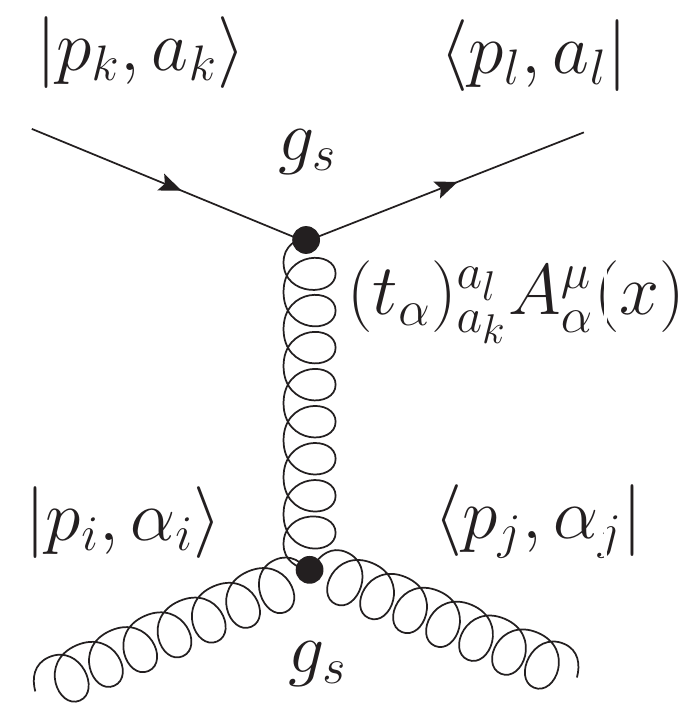}
\end{center}
\end{minipage}
\hfill
\begin{minipage}{0.3\textwidth}
\begin{center}
\includegraphics[width=0.85\textwidth]{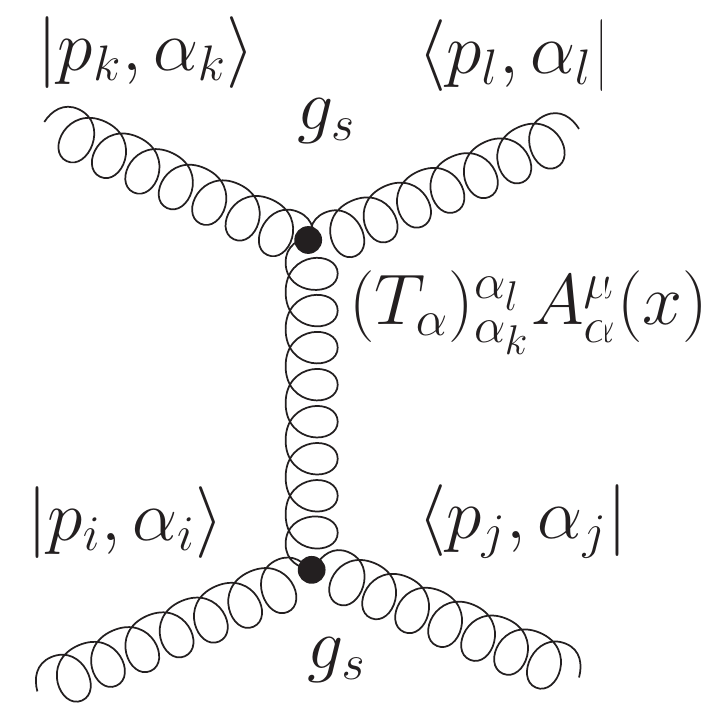}
\end{center}
\end{minipage}
\caption{Left: The $qq$ interaction. Center: The $qg$ interaction. Right: The
  $gg$ interaction.}
\label{fig:figure_4_1}
\end{figure}
\begin{align}
C_{R}^{qq}\equiv 
\frac{1}{d}\sum_{a_ja_i}\frac{1}{d}\sum_{a_la_k}& (t_\alpha)^{a_j}_{a_i}(t_\beta^\dag)^{a_i}_{a_j}
(t_\alpha)^{a_l}_{a_k}(t_\beta^\dag)^{a_k}_{a_l}=\frac{1}{d^2}\Tr^2(t_\alpha t_\beta)=\frac{N_c^2-1}{4N_c^2}=\frac{2}{9}.\label{qq_colorcharge}
\end{align}
Here we used the hermiticity of the $t_\alpha$ matrices and the normalization
$t_\alpha=\lambda_\alpha/2$, where $\lambda_\alpha$ are the Gell-Mann matrices
\cite{gellmann1962}, so $\Tr(t_\alpha t_\beta)=T_f\delta_{\alpha\beta}$ where
$T_f=1/2$ is the first invariant or Casimir. Similarly, for a \textit{qg} the
gluon emitted by a target gluon is given by the replacement
$(t_\alpha)^{a_j}_{a_i} \to (T_\alpha)^{\alpha_j}_{\alpha_i} =
-if_{\alpha\alpha_i}^{\alpha_j}$. We find an averaged charge of
\begin{align}
C_R^{qg}\equiv
\frac{1}{d}\sum_{\alpha_j\alpha_i}\frac{1}{d_A}\sum_{a_la_k}& (t_\alpha)^{a_l}_{a_k}(t_\beta^\dag)^{a_k}_{a_l}
(T_\alpha)^{\alpha_j}_{\alpha_i}(T_\beta^\dag)^{\alpha_i}_{\alpha_j}=\frac{1}{dd_A}\Tr(t_\alpha
t_\beta)\Tr(T_\alpha T_\beta)=T_f=\frac{1}{2},\label{gq_colorcharge}
\end{align}
where we used the adjoint representation normalization $\Tr(T_\alpha
T_\beta)=C_A\delta_{\alpha\beta}$, with $C_A=N_c$. And finally the case of a
gluon scattered by a target gluon produces using the same procedure
\begin{align}
C_R^{gg}\equiv
\frac{1}{d_A}\sum_{\alpha_j\alpha_i}\frac{1}{d_A}\sum_{\alpha_k\alpha_l}
(T_\alpha)^{\alpha_j}_{\alpha_i}(T_\beta^\dag)^{\alpha_i}_{\alpha_j}
(T_\alpha)^{\alpha_l}_{\alpha_k}(T_\beta^\dag)^{\alpha_k}_{\alpha_l}=\frac{1}{d_A^2}\Tr^2(T_\alpha
T_\beta)=\frac{N_c^2}{N_c^2-1}=\frac{9}{8}.\label{gg_colorcharge}
\end{align}
In Fig. \ref{fig:figure_4_1} these three different diagrams of scattering
scattering are shown. The above results can be reproduced also in the
classical and static field limit. For target quarks heavier than the average
momentum change scale $\mu_d$, or for moving quarks with a typical energy
$\langle p_i^0 \rangle$ much greater than $\mu_d$ but much smaller than the
energy of the traveling parton, quarks in the target can be considered almost
at rest, $p_i^0\simeq m_q$ or $p_i^0\simeq \langle p_i^0\rangle$
respectively. The recoil can be considered negligible $p_j\simeq p_i$ and we
find
\begin{align}
\sqrt{\frac{m_q}{p_j^0}}\bar{u}_{s_j}(p_j)\gamma_0
u_{s_i}(p_i)\sqrt{\frac{m_q}{p_i^0}}\approx
1, \medspace\medspace\medspace\medspace \sqrt{\frac{m_q}{p_j^0}}\bar{u}_{s_j}(p_j)\gamma_k
u_{s_i}(p_i)\sqrt{\frac{m_q}{p_i^0}}\approx \frac{(p_i)_k}{m_q}\ll 1,
\end{align}
and then the gluon field simplifies to
\begin{align}
A^\mu_\alpha(x) = \delta^{\mu}_{0} \frac{4\pi g_s
  t_\alpha}{-(p_j-p_i)^2+\mu_d^2}e^{+i(p_j-p_i)x}.
\end{align}
\begin{figure}
\begin{center}
\includegraphics[width=0.65\textwidth]{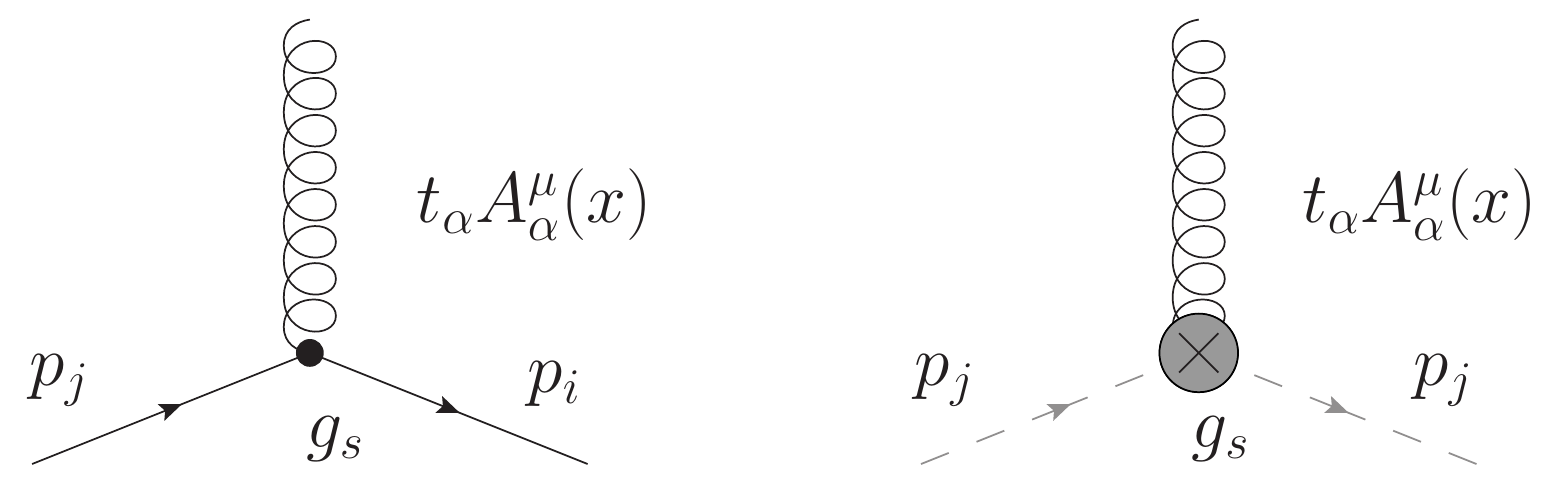}
\end{center}
\caption{Left: The field created by a quantum current $\langle
  i|j\rangle$. Right: The field as a classic and static current where $p_i\approx p_j$.}
\label{fig:colorfield}
\end{figure}
In this approximation the slow varying momentum dependence of the spinor terms
is assumed. After coherently forming the superposition in momentum and
integrating in time we find
\begin{align}
\langle A^\mu_\alpha(x) \rangle_{q,t} &\equiv \delta^{\mu}_{0}g_st_\alpha\int\frac{d^4q}{(2\pi)^4} \int dt  \frac{4\pi 
  }{-q^2+\mu_d^2}e^{iq\cdot x} =\delta^{\mu}_{0}g_st_\alpha\frac{1}{|\v{x}|}e^{-\mu_d|\v{x}|},\label{classicgluonfield}
\end{align}
where the last integral has been done with the help of Cauchy's theorem. So
the classical field of a particle of charge $g_s(t_\alpha)^{a_j}_{a_i}$ at
$\v{x}=0$ is recovered,
i.e. $J_\alpha^0(x)=g_st_\alpha\delta^{(3}(\v{x})$ and
$\v{J}_\alpha=\v{0}$. Under these simplifications our colored medium, then,
will be characterized from here onwards as the classical gluon field of a set
of $n$ static quarks, of the form
\begin{align}
A_\alpha^0(x)=\sum_{k=1}^n \frac{g_s t_\alpha}{|\v{x}-\v{r}_k|}e^{-\mu_d|\v{x}-\v{r}_k|}. \label{fielddefinition}
\end{align}
For a mixed scenario consisting of gluons and quarks the coupling of the gluon
content can be simply replaced with the adjoint representation $(t_\alpha)\to
(T_\alpha)$ in this classical approximation. In our following derivations, for
simplicity, we will restrict to a medium exclusively composed of quarks.
\section{Scattering amplitude}
\label{sec:section_4_2}
From \eqref{fielddefinition} a medium consisting of quarks and/or gluons will
be characterized as an external field with one non vanishing time independent
component, namely $A^0_\alpha(x)$. The quark state at high energies under this
interaction, assuming that the asymptotic state has momentum $p_i$ placed
along the $x_3$ direction, \textit{c.f} Section \ref{section_1_2}, is
given by \eqref{states}
\begin{align}
\psi^{(n)}(x)=\left(1+\frac{i}{2p_i^0}\v{\alpha}\cdot\v{\nabla}+\frac{1}{2p_i^0}\v{\alpha}\cdot\v{p}\right)\varphi_s^{(n)}(x),
\end{align}
where $\varphi_s^{(n)}(x)$ is a stationary and high energy, Schrodinger-like,
solution to the equation \eqref{highenergy_equation} with asymptotically free
initial condition. We write then
\begin{align}
\frac{\partial \varphi_s^{(n)}(x)}{\partial
  x_3}=\left(ip_i^3-\frac{i}{\beta_p}g_st_\alpha A^0_\alpha(\v{x})\right)\varphi_s^{(n)}(x),\medspace\medspace\medspace\medspace\medspace\medspace\medspace\medspace\lim_{x_3\to-\infty}\varphi_s^{(n)}(x)=\varphi_s^{(0)}(x),\label{highenergy_equation_quark}
\end{align}
where $\beta_p$ is the quark velocity and $\varphi_s^{(0)}(x)$ a free solution
of momentum $p_i$. The solution to this matrix equation
\eqref{highenergy_equation_quark} with that initial condition is given by the
ordered exponential
\begin{align}
\varphi_s^{(n)}(x)=\mathcal{P}\exp\left(-i\frac{g_s}{\beta_p}\int_{-\infty}^{x_3}
dx_3' \medspace t_\alpha A_\alpha^0(\v{x}) \right)\varphi_{s}^{(0)}(x).
\end{align}
Noticing that the term $\v{\alpha}\cdot\v{p}$ is canceled when the derivative
of the free part is taken, we find the quark equivalent of
\eqref{wave_integration}
\begin{align}
\psi^{(n)}(x)=\left\{\left(1+i\frac{\gamma_k\gamma_0}{2p_i^0}\partial_k\right)\mathcal{P}\exp\left(-i\frac{g_s}{\beta_p}\int_{-\infty}^{x_3}
dx_3'\medspace t_\alpha A_\alpha^0(\v{x}) \right)\right\}\psi^{(0)}(x).\label{quarkwave}
\end{align}
We will explicitely write the states with the color vector. For the
asymptotic initial quark in particular
\begin{align}
\psi^{(0)}(x)=e^{-ip_i\cdot x}\sqrt{\frac{m_q}{p_i^0}}u_{s_i}^{a_i}(p_i).
\end{align}
Let the eikonal phase in the high energy integration of the wave be denoted as
\begin{align}
W^{(n)}(\v{x},p_i)\equiv\mathcal{P}\exp\left(-i\frac{g_s}{\beta_p}\int^{x_3}_{-\infty}
dx_3' \medspace t_\alpha A_\alpha^0(\v{x})\right).
\end{align}
From here onwards path ordering is going to be implicitly assumed. The
scattering amplitude of finding the quark in a state with momentum $p_f$ due
to the external field actuation can be found \cite{itzykson1980} by using the
Lippmann-Schwinger relation
\begin{align}
\psi^{(n)}(x)&=\psi^{(0)}(x)+\int d^4y
S_F(x-y)\gamma_0g_st_\alpha A^0_\alpha(\v{y})\psi^{(n)}(y)=\psi^{(0)}(x)+\psi^{(n)}_{diff}(x),\nonumber\\
\end{align}
where the Feynman-Stueckelberg quark propagator is given by
\begin{align}
S_F^q(x) = \int \frac{d^4p}{(2\pi)^4}e^{ip\cdot x} &\frac{\slashed{p}+m_q}{p^2-m_q^2}=\frac{1}{i}\sum_{s,a}\int
\frac{d^3\v{p}}{(2\pi)^3}\\
&\times\left(\sqrt{\frac{m_q}{p^0}}u_{s}^{a}(p)\otimes
\bar{u}_{a}^{s}(p)\sqrt{\frac{m_q}{
p^0}}\right)e^{-i\v{p}\cdot\v{x}}\nonumber.
\end{align}
Here we restricted to energy positive solutions and we used the completeness
relation in spin and in color
\begin{align}
\sum_{s,a}u_{s}^a(p)\otimes \bar{u}_{a}^s(p) =
\frac{\slashed{p}+m_q}{2m_q}.
\end{align}
Then, the diffracted part in the Lippmann-Schwinger equation produces a
superposition of states of momentum $p_f$, spin $s_f$ and color $a_f$ of the form
\begin{align}
\psi_{diff}^{(n)}(x)=\sum_{s_f,a_f}\int \frac{d^3\v{p}_f}{(2\pi)^3}e^{-i\v{p}_f\cdot\v{x}} \sqrt{\frac{m_q}{p_f^0}}u_{s_f}^{a_f}&(p_f) \left(M_{s_fs_i}^{(n)}(p_f,p_i)\right)_{a_i}^{a_f},
\end{align}
so that we write for the amplitude of finding the quark in this final state
$f$ as 
\begin{align}
&\left(M_{s_fs_i}^{(n)}(p_f,p_i)\right)_{a_i}^{a_f} \equiv
\bigg\langle a_f\bigg|\int d^4y e^{+i(p_f-p_i) y}\\
\times&\sqrt{\frac{m_q}{p_f^0}}\bar{u}_{s_f}(p_f)\gamma_0\left( -ig_st_\alpha
A^0_\alpha(\v{y})\right)\left\{\left(1+i\frac{\gamma_k\gamma_0}{2p_i^0}\partial_k\right)W^{(n)}(\v{y},p_i)\right\}u_{s_i}(p_i)\sqrt{\frac{m_q}{p_i^0}}\bigg|a_i\bigg\rangle\nonumber.
\end{align}
The integration of the above equation in the high energy limit follows exactly
the same steps as those presented at Section \ref{sec:section_2_1}. In the
high energy limit, provided that the Glauber condition
\eqref{glaubercondition1} holds, we can neglect the $1/2p_i^0$ operator
correction. By defining $\v{q}=\v{p}_f-\v{p}_i$ and neglecting by now the
beyond eikonal corrections in $q_z$ one easily finds
\begin{align}
M_{s_fs_i}^{(n)}(p_f,p_i)= 2\pi&\delta(p_f^0-p_i^0) \beta_p
\sqrt{\frac{m_q}{p_f^0}}\bar{u}_{s_f}(p_f)\gamma_0
u_{s_i}(p_i)\sqrt{\frac{m_q}{p_i^0}} \label{scatteringamplitude_quark}\\
&\times \int d^2\v{y}_t
e^{-i\v{q}_t\cdot\v{y}_t}\left(\exp\left[-i\frac{g_s}{\beta_p}\int_{-\infty}^{+\infty}
  dy_3 \medspace t_\alpha A_\alpha^0(\v{y})\right]-1\right).\nonumber
\end{align}
The spinorial product can be reduced to a spin conservation delta in the
high energy limit
\begin{align}
\sqrt{\frac{m_q}{p_f^0}}\bar{u}_{s_f}(p_f)\gamma_0
u_{s_i}(p_i)\sqrt{\frac{m_q}{p_i^0}}\simeq\delta_{s_i}^{s_f}
\end{align}
For convenience we define the integral part of \eqref{scatteringamplitude_quark}
as
\begin{align}
F_{el}^{(n)}(\v{q}_t)\equiv \int d^2\v{y}_t
e^{-i\v{q}_t\cdot\v{y}_t}\left(\exp\left[-i\frac{g_s}{\beta_p}\int_{-\infty}^{+\infty}
  dy_3 \medspace t_\alpha A_\alpha^0(\v{y})\right]-1\right),\label{scatteringamplitudef_quark}
\end{align}
since spin and energy are preserved and $F_{el}^{(n)}(\v{q})$ contains the
relevant momentum distribution after the scattering. We introduce now the
notation shorthands
\begin{align}
i\frac{g_s}{\beta_p}\int^{+\infty}_{-\infty} dy_3 t_\alpha A^0_\alpha(\v{y})
=i\frac{g_s^2}{\beta_p}t_\alpha
t_\alpha\sum_{k=1}^{n}\chi_0^{(1)}(\v{y}_t-\v{r}_t^k)=i\frac{g_s^2}{\beta_p}t_\alpha
t_\alpha\sum_{k=1}^{n}\chi_0^{k}(\v{y}_t).
\end{align}
The color transitions of the target and traveling quarks are expected to be
evaluated in $F_{el}^{(n)}(\v{q})$ and usually color-averaged in its
square. For a color averaged single target quark ($n=1$) performing the color
transition $b_i\to b_f$ we obtain the following useful optical theorem
\begin{align}
\frac{1}{N_c}\int& \frac{d^2\v{q}_t}{(2\pi)^2}\Tr\left(F_{el}^{(1)\dag}(\v{q}_t)F_{el}^{(1)}(\v{q}_t)\right)= \frac{1}{N_c}\int
d^2\v{x}_t
\left(\exp\left(+i\frac{g_s^2}{\beta_p}t_\beta^\dag
t_\beta^\dag\chi_0^{1}(\v{x}_t)\right)-1\right)^{b_i}_{b_f}\nonumber\\
&\times\left(\exp\left(-i\frac{g_s^2}{\beta_p}t_\alpha
t_\alpha\chi_0^{1}(\v{x}_t)\right)-1\right)^{b_f}_{b_i}=-\frac{1}{N_c}\Tr\left(F_{el}^{(1)\dag}(\v{0})+F_{el}^{(1)}(\v{0})\right),\label{opticaltheorem_quark_single}
\end{align}
where the trace refers from here onwards to the colors in the target space
only. A symmetrical relation can be written for the color average in the
projectile or for both averages together, in general. We notice that the color
average of $|F_{el}^{(1)}(\v{q})|^2$ does not admit, however, a
reexponentiation. In order to see this fact we expand the amplitude using
\eqref{scatteringamplitudef_quark} and \eqref{fielddefinition}. We get
\begin{align}
F^{(1)}_{el}(\v{q})=-i\frac{g_s^2}{\beta_p}t_{\alpha_1}t_{\alpha_1}\hat{A}_0^{(1)}(\v{q})-\frac{g_s^4}{2\beta_p^2}t_{\alpha_2}t_{\alpha_2}t_{\alpha_1}t_{\alpha_1}\int\frac{d^2\v{k}}{(2\pi)^2}\hat{A}_0^{(1)}(\v{k})\hat{A}_0^{(1)}(\v{q}-\v{k})+\ldots,\label{scatteringamplitudef_perturbative_quark}
\end{align}
where the Fourier transform of the single field satisfies
$\hat{A}_0^{(1)}(\v{q})=4\pi/(\v{q}^2+\mu_d^2)$ for the Debye screened
interaction \eqref{fielddefinition}. The total color average of the squared
amplitude produces
\begin{align}
&\left(F_{el}^{(1)\dag}(\v{q}_t)\right)^{a_ib_i}_{a_fb_f}\left(F_{el}^{(1)}(\v{q}_t)\right)^{a_fb_f}_{a_ib_i}  =
  \frac{g_s^4}{\beta_p}^2C_R^{(2)}\left|\hat{A}_0^{(1)}(\v{q})\right|^2+\frac{g_s^6}{2\beta_p^3}C_R^{(3)}\hat{A}_0^{(1)}(\v{q})\\ &\times\int\frac{d^2\v{k}}{(2\pi)^2}\hat{A}_0^{(1)}(\v{k})\hat{A}_0^{(1)}(\v{q}-\v{k})+\frac{g_s^8}{4\beta_p^4}C_R^{(4)}\left(\int\frac{d^2\v{k}}{(2\pi)^2}\hat{A}_0^{(1)}(\v{k})\hat{A}_0^{(1)}(\v{q}-\v{k})\right)^2+\ldots.\nonumber
\end{align}
The required color charges are given by operations like
\eqref{qq_colorcharge}. Indeed for the simplest diagram we find the
aforementioned result \eqref{qq_colorcharge}
\begin{align}
C_R^{(2)}=\frac{1}{N_c^2}\Tr(t_{\alpha_1}t_{\beta_1})=\frac{N_c^2-1}{4N_c^2}=\frac{2}{9}.
\end{align}
The interference term $C_R^{(3)}$ has to cancel by symmetry under
conjugation. We obtain, of course
\begin{align}
C_R^{(3)}=\frac{1}{N_c^2}\left(\Tr^2(t_{\alpha_1}t_{\beta_2}t_{\beta_1})-\Tr^2(t_{\alpha_2}t_{\alpha_1}t_{\beta_1})\right)=0,
\end{align}
as expected. The next term requires a little more of work. It is given by
\begin{align}
C_R^{(4)}=\frac{1}{N_c^2}\left(\frac{1}{2}\delta^{a_f}_{b_1}\delta^{b_f}_{a_1}-\frac{1}{6}\delta^{a_f}_{a_1}\delta^{b_f}_{b_1}\right)\left(\frac{1}{2}\delta^{a_1}_{b_i}\delta^{b_1}_{a_i}-\frac{1}{6}\delta^{a_1}_{a_i}\right)(t_{\beta_2})^{b_1}_{b_i}
(t_{\beta_2})^{a_i}_{a_1}(t_{\beta_1})^{a_1}_{a_f}(t_{\beta_1})^{b_1}_{b_f}\nonumber\\ =\frac{1}{N_c^2}\left(\frac{10}{36}\Tr^2(t_{\beta_1}t_{\beta_2})-\frac{2}{12}\Tr(t_{\beta_1}t_{\beta_2}t_{\beta_1}t_{\beta_2})\right)=\frac{2}{27}.
\end{align}
It can be shown that the next order squared-diagram would produce
$C_R^{(6)}=22/729$. So that the average color charge of two kicks with the
same center $2/27$ is not anymore the square of the averaged color charge of
two single kicks $(2/9)^2$, and so on. Since a reexponentiation does not hold,
in the high energy formalism we will restrict to leading order evaluations in
the coupling of the single amplitudes. 

On the other hand, the scattering amplitude for the gluon is found by tracing
back the same approach with the gluon propagator in the Feynman gauge.  By
using the completeness and orthogonality relations
\begin{align}
\sum_{\lambda}\epsilon_\mu^\lambda(k) \epsilon_\nu^{\lambda *}(k)=g_{\mu\nu},\medspace\medspace\medspace\medspace\medspace\medspace\medspace\medspace\medspace\medspace\epsilon^{\lambda_f}_\mu(k_f)\epsilon_{\lambda_i}^\mu(k_i)=-\delta^{\lambda_f}_{\lambda_i},
\end{align}
one arrives at a very similar form to \eqref{scatteringamplitude_quark}
\begin{align}
M_{\lambda_f\lambda_i}^{(n)}(k_f,k_i)&= 2\pi\delta(\delta k^0) \beta_k
\epsilon^{\lambda_f*}_\mu(k_f)\epsilon_{\lambda_i}^\mu(k_i)\nonumber\\
&\times \int d^2\v{y}_t
e^{-i\delta\v{k}_t\cdot\v{y}_t}\left(\exp\left[-i\frac{g_s}{\beta_p}\int_{-\infty}^{+\infty}
  dy_3 \medspace T_\alpha
  A_\alpha^0(\v{y})\right]-1\right).\label{scatteringamplitude_gluon}
\end{align}
Similarly to the quark case, in the high energy limit we find $k_f\simeq k_i$
and thus we obtain a polarization conservation delta. The gluon dynamics in
this sense is similar to the quark except for the coupling definitions, and
contained in the function $F_{el}^{(n)}(\v{q})$.
\section{Multiple scattering effects}
\label{sec:section_4_3}
\begin{figure}[h]
\centering
\includegraphics[width=0.8\textwidth]{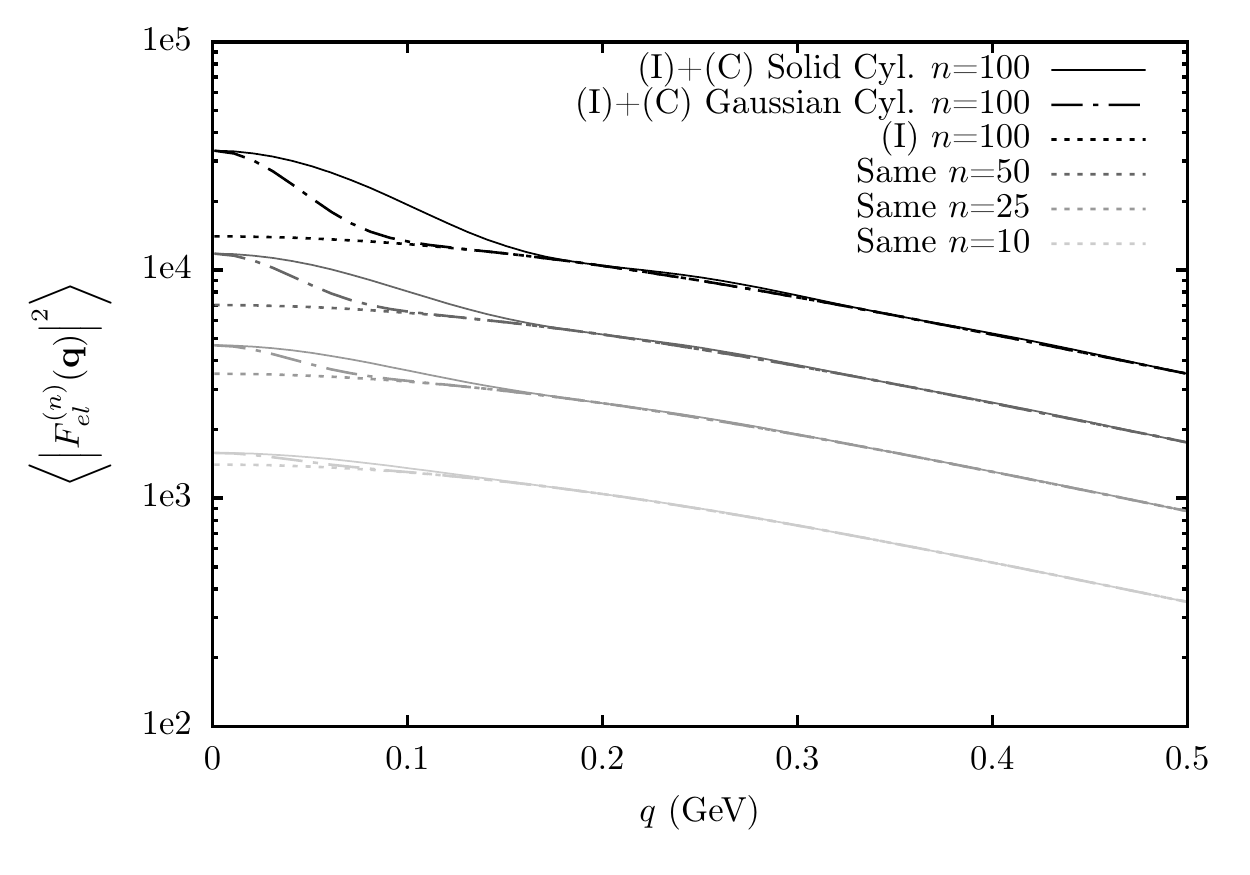}
\caption{Averaged squared elastic amplitude at leading order in the coupling
  and in the number of collisions as a function of the momentum change, of a
  quark emerging from a medium with $n$ = 100 quarks (black lines), with Debye
  screening mass of $\mu_d$ = 0.5 GeV and a medium radius of $R =$ 10
  $r_d$. Incoherent (I) contribution (diagonal term) is shown in dotted lines,
  and incoherent (I) and coherent (C) contributions (diagonal and non diagonal
  term) are shown for a cylinder with Gaussian decaying density (dot-dashed
  lines) and a solid cylinder (solid lines). Same results for $n$ = 50 (dark
  grey), $n$ = 25 (medium grey) and $n$ = 10 (lightest grey).}
\label{fig:figure_4_3}
\end{figure}

Amplitude \eqref{scatteringamplitude_quark} can be squared and averaged over
multiple scatterers configurations. As in the QED case we find an incoherent
and a coherent contribution, but with some differences related to the color
behavior. It results illustrative to obtain this transverse interference
behavior by means of an expansion in the number of collisions prior to a
direct evaluation of the full expression. An inspection of
\eqref{scatteringamplitudef_quark} suggests defining
\begin{align}
\Gamma_k\equiv\exp\left(-i\frac{g_s^2}{\beta_p}t_{\alpha}t_{\alpha}\chi_0^{k}(\v{y}_t)\right)-1.
\end{align}
Let us denote by $b_f^k$ and $b_i^k$ the final and initial color,
respectively, of the target quark $k$, then omitting the indices in the left
hand side
\begin{align} 
F_{el}^{(n)}(\v{q})=\int
d^2\v{x}e^{-i\v{q}\cdot\v{x}}\left(\prod_{k=1}^{n}\left((\Gamma_k)^{b_f^k}_{b_i^k}+\delta^{b_f^k}_{b_i^k}\right)-\prod_{k=1}^n\delta^{b_f^k}_{b_i^k}\right).\label{scatteringamplitudef_quark_explicitcolortarget}
\end{align}
It is now easy to expand the amplitude in the number of collisions. We notice
\begin{align}
F_{el}^{(n)}(\v{q})=\int d^2\v{x}
e^{-i\v{q}\cdot\v{x}}&\left(\sum_{k=1}^{n}(\Gamma_k)_{b_i^k}^{b_f^k}\prod_{j\neq
k}\delta^{b_f^j}_{b_i^j}+\sum_{k=1}^n\sum_{j=k+1}^{n}(\Gamma_k)^{b_f^k}_{b_i^k}(\Gamma_j)^{b_f^j}_{b_i^j}\prod_{l\neq
j\neq k}\delta_{b_i^l}^{b_f^l}+\ldots\right)\nonumber\\
&=\sum_{k=1}^nI_k+\sum_{k=1}^n\sum_{j=k+1}^nI_{kj}+\cdots.
\end{align}
The single collision contribution can be rewritten as
\begin{align}
\sum_{k=1}^nI_k&=\sum_{k=1}^n\left(\prod_{j\neq k}\delta^{b_f^j}_{b_i^j}\right)\int d^2\v{x}_te^{-i\v{q}_t\cdot\v{x}_t}\left(\exp\left(-i\frac{g_s^2}{\beta_p}t_\alpha
t_\alpha\chi_0^{k}(\v{y}_t)\right)-1\right)^{b_f^k}_{b_i^k}\nonumber\\
&=\sum_{k=1}^n\left(\prod_{j\neq k}\delta^{b_f^j}_{b_i^j}\right)e^{-i\v{q}_t\cdot\v{r}_t^k}\left(F_{el}^{(1)}(\v{q}_t)\right)^{b_f^k}_{b_i^k},
\end{align}
which is the expected result, consisting in the transition of the target quark
$k$ causing the collision and the no transition of the rest of quarks. Due to
the high energy limit any term of higher order can be always written as a
convolution
\begin{align}
I_{kj}=\left(\prod_{l\neq j \neq k}\delta^{b_f^l}_{b_i^k}\right)\int\frac{d^2\v{k}_t}{(2\pi)^2}e^{-i\v{k}_t\cdot\v{r}_t^k-i(\v{q}_t-\v{k}_t)\cdot\v{r}_t^j}\left(F_{el}^{(1)}(\v{k}_t)\right)^{b_f^k}_{b_i^k}\left(F_{el}^{(1)}(\v{q}_t-\v{k}_t)\right)^{b_f^j}_{b_i^j},
\end{align}
which is easily interpretable also in terms of single processes.  We can
consider now the squared amplitude, which with this notation is given by
\begin{align}
F_{el}^{(n)\dag}(\v{q})F_{el}^{(n)}(\v{q})= \sum_{k=1}^{n}\sum_{j=1}^n
I_j^\dag I_k+\sum_{k=1}^{n}\sum_{j=1}^{n}\sum_{l=j+1}^n\bigg(I_{jl}^\dag
I_k+I_{k}^\dag I_{jl}\bigg)+\cdots.
\end{align}
The first contribution is given by the square of the single collision
amplitudes, which is given by terms as
\begin{align}
I_j^\dag I_k=\left(\prod_{l\neq k}\delta^{b_f^l}_{b_i^l}\right)\left(\prod_{m\neq j}\delta^{b_i^m}_{b_f^m}\right)
\left(e^{+i\v{q}_t\cdot\v{r}_t^j}\left(F_{el}^{(1)\dag}(\v{q}_t)\right)_{b_f^j}^{b_i^j}\right)
\times
\left(e^{-i\v{q}_t\cdot\v{r}_t^k}\left(F_{el}^{(1)}(\v{q}_t)\right)^{b_f^k}_{b_i^k}\right).
\end{align}
The sum of this terms can be always arranged into a diagonal and a non
diagonal contribution following
\begin{align}
\sum_{k=1}^n\sum_{j=1}^n I_j^\dag I_k= \sum_{k=1}^n I_k^\dag
I_k+\sum_{k=1}^n\sum_{j\neq k}^n I_j^\dag I_k,
\end{align}
which produces for the diagonal contribution
\begin{align}
\sum_{k=1}^n I_k^\dag I_k = n\left(\prod_{j\neq
  k}^n\delta^{b_f^j}_{b_i^j}\delta^{b_i^j}_{b_f^j}\right)\bigg(F_{el}^{(1)\dag}(\v{q}_t)\bigg)^{b_i^k}_{b_f^k}\bigg(F_{el}^{(1)}(\v{q}_t)\bigg)^{b_f^k}_{b_i^k}=n
N_c^{n-1}\Tr\left(F_{el}^{(1)\dag}(\v{q}_t)F_{el}^{(1)}(\v{q}_t)\right),\label{diagonal_term}
\end{align}
where trace refers to the target color space. By dividing by the dimension of
the target color space and by perturbatively expanding the amplitudes up to
first order in the coupling \eqref{scatteringamplitudef_perturbative_quark} we
get
\begin{align}
\frac{1}{N_c^n}\sum_{k=1}^n I_k^\dag I_k \simeq
\frac{n}{N_c}\frac{g_s^4}{\beta_p^2}t_{\alpha}t_\beta \Tr(t_\alpha t_\beta)\left(\hat{A}_0^{(1)}(\v{q})\right)^2= n\frac{g_s^4}{\beta_p^2}\frac{N^2_c-1}{4N_c^2}\left(A_0^{(1)}(\v{q})\right)^2,\label{diagonal_term_firstorder}
\end{align}
which is $n$, the number of independent collisions, times the squared
amplitude of a single collision. Notice that the color average in the target
produces an average color conservation in the projectile
$\delta^{a_f}_{a_i}$. For the non-diagonal contribution we find
\begin{align}
\sum_{k=1}^n\sum_{j\neq k}^n I_j^\dag I_k&=\sum_{k=1}^{n}\sum_{j\neq
  k}^n\left(\prod_{l\neq k}^n\delta^{b_f^l}_{b_i^l}\right)\left(\prod_{m\neq
  j}^n\delta^{b_i^m}_{b_f^m}\right)e^{-i\v{q}_t\cdot(\v{r}_t^k-\v{r}_t^j)}\bigg(F_{el}^{(1)\dag}(\v{q}_t)\bigg)^{b_i^j}_{b_f^j}\bigg(F_{el}^{(1)}(\v{q}_t)\bigg)^{b_f^k}_{b_i^k}\nonumber\\ &=N_c^{n-2}\Tr\bigg(F_{el}^{(1)\dag}(\v{q}_t)\bigg)\Tr\bigg(F_{el}^{(1)}(\v{q}_t)\bigg)\left(\sum_{k=1}^{n}\sum_{j\neq
  k}^ne^{-i\v{q}_t\cdot(\v{r}_t^k-\v{r}_t^j)}\right).\label{nondiagonal_term}
\end{align}
Unlike to the QED case, by using \eqref{scatteringamplitudef_perturbative_quark}
we get that the first order in the coupling vanishes due to the traceless
$t_\alpha$ matrices, so at leading order in the coupling
\begin{align}
\Tr\bigg(F_{el}^{(1)\dag}(\v{q}_t)\bigg)\Tr\bigg(F_{el}^{(1)}(\v{q}_t)\bigg)=\left(-\frac{1}{2}\frac{g_s^4}{\beta_p^2}t_{\alpha}t_\beta
\Tr(t_\alpha t_\beta)\int\frac{d^2\v{k}}{(2\pi)^2}\hat{A}_0^{(1)}(\v{k})\hat{A}_0^{(1)}(\v{q}-\v{k})\right)^2.
\end{align}
By averaging over target spatial configurations, for simplicity over a
solid cylinder of section $\Omega=\pi R^2$ we have
\begin{align}
\left\langle \sum_{k=1}^{n}\sum_{j\neq
  k}^ne^{-i\v{q}_t\cdot(\v{r}_t^k-\v{r}_t^j)} \right\rangle_\Omega =
n(n-1)\left(\frac{1}{\pi R^2}\int_\Omega
d^2\v{r}_te^{-i\v{q}_t\cdot\v{r}_t}\right)^2=n(n-1)\left(\frac{w(\v{q},R)}{\pi R^2}\right)^2.
\end{align}
where $w(\v{q},R)$ is the window function given at
\eqref{window_function}. Then we get
\begin{align}
\frac{1}{N_c^n}\sum_{k=1}^n\sum_{j\neq k}^n \left\langle I_j^\dag
I_k\right\rangle_\Omega \simeq n^2
\left(\frac{1}{2}\frac{g_s^4}{\beta_p^2}\frac{N_c^2-1}{4N_c^2}\frac{w(\v{q},R)}{\pi
R^2}\int\frac{d^2\v{k}}{(2\pi)^2}\hat{A}_0^{(1)}(\v{k})\hat{A}_0^{(1)}(\v{q}-\v{k}) \right)^2,
\end{align}
where $n$ was assumed large so that $n(n-1)\simeq n^2$. Observe that the
diagonal contribution produced a term of order $n$ and $\alpha_s^2$, the
number of independent collisions times the leading order of a color averaged
single collision, whereas the non diagonal contribution produces a term of
order $n^2$ and $\alpha_s^4$, which measures the $n^2$ different interferences
between single collisions with different scattering centers. By performing the
last integral in momentums the final result can be written explicitely as
\begin{align}
\left\langle\left|F_{el}^{(n)}(\v{q})\right|^2\right\rangle\cong&
\medspace n\frac{N_c^2-1}{4N_c^2}\left(\frac{\alpha_s}{\beta_p}\frac{4\pi
}{\v{q}^2+\mu_d^2}\right)^2\nonumber\\
&+\left(n\frac{N_c^2-1}{4N_c^2}\frac{w(\v{q},R)}{\pi R^2}\frac{1}{2}\left(\frac{\alpha_s}{\beta_p}\right)^2\frac{16\pi
\arcsinh\left(\frac{q}{2\mu_d}\right)}{q\sqrt{q^2+4\mu_d^2}}\right)^2.\label{leadingorder_squaredamplitude_quark}
\end{align}
For convenience we give the window functions for a solid cylinder and for a
cylinder with normalized Gaussian decaying density, of a typical width $R^2$,
\begin{align}
\frac{w_{c}(q,R)}{\pi R^2}\equiv 2\frac{J_1(qR)}{qR},\medspace\medspace\medspace\medspace\medspace\medspace\medspace\medspace\medspace \frac{w_{g}(q,R)}{\pi R^2} \equiv\exp\left(-\frac{q^2R^2}{2}\right).
\end{align}
Notice that in the limit $R\to\infty$ we recover always the forward
propagation $\delta^2(\v{q})$ in the non-diagonal contribution, as expected
due to transverse homogeneity. In Figure \ref{fig:figure_4_3} the above result
is shown for a typical QCD medium. As it can be seen, although the
non-diagonal contribution is a higher order correction in the coupling to the
diagonal contribution, for $n$ = 10 and higher number of constituents its
contribution in the soft scattering zone $q\sim 1/R$ is the dominant one. For
mediums of large length the above discussion is not enough, since the
scattering typically involves more than one collision. For the next order
evaluations we refer to the procedure at Section \ref{sec:section_2_4}. Since
the expansion in the number of collisions is an alternated series, a
functional form is required. We define then the shorthand notations
\begin{align}
W_0^{(n)}(\v{y})\equiv \exp\left(-i\sum_{k=1}^n\frac{g_s^2}{\beta}t_\alpha t_\beta
\chi^{k}_0(\v{y})\right)=\prod_{k=1}^n\left(\exp\left(-i\frac{g_s^2}{\beta_p}t_\alpha t_\alpha
\chi^{k}_0(\v{x})\right)\right)=\prod_{k=1}^nW_0^k(\v{y}).\label{wdefinition_qcd}
\end{align}
Then the average over target colors and positions can be formally written as
\begin{align}
\frac{1}{N_c^n}\Tr\left\langle
M_{s_{f}s_i}^{(n)\dag}(p_f,p_i)M_{s_{f}s_i}^{(n)}(p_f,p_i)\right\rangle =
2\pi\delta(q^0)\beta_p\delta^{s_f}_{s_i}\frac{1}{N_c^n}\Tr\left\langle
F_{el}^{(n)\dag}(\v{q})F_{el}^{(n)}(\v{q})\right\rangle,\label{averaged_squared_amplitude_quarks}
\end{align}
where the trace refers to the independent color average/traces of the
different $n$ centers. We divided by an overall infinite $T=2\pi\delta(0)$
factor, related to time translation invariance, and by the quark incoming flux
$\beta_p$. We have to evaluate
\begin{align}
\frac{1}{N_c^n}\Tr\left(\left\langle F_{el}^{(n)\dag}(\v{q})F_{el}^{(n)}(\v{q})\right\rangle\right)=\int
d^2\v{x}&\int d^2\v{y}
e^{-i\v{q}\cdot(\v{x}-\v{y})}\\
&\times\frac{1}{N_c^n}\Tr\left\langle\left(W_0^{(n)\dag}(\v{y})-1\right)\left(W_0^{(n)}(\v{x})-1\right)\right\rangle\nonumber.\label{averaged_squared_amplitudef_quarks}
\end{align}
The average on the right hand side, by using our notation, factorizes as
\begin{align}
\frac{1}{N_c^n}\Tr\left\langle\left(W_0^{(n)\dag}(\v{y})-1\right)\right.&\left.\left(W_0^{(n)}(\v{x})-1\right)\right\rangle=\prod_{k=1}^n\left(\frac{1}{N_c}\Tr\left\langle
W^{k\dag}_0(\v{y})W_0^k(\v{x})\right\rangle\right)
\nonumber\\
-\prod_{k=1}^n&\left(\frac{1}{N_c}\Tr\left\langle W_0^{k\dag}(\v{y})\right\rangle\right)-\prod_{k=1}^n\left(\frac{1}{N_c}\Tr\left\langle W_0^{k}(\v{x})\right\rangle\right)+1.
\end{align}
As learned from the QED case and the statistical relation $\langle
x^2\rangle=\langle x\rangle^2+\sigma^2$, our previous discussion regarding to
a splitting into an incoherent and a coherent contributions suggests adding
and subtracting now the term
\begin{align}
\pm \prod_{k=1}^n\left(\frac{1}{N_c}\Tr\left\langle
W_0^{(1)\dag}(\v{y})\right\rangle\right)\prod_{k=1}^n\left(\frac{1}{N_c}\Tr\left\langle W_0^{(1)}(\v{x})\right\rangle\right).
\end{align}
Then we can pair the following terms as
\begin{align}
1-\prod_{k=1}^n\left(\frac{1}{N_c}\Tr\left\langle
W_0^{k\dag}(\v{y})\right\rangle\right)&-\prod_{k=1}^n\left(\frac{1}{N_c}\Tr\left\langle
W_0^{k}(\v{x})\right\rangle\right)\nonumber\\
&+\prod_{k=1}^n\left(\frac{1}{N_c}\Tr\left\langle
W_0^{k\dag}(\v{y})\right\rangle\right)\prod_{k=1}^n\left(\frac{1}{N_c}\Tr\left\langle
W_0^{(1)}(\v{x})\right\rangle\right)\nonumber\\
=\left(\prod_{k=1}^n\left(\frac{1}{N_c}\Tr\left\langle W_0^{k\dag}(\v{y})\right\rangle\right)-1\right)&\left(\prod_{k=1}^n\left(\frac{1}{N_c}\Tr\left\langle W_0^{(1)}(\v{x})\right\rangle\right)-1\right),
\end{align}
in such a way that we found the usual splitting into a coherent and an
incoherent contribution
\begin{align}
\frac{1}{N_c}\Tr\left(\left\langle F_{el}^{(n)\dag}(\v{q})F_{el}^{(n)}(\v{q})\right\rangle\right)=\hat{\Pi}_2^{(n)}(\v{q})+\hat{\Sigma}_2^{(n)}(\v{q}),
\end{align}
where the coherent contribution is given by the color/space
averaged amplitude squared
\begin{align}
\hat{\Pi}_2^{(n)}(\v{q})\equiv \left(\int
d^2\v{x}
e^{-i\v{q}\cdot\v{x}}\right.&\left.\left(\prod_{k=1}^n\left(\frac{1}{N_c}\Tr\left\langle
W_0^{k}(\v{x})\right\rangle\right)-1\right)\right)^\dag\label{coherent_contributionq_definition}\\
&\times\left(\int
d^2\v{x}
e^{-i\v{q}\cdot\v{x}}\left(\prod_{k=1}^n\left(\frac{1}{N_c}\Tr\left\langle
W_0^{k}(\v{x})\right\rangle\right)-1\right)\right),\nonumber
\end{align}
and the incoherent contribution, which admits an statistical and classical
interpretation as a deviation $\sigma^2$ from the quantum coherent
contribution, is given by
\begin{align}
\hat{\Sigma}_2^{(n)}(\v{q})\equiv \int
d^2\v{x}\int d^2\v{y}&
e^{-i\v{q}\cdot(\v{x}-\v{y})}\left(\prod_{k=1}^n\left(\frac{1}{N_c}\Tr\left\langle
W_0^{k\dag}(\v{y})W_0^{k}(\v{x})\right\rangle\right)\right.\label{incoherent_contributionq_definition}\\
&-\left.\prod_{k=1}^n\left(\frac{1}{N_c}\Tr\left\langle
W_0^{k\dag}(\v{y})
\right\rangle\right)\prod_{k=1}^n\left(\frac{1}{N_c}\Tr\left\langle W_0^{k}(\v{x})\right\rangle\right)\right)\nonumber.
\end{align}
The operation $\langle\star\rangle$ indicates an average over medium space
configuration in a cylinder of transverse area $\Omega=\pi R^2$ and length
$l$. The factorization in single averages provides a great simplification. For
the incoherent contribution the first term in
\eqref{incoherent_contributionq_definition} produces
\begin{align}
\prod_{k=1}^n\left(\frac{1}{N_c}\Tr\left\langle
W_0^{k\dag}(\v{y})W_0^{k}(\v{x})\right\rangle \right)\equiv \left(\frac{1}{\pi
R^2}\int_\Omega d^2\v{r}_t \frac{1}{N_c}\Tr
W_0^{k\dag}(\v{y})W_0^{k}(\v{x})\right)^n.
\end{align}
Since $\chi_0^k(\v{x})$ as a function of $\v{x}-\v{r}_k$ vanishes for
distances larger than $\mu_d$ we can take the exponential approximation even
for low $n$, \textit{c.f.}  Section \ref{sec:section_2_4}, provided that the
condition $\mu_d^2\ll R^2$ is satisfied. By adding and subtracting $1$ and
assuming a constant density $n_0$ then
\begin{align}
\lim_{R\to\infty}\left(1+\frac{1}{\pi
R^2}\int_\Omega
d^2\v{r}_t\right.&\left.\left[\frac{1}{N_c}\Tr\left(W_0^{1\dag}(\v{y})W_0^1(\v{x})\right)-1\right]\right)^n\\
&\simeq \lim_{R\to\infty}\exp\left\{n_0l\int_\Omega d^2\v{r}_t\left(\frac{1}{N_c}\Tr\left(W_0^{1\dag}(\v{y})W_0^1(\v{x})\right)-1\right)\right\}\nonumber.
\end{align}
By taking the $R\to \infty$ limit and rearranging the terms in the
exponential we obtain
\begin{align}
&\lim_{R\to\infty}\int_\Omega
d^2\v{r}_t\left(\frac{1}{N_c}\Tr\left(W_0^{1\dag}(\v{y})W_0^{1}(\v{x})\right)-1\right)=\\
&\frac{1}{N_c}\Tr\left( F_{el}^{(1)\dag}(\v{0})\right)+\frac{1}{N_c}\Tr\left(
F_{el}^{(1)}(\v{0})\right)+\int\frac{d^2\v{q}}{(2\pi)^2}e^{i\v{q}\cdot(\v{x}-\v{y})}\frac{1}{N_c}\Tr\left(F_{el}^{(1)\dag}(\v{\v{q}})F_{el}^{(1)}(\v{\v{q}})\right)\nonumber.
\end{align}
which by using the optical theorem \eqref{opticaltheorem_quark_single}
simplifies to
\begin{align}
\int\frac{d^2\v{q}}{(2\pi)^2}\bigg(e^{i\v{q}\cdot(\v{x}-\v{y})}-1\bigg)\frac{1}{N_c}\Tr\left(F_{el}^{(1)\dag}(\v{\v{q}})F_{el}^{(1)}(\v{\v{q}})\right),
\end{align}
Similarly for the second term in the incoherent contribution
\eqref{incoherent_contributionq_definition} we find
\begin{align}
&\prod_{k=1}^n\left(\frac{1}{N_c}\Tr\left\langle
  W_0^{k\dag}(\v{y})\right\rangle\right)\prod_{k=1}^n\left(\frac{1}{N_c}\Tr\left\langle W_0^{k}(\v{x})\right\rangle\right)\\
&\equiv \left(\frac{1}{\pi
  R^2}\int_\Omega d^2\v{r}_t\frac{1}{N_c}\Tr
\left(W_0^{1\dag}(\v{y})\right)\right)^n\left(\frac{1}{\pi
  R^2}\int_\Omega d^2\v{r}_t\frac{1}{N_c}\Tr \left(W_0^{1}(\v{x})\right)\right)^n\nonumber\\
&\simeq \exp\left\{n_0l\left(\int_\Omega d^2\v{r}_t\left(\frac{1}{N_c}\Tr\left(W_0^{1\dag}(\v{y})\right)-1\right)+\int_\Omega d^2\v{r}_t\left(\frac{1}{N_c}\Tr\left(W_0^{1}(\v{x})\right)-1\right)\right)\right\}\nonumber.
\end{align}
The term in the exponential, after identifying the single color-space averaged
elastic amplitudes and by using \eqref{opticaltheorem_quark_single}, can be
rewritten as
\begin{align}
\lim_{R\to\infty} \int_\Omega
d^2\v{r}_t\left(\frac{1}{N_c}\Tr\left(W_0^{1\dag}(\v{y})\right)-1\right)+\lim_{R\to\infty}\int_\Omega
d^2\v{r}_t\left(\frac{1}{N_c}\Tr\left(W_0^{1}(\v{x})\right)-1\right)\nonumber\\
=\frac{1}{N_c}\Tr\left(F_{el}^{(1)\dag}(\v{0})\right)+\frac{1}{N_c}\Tr\left(
F_{el}^{(1)}(\v{0})\right)=-\int\frac{d^2\v{q}}{(2\pi)^2}\frac{1}{N_c}\Tr\left(F_{el}^{(1)\dag}(\v{\v{q}})F_{el}^{(1)}(\v{\v{q}})\right).
\end{align}
Finally the incoherent contribution
\eqref{incoherent_contributionq_definition} can be written as
\begin{align}
\hat{\Sigma}_2^{(n)}(\v{q},l)&= \pi
R^2\exp\left[-n_0l\int\frac{d^2\v{q}}{(2\pi)^2}\frac{1}{N_c}\Tr\left(F_{el}^{(1)\dag}(\v{\v{q}})F_{el}^{(1)}(\v{\v{q}})\right)\right]\\
&\times\int d^2\v{x} e^{-i\v{q}\cdot\v{x}}\left\{\exp\left[n_0l\int\frac{d^2\v{q}}{(2\pi)^2}e^{i\v{q}\cdot\v{x}}\frac{1}{N_c}\Tr\left(F_{el}^{(1)\dag}(\v{\v{q}})F_{el}^{(1)}(\v{\v{q}})\right)\right]-1\right\}\nonumber,
\end{align}
which as expected is a solution to the Moliere transport equation
\cite{moliere1948,bethe1953} for QCD matter with boundary condition
$\Sigma_2^{(1)}(\v{q},0)=0$, i.e. in absence of matter the diffracted wave
vanishes. It is easy to show that this transport equation is given by
\begin{align}
\frac{\partial}{\partial l}\Sigma_2^{(n)}(\v{q},l) =n_0\int
\frac{d^2\v{k}_t}{(2\pi)^2}\frac{1}{N_c}\Tr\left(F_{el}^{(1)}(\v{k})F_{el}^{(1)}(\v{k})\right)\left(\Sigma_2^{(n)}(\v{q}-\v{k},l)-\Sigma_2^{(n)}(\v{q},l)\right)\nonumber\\
+\frac{n_0}{N_c}\Tr\left(F_{el}^{(1)}(\v{k})F_{el}^{(1)}(\v{k})\right)\pi R^2\exp\left[-n_0l\int\frac{d^2\v{q}}{(2\pi)^2}\frac{1}{N_c}\Tr\left(F_{el}^{(1)\dag}(\v{\v{q}})F_{el}^{(1)}(\v{\v{q}})\right)\right].\label{moliere_qcd}
\end{align}
The interpretation is straightforward: the number of states with momentum
transfer $\v{q}$ at a depth $l$ grows with the number of states with
$\v{q}-\v{k}$ at $l-\delta l$, experimenting a single collision acquiring
$\v{k}$, and decreases with the number of states with already $\v{q}$ at $l$
experimenting a single collision loosing arbitrary momentum. The extra
boundary term, correcting the Moliere result, accounts for the probability of
penetrating till $l$ without interacting and experimenting a single collision
of momentum transfer $\v{q}$. Notice also that an expansion in the number of
collisions leads to the diagonal contribution, as expected from an incoherent
average. Indeed
\begin{align}
\hat{\Sigma}_2^{(n)}&(\v{q},l)= \pi R^2 \int
d^2\v{x}e^{-i\v{q}\cdot\v{x}}\left(n_0l\int
\frac{d^2\v{k}}{(2\pi)^2}e^{+i\v{k}\cdot\v{x}}\frac{1}{N_c}\Tr\left(F_{el}^{(1)\dag}(\v{\v{k}})F_{el}^{(1)}(\v{\v{k}})\right)+\cdots\right)\nonumber\\
&=\frac{\pi R^2n_0l}{N_c}\Tr\left(F_{el}^{(1)\dag}(\v{\v{q}})F_{el}^{(1)}(\v{\v{q}})\right)+\cdots=\frac{n}{N_c}\Tr\left(F_{el}^{(1)\dag}(\v{\v{q}})F_{el}^{(1)}(\v{\v{q}})\right)+\cdots,
\end{align}
which is just the first diagonal term \eqref{diagonal_term} found in the
initial discussion. The Fourier transforms of the single differential elastic
cross sections can be given in closed form order to order in the coupling
$\alpha_s$. Using \eqref{scatteringamplitudef_perturbative_quark} we find
\begin{align}
&\sigma_{el}^{(1)}(\v{x})=\int\frac{d^2\v{q}}{(2\pi)^2}e^{i\v{q}\cdot\v{x}}\frac{1}{N_c}\Tr\left(F_{el}^{(1)\dag}(\v{\v{q}})F_{el}^{(1)}(\v{\v{q}})\right)\label{sigma_qq_qcd}\\
&=\frac{4\pi g_s^4}{N_c\beta_p^2}t_\beta t_\alpha
\Tr(t_\alpha
t_\beta)\int\frac{d^2\v{q}}{(2\pi)^2}e^{i\v{q}\cdot\v{x}}\left(\hat{A}_0^{(1)}(\v{q})\right)^2=\frac{4\pi
  g_s^4}{\beta_p^2\mu_d^2}\frac{N_c^2-1}{4N_c^2}\mu_d|\v{x}|K_1(\mu_d|\v{x}|).\nonumber
\end{align}
Observe that these functions still have the remaining and trivial matrix
structure in the color of the projectile. After averaging over the traveling
quark colors $\Tr\sigma_{el}^{(1)}(\v{0})/N_c\equiv \sigma_t^{(1)}$ we obtain
the (scalar) total single elastic cross section. Finally we write for the
incoherent term
\begin{align}
\hat{\Sigma}_2^{(n)}(\v{q},l)= \pi
R^2\exp\left[-n_0l\sigma_{el}^{(1)}(\v{0})\right]\int d^2\v{x} e^{-i\v{q}\cdot\v{x}}\bigg(\exp\left[n_0l\sigma_{el}^{(1)}(\v{x})\right]-1\bigg),\label{incoherent_scattering_qcd}
\end{align}
which preserves color in the traveling quark so the incoherent scattering in
QCD, for a color averaged medium behaves, except for the coupling strength,
exactly as in the QED scenario. For a medium length verifying $\delta
l\lesssim \lambda$, where the mean free path is given by
$\lambda=1/n_0\sigma_{el}^{(1)}(\v{0})\equiv 1/n_0\sigma_t^{(1)}$, a
perturbative truncation of the distribution is enough. The average momentum
transfer is then given by
\begin{align}
\left\langle \v{q}^2(\delta l)\right\rangle =\frac{\int d^2\v{q}\medspace
  \v{q}^2\Sigma_2^{(n)}(\v{q},\delta l)}{\int d^2\v{q}\medspace
  \Sigma_2^{(n)}(\v{q},\delta l)}\cong \frac{1}{\sigma_{el}^{(1)}(\v{0})} \int
  d^2\v{q}\v{q}^2\medspace\frac{1}{N_c}\Tr\left(F_{el}^{(1)}(\v{q})F_{el}^{(1)}(\v{q})\right),
\end{align}
which is just the normalized incoherent sum of the averages of the single
collision distributions. The above expression diverges at large $\v{q}$, as
expected. However, by using the energy conservation delta at
\eqref{averaged_squared_amplitude_quarks}, instead of only
\eqref{averaged_squared_amplitudef_quarks}, we find
\begin{align}
\left\langle \v{q}^2(\delta l)\right\rangle=\mu_d^2\left(2\log\left(\frac{2p_0^0}{\mu_d}\right)-1\right).
\end{align}
For mediums of larger size we can use the transport/Moliere equation
\eqref{moliere_qcd}. We then find an equivalent equation for the squared
momentum transfer ,
\begin{align}
\frac{\partial}{\partial l}\left\langle \v{q}^2(l)\right\rangle  =
n_0\sigma_{el}^{(1)}(\v{0})\left\langle \v{q}^2(\delta l)\right\rangle\equiv
2\hat{q},\label{transport_equation_pt2_qcd}
\end{align}
so that the medium transport parameter $\hat{q}$ reads
\begin{align}
\hat{q}=\frac{n_0}{2}
\left(\frac{4\pi
  g_s^4}{\beta_p^2\mu_d^2}\frac{N_c^2-1}{4N_c^2}+\cdots\right)\times \mu_d^2\left(2\log\left(\frac{2p_0^0}{\mu_d}\right)-1\right).
\end{align}
The coherent contribution \eqref{coherent_contributionq_definition}, on the
other hand, consists in the averaged amplitude
\eqref{scatteringamplitude_quark} squared, encoding the diffractive behavior
of the medium. For its evaluation we observe
\begin{align}
\prod_{k=1}^n\left(\frac{1}{N_c}\Tr\left\langle
W_0^{k}(\v{x})\right\rangle\right)=\left(1+\frac{1}{\pi R^2}\int_\Omega
d^2\v{r}_t\left( \frac{1}{N_c}\Tr\left(W_0^{1}(\v{x}_t)\right)-1\right)\right)^n\nonumber\\
\simeq\exp\left[n_0l\int_\Omega d^2\v{r}_t \left(\frac{1}{N_c}\Tr \left(W_0^{1}(\v{x})-1\right)\right)\right].
\end{align}
By using \eqref{scatteringamplitudef_quark} and \eqref{wtrick} the term in the
exponential can be written as a function $\pi_{el}^{(1)}(\v{x})$
\begin{align}
\int_\Omega d^2\v{r}_t \left(\frac{1}{N_c}\Tr \left(W_0^{(1)}(\v{x})-1\right)\right)
&=
  \int\frac{d^2\v{q}}{(2\pi)^2}e^{i\v{q}\cdot\v{x}}w(\v{q},R)\frac{1}{N_c}\Tr
  F_{el}^{(1)}(\v{q})\equiv
  \pi_{el}^{(1)}(\v{x}),
\end{align}
where the window function $w(\v{q},R)$ can be given in general for any
geometry. In particular for the cylinder or the Gaussian decaying cylinder, as
we noted in \eqref{window_function}. Since at all orders the color traces
produce color conservation in the projectile, the coherent contribution is
given by a perfect square times the unit matrix
\begin{align}
\hat{\Pi}_2^{(1)}(\v{q})=\left|\int
d^2\v{x} e^{-i\v{q}\cdot\v{x}}\left(\exp\bigg(n_0l\pi_{el}^{(1)}(\v{x})\bigg)-1\right)\right|^2.
\end{align}
Observe that an expansion in the number of collisions produces
\begin{align}
\int d^2\v{x}_t
e^{-i\v{q}\cdot\v{x}}\left(\exp\left(n_0l\pi_{el}^{(1)}(\v{x})\right)-1\right)= n_0lw(\v{q},R)\frac{1}{N_c}\Tr
  F_{el}^{(1)}(\v{q})+\cdots,
\end{align}
and then at first order we obtain
\begin{align}
\hat{\Pi}_2^{(1)}(\v{q})= n^2\left(\frac{w(\v{q},R)}{\pi
  R^2}\right)^2\frac{1}{N_c}\Tr \left(F_{el}^{(1)\dag}(\v{q})\right)\frac{1}{N_c}\Tr \left(F_{el}^{(1)}(\v{q})\right)+\cdots,
\end{align}
which is the non-diagonal contribution \eqref{nondiagonal_term} we have found
in the single scattering regime. The typical width of the function
$\pi_{el}^{(1)}(\v{x})$ is $R$, instead of the width $r_d=1/\mu_d$, the
dimensions of a single scattering center, of the function
$\sigma_{el}^{1}(\v{x})$. Consequently, the typical momentum change in a
coherent scattering is always smaller than $\mu_d$ provided that
$R>r_d=1/\mu_d$, as we already encountered in the expansion in the number of
collisions. For $R\gg r_d$, a leading order in $\alpha_s$ evaluation of this
function is possible
\begin{align}
\pi_{el}^{(1)}(\v{x})=-\frac{\alpha^2_s}{2\beta_p^2}\frac{N_c^2-1}{4N_c^2}1\int \frac{d^2\v{q}}{(2\pi)^2}e^{i\v{q}\cdot\v{x}}w(\v{q},R)\frac{16\pi\arcsinh\left(\frac{q}{2\mu_d}\right)}{q\sqrt{q^2+\mu_d^2}}+\cdots.
\end{align}
In particular we find for a cylinder with Gaussian decaying density, in the
saddle point approximation,
\begin{align}
\pi_{el}^{(1)}(\v{x})=-\frac{\alpha^2_s}{2\beta_p^2}\frac{N_c^2-1}{4N_c^2}\frac{\pi}{\mu_d^2}\exp\left(-\frac{\v{x}^2}{2R^2}\right)+\cdots,
\end{align}
which has less than a $\sim 2\%$ of deviation of the actual value for
$R/r_d>10$ and for $x/R<4$. We also notice that, as expected, when
$R\to\infty$, $\pi_{el}^{(1)}(\v{x})$ is independent on $\v{x}$ and we recover
the pure forward contribution. An evolution equation at the level of the
amplitude can be written 
\begin{align}
\frac{\partial}{\partial l}\Pi_1^{(n)}(\v{q})=n_0\int
\frac{d^2\v{k}_t}{(2\pi)^2}w(\v{k},R)\frac{1}{N_c}\Tr\left(F_{el}^{(1)}(\v{k})\right)&\Pi_1^{(n)}(\v{q-k})\\
&+w(\v{q},R)\frac{1}{N_c}\Tr\left(F_{el}^{(1)}(\v{q})\right)\nonumber.
\end{align}
which can not be interpreted in probabilistic terms since
$\Pi_2^{(n)}(\v{q})=\Pi_1^{(n)\dag}(\v{q})\Pi_1^{(n)}(\v{q})$. 
\section{Beyond eikonal scattering}
\label{sec:section_4_4}
We will now assume that the $q_z$ component of the momentum change cannot be
omitted, either in order to take into account the $1/p_0$ corrections of the
preceding pure eikonal results or in order to consider the interference
between amplitudes with different on-shell states, as is the case of inelastic
processes. In that case the phase accumulated by the dropped term $q_zz$ can
be large at large distances, and as we have seen in the QED case, lead to
interference phenomena such as the LPM effect. The amplitude for a set of $n$
quarks placed sharing equal $z$ coordinate can be written as
\begin{align}
M_{s_fs_i}^{(n)}(p_f,p_i)=2\pi\delta(\delta p_i^0)\delta^{s_f}_{s_i}\beta_p\int
d^3\v{y} e^{-i\v{q}\cdot\v{y}}\frac{\partial}{\partial y_3}\exp\left(-i\frac{g_s}{\beta_p}\int^{y_3}_{-\infty} dy_3't_\alpha A_\alpha^0(\v{y})\right),
\end{align}
where the eikonal phase reads, using \eqref{fielddefinition},
\begin{align}
\int_{-\infty}^{y_3} dy_3' \medspace t_\alpha
A_\alpha^0(\v{y})=g_st_\alpha t_\alpha\sum_{i=k}^n \int^{y_3}_{-\infty}
dy_3' \int\frac{d^3\v{q}}{(2\pi)^3}\frac{4\pi}{\v{q}^2+\mu_d^2}e^{-i\v{q}\cdot(\v{y}-\v{r}_k)}.
\end{align}
We will assume that $\v{q}^2\simeq \v{q}_t^2$ so that the longitudinal
structure of each center is neglected, but we keep $q_z$ in the phase since
$y_3$ can be arbitrarily large and we want to keep the longitudinal structure
of the medium. This approximation produces
\begin{align}
 \int^{y_3}_{-\infty}
dy_3'&
\int\frac{d^3\v{q}}{(2\pi)^3}\frac{4\pi}{\v{q}^2+\mu_d^2}e^{-i\v{q}\cdot(\v{y}-\v{r}_k)}\simeq\int^{y_3}_{-\infty}
dy_3'\delta(y_3'-r_3^k)\\
&\times\int
\frac{d^2\v{q}_t}{(2\pi)^2}\frac{4\pi}{\v{q}^2_t+\mu_d^2}e^{-i\v{q}_t\cdot(\v{y}_t-\v{r}_t^k)}
=\Theta(y_3-r_3^k)\chi^{(1)}_0(\v{y}_t-\v{r}_t^k)\equiv\Theta^k(y_3)\chi^{k}_0(\v{y}_t)\nonumber.
\end{align}
Since we placed the quarks at the same $z$ coordinate we can easily integrate
by parts the amplitude, which takes the simpler form
\begin{align}
M_{s_fs_i}^{(n)}(p_f,p_i)=\bigg(S_{s_fs_i}^{(n)}(p_f,p_i)-S_{s_fs_i}^{(0)}(p_f,p_i)\bigg)\label{elastic_amplitudem_1layer_qcd},
\end{align}
where the eikonal amplitude, including the no collision amplitude, is given as
usual by
\begin{align}
S_{s_fs_i}^{(n)}(p_f,p_i)\equiv&2\pi\beta_p\delta(\delta
p_i^0)\delta^{s_f}_{s_i}\int
d^2\v{y}_te^{-i\v{q}\cdot\v{y}}\exp\left(-i\frac{g_s^2}{\beta_p}t_\alpha
t_\alpha \sum_{k=1}^n\chi^k_0(\v{y})\right)\label{elastic_amplitudes_1layer_qcd},
\end{align}
where the total three momentum change $\v{q}\cdot\v{y}$ appears now in the
phase. We can consider the medium as a set of $n$ layers of quarks at $z_1$,
$z_2$ ... $z_n$, of $n(z_i)$ quarks respectively. The total number of quarks
in the medium is defined as
\begin{align}
\sum_{i=1}^nn(z_i)\equiv N.
\end{align}
At high energies the amplitude of finding the quark with momentum $p_n$ after
the passage through these layers is given by the convolution of single layer
amplitudes, then
\begin{align}
S_{s_ns_0}^{(N)}(p_n,p_0)=\sum_{s}\left(\prod_{i=1}^{n-1}\int
\frac{d^3\v{p}_i}{(2\pi)^3}\right)\left(\prod_{i=1}^n
S_{s_is_{i-1}}^{n(z_i)}(p_i,p_{i-1})\right)\label{elastic_amplitudes_nlayers_qcd},
\end{align}
where sum over intermediate spins and ordering is assumed. So that we find
\begin{align}
M_{s_ns_0}^{(N)}(p_n,p_0)=S_{s_ns_0}^{(N)}(p_n,p_0)-S_{s_ns_0}^{(0)}(p_n,p_0).
\end{align}
At this point we notice that a path integral expression for the amplitude
exists by taking the separation between layers $\delta z_k\to 0$. The
resulting expression is, however, uninteresting out of a formal context, since
the interaction carries the matrix structure in the traveling and target color
spaces. We instead take the intensity and average over target colors, which
leads to color preservation in the projectile. By splitting as before in a
incoherent and a coherent contribution we obtain
\begin{align}
\left\langle
M_{s_ns_0}^{(N)\dag}(p_n,p_0)M_{s_ns_0}^{(N)}(p_n,p_0)\right\rangle = \Sigma_2^{(N)}(p_n,p_0)+\Pi_s^{(n)}(p_n,p_0),
\end{align}
where the coherent contribution is given by the averaged amplitudes squared
\begin{align}
\Pi_2^{(N)}(p_n,p_0)=\left\langle
S_{s_ns_0}^{(N)\dag}(p_n,p_0)-S_{s_ns_0}^{(0)\dag}(p_n,p_0)\right\rangle\left\langle
S_{s_ns_0}^{(N)}(p_n,p_0)-S_{s_ns_0}^{(0)}(p_n,p_0)\right\rangle ,
\end{align}
and the incoherent contribution instead by
\begin{align}
\Sigma_2^{(N)}(p_n,p_0)=\left\langle
S_{s_ns_0}^{(N)\dag}(p_n,p_0)S_{s_ns_0}^{(N)}(p_n,p_0)\right\rangle-\left\langle
S_{s_ns_0}^{(N)\dag}(p_n,p_0)\right\rangle\left\langle S_{s_ns_0}^{(N)}(p_n,p_0)\right\rangle.
\end{align}
The first term of the incoherent contribution can be written, after dividing
by the incoming quark flux and the infinite factor $T=2\pi\delta(0)$
accounting for time translation invariance, as the product
\begin{align}
&\left\langle
S_{s_ns_0}^{(N)\dag}(p_n,p_0)S_{s_ns_0}^{(N)}(p_n,p_0)\right\rangle=2\pi\beta_p\delta(p_n^0-p_0^0)\delta^{s_n}_{s_0}\\
&\times\prod_{i=1}^{n-1}\left(\int\frac{d^2\v{p}_i^t}{(2\pi)^2}\int\frac{d^2\v{u}_i^t}{(2\pi)^2}\exp\left(-i\frac{\big(\v{p}_i^t\big)^2-\big(\v{u}_i^t\big)^2}{2p_0^0}\delta
z_i\right)\right)\nonumber\\
&\times\prod_{i=1}^n\left(\int d^2\v{x}_i^t\int
d^2\v{y}_i^te^{-i\delta\v{p}_i^t\cdot\v{x}_i^t+i\delta\v{u}_i^t\cdot\v{y}_i^t}\prod_{k=1}^{n(z_i)}\left(\frac{1}{N_c}\Tr\left\langle
W_0^{k\dag}(\v{y}_t)W_0^k(\v{x}_t)\right\rangle\right)\right)\nonumber,
\end{align}
where $\delta z_i\equiv z_{i+1}-z_i$ and $\delta \v{p}_i=\v{p_i}-\v{p}_{i-1}$
and the internal momentum in the conjugated amplitude has been denoted by
$\v{u}_i$. The bottom line is just a product of averages at each layer of
length $\delta z_i$ and density $n_0(z_i)$ like the one computed in the
previous section, thus we write
\begin{align}
\int&d^2\v{x}_i^t\int
d^2\v{y}_i^te^{-i\delta\v{p}_i^t\cdot\v{x}_i^t+i\delta\v{u}_i^t\cdot\v{y}_i^t}\prod_{k=1}^{n(z_i)}\left(\frac{1}{N_c}\Tr\left\langle
W_0^{k\dag}(\v{y}_t)W_0^k(\v{x}_t)\right\rangle\right)\\
&=(2\pi)^2\delta^2(\delta\v{p}_i^t-\delta\v{u}_i^t)\int d^2\v{x}_t^i
e^{-i\delta\v{p}_i^t\cdot \v{x}_i^t}\exp\bigg(\delta
z_in_0(z_i)\left(\sigma_{el}^{(1)}(\v{x}_i^t)-\sigma_{el}^{(1)}(\v{0})\right)\bigg)\nonumber.
\end{align}
Similarly for the second term in the incoherent contribution we find
\begin{align}
&\left\langle
S_{s_ns_0}^{(N)\dag}(p_n,p_0)\right\rangle\left\langle S_{s_ns_0}^{(N)}(p_n,p_0)\right\rangle=2\pi\beta_p\delta(p_n^0-p_0^0)\delta^{s_n}_{s_0}\\
&\medspace\medspace\times\prod_{i=1}^{n-1}\left(\int\frac{d^2\v{p}_i^t}{(2\pi)^2}\int\frac{d^2\v{u}_i^t}{(2\pi)^2}\exp\left(-i\frac{\big(\v{p}_i^t\big)^2-\big(\v{u}_i^t\big)^2}{2p_0^0}\delta
z_i\right)\right)\nonumber\\
&\medspace\medspace\times\prod_{i=1}^n\left(\int d^2\v{x}_i^t\int
d^2\v{y}_i^te^{-i\delta\v{p}_i^t\cdot\v{x}_i^t+i\delta\v{u}_i^t\cdot\v{y}_i^t}\prod_{k=1}^{n(z_i)}\left(\frac{1}{N_c}\Tr\left\langle
W_0^{k\dag}(\v{y}_t)\right\rangle\Tr\left\langle W_0^k(\v{x}_t)\right\rangle\right)\right)\nonumber,
\end{align}
where as before by using the results of the preceding section we obtain for
each of the single layer averages
\begin{align}
\int&d^2\v{x}_i^t\int
d^2\v{y}_i^te^{-i\delta\v{p}_i^t\cdot\v{x}_i^t+i\delta\v{u}_i^t\cdot\v{y}_i^t}\prod_{k=1}^{n(z_i)}\left(\frac{1}{N_c}\Tr\left\langle
W_0^{k\dag}(\v{y}_t)\right\rangle\left\langle W_0^k(\v{x}_t)\right\rangle\right)\\
&=(2\pi)^2\delta^2(\delta\v{p}_i^t-\delta\v{u}_i^t)\int d^2\v{x}_t^i
e^{-i\delta\v{p}_i^t\cdot \v{x}_i^t}\exp\bigg(-\delta
z_in_0(z_i)\sigma_{el}^{(1)}(\v{0})\bigg)\nonumber.
\end{align}
We can now proceed to integrate in the conjugate momentums $\v{u}_i^t$
observing that since $\v{u}_0^t\equiv \v{p}_0^t$ and, by using the transverse
momentum deltas, the longitudinal phase vanishes
\begin{align}
\prod_{i=1}^{n-1}\left(\int\frac{d^2\v{u}_i^t}{(2\pi)^2}\exp\left(-i\frac{\big(\v{p}_i^t\big)^2-\big(\v{u}_i^t\big)^2}{2p_0^0}\delta
z_i\right)\right)\prod_{i=1}^n\left((2\pi)^2\delta^2(\delta\v{p}_i^t-\delta\v{u}_i^t)\right)=(2\pi)^2\delta^2(\v{0}),
\end{align}
as expected by symmetry arguments for a medium of infinite transverse
size. Then the incoherent contribution is given simply by
\begin{align}
\Sigma_2^{(N)}(p_n,p_0)=2\pi\delta(p_n^0-p_0^0)\beta_p \delta^{s_n}_{s_0}\pi
R^2\prod_{i=1}^{n-1}\left(\int \frac{d^2\v{p}_i^t}{(2\pi)^2}\right)\prod_{i=1}^{n}\Bigg(\int d^2\v{x}_i^te^{-i\delta\v{p}_i^t\cdot\v{x}_i^t}\Bigg)\nonumber\\
\times\left\{\prod_{i=1}^n\exp\bigg(\delta
z_in_0(z_i)\left(\sigma_{el}^{(1)}(\v{x}_i^t)-\sigma_{el}^{(1)}(\v{0})\right)\bigg)-\prod_{i=1}^n\exp\bigg(-\delta
z_in_0(z_i)\sigma_{el}^{(1)}(\v{0})\bigg)\right\}.
\end{align}
We notice that the integration in the internal momentum variables is trivial
and the internal scattering structure is lost. The joint action of the squared
averaged amplitudes at each layer simply convolute without phase and the pure
eikonal limit remains valid. By taking the $\delta z_i\to 0$ limit we obtain
\begin{align}
\Sigma_2^{(N)}(p_n,p_0)=2\pi\delta(p_n^0-p_0^0)\beta_p \delta^{s_n}_{s_0}\pi
R^2\exp\left(-\sigma_{el}^{(1)}(\v{0})\int_{z_1}^{z_n} dz \medspace
n_0(z)\right)\nonumber\\
\times\int d^2\v{x}_te^{-i(\v{p}_n^t-\v{p}_0^t)\cdot\v{x}_t}\left(\exp\left(\sigma_{el}^{(1)}(\v{x}_t)\int_{z_1}^{z_n} dz \medspace n_0(z)\right)-1\right),
\end{align}
which except for the varying local density is our previous result. For the
coherent contribution the scenario is however different if we do not take the
$R\to\infty$ limit. We have
\begin{align}
\left\langle
S_{s_ns_0}^{(N)}(p_n,p_0)\right\rangle& =
2\pi\delta(p_n^0-p_0^0)\beta_p\delta^{s_n}_{s_0}\nonumber\\
&\times\prod_{i=1}^{n-1}\left(\int\frac{d^2\v{p}_i^t}{(2\pi)^2}\right)\prod_{i=1}^{n}\left(\int
d^2\v{x}_i^te^{-i\delta\v{p}_i^t\cdot\v{x}_i^t}\prod_{k=1}^{n_0(z_i)}\left(\frac{1}{N_c}\Tr\left\langle
W_0^k(\v{x}_k^t)\right\rangle\right)\right)\nonumber\\
&\times\exp\left(+i\frac{(\v{p}_n^t)^2}{2p_0^0}z_n-i\sum_{i=1}^{n-1}\delta z_i\frac{\big(\v{p}_i^t\big)^2}{2p_0^0}-i\frac{\big(\v{p}_0^t\big)^2}{2p_0^0}z_1\right).
\end{align}
The total kinematical phase must be rearranged as follows
\begin{align}
-i\v{p}_n^t\cdot\v{x}_n^t+i\frac{(\v{p}_n^t)^2}{2p_0^0}z_n+i\sum_{i=1}^{n-1}\v{p}_i^t\cdot\frac{\delta\v{x}_i^t}{\delta
z_i}\delta z_i-i\sum_{i=1}^n\frac{(\v{p}_i^t)^2}{2p_0^0}\delta z_i-i\frac{\big(\v{p}_0^t\big)^2}{2p_0^0}z_1+i\v{p}_0^t\v{x}_1^t,
\end{align}
and the medium contribution is given by the results of the previous section
\begin{align}
\prod_{k=1}^{n_0(z_i)}\left(\frac{1}{N_c}\Tr\left\langle
W_0^k(\v{x}_k^t)\right\rangle\right)=\exp\left(\delta z_i n_0(z_i)\pi_{el}^{(1)}(\v{x}_i^t)\right).
\end{align}
Then we can perform now the integral in internal momenta. If we omit the two
boundary terms in $z_n$ and $z_0$, which otherwise will cancel when taking the
square of the amplitude, we find
\begin{align}
&\left\langle S_{s_ns_0}^{(N)}(p_n,p_0)\right\rangle =
2\pi\delta(p_n^0-p_0^0)\beta_p\delta^{s_n}_{s_0}\prod^{n-1}_{i=1}\left(\frac{i\delta
z_i}{2\pi p_0^0}\right)\prod_{i=1}^n\left(\int d^2\v{x}_t^i\right)\\
&\times\exp\left(-i\v{p}_n^t\cdot\v{x}_n^t+i\sum_{i=1}^{n-1}\frac{p_0^0}{2}\left(\frac{\delta \v{x}_i^t}{\delta
z_i}\right)^2\delta z_i+\sum_{i=1}^{n}\delta z_i n_0(z_i)\pi_{el}^{(1)}(\v{x}_i^t)+i\v{p}_0^t\v{x}_1^t\right)\nonumber,
\end{align}
which by taking the $\delta z\to 0$ limit transforms into a path integral in
the transverse plane with time variable the $z$ position as
\begin{align}
\left\langle S_{s_ns_0}^{(N)}(p_n,p_0)\right\rangle=
2\pi\delta(p_n^0-p_0^0)\beta_p\delta^{s_n}_{s_0}\int d^2\v{x}_n^te^{-i\v{p}_n^t\cdot\v{x}_n^t}\int d^2\v{x}_1^t
e^{+i\v{p}_0^t\v{x}_1^t}\nonumber\\
\int \mathcal{D}^2\v{x}_t(z)\exp\left(i\int^{z_n}_{z_1}
dz\left(\frac{p_0^0}{2}\dot{\v{x}}_t^2(z)-in_0(z)\pi_{el}^{(1)}\Big(\v{x}_t(z)\Big)\right)\right).
\end{align}
Similarly the second term in the coherent contribution is simply found by
doing $N$=0 in the above relation, thus by dividing by the incoming quark flux
and a factor $T=2\pi\delta(0)$ accounting time translation invariance, we
finally write
\begin{align}
&\Pi_2^{(N)}(p_{n},p_0)=2\pi\delta(p_n^0-p_0^0)\beta_p\delta^{s_n}_{s_0}\Bigg|\int d^2\v{x}_n^te^{-i\v{p}_n^t\cdot\v{x}_n^t}\int d^2\v{x}_1^t
e^{+i\v{p}_0^t\v{x}_1^t}\\&\times\int \mathcal{D}^2\v{x}_t(z)\exp\left(i\int^{z_n}_{z_1}
dz\medspace\frac{p_0^0}{2}\dot{\v{x}}_t^2(z)\right)\left(\exp\left(\int_{z_1}^{z_n}
dz \medspace n_0(z)\pi_{el}^{(1)}\Big(\v{x}_t(z)\Big)\right)-1\right)\Bigg|^2\nonumber,
\end{align}
which either in the $p_0^0\to\infty$ or the $R\to\infty$ limits transforms
into our previous eikonal result.

%% file: chapter5/chapter5.tex
\chapter{High energy emission in QCD}
The LPM effect in a QCD multiple scattering scenario has become an useful tool
of indirectly observing the properties of the hadronic matter at extremely
high temperatures. High energy collisions of heavy nuclei at the RHIC
\cite{gyulassy2004,adare2008a,adare2008b,adcox2001a} and the LHC
\cite{abelev2012a,cms2012a,aad2015} hadron colliders are expected to reproduce
the required extreme conditions for the QGP formation. Among other probes, the
energy loss of hard jets constituents
\cite{bjorken1982a,ryskin1990a,gavin1992a,gyulassy1990a,gyulassy1990b}
produced in these collisions can be used to trace back the relevant
characteristics of the formed QCD medium. While traveling through the QGP, a
high energetic quark or gluon may undergo a collisional process with the
medium constituents. The leading contribution to the energy loss suffered in
this multiple collision scenario is of radiative nature
\cite{gyulassy1990a}. As we have already seen, the radiation of quanta leading
to this energy loss is going to be severely modified by the existence of
multiple sources of scattering \cite{termikaelian1953,landau1953a,migdal1956},
compared to an emission scenario consisting in an incoherent sum of single
collision intensities \cite{bethe1934,gunion1982}. Hence a systematic study of
the intensity of gluon bremsstrahlung in media and the related parton energy
loss shall provide indicative signs of the QGP formation and its
characteristics.

Early studies of the intensity of gluon bremsstrahlung in a multiple
scattering context exist \cite{sorensen1992a,gavin1992a,ryskin1990a}. The main
difficulties introduced with respect to the QED case we have reviewed in
Chapter 3 are related to the non abelian nature of QCD, namely the color
structure and the gluon ability to reinteract with the medium. As we will
show, these circumstances lead to a LPM interference in the soft limit
dominated by the gluon rescattering, in analogy with the dielectric and
transition radiation effects occurring in QED and shown at Figure
\ref{fig:figure_3_1}. A first attempt for a QCD evaluation was given at
\cite{sorensen1992a} directly using the Migdal prediction \cite{migdal1956}
but without including gluon rescattering, and a more detailed evaluation was
then given at \cite{gyulassy1994a,wang1994a} without gluon rescattering effect
in the longitudinal phase. Semi-infinite medium calculations in the
Fokker-Planck approximation for the angle integrated intensity, either
following a transport approach \cite{baier1995} or a path integral formalism
\cite{zakharov1996a,zakharov1998} were soon presented including the full gluon
rescattering effect. These works can be shown to be the QCD equivalent to the
Boltzmann transport approach \cite{bell1958} used by Migdal \cite{migdal1956}
with the adequate approximations \cite{baier1998a}. Further extensions
including the angular dependence, finite, cold or expanding target scenarios
were later developed
\cite{zakharov1996b,zakharov1997,zakharov1999a,zakharov2000a,baier2001b,baier1997b,baier1997a,baier1998b}
although usually lacking a general formulation, until a finite size evaluation
of the angle dependence of the spectrum was given
\cite{wiedemann1999,wiedemann2000a,wiedemann2000b,wiedemann2004,salgado2003a}
generalizing the previous results. The intensity of gluon bremsstrahlung
within this framework
\cite{kovner2001,dokshitzer2001,baier2001a,salgado2002,salgado2003a,salgado2003b,zakharov2004a,armesto2004a,arnold2009a,armesto2009a,zapp2008a,caronhuot2010,zapp2011,armesto2011a,armesto2012a,apolinario2012a,apolinario2012b,zapp2012a,zapp2012b,armesto2013a,apolinario2015a,apolinario2014a,apolinario2014b,arleo2017,altinoluk2014a,altinoluk2018a}
was then widely used to make predictions of the quenching behavior of the
hadron spectra due to a multiple collision process with confined or deconfined
QCD matter. Several other distinct frameworks were also developed at the same
time or soon after the previous works, among them a reaction operator
formalism for finite media
\cite{gyulassy1999b,gyulassy1999a,gyulassy2000,gyulassy2001} and for QGP
\cite{zhang2004,djordjevic2004}, a high twist formalism for finite nuclei
\cite{wang2001a,guo2006} and a finite field theory framework
\cite{arnold2000,arnold2001,turbide2005}. Applications of these results were
also presented to make finite plasma, cold nuclear matter and heavy flavor
suppression predictions at RHIC and LHC, see for instance
\cite{vitev2002,chang2014,djordjevic2015}. An excellent comparison of all
these frameworks was given at \cite{bass2008a} and very good and detailed
reviews on the subject have been written
\cite{blaizot2001,casalderreysolana2007,wiedemann2009a,armesto2011b,gyulassy2003a,denterria2009a}.

The aim of this chapter is to develop a formalism \cite{feal2018a} in which
the intensity of gluons in a multiple scattering scenario can be evaluated for
a general interaction with the constituents of a finite or structured medium,
and in which the angular spectrum is also considered. We will basically follow
the steps we took at Chapter 3 for the QED case, to construct a
non-perturbative description of the emission amplitude. At the level of the
amplitude the diagrammatic structure becomes clear and the soft photon theorem
\cite{weinberg1965,bassetto1984a,mueller1989a} for finite media can be
understood, together with the LPM effect, as part of the same coherence
phenomena. The arising gluon intensity, which is the central result of this
chapter, will lead, under the adequate color averaged interaction of the gluon
and its Fokker-Planck approximation, to the well known results at
\cite{baier1995,zakharov1996a,wiedemann1999}. In order to make a direct
connection with the previous results, the results obtained for the multiple
soft scattering case are easily adapted to implement an approximation for
the emission scenario after a first hard collision.

In Section \ref{sec:section_5_1} we define amplitude of emission of a gluon
under the effect of the medium as a set of static and classical Debye screened
sources, as we explained at Section \ref{sec:section_4_1}. The diagrammatic
structure of the bremsstrahlung amplitude for multiple collisions is explained
and the Bertsch-Gunion limit \cite{gunion1982} recovered for the single
collision scenario. In Section \ref{sec:section_5_2} we compute the intensity
of gluons for a color and space averaged target with finite length $l$ but for
simplicity with macroscopical transverse size $R\to\infty$. In this
approximation the transverse coherent contributions to the multiple scattering
distributions are neglected. A color averaged effective interaction for the
gluon is used which translates into Debye screened single interactions
agreeing with previous results
\cite{nikolaev1994b,zakharov1996a,wiedemann1999}. The resulting intensity is
shown to satisfy a QED like form, where the LPM interfering phase, however, is
governed by the gluon rescattering, and then the role of the traveling fermion
is played by the gluon. Emission in a multiple soft scattering scenario will
be shown to be dominated by a longitudinal coherent contribution,
corresponding to the first and last leg gluons of the process, and lightly
enhanced for small gluon energies as a result of the LPM effect, and then
suppressed in the soft regime due to the mass suppression effect. An
energy-loss formalism for that scenario will be developed using energy
conservation constraints. Finally, the radiation due to a multiple soft
collision scenario after a hard collision will be approximated as a particular
subset of terms. The intensity in that case, except for the suppression in the
soft limit related to a non vanishing gluon mass, will be shown to reproduce
the well known results \cite{baier1995,wiedemann1999,salgado2003a} if the
Fokker-Planck approximation is used.
\section{Amplitude}
\label{sec:section_5_1}
\begin{figure}
\begin{minipage}{0.33\textwidth}
\begin{center}
\includegraphics[width=1.\textwidth]{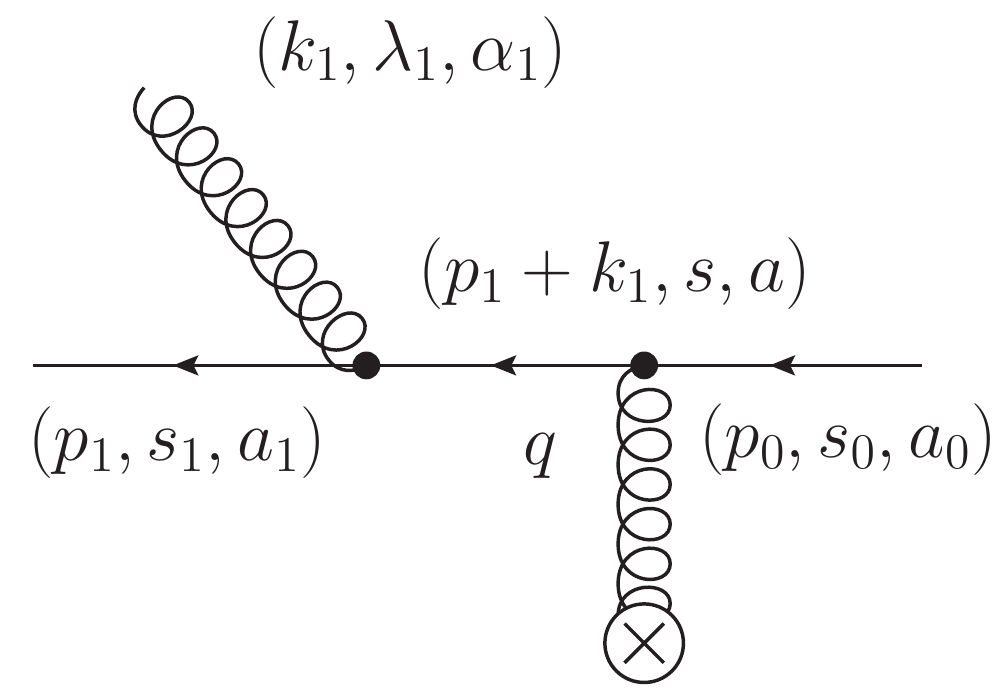}
\end{center}
\end{minipage}
\hfill
\begin{minipage}{0.33\textwidth}
\begin{center}
\includegraphics[width=1.\textwidth]{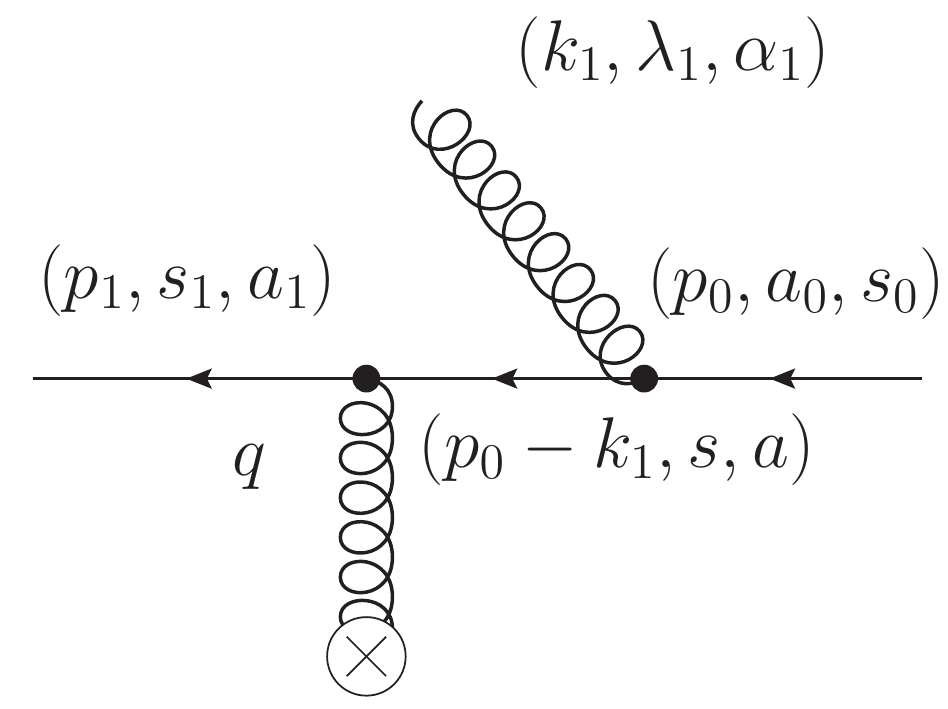}
\end{center}
\end{minipage}
\hfill
\begin{minipage}{0.28\textwidth}
\begin{center}
\includegraphics[width=1.\textwidth]{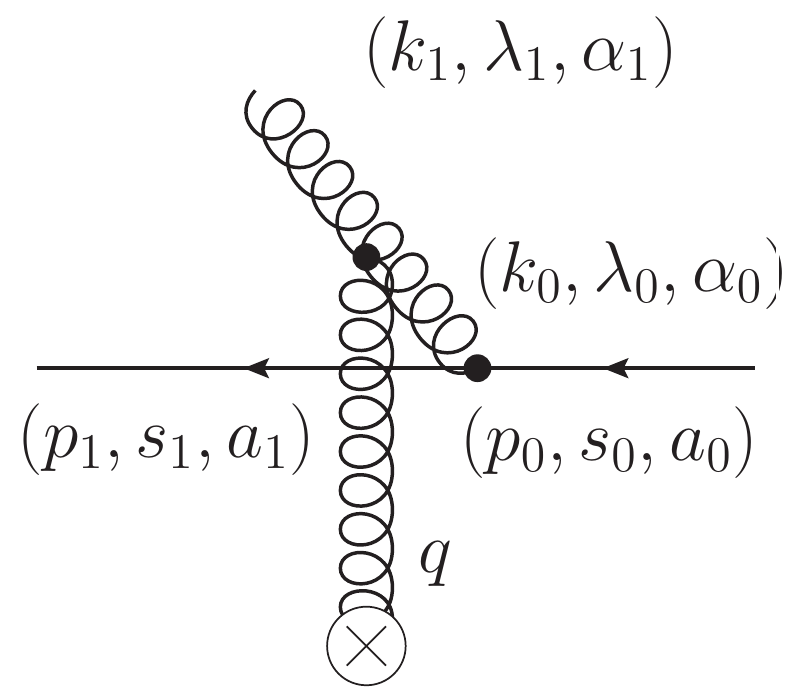}
\end{center}
\end{minipage}
\caption{The Feynman diagrams appearing in the leading order
  expansion in the coupling constant $g_s^2$ of
  $\mathcal{M}_{em}^{(n)}-\mathcal{M}_{em}^{(0)}$ in the single collision
  regime \eqref{emission_amplitude_1layer_qcd}.}
\label{fig:figure_5_1}
\end{figure}

Let us consider an ideal scenario for an on-shell quark coming from the
infinity and going to the infinity, while undergoing a multiple scattering
process with a QCD medium. The medium will be composed of $N$ static and
classical sources characterized by \eqref{fielddefinition}, extending from
$z_1$ to $z_n$ in the initial quark direction. In QCD the amplitude of
emission \eqref{amplitude_emission_m} of a single gluon, \textit{c.f.}
Sections \ref{sec:section_3_1} and \ref{sec:section_4_2}, is given by
\begin{align}
\mathcal{M}_{em}^{(N)}\equiv -ig_s\int d^4x
\psi_{f}^{(N)}(x)A_{\mu,\alpha}^{(N)}(x)t_\alpha\gamma^\mu \psi_i^{(N)}(x),\label{emission_amplitude_qcd_definition}
\end{align}
where $\psi_f^{(N)}(x)$, $A_{\mu,\alpha}^{(N)}(x)$ and $\psi_i^{(N)}(x)$ are
the states of the outcoming quark, the emitted gluon and the incoming quark,
respectively. The initial and final quark asymptotic states are denoted by
$\psi_i^{(0)}(x)$ and $\psi_f^{(0)}(x)$, with momentum, spin and color
$(p_0,s_0,a_0)$ and $(p_n,s_n,a_n)$, respectively
\begin{align}
\psi_{i}^{(0)}(x)\equiv\sqrt{\frac{m_q}{p_0^0}}u_{s_0}^{a_0}(p_0)e^{-ip_0\cdot
  x},\medspace\medspace\medspace\medspace \medspace\medspace\medspace\medspace\psi_{f}^{(0)}(x)\equiv\sqrt{\frac{m_q}{p_n^0}}u_{s_n}^{a_n}(p_n)e^{-ip_n\cdot
  x},
\end{align}
where we used spinor conventions given at Appendix \ref{appendix1} and $m_q$
is the quark mass. The gluon final asymptotic state is written
$A_{\mu,\alpha_n}^{(0)}(x)$, with momentum, polarization and color
$(k_n,\lambda_n,\alpha_n)$
\begin{align}
A_{\mu,\alpha_n}^{(0)}(x)=\mathcal{N}_k\epsilon_{\mu,\alpha_n}^{\lambda_n}(k_n)e^{-ik_n\cdot
x},
\end{align}
where the normalization, given in Gaussian units, reads
$\mathcal{N}_k=\sqrt{4\pi/2\omega}$, $\omega$ is the gluon energy and the
polarization vector reads $\epsilon^{\lambda_n}(k_n)$. The high energy states
at any coordinate $x$ under the effect of the $N$ sources of the medium can be
written as a superposition of diffracted states of the form
\begin{align}
\psi_i^{(N)}(x)=\psi_i^{(0)}(x)+\int\frac{d^3\v{p}_k}{(2\pi)^3}\sqrt{\frac{m_q}{p_k^0}}u_{s_k}^{a_k}(p_k)e^{-ip_k\cdot
  x}\left(M_{q}^{(N)}(p_k,p_0;z_k,z_1)\right)_{a_0s_0}^{a_ks_k},\label{quark_incoming_state}
\end{align}
sum over repeated indices assumed. The matrix elements of
$M^{(N)}_{q}(p_k,p_0;z_k,z_1)$ are the beyond eikonal elastic amplitude
\eqref{elastic_amplitudem_1layer_qcd} of finding the quark in the state
$(p_k,s_k,a_k)$ after traversing the colored matter in the interval
$(z_1,z_k)$. These amplitudes carry local and ordered longitudinal phases,
responsible of regulating the coherence in the squared amplitude leading to
the LPM effect. Similarly for the final quark we write
\begin{align}
\psi_f^{(N)}(x)=\psi_f^{(0)}(x)+\int\frac{d^3\v{p}_k}{(2\pi)^3}\sqrt{\frac{m_q}{p_k^0}}u^{a_k}_{s_k}(p_k)e^{-ip_k\cdot
  x}\left(M_{q}^{(N)}(p_k,p_n;z_k,z_n)\right)^{a_ks_k}_{a_ns_n}.\label{quark_outcoming_state}
\end{align}
On contrary to QED case \eqref{amplitude_emission_m}, in QCD the gluon is
allowed to reinteract with the medium once it is emitted, then we write also
for its state
\begin{align}
A_{\mu,\alpha}^{(N)}(x)=A_\mu^{(0)}(x)\delta_{\alpha}^{\alpha_n}+\int\frac{d^3\v{k}_k}{(2\pi)^3}\mathcal{N}_k\epsilon_\mu^{\lambda_k}(k)e^{-ik\cdot x}\left(M_g^{(N)}(k_k,k_n;z_k,z_n)\right)_{a_f\lambda_f}^{a_k\lambda_k},\label{gluon_state}
\end{align}
where the matrix elements of $M^{(N)}_{g}(k_k,k_n;z_k,z_n)$ are the beyond
eikonal elastic amplitude \eqref{elastic_amplitudem_1layer_qcd} of finding the
gluon in the state $(k_n,\lambda_n,\alpha_n)$ after traversing the colored
matter in the interval $(z_k,z_n)$. The emitted gluon will be considered
massive, with a plasma frequency given by $\omega_p=m_g$ to account for medium
effects in the dispersion relation. Then the gluon 4-momentum is denoted by
$k=\omega(1,\beta_k\v{k})$, where the gluon velocity is given by
\begin{align}
\beta_k=\sqrt{1-m_g^2/\omega^2}.
\end{align}
The longitudinal polarization related to $m_g\neq$0 will be, however,
neglected. The effective gluon mass can be considered of the order of $\mu_d$,
the effective screening mass of the external field of the medium
\cite{arnold1995a,blaizot2001,nakamura2004,kaczmarek2005a,panero2013a}. As
with the QED analogous \eqref{amplitude_emission_m}, the insertion of
\eqref{quark_incoming_state}, \eqref{quark_outcoming_state} and
\eqref{gluon_state} in \eqref{emission_amplitude_qcd_definition} produces a
term containing the three unscattered states, representing the emission
amplitude in the vacuum
\begin{align}
\mathcal{M}_{em}^{(0)}&=-ig_s\mathcal{N}_k\int d^4x
\psi_{f}^{(0)}(x)A_{\mu,\alpha}^{(0)}(x)t_\alpha\gamma^\mu
\psi_i^{(0)}(x)\label{vacuum_diagram_qcd}\\ &=-ig_s\mathcal{N}_k(2\pi)^4\delta^4(p_n+k_n-p_i)\Big(t_{\alpha_n}\Big)^{a_n}_{a_0}\sqrt{\frac{m_q}{p_n^0}}\bar{u}_{s_n}(p_n)\gamma^\mu
\epsilon_\mu^{\lambda_n}(k)u_{s_0}(p_0)\sqrt{\frac{m_q}{p_0^0}},\nonumber
\end{align}
which vanishes due to energy momentum conservation. This diagram has to be
subtracted from the full amplitude in order to avoid spurious divergences in
the evaluation of the non-perturbative scattering amplitudes. Then we define
the amplitude of emission \textit{while} interacting with the medium. By using
$S=1+M$ and inserting \eqref{quark_incoming_state},
\eqref{quark_outcoming_state} and \eqref{gluon_state} in
\eqref{emission_amplitude_qcd_definition} and subtracting
\eqref{vacuum_diagram_qcd} we obtain
\begin{align}
\mathcal{M}_{em}^{(N)}-&\mathcal{M}_{em}^{(0)}=-ig_s\mathcal{N}_k\int
\frac{d^3\v{p}_k}{(2\pi)^3}\int\frac{d^3\v{k}_k}{(2\pi)^3}\int_{-\infty}^{+\infty}
dz\exp\Big(-iq(p,p+k)z_k\Big)\nonumber\\
&\times f_{s'_ks_k}^{\lambda_k}(p_k,p_k+k_k)\left(S_g^{(N)}(k_n,k;z_n,z_k)\right)^{\alpha_n}_{\alpha_k}\label{amplitude_emission_momentum_m_qcd}\\
&\times\left(S_q^{(N)}(p_n,p;z_n,z_k)\right)^{a_ns_n}_{a'_ks'_k}\Big(t_{\alpha_k}\Big)^{a'_k}_{a_k}\left(S_q^{(N)}(p_k+k_k,p_0;z_k,z_1)\right)^{a_ks_k}_{a_0s_0}-\mathcal{M}_{em}^{(0)},\nonumber
\end{align}
where the quark and gluon elastic amplitudes $S_q$ and $S_g$ are the
respective beyond eikonal and ordered evaluations outlined at
\eqref{elastic_amplitudes_nlayers_qcd}. The shorthand notation
$f_{s'_ks_k}^{\lambda_k}(p_k,p_k+k_k)$ refers to the emission vertex, which
factorizes and reads
\begin{align}
f_{s'_ks_k}^{\lambda_k}(p_k,p_k+k_k)\equiv \sqrt{\frac{m_q}{p_0^0-\omega}}\bar{u}_{s'_k}(p_k)\epsilon_\mu^{\lambda_k}(k_k) \gamma^\mu
u_{s_k}(p_k+k_k)\sqrt{\frac{m_q}{p_0^0}}.\label{emission_vertex_qcd}
\end{align}
If $p_{p_0^0}^z(\v{p}^t)$ denotes the longitudinal momentum of a parton state
with energy $p_0^0$ and transverse momentum $\v{p}^t$, the local longitudinal
phase arising at the emission point is given by
\begin{align}
q(p_k,p_k+k_k)&\equiv
p_{p_0^0-\omega}^z(\v{p}_k^t)+k^z_\omega(\v{k}_k^t)-p_{p_0^0}^z(\v{p}_k^t+\v{k}_k^t),\nonumber\\
q(p_k-k_k,p_k)&\equiv p_{p_0^0-\omega}^z(\v{p}_k^t-\v{k}_k^t)+k^z_\omega(\v{k}_k^t)-p_{p_0^0}^z(\v{p}_k^t),\label{longitudinal_momentum_change_vertex_qcd}
\end{align}
depending on which of the intermediate quark momenta $\v{p}_k$, either at
\eqref{quark_incoming_state} or \eqref{quark_outcoming_state}, is chosen to
perform the remaining integral in
\eqref{amplitude_emission_momentum_m_qcd}. The relation of
\eqref{longitudinal_momentum_change_vertex_qcd} with the pole of an off-shell
quark propagator for the leg after or before the emission can be seen by
taking the high energy limit, $\beta_k\cong 1$ and $\beta_p\cong 1$,
respectively, where $\beta_p$ is the velocity of the quark in the QCD
medium. Indeed we find
\begin{align}
q_z(p,p+k)\simeq 
-\frac{\omega}{2p_0^0(p_0^0-\omega)}\left(m_q^2+\left(\v{p}_t-\frac{p_0^0-\omega}{\omega}\v{k}_t\right)^2\right)-\frac{m_g^2}{2\omega}\simeq-\frac{k_\mu
p^\mu}{p_0^0},\label{pole_after_qcd}
\end{align}
if we carry the integration with $\v{p}$, the momentum of the leg after the
emission, of modulus $\beta_p p^0=\beta_p(p_0^0-\omega)$, or
\begin{align}
q_z(p-k,p)\simeq 
-\frac{\omega}{2p_0^0(p_0^0-\omega)}\left(m_q^2+\left(\v{p}_t-\frac{p_0^0}{\omega}\v{k}_t\right)^2\right)-\frac{m_g^2}{2\omega}\simeq-\frac{k_\mu
p^\mu}{p_0^0-\omega},\label{pole_before_qcd}
\end{align}
if $\v{p}$ corresponds to the quark momentum at the leg before the emission,
of modulus $\beta_pp^0=\beta_p p_0^0$. It is interesting to compare
\eqref{pole_after_qcd} and \eqref{pole_before_qcd} with the QED analogous
\eqref{pole_after} and \eqref{pole_before}, and observe that in the soft
$\omega\ll p_0^0$ limit the resonances \eqref{pole_after_qcd} and
\eqref{pole_before_qcd} are dominated by the gluon momentum change and
mass. The interpretation of the emission amplitude
\eqref{amplitude_emission_momentum_m_qcd} is straightforward from right to
left. The quark can be found at a depth $z_k$ with energy $p_0^0$ and a given
amplitude of being with deflected state $(p_k,s_k,a_k)$. At that point it
emits a gluon of color $\alpha_k$, loosing momentum $k_k$, changing spin from
$s_k$ to $s'_k$ and color from $a_k$ to $a'_k$. The amplitude of this emission
vertex \eqref{emission_vertex_qcd} factorizes from the elastic propagation as
\begin{align}
\left\{\sqrt{\frac{m_q}{p_0^0-\omega}}\bar{u}_{s'_k}(p_k)\epsilon_\mu^{\lambda_k}(k_k) \gamma^\mu
u_{s_k}{k}(p_k+k_k)\sqrt{\frac{m_q}{p_0^0}}\right\}\Big(t_{\alpha_k}\Big)^{a'_k}_{a_k}.
\end{align}
From there onwards the quark and the gluon continue propagating till the end
of the medium $z_n$ with a given amplitude of ending at the states
$(p_n,s_n,a_n)$ and $(k_n,\lambda_n,\alpha_n)$, respectively. Finally we sum
(integrate) over the different points in which the gluon can be emitted, which
adequately inserts the corresponding off-shell propagators
\eqref{pole_after_qcd} and \eqref{pole_before_qcd}. As an illustrative example
we consider the single case, i.e. just $N=n(z_1)=n_1$ centers distributed at
equal coordinate $z_1=0$. Then the elastic amplitudes
\eqref{elastic_amplitudes_nlayers_qcd} reduce to the single layer case
\eqref{elastic_amplitudes_1layer_qcd} and we obtain from
\eqref{amplitude_emission_momentum_m_qcd}
\begin{align}
&\mathcal{M}^{(n_1)}_{em}=-g_s\mathcal{N}\Bigg\{\int\frac{d^3\v{k}_0}{(2\pi)^3}\frac{p_0^0-\omega}{k_\mu^0p_0^\mu}\left(S_g^{(n_1)}(k_1,k_0)\right)^{\alpha_1}_{\alpha_0}\left(S_q^{n_1}(p_1,p_0-k_0)\right)^{a_1s_1}_{a_ks_k}\Big(t_{\alpha_0}\Big)^{a_k}_{a_0}\nonumber\\
&\times f_{s_ks_0}^{\lambda_0}(p_0-k_0,p_0)-\frac{p_0^0}{k_\mu^1p_1^\mu}f_{s_1s_k}^{\lambda_1}(p_1,p_1+k_1)\Big(t_{\alpha_1}\Big)^{a_1}_{a_k}\left(S_q^{(n_1)}(p_1+k_1,p_0)\right)^{a_ks_k}_{a_0s_0}\Bigg\}.\label{amplitude_emission_mometum_m_qcd_singlelayer}
\end{align}
We observe that in order to obtain $\mathcal{M}_{em}^{(0)}$ we simply have to
evaluate the above expression in $n_1=0$. The leading order of
\eqref{elastic_amplitudes_1layer_qcd} in $g_s^2$ produces for the quark
\begin{align}
\Big(S_q^{(n_1)}(p_1,p_0)\Big)^{a_1s_1}_{a_0s_0}&=(2\pi)^3\delta^3(\v{q}_q)\delta^{s_1}_{s_0}\delta^{a_1}_{a_0}+2\pi\delta(q_q^0)\label{quark_amplitude_leading_order}\\\times\sqrt{\frac{m_q}{p_1^0}}&\bar{u}_{s_1}(p_1)
\gamma_0u_{s_0}(p_0)\sqrt{\frac{m_q}{p_0^0}}\left(-ig_s^2\Big(t_{\alpha}\Big)^{a_{1}}_{a_0}t_{\alpha}A_0^{(1)}\Big(\v{q}_q^t\Big)\sum_{i=1}^{n_1}e^{-i\v{q}\cdot\v{r}_i}+\cdots\right),\nonumber
\end{align}
where the momentum change is $\v{q}_q=\v{p}_1-\v{p}_0$ and
$A_0^{(1)}(\v{q})=4\pi/(\v{q}^2+\mu_d^2)$ is the Fourier transform of the
external field \eqref{fielddefinition}. Similarly for the gluon we find, by
changing to the adjoint representation, \textit{c.f.} Section
\ref{sec:section_4_2},
\begin{align}
\Big(S_g^{(n_1)}(k_1,k_0)\Big)^{\alpha_1\lambda_1}_{\alpha_0\lambda_0}&=(2\pi)^3\delta^3(\v{q}_g)\delta^{\lambda_1}_{\lambda_0}\delta^{\alpha_1}_{\alpha_0}+2\pi\delta(q_g^0)\label{gluon_amplitude_leadingorder}\\&\times
\epsilon_\mu^{\lambda_1}(k_1)\epsilon^{\mu}_{\lambda_0}(k_0)\left(-ig_s^2\Big(T_{\alpha}\Big)^{\alpha_1}_{\alpha_0}t_{\alpha}A_0^{(1)}\Big(\v{q}_g^t\Big)\sum_{i=1}^{n_1}e^{-i\v{q}\cdot\v{r}_i}+\cdots\right),\nonumber
\end{align}
where the momentum change is denoted $\v{q}_g=\v{k}_1-\v{k}_0$. The
subtraction of $\mathcal{M}_{em}^{(0)}$ removes the trivial amplitude,
i.e. no collision at any point. In the pure high energy limit the spinor and
polarization content of the elastic amplitudes reduce to conservation
deltas. In that case we find after inserting
\eqref{quark_amplitude_leading_order} and \eqref{gluon_amplitude_leadingorder}
in \eqref{amplitude_emission_mometum_m_qcd_singlelayer} and using
$\Big(T_{\alpha}\Big)^{\alpha_1}_{\alpha_0}t_{\alpha_0}=[t_{\alpha_1},t_\alpha]$
\begin{align}
\mathcal{M}_{em}^{(n_1)}-\mathcal{M}_{em}^{(0)}=-ig_s^3\mathcal{N}_kA_0^{(1)}(\v{q}_g)\sum_{i=1}^{n_1}e^{-i\v{q}\cdot\v{r}_i}\left(\frac{p_0^0}{k_\mu^1p_1^\mu}f_{s_1s_0}^{\lambda_1}(p_1,p_1+k_1)\Big(t_{\alpha_1}t_{\alpha}\Big)^{a_1}_{a_0}\right.\nonumber\\ -\left.\frac{p_0^0-\omega}{k_\mu^1p_0^\mu}f_{s_1s_0}^{\lambda_1}(p_0-k_1,p_0)\Big(t_{\alpha}t_{\alpha_1}\Big)^{a_1}_{a_0}-\frac{p_0^0-\omega}{k_\mu^0p_0^\mu}f_{s_1s_0}^{\lambda_0}(p_0-k_0,p_0)\Big[t_{\alpha_1},t_\alpha\Big]^{a_1}_{a_0}\right)t_\alpha,\label{emission_amplitude_1layer_qcd}
\end{align}
with $\v{q}\equiv \v{p}_1+\v{k}_1-\v{p}_0$ and then
$\v{q}=\v{k}_1-\v{k}_0$. Equation \eqref{emission_amplitude_1layer_qcd} will
become more familiar assuming that $\omega\ll p_0^0$. In that case we can
omit the transverse components $\v{p}_1^t$ and $\v{p}_0^t$ in
\eqref{pole_after_qcd} and \eqref{pole_before_qcd}. By expanding the
denominators \eqref{longitudinal_momentum_change_vertex_qcd} and the vertex in
transverse and longitudinal components we obtain, in the $\omega\to 0$ limit,
\begin{align}
\mathcal{M}_{em}^{(n)}-\mathcal{M}_{em}^{(0)}=ig_s^3\mathcal{N}_k\left(\frac{2\v{\epsilon}^t_\lambda\cdot\v{k}_1^t}{m_g^2+(\v{k}_1^t)^2}-\frac{2\v{\epsilon}^t_\lambda\cdot\v{k}_0^t}{m_g^2+(\v{k}_0^t)^2}\right)A_0^{(1)}(\v{q}_t) \Big[t_{\alpha},t_{\alpha_1}\Big]^{a_1}_{a_0}t_\alpha\sum_{i=1}^{n_1}e^{-i\v{q}\cdot\v{r}_i},\label{bertsch_gunion_limit}
\end{align}
whose square is the massive version of the Bertsch-Gunion result
\cite{gunion1982}. A diagrammatic representation of
\eqref{emission_amplitude_1layer_qcd} is shown in Figure \ref{fig:figure_5_1}.
For mediums of larger thickness the above formula does not hold. In that case
we can discretize the medium into a set of $n$ layers from $z_1$ to $z_n$
verifying
\begin{align}
N= \sum_{i=1}^nn(z_i) \equiv \sum_{i=1}^nn_i.
\end{align}
Then the elastic amplitudes for the quark and the gluon are given as
convolutions in the form of \eqref{elastic_amplitudes_nlayers_qcd}. The
integration in $z$ of \eqref{amplitude_emission_momentum_m_qcd} produces three
different zones so that we write
\begin{align}
\mathcal{M}_{em}^{(N)}=g_s\mathcal{N}_k\left(\prod_{i=1}^{n-1}\frac{d^3\v{p}_i}{(2\pi)^3}\right)\left(\prod_{i=k}^{n-1}\frac{d^3\v{k}_i}{(2\pi)^3}\right)\left(\mathcal{M}_f+\mathcal{M}_{int}+\mathcal{M}_{i}\right).
\end{align}
The first term corresponds to the integration of
\eqref{amplitude_emission_momentum_m_qcd} in the $(z_n,+\infty)$ region and
contains the gluon emitted in the last leg. It reads
\begin{align}
\mathcal{M}_{f}\equiv\bigg(t_{\alpha_n}\bigg)^{a_n}_{a'_n}\frac{f_{s_ns'_n}^{\lambda_n}(p_n,p_n+k_n)}{k^n_\mu
p^\mu_n/p_0^0}\exp\left(+i\frac{k_\mu^np^\mu_n}{p_0^0}z_n\right)\bigg(S^{(n_n)}_{s'_ns_{n-1}}(\delta
p_n+k_n)\bigg)^{a'_n}_{a_{n-1}}\nonumber\\
\times\left(\prod_{i=1}^{n-1}\bigg(S_{s_is_{i-1}}^{(n_i)}(\delta p_i)\bigg)^{a_i}_{a_{i-1}}\right),\label{last_gluon_amplitude}
\end{align}
where $\delta p_i\equiv p_i-p_{i-1}$ is the momentum change of the quark at
the layer at $z_i$ and the abbreviations $S_{s_is_{i-1}}^{n(z_i)}(\delta
p_i)\equiv S_{s_{i}s_{i-1}}^{n(z_i)}(p_i,p_{i-1})$ have been used. The
integration of \eqref{amplitude_emission_momentum_m_qcd} in the inner region
$(z_0,z_n)$ contains all the gluons emitted from internal legs. Indeed we find
\begin{align}
&\mathcal{M}_{int}\equiv-\sum_{k=1}^{n-1}\left(\prod_{i=k+1}^n\bigg(S_{s_is_{i-1}}^{(n_i)}(\delta
p_i)\bigg)^{a_i}_{a_{i-1}}\bigg(S_{\lambda_i\lambda_{i-1}}^{(n_i)}(\delta
k_i)\bigg)^{\alpha_i}_{\alpha_{i-1}}\right)\bigg(t_{\alpha_k}\bigg)^{a_k}_{a'_k}\frac{f_{s_ks'_k}^{\lambda_k}(p_k,p_k+k_k)}{k_\mu^kp^\mu_k/p_0^0}\times\nonumber\\
&\left(\exp\left(i\frac{k_\mu^kp^\mu_k}{p_0^0}z_{k+1}\right)-\exp\left(i\frac{k_\mu^kp^\mu_k}{p_0^0}z_k\right)\right)\bigg(S^{(n_k)}_{s'_ks_{k-1}}(\delta
p_k+k_k)\bigg)_{a_{k-1}}^{a'_k}\left(\prod_{i=1}^{k-1}\bigg(S_{s_is_{i-1}}^{(n_i)}(\delta p_i)\bigg)^{a_i}_{a_{i-1}}\right).\label{internal_gluons_amplitude}
\end{align}
\begin{figure}
\begin{center}
\includegraphics[width=1.\textwidth]{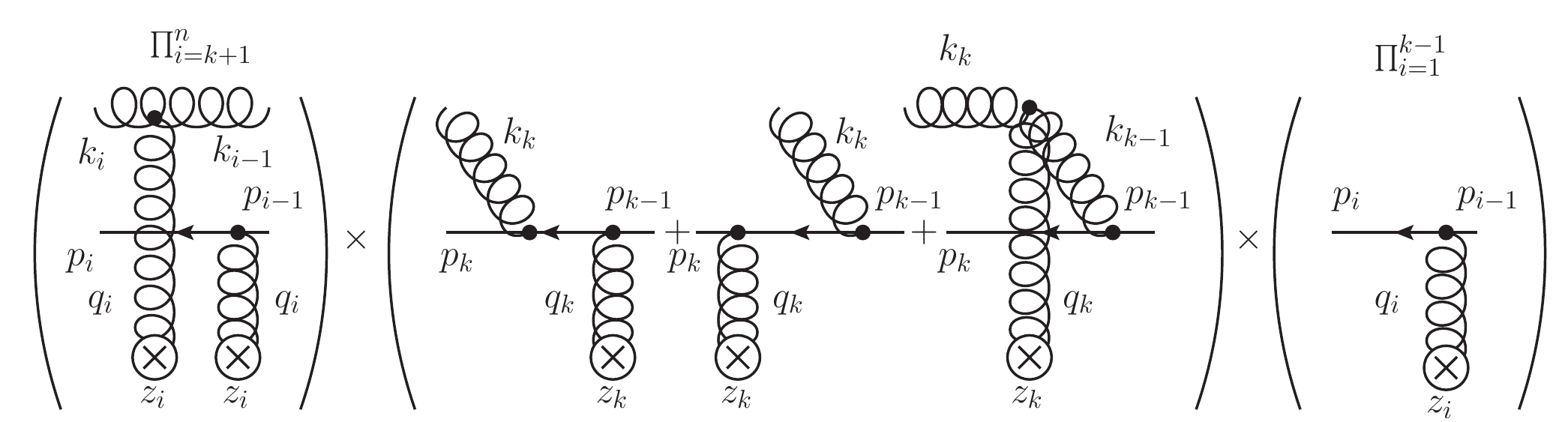}
\end{center}
\caption{The Feynman diagrams \cite{feal2018a} appearing in an expansion in
  the coupling constant $g_s^2$ of
  $\mathcal{M}_{em}^{(n)}-\mathcal{M}_{em}^{(0)}$ in the multiple collision
  regime \eqref{amplitude_emission_sum_mk_qcd}.}
\label{fig:figure_5_2}
\end{figure}%
Finally the term containing the gluon emitted from the first leg is given by
the integration of  \eqref{amplitude_emission_momentum_m_qcd} in the region
$(-\infty,z_0)$. It produces
\begin{align}
&\mathcal{M}_{i}\equiv -\left(\prod_{i=2}^{n}
\bigg(S_{s_is_{i-1}}^{(n_i)}(\delta
p_i)\bigg)^{a_i}_{a_{i-1}}\bigg(S_{\lambda_i\lambda_{i-1}}^{(n_i)}(\delta
k_i)\bigg)^{\alpha_i}_{\alpha_{i-1}}\right)\label{first_gluon_amplitude}\\
&\times\left(S_{s_1s'_0}^{(n_1)}(\delta
p_1+k_1)\right)_{a'_0}^{a_1}\bigg(S_{\lambda_1\lambda_0}^{(n_1)}(\delta k_1)\bigg)_{\alpha_0}^{\alpha_1}\bigg(t_{\alpha_0}\bigg)^{a'_0}_{a_0}\frac{f_{s'_0s_0}^{\lambda_0}(p_0-k_0,p_0)}{k_\mu^0p^\mu_0/(p_0^0-\omega)}\exp\left(+i\frac{k_\mu^0p^\mu_0}{p_0^0-\omega}z_1\right).\nonumber
\end{align}
Similarly to the QED case, in the soft gluon approximation $\omega\ll p_0^0$
the first term \eqref{last_gluon_amplitude} and the last term
\eqref{first_gluon_amplitude} are the only ones surviving when $\omega\to 0$,
since the phases, which are of the form \eqref{pole_after_qcd} or
\eqref{pole_before_qcd}, vanish, and then the internal sum
\eqref{internal_gluons_amplitude} cancels
\cite{weinberg1965,weinberg1995,bassetto1984a,mueller1989a}. Notice, however,
that previously to this cancellation the phase has an enhancement $1/\omega$
due to the $m_g$ and $\v{k}_k^t$ terms, which means that incoherence may occur
in the sum of the amplitudes in the soft regime, previous to the phase
cancellation, in analogy with the dielectric and transition radiation effects
of QED. In the regime of vanishing phases the coherent plateau is found, and
the medium emits as a single entity of equivalent charge the amount of colored
matter in $(z_1,z_n)$. The three terms \eqref{last_gluon_amplitude},
\eqref{internal_gluons_amplitude} and \eqref{first_gluon_amplitude} can be
reunited into a single expression. For that purpose we simplify now the
notation by extracting the energy, spin and polarization conservation deltas
from the elastic amplitudes as follows
\begin{align}
\left(S_{s_{i}s_{i-1}}^{(n_i)}(\delta
p_i)\right)^{a_i}_{a_{i-1}}&=\delta^{s_i}_{s_{i-1}}2\pi\beta_p\delta(\delta p_i^0)\left(S_q^{i}(\delta\v{p}_i)\right)^{a_i}_{a_{i-1}}\nonumber\\
\left(S_{\lambda_{i}\lambda_{i-1}}^{(n_i)}(\delta
k_i)\right)^{\alpha_i}_{\alpha_{i-1}}&=\delta^{\lambda_i}_{\lambda_{i-1}}2\pi\beta_k\delta(\delta k_i^0)\left(S_g^{i}(\delta\v{k}_i)\right)^{\alpha_i}_{\alpha_{i-1}}.
\end{align}
We define also the $f_k^+$ and $f_k^-$, $\varphi_k^+$ and $\varphi_k^{-}$
shorthands for the advanced and retarded propagator together with the spinorial
vertices, and the local phases, respectively, at the emission point $z_k$
\begin{align}
f_k^{+}&\equiv
\frac{f_{s_ns_0}^{\lambda_n}(p_k,p_k+k_k)}{k_\mu^{k}p^\mu_k/p_0^0},\medspace
f_k^{-}\equiv\frac{f^{\lambda_{n}}_{s_{n}s_{0}}(p_{k}-k_{k},p_{k})}{k_\mu^{k}p^\mu_{k}/(p_0^0-\omega)}, \varphi_k^+=\frac{k_\mu^kp^\mu_k}{p_0^0}z_k,
\varphi_k^-=\frac{k_\mu^{k}p^\mu_{k}}{p_0^0-\omega}z_k.
\end{align}
With this notation the sum of all the gluons intervening in the amplitude can
be written as
\begin{align}
\mathcal{M}_{em}^{(N)}=g_s\mathcal{N}_k2\pi\beta_p\delta(p_n^0+\omega-p_0^0)\left(\prod_{i=1}^{n-1}\int\frac{d^2\v{p}_i^t}{(2\pi)^2}\right)\left(\prod_{i=k}^{n-1}\int\frac{d^2\v{k}_i^t}{(2\pi)^2}\right)\times\sum_{k=1}^n\mathcal{M}_k,
\end{align}
where a single Bethe-Heitler element of gluon bremsstrahlung is given by
\begin{align}
\mathcal{M}_k&=\prod_{i=k+1}^{n}\left(\left(S_q^{i}(\delta\v{p}_i)\right)^{a_i}_{a_{i-1}}\left(S_g^{i}(\delta\v{k}_i)\right)^{\alpha_i}_{\alpha_{i-1}}\right)\bigg(f_k^{+}e^{i\varphi_k^+}\Big(t_{\alpha_k}\Big)^{a_k}_{a'_k}\left(S_q^{k}(\delta\v{p}_k+\v{k}_k)\right)^{a'_k}_{a_{k-1}}\nonumber\\
&-f_{k-1}^-e^{i\varphi_k^-}\left(S_g^k(\delta\v{k}_k)\right)^{\alpha_k}_{\alpha_{k-1}}\left(S_q^k(\delta\v{p}_k+\v{k}_{k-1})\right)^{a_k}_{a'_{k-1}}\Big(t_{\alpha_{k-1}}\Big)^{a'_{k-1}}_{a_{k-1}}\bigg)
\prod_{i=1}^{k-1}\left(S_q^i(\delta\v{k}_k)\right)^{a_i}_{a_{i-1}}.\label{amplitude_emission_mk_qcd_v2}
\end{align}
We observe that the vacuum term is given by the evaluation of the former
expression in $(n_i)=0$ for all $i$. Thus we finally write
\begin{align}
\mathcal{M}_{em}^{(N)}-\mathcal{M}_{em}^{(0)}&=g_s\mathcal{N}_k2\pi\beta_p\delta(p_n^0+\omega-p_0^0)\nonumber\\
&\times\left(\prod_{i=1}^{n-1}\int\frac{d^2\v{p}_i^t}{(2\pi)^2}\right)\left(\prod_{i=k}^{n-1}\int\frac{d^2\v{k}_i^t}{(2\pi)^2}\right)\times\sum_{k=1}^n\left(\mathcal{M}_k-\mathcal{M}_k^{(0)}\right).\label{amplitude_emission_sum_mk_qcd}
\end{align}
Amplitude \eqref{amplitude_emission_sum_mk_qcd} is just a sum of single
Bethe-Heitler/Bertsch-Gunion elements in a QCD scenario. A diagrammatic
representation is depicted in Figure \ref{fig:figure_5_2}. The internal phases
in the scattering amplitudes are responsible of modulating the interference
behavior leading to the LPM effect in the square of
\eqref{amplitude_emission_sum_mk_qcd}.
\section{Intensity and energy loss}
\label{sec:section_5_2}
\begin{figure}
\centering
\includegraphics[width=0.6\textwidth]{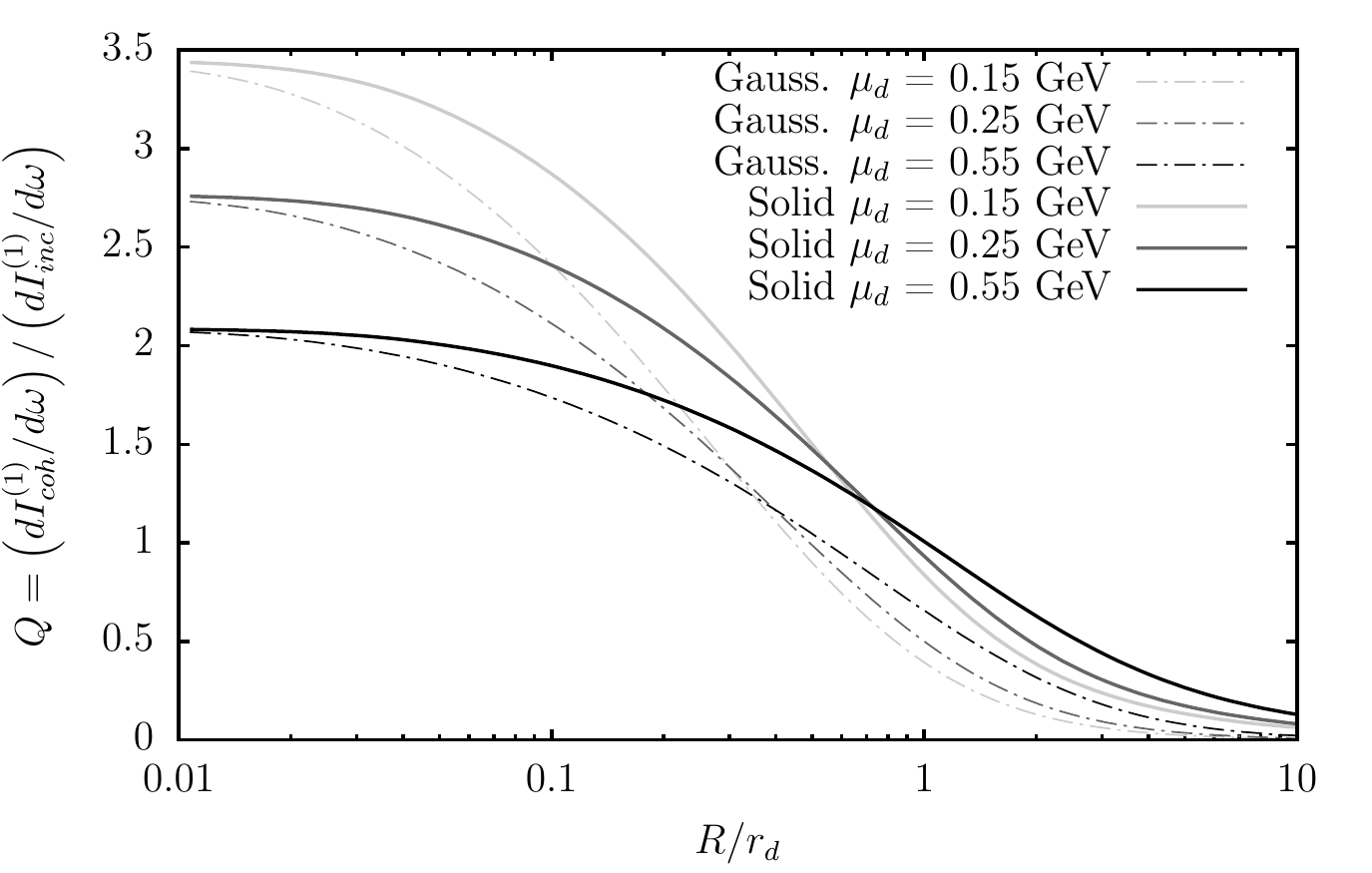}
\caption{Quotient between the transverse coherent and transverse incoherent
  intensities for the single scattering regime out of the suppression zone
  $\omega\gg m_g$ as a function of the transverse size of the QCD medium $R$
  in units of the Debye radius of the plasma $r_d=\mu_d^{-1}$, for a medium of
  $n_0T_f$ = 8 fm$^{-3}$ corresponding to a transport parameter $\hat{q}$ =
  0.98 GeV$^2$/fm, and a gluon mass of $m_g$ = 0.15 GeV, for a Gaussian
  decaying density (dot-dashed lines) and an uniform density in a solid
  cylinder (solid lines) for different Debye screenings, as marked.}
\label{fig:figure_5_12}
\end{figure}

The amplitude \eqref{amplitude_emission_sum_mk_qcd} must be squared and
averaged over medium configurations, and summed and averaged over final and
initial states respectively. The medium will be chosen as a solid cylinder
with transverse area $\pi R^2$ and length $l=z_n-z_1$. The transverse
dimensions will be assumed much larger than the dimensions of a single
scatterer, i.e. $R\gg \mu_d^{-1}=r_d$, where $r_d$ is the Debye radius of the
plasma, in such a way that the transverse boundary effects can be
neglected. The unpolarized differential intensity of gluons in the energy
interval $\omega$ and $\omega+d\omega$, in the solid angle $\Omega_k$ and
$\Omega_k+d\Omega_k$, per unit of quark incoming flux, time and transverse
area is given by
\begin{align}
dI\equiv \frac{\omega^2d\omega d\Omega_k}{(2\pi)^3}\frac{1}{\beta_pT\pi
  R^2}&\frac{1}{2}\sum_{s_ns_0}\sum_{\lambda_n}\frac{1}{N_c}\sum_{a_na_0}\sum_{\alpha_n}\nonumber\\
\times&\int\frac{d^3\v{p}_n}{(2\pi)^3}\left\langle \bigg(\mathcal{M}_{em}^{(N)}-\mathcal{M}_{em}^{(0)}\bigg)^*\bigg(\mathcal{M}_{em}^{(N)}-\mathcal{M}_{em}^{(0)}\bigg)\right\rangle,\label{intensity_definition_qcd}
\end{align}
where the operation $\langle \star \rangle$ denotes an average over target
configurations, both in color and spatial dimensions. The form of
\eqref{amplitude_emission_sum_mk_qcd} as $M-1$ suggests, as with the elastic
case, an splitting into a incoherent and a coherent contribution. We then
proceed as before by adding and subtracting the term $\left\langle
\mathcal{M}_{em}^{(N)}\right\rangle^*\left\langle
\mathcal{M}_{em}^{(N)}\right\rangle$ in which case we obtain
\begin{align}
\frac{1}{2}\sum_{s_ns_0}\sum_{\lambda_n}\frac{1}{N_c}\sum_{a_na_0}\sum_{\alpha_n}\left\langle \bigg(\mathcal{M}_{em}^{(N)}-\mathcal{M}_{em}^{(0)}\bigg)^*\bigg(\mathcal{M}_{em}^{(N)}-\mathcal{M}_{em}^{(0)}\bigg)\right\rangle=\Sigma_{em}^{(N)}+\Pi_{em}^{(N)},
\end{align}
where the coherent part, which contains the diffractive and quantum
contribution of the medium boundaries to the emission intensity, is given by
the averaged emission amplitude squared,
\begin{align}
\Pi_{em}^{(N)}\equiv \frac{1}{2}\sum_{s_ns_0}\sum_{\lambda_n}\frac{1}{N_c}\sum_{a_na_0}\sum_{\alpha_n}\left\langle\mathcal{M}_{em}^{(n)}-\mathcal{M}_{em}^{(0)}\right\rangle^*\left\langle\mathcal{M}_{em}^{(n)}-\mathcal{M}_{em}^{(0)}\right\rangle,\label{pi_em_qcd_definition}
\end{align}
whereas the incoherent part, which contains a contribution which in the
macroscopic limit $R\gg \mu_d^{-1}$ can be interpreted in probabilistic terms,
is given by
\begin{align}
\Sigma_{em}^{(N)}\equiv \frac{1}{2}\sum_{s_ns_0}\sum_{\lambda_n}\frac{1}{N_c}\sum_{a_na_0}\sum_{\alpha_n}\left\langle
\left(\mathcal{M}_{em}^{(N)}\right)^*\mathcal{M}_{em}^{(N)}\right\rangle-\left\langle\mathcal{M}_{em}^{(N)}\right\rangle^*\left\langle\mathcal{M}_{em}^{(N)}\right\rangle.\label{sigma_em_qcd_definition}
\end{align}
For simplicity we take then $R\to\infty$ and neglect the boundary contribution
of the coherent part \eqref{pi_em_qcd_definition}. In that limit the coherent
part produces a set of pure forward scatterings so that each $\mathcal{M}_k$
vanishes and so it does the sum \eqref{amplitude_emission_sum_mk_qcd}. This
effect can be seen in Fig. \ref{fig:figure_5_12}, where the ratio between the
transverse-coherent and the transverse-incoherent intensities is shown, for
the single scattering regime, as a function of the transverse size of the
medium.
Then the relevant contribution \eqref{sigma_em_qcd_definition} can be
rewritten, using \eqref{amplitude_emission_sum_mk_qcd}, as
\begin{align}
&\Sigma_{em}^{(N)}=g_s^2\mathcal{N}_k^2\beta_pT2\pi\beta_p\delta(p_n^0+\omega-p_0^0)\frac{1}{2}\sum_{s_ns_0}\sum_{\lambda_n}\frac{1}{N_c}\sum_{a_na_0}\sum_{\alpha_n}\left(\prod_{i=k-1}^{n-1}\int\frac{d^2\v{k}_i^t}{(2\pi)^2}\right)\nonumber\\
&\times\left(\prod_{i=j-1}^{n-1}\int\frac{d^2\v{v}_i^t}{(2\pi)^2}\right)\left(\prod_{i=1}^{n-1}\int\frac{d^2\v{p}_i^t}{(2\pi)^2}\int\frac{d^2\v{u}_i^t}{(2\pi)^2}\right)\sum_{j,k=1}^n\left(\left\langle\mathcal{M}_{j}^*\mathcal{M}_k\right\rangle-\left\langle\mathcal{M}_j^*\right\rangle\Big\langle\mathcal{M}_k\Big\rangle\right).\label{incoherent_emission_qcd_definition}
\end{align}
The quark and gluon momenta in the amplitude have been denoted by $\v{p}_i$
and $\v{k}_i$, respectively, and by $\v{u}_i$ and $\v{v}_i$ in the conjugated
amplitude. Similarly the quark and gluon color index will be denoted by $a_i$
and $\alpha_i$ in the amplitude and by $b_i$ and $\beta_i$ in the conjugated
amplitude. With this notation we notice that $b_0=a_0$ and $b_n=a_n$,
$\v{p}_0=\v{u}_0$ and $\v{p}_n=\v{u}_n$ for the observed initial and final
states of the quark, respectively, and $\alpha_n=\beta_n$ and
$\v{k}_n=\v{v}_n$ for the observed final state of the gluon. The averaged
squared terms on the right hand side of
\eqref{incoherent_emission_qcd_definition} can be written as
\begin{align}
\left\langle\mathcal{M}_j^*\mathcal{M}_k\right\rangle = A_{j+1}^n\Bigg(f_j^{+*}e^{-i\varphi_j^+}&\Big(t_{\beta_j}\Big)^{b'_j}_{b_j}A_j^+
-A_j^+
f_{j-1}^{-*}e^{-i\varphi_j^-}\Big(t_{\beta_{j-1}}\Big)^{b_{j-1}}_{b'_{j-1}}\Bigg)A_{j+1}^{k-1}\label{m_k_m_j_incoherent}\\
&\times\Bigg(f_k^+e^{i\varphi_k^+}\Big(t_{\alpha_k}\Big)^{a_k}_{a'_{k}}A_k^+-A_k^-
f_{k-1}^-e^{i\varphi_k^-}\Big(t_{\alpha_{k-1}}\Big)^{a'_{k-1}}_{a_{k-1}}\Bigg)A_1^{k-1},\nonumber
\end{align}
where the $A_k^j$ are abbreviations denoting the averaged squared elastic
amplitudes for the layers from $z_k$ to $z_j$. The squared average terms in
the right hand side of \eqref{incoherent_emission_qcd_definition} are written
similarly as
\begin{align}
\left\langle\mathcal{M}_j^*\right\rangle\Big\langle\mathcal{M}_k\Big\rangle = B_{j+1}^n\Bigg(f_j^{+*}e^{-i\varphi_j^+}&\Big(t_{\beta_j}\Big)^{b'_j}_{b_j}B_j^+
-B_j^+
f_{j-1}^{-*}e^{-i\varphi_j^-}\Big(t_{\beta_{j-1}}\Big)^{b_{j-1}}_{b'_{j-1}}\Bigg)B_{j+1}^{k-1}\label{m_k_m_j_coherent}\\
&\times\Bigg(f_k^+e^{i\varphi_k^+}\Big(t_{\alpha_k}\Big)^{a_k}_{a'_{k}}B_k^+-B_k^-
f_{k-1}^-e^{i\varphi_k^-}\Big(t_{\alpha_{k-1}}\Big)^{a'_{k-1}}_{a_{k-1}}\Bigg)B_1^{k-1},\nonumber
\end{align}
where $B_k^{j}$ denote the averaged amplitudes squared for the layers from
$z_k$ to $z_j$. Both at \eqref{m_k_m_j_incoherent} and
\eqref{m_k_m_j_coherent} we have used the fact that the averages factorize
layer to layer due to the z-ordered form of the beyond eikonal elastic
amplitudes \eqref{elastic_amplitudes_nlayers_qcd}. The first set of
scatterings corresponds to the passage of the quark state ($q$) and its
conjugate state ($\bar{q}$) from the layer at $z_1$ to the on at $z_k$,
\begin{align}
A^{k-1}_{1}=\prod_{i=1}^{k-1}\left(\frac{1}{N_c^{n_i}}\Tr\left\langle\left(S_q^{i\dag}(\delta
\v{u}_i)\right)^{b_{i-1}}_{b_{i}}\left(S_{q}^i(\delta
\v{p}_i)\right)^{a_i}_{a_{i-1}}\right\rangle\right)=\prod_{i=1}^{k-1}A_i.
\end{align}
We refer to the Section \ref{sec:section_4_3} for the details of the averaging
process and leave implicit the color indices and the ordering. For a single
layer average we find
\begin{align}
A_i=&(2\pi)^2\delta^2(\delta \v{p}_i^t-\delta\v{u}_i^t)\exp\bigg(-i\Big(\delta
p_i^z-\delta u_i^z\Big)z_i\bigg)\medspace \times\phi^{(n_i)}_{q\bar{q}}(\delta
\v{p}_i^t),\label{single_layer_average_qq}
\end{align}
where $\delta p_i$ and $\delta u_i$ are the momentum change of the $q$ and
$\bar{q}$ states and the elastic distribution
$\phi_{q\bar{q}}^{(n_i)}(\delta\v{p}_i^t)$ satisfies
\begin{align}
\phi_{q\bar{q}}^{(n_i)}(\delta\v{p}_i^t)=\int
d^2\v{x}_i^{q\bar{q}}\exp\Bigg(-i\delta\v{p}_i^t\cdot\v{x}_i^{q\bar{q}}+n_0(z_i)\delta
z_i\Big(\sigma_{q\bar{q}}^{(1)}(\v{x}_i^{q\bar{q}})-\sigma_{q\bar{q}}^{(1)}(\v{0})\Big)\Bigg).\label{phi_qbarq}
\end{align}
Here $n_0(z_i)$ and $\delta z_i$ are the density and thickness of the target
quark layer, respectively, and $\sigma_{q\bar{q}}^{(1)}(\v{0})$ the
Fourier transform of the single elastic amplitude of a quark given at
\eqref{sigma_qq_qcd}. For the interaction \eqref{fielddefinition} we find at
leading order
\begin{align}
\sigma^{(1)}_{q\bar{q}}(\v{x})\equiv \int \frac{d^2\v{q}}{(2\pi)^2}
e^{i\v{q}\cdot\v{x}}\frac{1}{N_c}\Tr\left(F_q^{(1)\dag}(\v{q})F_q^{(1)}(\v{q})\right)=\frac{4\pi
g_s^4}{\beta_p^2\mu_d^2}\frac{N_c^2-1}{4N_c^2}\mu_d|\v{x}|K_1(\mu_d|\v{x}|),\label{sigma_qbarq_qcd}
\end{align}
where $K_1(x)$ is the modified Bessel function. We observe that
\eqref{sigma_qbarq_qcd} preserves the color of the quark states $q$ and
$\bar{q}$ hence any multiple interaction like \eqref{phi_qbarq} and any
convolution of \eqref{phi_qbarq} does too. The macroscopic limit $R\to\infty$
and the condition $\v{u}_0^t=\v{p}_0^t$ leads to an observed quark trajectory
satisfying $\v{u}_i^t=\v{p}_i^t$ for $i=1,\ldots,k-1$. This fact together with
energy conservation $u_i^0=p_i^0=p_0^0$ leads to a vanishing longitudinal
phase in all the interval
\begin{align}
\delta p_i^z-\delta u_i^z =
\bigg(p^z_{p_0^0}(\v{p}_i^t)-p_{p_0^0}^z(\v{p}_{i-1}^t)\bigg)-\bigg(p^z_{p_0^0}(\v{u}_i^t)-p_{p_0^0}^z(\v{u}_{i-1}^t)\bigg)=0.
\end{align}
Correspondingly for the entire passage from $z_1$ to $z_{k-1}$ we find,
using reiteratively \eqref{single_layer_average_qq},
\begin{align}
A^{k-1}_1=\prod_{i=1}^{k-1}A_i=\prod_{i=1}^{k-1}\left((2\pi)^2\delta^2(\v{p}_{i-1}^t-\v{u}_{i-1}^t)\times\phi_{q\bar{q}}^{(n_i)}(\delta\v{p}_i^t)\right)\label{a_k-1_1_qcd}.
\end{align}
The next average corresponds to the layer at $z_k$ where the gluon is emitted
in the amplitude and then two possibilities arise. The first one corresponds
to an advanced collision previous to the emission, and it is given by
\begin{align}
A_k^+\equiv\frac{1}{N_c^{n_k}}\Tr\left\langle\left(S_q^{k\dag}(\delta \v{u}_k)\right)^{b_{k-1}}_{b_k}\left(S_q^{k}(\delta
\v{p}_k+\v{k}_k)\right)^{a'_k}_{a_{k-1}}\right\rangle.
\end{align}
Both amplitudes are yet on-shell with the initial quark
$p_{k-1}^0=u_{k-1}^0=p_0^0$ and from \eqref{a_k-1_1_qcd} we inherit
$\v{u}_{k-1}^t=\v{p}_{k-1}^t$, so we find, using
\eqref{single_layer_average_qq}, that $\v{p}_k^t+\v{k}_k^t=\v{u}_k^t$ so that
the longitudinal phase at \eqref{single_layer_average_qq} vanishes again
\begin{align}
\delta p_k^z-\delta u_k^z = \bigg(p^z_{p_0^0}(\v{p}_k^t+\v{k}_k^t)-p_{p_0^0}^z(\v{p}_{k-1}^t)\bigg)-\bigg(p^z_{p_0^0}(\v{u}_k^t)-p_{p_0^0}^z(\v{u}_{k-1}^t)\bigg)=0.
\end{align}
Then the averaged squared amplitude of the advanced collision produces
\begin{align}
A_k^+=(2\pi)^2&\delta^2(\v{p}_k^t+\v{k}_k^t-\v{u}_k^t)\times\phi_{q\bar{q}}^{(n_i)}(\delta \v{p}_k^t+\v{k}_k^t).\label{a_k+_qcd}
\end{align}
The other possibility at the emission layer at $z_k$ is that the collision
occurs just after the emission. Since the gluon can reinteract with the target
quarks for this contribution we get the average of three elastic amplitudes,
of the form
\begin{align}
A_k^-\equiv
\frac{1}{N_c^{n_k}}\Tr\left\langle\left(S_q^{k\dag}(\delta \v{u}_k)\right)^{b_{k-1}}_{b_k}\left(S_q^k(\delta
\v{p}_k+\v{k}_{k-1})\right)^{a_k}_{a'_{k-1}}\left(S_g^k(\delta
\v{k}_k)\right)^{\alpha_k}_{\alpha_{k-1}}\right\rangle.
\end{align}
From \eqref{a_k-1_1_qcd} we obtain $\v{u}_{k-1}^t=\v{p}_{k-1}^t$ but now one
amplitude is on-shell with the initial quark and the other with the final
quark, $p_{k-1}^0+\omega=u_{k-1}=p_0^0$. From the averaging process it can be
shown that $\v{u}_{k}^t=\v{p}_k^t+\v{k}_k^t$ and then the longitudinal phase
is given by
\begin{align}
\delta p_k^z+\delta k_k^z-\delta u_k^z =
\bigg(p^z_{p_0^0-\omega}(\v{p}_k^t)+p_\omega(\v{k}_{k}^t)&-p_{p_0^0}^z(\v{p}_k^t+\v{k}_k^t)\bigg)\\
-\bigg(p_{p_0^0-\omega}^z(\v{p}_{k-1}^t-\v{k}_{k-1}^t)+p^z_{\omega}(\v{k}_{k-1}^t)&-p_{p_0^0}^z(\v{p}_{k-1}^t)\bigg)=-\frac{k_\mu^kp^\mu_{k}}{p_0^0}+\frac{k_\mu^{k-1}p^\mu_{k-1}}{p_0^0-\omega},\nonumber
\end{align}
where we have used \eqref{pole_after_qcd} and \eqref{pole_before_qcd}. The average procedure follows the same steps, leading to
\begin{align}
A_k^-&=(2\pi)^2\delta^2(\v{p}_{k}^t+\v{k}_k^t-\v{u}_{k}^t)\exp\left(i\frac{k_\mu^kp^\mu_{k}}{p_0^0}z_k-i\frac{k_\mu^{k-1}p^\mu_{k-1}}{p_0^0-\omega}z_k\right)\phi_{q\bar{q}g}^{(n_k)}(\delta\v{p}_k^t+\v{k}_{k-1}^t,\delta\v{k}_{k}^t),\label{a_k-_qcd}
\end{align}
where the elastic distribution for the three states
$\phi_{q\bar{q}g}^{(n_k)}(\delta\v{p}_k^t,\delta\v{k}_{k}^t)$ can be shown to
satisfy
\begin{align}
&\phi_{q\bar{q}g}^{(n_k)}(\delta\v{p}_k^t,\delta\v{k}_{k}^t)\equiv\int
d^2\v{x}_k^{q\bar{q}}\int d^2\v{x}_{k}^{g\bar{q}}\exp\Bigg(-i\delta\v{p}_k^t\cdot\v{x}_k^{q\bar{q}}-i\delta\v{k}_k^t\cdot\v{x}_k^{g\bar{q}}\Bigg)\label{phi_qbarqg}\\
\times&\exp\Bigg(n_0(z_i)\delta
z_i\bigg(\varphi_g^{(1)}(\v{0})+\left(\sigma_{q\bar{q}}\Big(\v{x}_{k}^{q\bar{q}}\Big)-\sigma_{q\bar{q}}(\v{0})\right)+\sigma_{g\bar{q}}\Big(\v{x}_{k}^{g\bar{q}}\Big)+\sigma_{gq}\Big(\v{x}_k^{g\bar{q}}-\v{x}_{k}^{q\bar{q}}\Big)\bigg)\Bigg)\nonumber.
\end{align}
Here the Fourier transform of the single squared amplitude of a quark
$\sigma_{q\bar{q}}^{(1)}(\v{x})$ is given by \eqref{sigma_qbarq_qcd} whereas
the Fourier transform of the crossed amplitudes are given by
\begin{align}
\sigma^{(1)}_{gq}(\v{x})\equiv \int \frac{d^2\v{q}}{(2\pi)^2}
&e^{i\v{q}\cdot\v{x}}\frac{1}{N_c}\Tr\left(
F_g^{(1)}(\v{q})F_q^{(1)}(\v{q})\right)\nonumber\\
&=\frac{4\pi g_s^4}{\beta_p\beta_k\mu_d^2}\frac{1}{2N_c}T_\alpha
t_\alpha \mu_d|\v{x}|K_1(\mu_d|\v{x}|)+\cdots\label{sigma_gq_qcd},
\end{align} 
which represents a collision affecting the $(q)$ and $(g)$ states with a
single quark, leaving $(\bar{q})$ unaltered, and similarly a single collision
affecting the $(g)$ and $(\bar{q})$ states leaving the state $(q)$ unaltered
is given by
\begin{align}
\sigma^{(1)}_{g\bar{q}}(\v{x})\equiv \int \frac{d^2\v{q}}{(2\pi)^2}
&e^{i\v{q}\cdot\v{x}}\frac{1}{N_c}\Tr\left(
F_g^{(1)}(\v{q})F_q^{(1)\dag}(\v{q})\right)\nonumber\\
&=-\frac{4\pi g_s^4}{\beta_p\beta_k\mu_d^2}\frac{1}{2N_c}T_\alpha
t_\alpha \mu_d|\v{x}|K_1(\mu_d|\v{x}|)+\cdots\label{sigma_gbarq_qcd}.
\end{align}
We observe that while \eqref{sigma_qbarq_qcd} preserves the color of the
traveling $(q)$ and $(\bar{q})$ states, both \eqref{sigma_gq_qcd} and
\eqref{sigma_gbarq_qcd} allow arbitrary and unobserved color rotations of the
intermediate states. The unpaired single elastic cross section appearing in
\eqref{phi_qbarqg} has been denoted by
\begin{align}
\varphi_g^{(1)}(\v{0})\equiv \frac{1}{N_c}\Tr\left(F_g^{(1)}(\v{0})\right)=-\frac{g_s^4}{2\beta_k^2}T_f\frac{4\pi}{\mu_d^2}+\cdots,
\end{align}
which satisfies $\varphi_{g}^{(1)}(\v{0})=-\sigma_{g\bar{g}}^{(1)}(\v{0})/2$
at leading order, where $\sigma_{g\bar{g}}^{(1)}(\v{0})$ is the single elastic
cross section of a gluon. The next passage is given by the intermediate
scatterings of $(q)$, $(\bar{q})$ and $(g)$ states through the layers from
$z_{k+1}$ to $z_{j-1}$. It is given by
\begin{align}
A_{k+1}^{j-1}\equiv
\prod_{i=k+1}^{j-1}\left(\frac{1}{N_c^{n_i}}\Tr\left\langle\left(S_q^{i\dag}(\delta \v{u}_i)\right)^{b_{i-1}}_{b_i}\left(S_q^{i}(\delta
\v{p}_i)\right)^{a_i}_{a_{i-1}}\left(S_g^i(\delta
\v{k}_i)\right)^{\alpha_i}_{\alpha_{i-1}}\right\rangle\right)=\prod_{i=k+1}^{j-1} A_i.
\end{align}
Either from \eqref{a_k+_qcd} or \eqref{a_k-_qcd} we inherit the condition
$\v{p}_k^t+\v{k}_k^t=\v{u}_k^t$ and the averaging produces then
$\v{p}_i^t+\v{k}_i^t=\v{u}_i^t$ for $i=k+1,\ldots,j-1$. In this passage the
quark state $(q)$ is on-shell with the final quark but $(\bar{q})$ is on-shell
yet with the initial quark, thus we find a longitudinal phase of the form
\begin{align}
\delta p_i^z+\delta k_i^z&-\delta u_i^z =
\bigg(p^z_{p_0^0-\omega}(\v{p}_i^t)+p_\omega(\v{k}_{i}^t)-p_{p_0^0}^z(\v{p}_i^t+\v{k}_i^t)\bigg)\\
&-\bigg(p_{p_0^0-\omega}^z(\v{p}_{i-1}^t)+p^z_{\omega}(\v{k}_{i-1}^t)-p_{p_0^0}^z(\v{p}_{i-1}^t+\v{k}_{i-1}^t)\bigg)=-\frac{k_\mu^ip^\mu_{i}}{p_0^0}+\frac{k_\mu^{i-1}p^\mu_{i-1}}{p_0^0}.\nonumber
\end{align}
Making an analogy with the QED classical phase \eqref{phase_difference}
obtained in the $R\to\infty$ limit, this suggests a reorganization of the
total contribution of the interval as follows
\begin{align}
-i\sum_{i=k+1}^{j-1}\bigg(\delta p_i^z+\delta k_i^z-\delta
u_i^z\bigg)z_i=+i\frac{k_\mu^{j-1}p^\mu_{j-1}}{p_0^0}z_{j-1}-i\sum_{i=k+1}^{j-2}\frac{k_\mu^ip^\mu_i}{p_0^0}\delta
z_i-i\frac{k_\mu^{k+1}p^\mu_{k+1}}{p_0^0}z_{k+1},
\end{align}
where we denoted $\delta z_i\equiv z_{i+1}-z_i$. The average of the set of
interactions of this passage can be rewritten then as
\begin{align}
A_{k+1}^{j-1}=&\left(\prod_{i=k+1}^{j-1}(2\pi)^2\delta^2(\v{p}_i^t+\v{k}_i^t-\v{u}_i^t)\times
\phi_{q\bar{q}g}^{(n_i)}(\delta
\v{p}_i^t,\delta\v{k}_i^t)\right)\label{a_j-1_k+1_qcd}\\
&\times\exp\left(+i\frac{k_\mu^{j-1}p^\mu_{j-1}}{p_0^0}z_{j-1}-i\sum_{i=k+1}^{j-2}\frac{k_\mu^ip^\mu_i}{p_0^0}\delta
z_i-i\frac{k_\mu^{k+1}p^\mu_{k+1}}{p_0^0}z_{k+1}\right).\nonumber
\end{align}
At $z_j$ a gluon is emitted in the conjugated amplitude. Two possibilities can
be contemplated. The first one corresponds to a collision before the emission,
which affects then only to the $(q)$, $(\bar{q})$ and $(g)$ states and is
given by the average
\begin{align}
A_j^+\equiv\frac{1}{N_c^{n_j}}\Tr\left\langle\left(S_q^{j\dag}(\delta
\v{u}_j+\v{v}_j)\right)^{b_{j-1}}_{b'_j}\left(S_q^j(\delta
\v{p}_j)\right)^{a_j}_{a_{j-1}}\left(S_g^j(\delta \v{k}_j)\right)^{\alpha_j}_{\alpha_{j-1}}\right\rangle.
\end{align}
From \eqref{a_j-1_k+1_qcd} we inherit the condition
$\v{u}_{j-1}^t=\v{p}_{j-1}^t+\v{k}_{j-1}^t$ so from the elastic average we obtain
$\v{u}_{j}^t+\v{v}_j^t=\v{p}_{j}^t+\v{k}_{j}^t$. Since
$p_{j-1}^0+\omega=u_{j-1}^0=p_0^0$ the longitudinal phase reads for this case
\begin{align}
\delta p_j^z+\delta k_j^z&-\delta u_j^z =
\bigg(p^z_{p_0^0-\omega}(\v{p}_j^t)+p_\omega(\v{k}_{j}^t)-p_{p_0^0}^z(\v{p}_j^t+\v{k}_j^t)\bigg)\\
&-\bigg(p_{p_0^0-\omega}^z(\v{p}_{j-1}^t)+p^z_{\omega}(\v{k}_{j-1}^t)-p_{p_0^0}^z(\v{p}_{j-1}^t+\v{k}_{j-1}^t)\bigg)=-\frac{k_\mu^jp^\mu_{j}}{p_0^0}+\frac{k_\mu^{j-1}p^\mu_{j-1}}{p_0^0}.\nonumber
\end{align}
Then the average of square advanced interaction at $z_j$ produces a term of
the form
\begin{align}
A_j^+=(2\pi)^2\delta^2(\v{p}_j^t+\v{k}_j^t-\v{u}_{j}^t-\v{v}_j^t)\phi_{q\bar{q}g}(\delta\v{p}_j,\delta\v{k}_j)\exp\left(+i\frac{k_\mu^jp^\mu_{j}}{p_0^0}z_j-i\frac{k_\mu^{j-1}p^\mu_{j-1}}{p_0^0}z_j\right)\label{a_j+_qcd}.
\end{align}
The alternative case at $z_j$ corresponds to a collision just after the
emission, which thus affects the four states and is given instead by the
average of four amplitudes
\begin{align}
A_j^-\equiv\frac{1}{N_c}\Tr\left\langle\left(S_q^{j\dag}(\delta
\v{u}_j+\v{v}_{j-1})\right)^{b'_{j-1}}_{b_j}\left(S_g^{j\dag}(\delta
\v{v}_j)\right)^{\beta_{j-1}}_{\beta_j}\left(S_q^j(\delta
\v{p}_j)\right)^{a_j}_{a_{j-1}}\left(S_g^j(\delta \v{k}_j)\right)^{\alpha_j}_{\alpha_{j-1}}\right\rangle.
\end{align}
From \eqref{a_j-1_k+1_qcd} we find the condition
$\v{u}_{j-1}^t=\v{p}_{j-1}^t+\v{k}_{j-1}^t$ and then
$\v{u}_{j}^t+\v{v}_j^t=\v{p}_{j}^t+\v{k}_{j}^t$. The phase arising in the collision
can be reorganized using \eqref{pole_after_qcd} and \eqref{pole_before_qcd} as
\begin{align}
\delta p_j^z&+\delta k^z_j-\delta u_j^z-\delta
v^z_j=\bigg(p^z_{p_0^0-\omega}(\v{p}_j^t)+p^z_\omega(\v{k}_j^t)-p^z_{p_0^0}(\v{p}_j^t+\v{k}_j^t)\bigg)\\
&-\bigg(p_{p_0^0-\omega}^z(\v{p}_{j-1}^t)+p_\omega^z(\v{k}_{j-1}^t)-p_{p_0^0}^z(\v{p}_{j-1}^t+\v{k}_{j-1}^t)\bigg)-\bigg(p_{p_0^0-\omega}^z(\v{p}_j^t+\v{k}_j^t-\v{v}_j^t)\nonumber\\
&+p_\omega^z(\v{v}_j^t)-p_{p_0^0}^z(\v{p}_j^t+\v{k}_j^t)\bigg)+\bigg(p_{p_0^0-\omega}^z(\v{p}_{j-1}^t+\v{k}_{j-1}^t-\v{v}_{j-1}^t)+p_\omega^z(\v{v}_{j-1}^t)\nonumber\\
&-p_{p_0^0}^z(\v{p}_{j-1}^t+\v{k}_{j-1}^t)\bigg)=-\frac{k_\mu^j
  p^\mu_j}{p_0^0}+\frac{k_\mu^{j-1}p^\mu_{j-1}}{p_0^0}+\frac{v_\mu^ju^\mu_{j}}{p_0^0}-\frac{v_\mu^{j-1}u^\mu_{j-1}}{p_0^0}.\nonumber
\end{align}
Notice that the last two terms can also be defined, if desired in terms of the
momentum of the $q$ state, as
\begin{align}
\frac{v_\mu^ju^\mu_{j}}{p_0^0}-\frac{v_\mu^{j-1}u^\mu_{j-1}}{p_0^0}=\frac{v_\mu^j(p+k)^\mu_{j}}{p_0^0-\omega}-\frac{v_\mu^{j-1}(p+k)^\mu_{j-1}}{p_0^0-\omega},
\end{align}
provided $(p+k)$ refers to a quark 4-momentum of modulus $p_0^0$ with transverse
component $\v{p}^t+\v{k}^t$. The elastic average at this step can be shown to
satisfy then
\begin{align}
A_j^{-}&=(2\pi)^2\delta^2(\v{p}_j^t+\v{k}_j^t-\v{u}_j^t-\v{v}_j^t)\times\phi_{q\bar{q}g\bar{g}}^{(n_j)}(\delta\v{p}_j^t,\delta\v{k}_j^t,\delta\v{v}_j^t)\label{a_j-_qcd}\\
&\times\exp\left(+i\frac{k_\mu^j p^\mu_j}{p_0^0}z_j-i\frac{k_\mu^{j-1}p^\mu_{j-1}}{p_0^0}z_j-i\frac{v_\mu^ju^\mu_{j}}{p_0^0}z_j+i\frac{v_\mu^{j-1}u^\mu_{j-1}}{p_0^0}z_j\right)\nonumber,
\end{align}
where the elastic distribution of the momentum of the three states produces
\begin{align}
&\phi_{q\bar{q}g\bar{g}}^{(n_j)}(\delta\v{p}_j^t,\delta\v{k}_j^t,\delta\v{v}_j^t)\equiv
\int d^2\v{x}_j^{q\bar{q}}\int d^2\v{x}_j^{g\bar{q}}\int
d^2\v{x}_j^{\bar{g}\bar{q}}\exp\bigg[-i\delta\v{p}_j^t\cdot\v{x}^{q\bar{q}}_{j}\label{phi_qbarqgbarg}\\
&-i\delta\v{k}_j^t\cdot\v{x}_j^{g\bar{q}}+i\delta\v{v}_{j}^{t}\cdot\v{x}_j^{\bar{g}\bar{q}}+n_0(z_j)\delta
z_j\bigg(\sigma_{g\bar{q}}^{(1)}\Big(\v{x}_j^{g\bar{q}}\Big)+\Big(\sigma_{q\bar{q}}^{(1)}\Big(\v{x}^{q\bar{q}}_j\Big)-\sigma_{q\bar{q}}^{(1)}(\v{0})\Big)\nonumber\\
&+\Big(\sigma_{g\bar{g}}^{(1)}\Big(\v{x}^{g\bar{g}}_j\Big)-\sigma_{g\bar{g}}^{(1)}(\v{0})\Big)+\sigma_{gq}^{(1)}\Big(\v{x}_j^{q\bar{q}}-\v{x}_j^{g\bar{q}}\Big)+\sigma_{q\bar{g}}^{(1)}\Big(\v{x}_j^{q\bar{q}}-\v{x}_j^{\bar{g}\bar{q}}\Big)+\sigma_{\bar{q}\bar{g}}^{(1)}\Big(\v{x}_j^{\bar{g}\bar{q}}\Big)\bigg)\bigg]\nonumber.
\end{align}
Here the Fourier transform of the quark single squared amplitude
$\sigma_{q\bar{q}}^{(1)}(\v{x})$ is given by \eqref{sigma_qbarq_qcd}, the
cross products $\sigma_{gq}^{(1)}(\v{x})$ and $\sigma_{g\bar{q}}^{(1)}(\v{x})$
are given at \eqref{sigma_gq_qcd} and \eqref{sigma_gbarq_qcd}, respectively,
and the new cross squared amplitudes appearing in this average are given at
leading order for \eqref{fielddefinition} by
\begin{align}
\sigma_{g\bar{g}}^{(1)}(\v{x})=\int\frac{d^2\v{q}}{(2\pi)^2}e^{i\v{q}\cdot\v{x}}\frac{1}{N_c}\Tr\left(F_g^{(1)\dag}(\v{q})F_g^{(1)}(\v{q})\right)=\frac{4\pi
g_s^4}{\beta_k^2\mu_d^2}T_f\mu_d|\v{x}|K_1(\mu_d|\v{x}|),\label{sigma_gbarg_qcd}
\end{align}
which represents the collision with the gluon state $(g)$ and the conjugate
state $(\bar{g})$ leaving the rest of states unaltered, and the other two
verify $\sigma_{q\bar{g}}^{(1)}(\v{x})=\sigma_{g\bar{q}}^{(1)}(\v{x})$ acting
however on the $q$ and $\bar{g}$ states, and
$\sigma_{\bar{q}\bar{g}}^{(1)}(\v{x})=\sigma_{gq}^{(1)}(\v{x})$, but acting on
the $(\bar{q})$ and $(\bar{g})$ states. Finally the passage of the quark state
$(q)$ and its conjugate $(\bar{q})$, and of the gluon state $(g)$ and its
conjugate state $(\bar{g})$, from the layer $z_{j+1}$ to the end $z_n$ is
given by the average
\begin{align}
A^n_{j+1}\equiv
\prod_{j+1}^n\left(\frac{1}{N_c^{n_i}}\Tr\left\langle\left(S_{q}^{i\dag}(\delta
\v{u}_i)\right)^{b_{i-1}}_{b_{i}}\left(S_g^{i\dag}(\delta
\v{v}_i)\right)^{\beta_{i-1}}_{\beta_i}\left(S_q^i(\delta
\v{p}_i)\right)^{a_i}_{a_{i-1}}\left(S_g^i(\delta \v{k}_i)\right)^{\alpha_i}_{\alpha_{i-1}}\right\rangle\right).
\end{align}
As before, the total longitudinal phase in the interval suggests a
rearrangement of the form
\begin{align}
-i\sum_{i=j+1}^{n}\Big(\delta p_i^z&-\delta u_i^1+ \delta k_i^z-\delta
v_i^z\Big)z_i=+i\frac{k_\mu^np^\mu_n}{p_0^0}z_n-i\sum_{i=j+1}^{n-1}\frac{k_\mu^ip^\mu_i}{p_0^0}\delta
z_i-i\frac{k_\mu^{j}p^\mu_j}{p_0^0}z_{j+1}\nonumber\\
&-i\frac{v_\mu^nu^\mu_{n}}{p_0^0}z_n+i\sum_{i=j+1}^{n-1}\frac{v_\mu^iu^\mu_i}{p_0^0}\delta z_i+i\frac{v_\mu^{j}u^\mu_j}{p_0^0}z_{j+1}.
\end{align}
The averages of the squared amplitude are of the same kind as the one
explained for the previous scattering \eqref{a_j-_qcd}, and we obtain
\begin{align}
A_{j+1}^n=\prod_{i=j+1}^n\bigg((2\pi)^2\delta^2(\v{p}_i^t+\v{k}_i^t-\v{u}_i^t&-\v{v}_i^t)\times\phi_{q\bar{q}g\bar{g}}^{(n_j)}(\delta\v{p}_j^t,\delta\v{k}_j^t,\delta\v{v}_j^t)\bigg)\label{a_n_j+1_qcd}\\
\times\exp\left(-i\sum_{i=j+1}^{n-1}\frac{k_\mu^ip^\mu_i}{p_0^0}\delta
z_i\right.&-\left.i\frac{k_\mu^{j}p^\mu_j}{p_0^0}z_{j+1}\right.\left.+i\sum_{i=j+1}^{n-1}\frac{v_\mu^iu^\mu_i}{p_0^0}\delta z_i+i\frac{v_\mu^{j}u^\mu_j}{p_0^0}z_{j+1}\right).\nonumber
\end{align}
Now, from \eqref{m_k_m_j_incoherent} the average over initial spins and the sum
over final spins and polarizations provides further simplifications. We 
notice, \textit{c.f.} Appendix \ref{appendix1},
\begin{align}
\frac{1}{2}&\sum_{s_ns_0}\sum_{\lambda_n}
\bigg(\Big(f_j^{+}\Big)^*t_{\beta_j}\phi_{q\bar{q}g}^{(n_j)}\Big(\delta\v{p}_j,\delta\v{k}_j\bigg)-\Big(f_{j-1}^{-}\Big)^*\phi_{q\bar{q}g\bar{g}}^{(n_j)}\Big(\delta\v{p}_j^t,\delta\v{k}_j^t,\delta\v{v}_j^t\Big)\medspace
t_{\beta_{j-1}}\bigg)\\
\times&\medspace\bigg(f_k^{+}t_{\alpha_k}\phi_{q\bar{q}}^{(0)}\Big(\delta\v{p}_k^t\Big)-f_{k-1}^-\phi^{(0)}_{q\bar{q}g}\Big(\delta\v{p}_k^t,\delta\v{k}_k^t\Big)\medspace
t_{\alpha_{k-1}}\bigg)=\frac{1}{\omega^2}\left(h^n(y)\v{\delta}_j^n\cdot\v{\delta}_k^n+h^s(y)\delta_j^s\delta_k^s\right),\nonumber
\end{align}
where $y=\omega/p_0^0$ is the fraction of energy carried by the gluon and the
functions $h^n(y)=(1+(1-y)^2)/2$ and $h^s(y)=y^2/2$ are the kinematical
weights of the spin no flip and spin flip contributions, which are given by
the currents
\begin{align}
\v{\delta}_k^n\equiv\frac{\v{k}_k\times\v{p}_k}{k_\mu^kp^\mu_k}t_{\beta_k}\phi_{q\bar{q}g}^{(n_k)}\Big(\delta\v{p}_k,\delta\v{k}_k\Big)-\frac{\v{k}_{k-1}\times\v{p}_{k-1}}{k_\mu^{k-1}p^\mu_{k-1}}\phi_{q\bar{q}g\bar{g}}^{(n_k)}\Big(\delta\v{p}_k^t,\delta\v{k}_k^t,\delta\v{v}_k^t\Big)t_{\beta_{k-1}},\label{qcd_current_pluselastic}
\end{align}
and
\begin{align}
\delta_k^s\equiv \frac{\omega
    p_0^0}{k_\mu^kp^\mu_k}t_{\beta_k}\phi_{q\bar{q}g}^{(n_k)}\Big(\delta\v{p}_k,\delta\v{k}_k\Big)-\frac{\omega p_0^0}{k_\mu^{k-1}p^\mu_{k-1}}\phi_{q\bar{q}g\bar{g}}^{(n_k)}\Big(\delta\v{p}_k^t,\delta\v{k}_k^t,\delta\v{v}_k^t\Big)t_{\beta_{k-1}}.
\end{align}
In the following we will assume that $\omega \ll p_0^0$. In this soft limit
the spin flip contribution can be neglected and the kinematical weight of the
non flip contribution is given simply by $h^n(y)\simeq 1$. Hence, after
inserting equations \eqref{a_k-1_1_qcd}, \eqref{a_k+_qcd}, \eqref{a_k-_qcd},
\eqref{a_j-1_k+1_qcd}, \eqref{a_j+_qcd}, \eqref{a_j-_qcd} and
\eqref{a_n_j+1_qcd} in \eqref{m_k_m_j_incoherent}, integrating in the momentum
trajectory of the $(\bar{q})$ states, summing over final spin and polarization
and average over initial spin we find
\begin{align}
\frac{1}{2}&\sum_{\lambda,
    s}\left(\prod_{i=1}^{n-1}\int\frac{d^2\v{u}_i^t}{(2\pi)^2}\right)\left\langle\mathcal{M}_j^\dag\mathcal{M}_k\right\rangle
  \label{m_k_m_j_incoherent_result}\\
&= (2\pi)^2\delta^2(\v{0})\exp\left(-i\sum_{i=j}^{n-1}\delta z_i\Bigg(\frac{k^i_\mu
  p^\mu_i}{p_0^0-\omega}-\frac{v_\mu^ip^\mu_i}{p_0^0-\omega}\Bigg)-i\sum_{i=k}^{j-1}\delta z_i\frac{k_\mu^i
  p^\mu_i}{p_0^0-\omega}\right)\nonumber\\
&\left(\prod_{i=j+1}^n\phi_{q\bar{q}g\bar{g}}^{(n_i)}\Big(\delta\v{p}_i,\delta\v{k}_i,\delta\v{v}_i\Big)\right)
\v{\delta}_j^n
\left(\prod_{i=k+1}^{j-1}\phi_{q\bar{q}g}
^{(n_i)}\Big(\delta\v{p}_i^t,\delta\v{k}_i^t\Big)\right)\v{\delta}_k^n\left(\prod_{i=1}^{k-1}\phi_{q\bar{q}}^{(n_i)}\Big(\delta\v{p}_i^t\Big)\right).\nonumber
\end{align}
The extra delta $(2\pi)^2\delta^2(\v{0})=\pi R^2$, which accounts the space
translation invariance in the transverse plane, appears due to the conditions
$\v{p}_n^t=\v{u}_n^t$ and $\v{k}_n^t=\v{v}_n^t$. The integration in quark
momenta has been shifted by $\v{k}^t$. The averaged elastic amplitudes squared
$B_j^k$ appearing at \eqref{m_k_m_j_coherent} follow the same steps as the
previous calculation. Since the same kinematical conditions hold, it can be
shown that the elastic distributions \eqref{phi_qbarq}, \eqref{phi_qbarqg} and
\eqref{phi_qbarqgbarg} have to be replaced with the squared averaged
amplitudes. An elastic average squared of the state $(q)$ and $(\bar{q})$ is
given by
\begin{align}
\phi_{q\bar{q}}^{(0)}\Big(\delta\v{p}_i^t\Big)\equiv \exp\Bigg(-n_0(z_i)\delta z_i\sigma_{q\bar{q}}^{(1)}(\v{0})\Big)\Bigg)\times(2\pi)^2\delta^2(\delta\v{p}_i^t).\label{phi_qbarq_vac}
\end{align}
Similarly, an elastic average squared of the states $(q)$, $(g)$ and $\bar{q}$
is given instead by
\begin{align}
\phi_{q\bar{q}g}^{(0)}(\delta\v{p}_i^t,\delta\v{k}_{i}^t)\equiv&\int
d^2\v{x}_k^{q\bar{q}}\int d^2\v{x}_{k}^{g\bar{q}}\exp\Bigg(-i\delta\v{p}_i^t\cdot\v{x}_i^{q\bar{q}}-i\delta\v{k}_i^t\cdot\v{x}_i^{g\bar{q}}\Bigg)\\
&\times\exp\Bigg(n_0(z_i)\delta
z_i\bigg(\varphi_g^{(1)}(\v{0})-\sigma_{q\bar{q}}^{(1)}(\v{0})+\sigma_{gq}^{(1)}\Big(\v{x}_i^{g\bar{q}}-\v{x}_{k}^{q\bar{q}}\Big)\bigg)\Bigg)\nonumber,
\end{align}
which produces $\delta\v{p}_i^t=\delta\v{k}_i^t$. And finally an elastic
average squared of the states $(q)$, $(g)$, $(\bar{q})$ and $(\bar{g})$
produces
\begin{align}
&\phi_{q\bar{q}g\bar{g}}^{(0)}(\delta\v{p}_i^t,\delta\v{k}_i^t,\delta\v{v}_i^t)\equiv
\int d^2\v{x}_i^{q\bar{q}}\int d^2\v{x}_i^{g\bar{q}}\int
d^2\v{x}_i^{\bar{g}\bar{q}}e^{-i\delta\v{p}_i^t\cdot\v{x}^{q\bar{q}}_{i}-i\delta\v{k}_i^t\cdot\v{x}_i^{g\bar{q}}+i\delta\v{v}_{i}^{t}\cdot\v{x}_i^{\bar{g}\bar{q}}}\nonumber\\
&\times\exp\Bigg(n_0(z_i)\delta
z_i\bigg(-\sigma_{q\bar{q}}^{(1)}(\v{0})-\sigma_{g\bar{g}}^{(1)}(\v{0})+\sigma_{gq}^{(1)}\Big(\v{x}_i^{q\bar{q}}-\v{x}_i^{g\bar{q}}\Big)+\sigma_{\bar{g}\bar{q}}^{(1)}\Big(\v{x}_i^{\bar{g}\bar{q}}\Big)\bigg)\Bigg),
\end{align}
which produces $\delta\v{p}_i^t=\delta\v{k}_i^t$. Using these elastic
distributions, from \eqref{m_k_m_j_coherent} we get
\begin{align}
\frac{1}{2}&\sum_{\lambda,
    s}\left(\prod_{i=1}^{n-1}\int\frac{d^2\v{u}_i^t}{(2\pi)^2}\right)\left\langle\mathcal{M}_j^\dag\right\rangle\Big\langle\mathcal{M}_k\Big\rangle
 \label{m_k_m_j_coherent_result}\\
&= (2\pi)^2\delta^2(\v{0})\exp\left(-i\sum_{i=j}^{n-1}\delta z_i\Bigg(\frac{k^i_\mu
  p^\mu_i}{p_0^0-\omega}-\frac{v_\mu^ip^\mu_i}{p_0^0-\omega}\Bigg)-i\sum_{i=k}^{j-1}\delta z_i\frac{k_\mu^i
  p^\mu_i}{p_0^0-\omega}\right)\nonumber\\
&\left(\prod_{i=j+1}^n\phi_{q\bar{q}g\bar{g}}^{(0)}\Big(\delta\v{p}_i,\delta\v{k}_i,\delta\v{v}_i\Big)\right)
\v{\delta}_j^n
\left(\prod_{i=k+1}^{j-1}\phi_{q\bar{q}g}
^{(0)}\Big(\delta\v{p}_i^t,\delta\v{k}_i^t\Big)\right)\v{\delta}_k^n\left(\prod_{i=1}^{k-1}\phi_{q\bar{q}}^{(0)}\Big(\delta\v{p}_i^t\Big)\right).\nonumber
\end{align}
The structure of the intensities \eqref{m_k_m_j_incoherent_result} and
\eqref{m_k_m_j_coherent_result} has substantial differences with the QED
scenario, \eqref{incoherent_kj_1_result} and \eqref{incoherent_kj_2_result},
related to the ability of the gluon to reinteract with the medium
constituents. First, due to the form of \eqref{qcd_current_pluselastic} it is
not possible yet to factorize the elastic distributions independently of $z_k$
and $z_j$. Second, the accumulated phase between $z_k$ and $z_j$, regulating
the interference and hence the LPM effect, is now dominated for soft gluons by
the accumulated squared momentum change of the gluon and its mass. Indeed, for
$\omega\ll p_0^0$ either looking at \eqref{pole_before_qcd} or from $k_\mu
p^\mu=\omega p_0^0(1-\beta_k\beta_p\hat{\v{k}}\cdot\hat{\v{p}})$ the quark can
be considered frozen with respect to the gluon rescattering in the initial
direction $p^\mu(z)=p^\mu(0)$. Then if we set $\beta_p=1$ the phase becomes
\begin{align}
\sum_{i=k}^{j-1}\frac{ k_\mu^ip^\mu_i}{p_0^0-\omega}\cong\sum_{i=k}^{j-1}\frac{ k_\mu^ip^\mu_0}{p_0^0}=\frac{m_g^2}{(1+\beta_k)\omega}(z_j-z_k) +\sum_{i=k}^{j-1}\frac{\delta
  \v{k}^2_i}{2\beta_k\omega}\delta z_i,\label{softgluon_phaseapproximation}
\end{align}
where $\delta\v{k}_i^2$ is the accumulated gluon squared momentum change with
respect to the initial quark direction. Third, the unobserved intermediate
states of the gluon leads to an extra phase in the interval from $z_j$ to
$z_n$ of the form
\begin{align}
&\sum_{i=j}^{n-1}\delta z_i\Bigg(\frac{k^i_\mu
  p^\mu_i}{p_0^0-\omega}-\frac{v_\mu^ip^\mu_i}{p_0^0-\omega}\Bigg)  \neq 0,
\end{align}
since $v_i\neq k_i$. And third, color rotation is allowed in the states from
$z_k$ onwards, since the elastic distributions \eqref{phi_qbarqg} and
\eqref{phi_qbarqgbarg} carry a non trivial matrix structure. We will assume,
however, that a color averaged effective interaction for the gluon exists
which leads to the same momentum transport as the elastic distributions
\eqref{phi_qbarqg} and \eqref{phi_qbarqgbarg} and thus the same phase
\eqref{softgluon_phaseapproximation}. For that purpose we take as an ansatz
the color average of \eqref{m_k_m_j_incoherent_result} and
\eqref{m_k_m_j_coherent_result} for the single layer $n$ = 1 case. Since an
expansion in $\delta z$ holds, using the relation $n_0(z_1)\delta z_1
(2\pi)^2\delta^2(\v{0})\equiv n_1$ we find from
\eqref{m_k_m_j_incoherent_result} a crossed term of the form
\begin{align}
(2\pi)^2\delta^{2}(\v{0})\frac{g_s^2}{N_c}\Tr\Big(t_{\alpha_1}\phi_{q\bar{q}g}^{(n_1)}\Big(\delta\v{p},\delta\v{k}\Big)t_{\alpha_0}\Big)\simeq
  n_1&\int
d^2\v{x}_1^{q\bar{q}}e^{-i\delta\v{p}\cdot\v{x}_1^{q\bar{q}}}
\int d^2\v{x}_1^{g\bar{q}}e^{-i\delta\v{k}\cdot\v{x}_1^{g\bar{q}}}\nonumber\\
\times\Bigg\{\frac{g_s^2}{N_c}\Tr\left(t_{\alpha_1}\Big(\varphi_g^{(1)}(\v{0})-\sigma_{q\bar{q}}^{(1)}(\v{0})\Big)t_{\alpha_0}\right)&+\frac{g_s^2}{N_c}\Tr\left(t_{\alpha_1}\sigma_{q\bar{q}}^{(1)}\Big(\v{x}_1^{q\bar{q}}\Big)t_{\alpha_0}\right)\\
+\frac{g_s^2}{N_c}\Tr\left(t_{\alpha_1}\sigma_{g\bar{q}}^{(1)}\Big(\v{x}_1^{g\bar{q}}\Big)t_{\alpha_0}\right)&+\frac{g_s^2}{N_c}\Tr\Big(t_{\alpha_1}\sigma_{gq}^{(1)}\Big(\v{x}_1^{g\bar{q}}-\v{x}_1^{q\bar{q}}\Big)t_{\alpha_0}\Big)\Bigg\}\nonumber.
\end{align}
The first contribution represents the passage without momentum changes. We are
interested instead in the last three terms.  Using \eqref{sigma_qbarq_qcd} the
first of these contributions can be rewritten as
\begin{align}
\frac{g_s^2}{N_c}\Tr&\left(t_{\alpha_1}\sigma_{q\bar{q}}^{(1)}\Big(\v{x}_1^{q\bar{q}}\Big)
t_{\alpha_0}\right)=\frac{g_s^2}{N_c}\Tr\Big(t_{\alpha_1}t_\alpha
t_\beta
 t_{\alpha_1}\Big)\frac{\delta_{\alpha\beta}}{2N_c}\frac{4\pi
   g_s^4}{\beta_p^2\mu_d^2}\mu_d\left|\v{x}_1^{q\bar{q}}\right|K_1\Big(\mu_d\left|\v{x}_1^{q\bar{q}}\right|\Big)\nonumber\\
&=-\frac{1}{N_c^2-1}\left(g_s^2\frac{N_c^2-1}{2N_c}\right)\left(\frac{N_c^2-1}{4N_c^2} \frac{4\pi g_s^4}{\beta_p^2\mu_d^2}\mu_d\left|\v{x}_1^{q\bar{q}}\right|K_1\Big(\mu_d\left|\v{x}_1^{q\bar{q}}\right|\Big)\right),
\end{align}
that is, $1/8$ times the independent color average of the emission vertex
$C_f=(N_c^2-1)/2N_c$ times the independent color average of the squared
scattering amplitude of a quark with a single quark. The second contribution
produces, using \eqref{sigma_gbarq_qcd},
\begin{align}
\frac{g_s^2}{N_c}&\Tr\left(t_{\alpha_1}\sigma_{g\bar{q}}^{(1)}\Big(\v{x}_1^{g\bar{q}}\Big)t_{\alpha_0}\right)=\frac{g_s^2}{N_c}\Tr\Big(t_\alpha
t_{\alpha_1}t_{\alpha_0}\Big)\Big(T_\alpha\Big)^{\alpha_1}_{\alpha_0}\frac{\delta_{\alpha\beta}}{2N_c}
\frac{4\pi
   g_s^4}{\beta_p\beta_k\mu_d^2}\mu_d\left|\v{x}_1^{g\bar{q}}\right|K_1\Big(\mu_d\left|\v{x}_1^{g\bar{q}}\right|\Big)\nonumber\\
&=\frac{N_c^2}{N_c^2-1}\left(g_s^2\frac{N_c^2-1}{2N_c}\right)\left(\frac{N_c^2-1}{4N_c^2} \frac{4\pi g_s^4}{\beta_p\beta_k\mu_d^2}\mu_d\left|\v{x}_1^{g\bar{q}}\right|K_1\Big(\mu_d\left|\v{x}_1^{g\bar{q}}\right|\Big)\right),
\end{align}
which is $9/8$ times the independent color average of the emission vertex
times the color average of the squared scattering amplitude of a quark with a
single quark. For the third contribution we get, using \eqref{sigma_gq_qcd},
\begin{align}
\frac{g_s^2}{N_c}&\Tr\left(t_{\alpha_1}\sigma_{gq}^{(1)}\Big(\v{x}_1^{gq}\Big)t_{\alpha_0}\right)=-\frac{g_s^2}{N_c}\Tr\Big(t_{\alpha_1}
t_{\alpha}t_{\alpha_0}\Big)\Big(T_\alpha\Big)^{\alpha_1}_{\alpha_0}\frac{\delta_{\alpha\beta}}{2N_c}
\frac{4\pi
   g_s^4}{\beta_p\beta_k\mu_d^2}\mu_d\left|\v{x}_1^{g{q}}\right|K_1\Big(\mu_d\left|\v{x}_1^{gq}\right|\Big)\nonumber\\
&=\frac{N_c^2}{N_c^2-1}\left(g_s^2\frac{N_c^2-1}{2N_c}\right)\left(\frac{N_c^2-1}{4N_c^2} \frac{4\pi g_s^4}{\beta_p\beta_k\mu_d^2}\mu_d\left|\v{x}_1^{gq}\right|K_1\Big(\mu_d\left|\v{x}_1^{gq}\right|\Big)\right),
\end{align}
which is again the same result as before. Then if we assume the high energy
limit both for the gluon and the quark $\beta_k\simeq \beta_p\simeq 1$ we can
write an effective elastic distribution for \eqref{phi_qbarqg} given by the
replacement
\begin{align}
&\left(\sigma_{q\bar{q}}^{(1)}\Big(\v{x}_i^{q\bar{q}}\Big)-\sigma_{q\bar{q}}^{(1)}(\v{0})\right)+\sigma_{g\bar{q}}^{(1)}\Big(\v{x}_i^{g\bar{q}}\Big)+\sigma_{gq}^{(1)}\Big(\v{x}_i^{g\bar{q}}-\v{x}_{i}^{q\bar{q}}\Big)\nonumber\\
&\to-\frac{1}{8}\left(\sigma_{q\bar{q}}^{(1)}\Big(\v{x}_i^{q\bar{q}}\Big)-\sigma_{q\bar{q}}^{(1)}(\v{0})\right)+\frac{9}{8}\sigma_{q\bar{q}}^{(1)}\Big(\v{x}_i^{g\bar{q}}\Big)+\frac{9}{8}\sigma_{q\bar{q}}^{(1)}\Big(\v{x}_i^{g\bar{q}}-\v{x}_{i}^{q\bar{q}}\Big),\label{sigma_gbarq_effective}
\end{align} 
where the (trivial) color structure of $\sigma_{q\bar{q}}^{(1)}(\v{x})$ at
\eqref{sigma_qq_qcd} has been dropped. Further simplifications produces the
soft limit of the phase. In that case the integrals in the quark momenta can
be performed producing $\v{x}_{i}^{q\bar{q}}=0$ in all the distributions so that
$\phi_{q\bar{q}}(\v{x}_i^{q\bar{q}})\to 1$. Indeed for the first set of
scatterings we obtain, using \eqref{phi_qbarq},
\begin{align}
\int\frac{d^2\delta\v{p}_i}{(2\pi)^2}\phi_{q\bar{q}}^{(n_i)}(\delta\v{p}_i^t)=1.
\end{align}
For the intermediate step of scatterings we find, using \eqref{phi_qbarqg},
\begin{align}
\int&\frac{d^2\delta
  \v{p}_i^t}{(2\pi)^2}\phi_{q\bar{q}g}^{(n_i)}(\delta\v{p}_i^t,\delta\v{k}_i^t)\\ &=\int
d^2\v{x}_i^{g\bar{q}}\exp\left(-i\delta\v{k}_i^t\cdot\v{x}_i^{g\bar{q}}\right)\exp\left(\varphi_{g}^{(1)}(\v{0})+\frac{9}{8}\sigma_{q\bar{q}}^{(1)}\Big(\v{x}_i^{g\bar{q}}\Big)+\frac{9}{8}\sigma_{qq}^{(1)}\Big(\v{x}_i^{g\bar{q}}\Big)\right).\nonumber
\end{align}
By summing the two last contributions the above term can be rewritten as the
squared scattering amplitude of a gluon with a single target quark
\begin{align}
\frac{9}{8}\sigma_{q\bar{q}}^{(1)}\Big(\v{x}_i^{g\bar{q}}\Big)+\frac{9}{8}\sigma_{q\bar{q}}^{(1)}\Big(\v{x}_i^{g\bar{q}}\Big)=\left(\frac{9}{8}+\frac{9}{8}\right)\frac{2}{9}\frac{4\pi
g_s^4}{\beta^2\mu_d^2}\mu_d\left|\v{x}_i^{g\bar{q}}\right|K_1\left(\mu_d
\left|\v{x}_i^{g\bar{q}}\right|\right)=\sigma_{g\bar{g}}^{(1)}(\v{x}^{g\bar{q}}_i),\label{sigma_gbarg_qcd_scalar}
\end{align}
where as before the color structure of
$\sigma_{g\bar{g}}^{(1)}(\v{x}^{g\bar{q}}_i)$ at \eqref{sigma_gbarg_qcd} has
been dropped \cite{nikolaev1994b,nikolaev1994a,wiedemann1999}. For the last set of scatterings the
integration in the quark momenta produces, using \eqref{phi_qbarqgbarg},
\begin{align}
\int\frac{d^2\delta\v{p}_i^t}{(2\pi)^2}\phi_{q\bar{q}g\bar{g}}^{(n_i)}(\delta\v{p}_i^t,\delta\v{k}_i^t,\delta\v{v}_i^t)&=(2\pi)^2\delta^2(\delta\v{v}_i^t-\delta\v{k}_i^t)\\&\times\int d^2\v{x}_i^{g\bar{q}}e^{-i\delta\v{k}_i^t\cdot\v{x}_i^{g\bar{q}}}\exp\left(\sigma_{g\bar{g}}^{(1)}\Big(\v{x}_i^{g\bar{g}}\Big)-\sigma_{g\bar{g}}^{(1)}(\v{0})\right)\nonumber,
\end{align}
which makes the phase vanish in the interval $z_j$ to $z_n$, since
$\v{v}_i^t=\v{k}_i^t$ for $i=j,\ldots,n$. Under these approximations we assume
that the gluon effective interaction is given by the solution of a
Moliere/transport equation \eqref{moliere_qcd} where the single collision
distribution is given by the Fourier transform of
$\sigma_{g\bar{g}}^{(1)}(\v{x}^{g\bar{q}}_i)$. From
\eqref{m_k_m_j_incoherent_result} we obtain then
\begin{align}
&\frac{1}{2N_c}\sum_{\lambda,
    s,a_0}\left(\prod_{i=1}^{n-1}\int\frac{d^2\v{p}_i^t}{(2\pi)^2}\int\frac{d^2\v{u}_i^t}{(2\pi)^2}\right)\left(\prod_{i=j-1}^{n-1}\int\frac{d^2\v{v}_i^t}{(2\pi)^2}\right)\left\langle\mathcal{M}_j^\dag\mathcal{M}_k\right\rangle
  \simeq(2\pi)^2\delta^2(\v{0})\frac{C_f}{\omega^2}\nonumber\\
&\exp\left(-i\sum_{i=k}^{j-1}\delta z_i\frac{k_\mu^i
  p^\mu_0}{p_0^0-\omega}\right)\left(\prod_{i=j+1}^n\phi_{g\bar{g}}^{(n_i)}\Big(\delta\v{k}_i\Big)\right)
\v{\delta}_j^n
\left(\prod_{i=k+1}^{j}\phi_{g\bar{g}}
^{(n_i)}\Big(\delta\v{k}_i^t\Big)\right)\v{\delta}_k^n\left(\prod_{i=1}^{k}\phi_{g\bar{g}}
^{(n_i)}\Big(\delta\v{k}_i^t\Big)\right),\label{m_k_m_j_incoherent_result2}
\end{align}
where we extracted $C_f=(N_c^2-1)/2N_c$, the color average of the emission
vertex, and for convenience we multiplied by one with a set of normalized
gluon distributions for the range $i=1,\ldots,k-1$. After summing in $j$ and
$k$ from \eqref{m_k_m_j_incoherent_result2} we obtain
\begin{align}
&\frac{1}{2N_c}\sum_{\lambda,
    s,a_0}\left(\prod_{i=1}^{n-1}\int\frac{d^2\v{p}_i^t}{(2\pi)^2}\int\frac{d^2\v{u}_i^t}{(2\pi)^2}\right)\left(\prod_{i=j-1}^{n-1}\int\frac{d^2\v{v}_i^t}{(2\pi)^2}\right)\sum_{j,k}^n\left\langle\mathcal{M}_j^\dag\mathcal{M}_k\right\rangle\nonumber\\
&=(2\pi)^2\delta^2(\v{0})\times\frac{C_f}{\omega^2}\left|\sum_{k=1}^n\exp\left(-i\sum_{i=k}^{n-1}\delta z_i\frac{k_\mu^i
  p^\mu_0}{p_0^0-\omega}\right)
\v{\delta}_k^n\right|^2\left(\prod_{i=1}^{n}\phi_{g\bar{g}}
^{(n_i)}\Big(\delta\v{k}_i^t\Big)\right).\label{m_k_m_j_incoherent_result3}
\end{align} 
Under the new color averaged effective elastic distributions the averages
appearing in the non flip emission currents \eqref{qcd_current_pluselastic}
can be factorized so that
\begin{align}
\v{\delta}_k^n\equiv\frac{\v{k}_k\times\v{p}_0}{k_\mu^kp^\mu_0}-\frac{\v{k}_{k-1}\times\v{p}_{0}}{k_\mu^{k-1}p^\mu_{0}}.\label{qcd_current}
\end{align}
Similarly for the elastic averaged amplitudes squared
\eqref{m_k_m_j_coherent_result} we obtain, after summing in $j$ and $k$,
\begin{align}
&\frac{1}{2N_c}\sum_{\lambda,
    s,a_0}\left(\prod_{i=1}^{n-1}\int\frac{d^2\v{p}_i^t}{(2\pi)^2}\int\frac{d^2\v{u}_i^t}{(2\pi)^2}\right)\left(\prod_{i=j-1}^{n-1}\int\frac{d^2\v{v}_i^t}{(2\pi)^2}\right)\sum_{j,k}^n
  \left\langle\mathcal{M}_j^\dag\right\rangle\Big\langle\mathcal{M}_k\Big\rangle\nonumber\\ &=(2\pi)^2\delta^2(\v{0})\times\frac{C_f}{\omega^2}\left|\sum_{k=1}^n\exp\left(-i\sum_{i=k}^{n-1}\delta
  z_i\frac{k_\mu^i p^\mu_0}{p_0^0-\omega}\right)
  \v{\delta}_k^n\right|^2\left(\prod_{i=1}^{n}\phi_{g\bar{g}}
  ^{(0)}\Big(\delta\v{k}_i^t\Big)\right).\label{m_k_m_j_coherent_result3}
\end{align} 
Correspondingly, after identifying $(2\pi)^2\delta^2(\v{0})\equiv\pi R^2$ and
$2\pi\beta_p\delta(0)\equiv \beta_pT$, inserting
\eqref{m_k_m_j_incoherent_result3} and \eqref{m_k_m_j_coherent_result3} in
\eqref{incoherent_emission_qcd_definition}, and using
\eqref{intensity_definition_qcd}, we finally find an intensity of the form
\begin{align}
\omega \frac{dI}{d\omega
  d\Omega_k}=\frac{g_s^2C_f}{(2\pi)^2}\left(\prod_{i=1}^{n-1}\frac{d^3\v{k}_i}{(2\pi)^3}\right)\left\{\left(\prod_{i=1}^n\phi_{g\bar{g}}^{(n_i)}(\delta\v{k}_i)\right)-\left(\prod_{i=1}^n\phi_{g\bar{g}}^{(0)}(\delta\v{k}_i)\right)\right\}\nonumber\\
\times \left|\sum_{k=1}^n\exp\left(-i\sum_{i=k}^{n-1}\delta z_i\frac{k_\mu^i
  p^\mu_0}{p_0^0-\omega}\right)
\v{\delta}_k^n\right|^2.\label{central_equation_qcd}
\end{align}
Equation \eqref{central_equation_qcd} is the central result of this
section. The resulting intensity is the evaluation, over the elastic
transports at each of the single layers, of a sum of single Bethe-Heitler
currents for the gluon \eqref{qcd_current} carrying each one a phase
\eqref{softgluon_phaseapproximation} responsible of modulating the LPM effect
in a QCD scenario. The elastic distributions appearing in
\eqref{central_equation_qcd} have been completed to the original three
dimensional elastic distributions and are given by
\begin{align}
\phi_{g\bar{g}}^{(n_i)}(\delta\v{k}_i)\equiv \exp\left(-n_0(z_i)\delta
z_i\sigma_{g\bar{g}}^{(1)}(\v{0})\right)\times(2\pi)^3\delta^3(\delta\v{k}_i)+2\pi\beta_k\delta
(\delta k^0)\Sigma_2^{(n_i)}(\delta\v{k}_i^t,\delta z_i).\label{phi_gbarg}
\end{align}
Equation \eqref{phi_gbarg} can be interpreted as the probability of no
colliding with the layer of density $n_0(z_i)$ and thickness $\delta z_i$,
given by $\exp\Big(-n_0(z_i)\delta z_i\sigma_{g\bar{g}}^{(1)}(\v{0})\Big)$,
times the forward distribution $(2\pi)^3\delta^{3}(\delta\v{k}_i)$, or the
collisional distribution in case of collision, which is given by
\begin{align}
\Sigma_{g\bar{g}}^{(n_i)}(\v{q},\delta z)\equiv
\exp\left(-n_0(z)\delta z\sigma_{g\bar{g}}^{(1)}(\v{0})\right)\int
d^2\v{x}e^{-i\v{q}\cdot\v{x}}\left(\exp\bigg(n_0(z)\delta
z\sigma_{g\bar{g}}^{(1)}(\v{x})\bigg)-1\right).\label{sigma_effective_qcd}
\end{align}
The Fourier transform of the squared scattering amplitude of a gluon with a
single quark of the QCD medium $\sigma_{g\bar{g}}^{(1)}(\v{x})$ is given by
\eqref{sigma_gbarg_qcd} but without the color structure. Its color averaged
charge $T_f$ equals the sum of the charges of the $gq$ and $g\bar{q}$
interactions \eqref{sigma_gbarq_qcd} and \eqref{sigma_gq_qcd}, which govern
the gluon interaction between $z_k$ and $z_j$. In \eqref{central_equation_qcd}
the overall no collision case is removed at the end. Its contribution is given
by the effective elastic distributions substituting the averaged elastic
amplitudes squared \eqref{m_k_m_j_coherent}
\begin{align}
\phi_{g\bar{g}}^{(0)}(\delta\v{k}_i)\equiv \exp\left(-n_0(z_i)\delta
z_i\sigma_{g\bar{g}}^{(1)}(\v{0})\right)\times(2\pi)^3\delta^3(\delta\v{k}_i)\label{phi_gbarg_nocollision}
\end{align}
Following \eqref{phi_gbarg} or \eqref{phi_gbarg_nocollision} then the gluon
mean free path in the medium $\lambda_{g\bar{g}}$, if we assume an uniform
density $n_0\equiv n_0(z_i)$, is given by
\begin{align}
\lambda_{g\bar{g}}\equiv\frac{1}{n_0\sigma_{g\bar{g}}^{(1)}(\v{0})}
\end{align}
For a medium of vanishing length $l\ll \lambda_{g\bar{g}}$ the cancellation of
the elastic distribution \eqref{phi_gbarg} with the no collision case
\eqref{phi_gbarg_nocollision} at \eqref{central_equation_qcd} guarantees that
the intensity vanishes. For mediums of very large length $l\gg
\lambda_{g\bar{g}}$ the subtraction of the overall no collision case
\eqref{phi_gbarg_nocollision} at \eqref{central_equation_qcd} can be
neglected. The motion of the gluon under the effective distributions
\eqref{phi_gbarg} satisfies an additivity rule for the squared momentum
change. Indeed, for a step such that $\delta l \lesssim \lambda_{g\bar{g}}$
then the total elastic distribution is given by the incoherent superposition
of the single elastic distributions of the centers at $\delta l$. The squared
momentum change in $\delta l$ equals, then, the momentum change in a single
collision
\begin{align}
\left\langle\delta\v{k}^2(\delta l)\right\rangle \equiv
\frac{\lambda_{g\bar{g}}}{\delta l}\int\frac{d^3\v{k}}{(2\pi)^3}\delta
\v{k}^2\phi_{g\bar{g}}^{(n)}(\delta\v{k})\simeq 2\mu_d^2\left(\log\left(\frac{2\beta_k\omega}{\mu_d}\right)-\frac{1}{2}\right)=\eta(\omega)\mu_d^2.\label{momentumchange_gluon_singlecollision}
\end{align}
where the function $\eta(\omega)$ accounts for the long tail of the Debye
interaction. For larger distances $l$ we can use the transport equation
\cite{moliere1948,bethe1953} satisfied by \eqref{sigma_effective_qcd} given at
\eqref{moliere_qcd}. We then find a gluon analogous of the quark transport
\eqref{transport_equation_pt2_qcd}, a transport equation for the gluon squared
momentum change
\begin{align}
\frac{\partial}{\partial l}\left\langle\delta\v{k}^2(l)\right\rangle =
n_0\sigma_{g\bar{g}}^{(1)}\left\langle \delta\v{k}^2(\delta
l)\right\rangle\equiv 2\hat{q}. \label{momentumchange_gluon_equation}
\end{align}
where we defined the transport parameter of the medium $\hat{q}$. The presence
of the long tail correction $\eta(\omega)$ in $\left\langle
\delta\v{k}^2(\delta l)\right\rangle$ at
\eqref{momentumchange_gluon_singlecollision} makes $\hat{q}$ slowly dependent
on the gluon energy. An accurate form of $\eta(\omega)$ under gluon
bremsstrahlung would require the effect of the convolution of the elastic
distributions \eqref{phi_gbarg} with the $\v{\delta}_k$ currents
\eqref{qcd_current}. Using our previous QED estimates, in a single collision
the functions $\v{\delta}_k$ substantially reduce the gluon maximal momentum
change \eqref{momentumchange_gluon_singlecollision} from the kinematical limit
$2\beta_k\omega$ to 2-3 times the gluon mass $m_g$ \cite{bethe1934}. For a
multiple collision scenario the LPM effect modulating the intensity
\eqref{central_equation_qcd} has to be taken into account in order to obtain
an accurate $\omega$ dependence of the medium transport parameter $\hat{q}$ as
measured through gluon bremsstrahlung \cite{zakharov1996a}.
\begin{figure}
\centering
\includegraphics[width=0.92\textwidth]{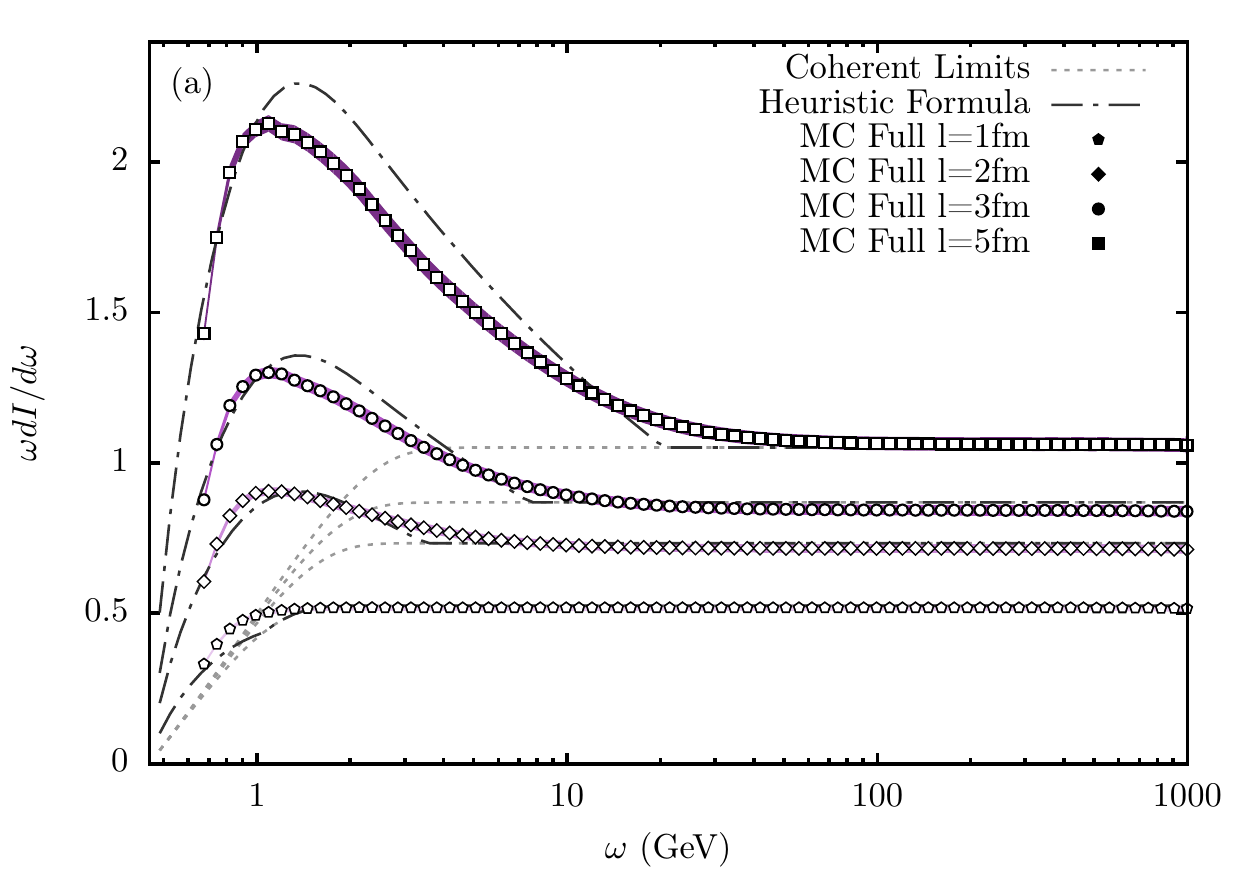}
\includegraphics[width=0.92\textwidth]{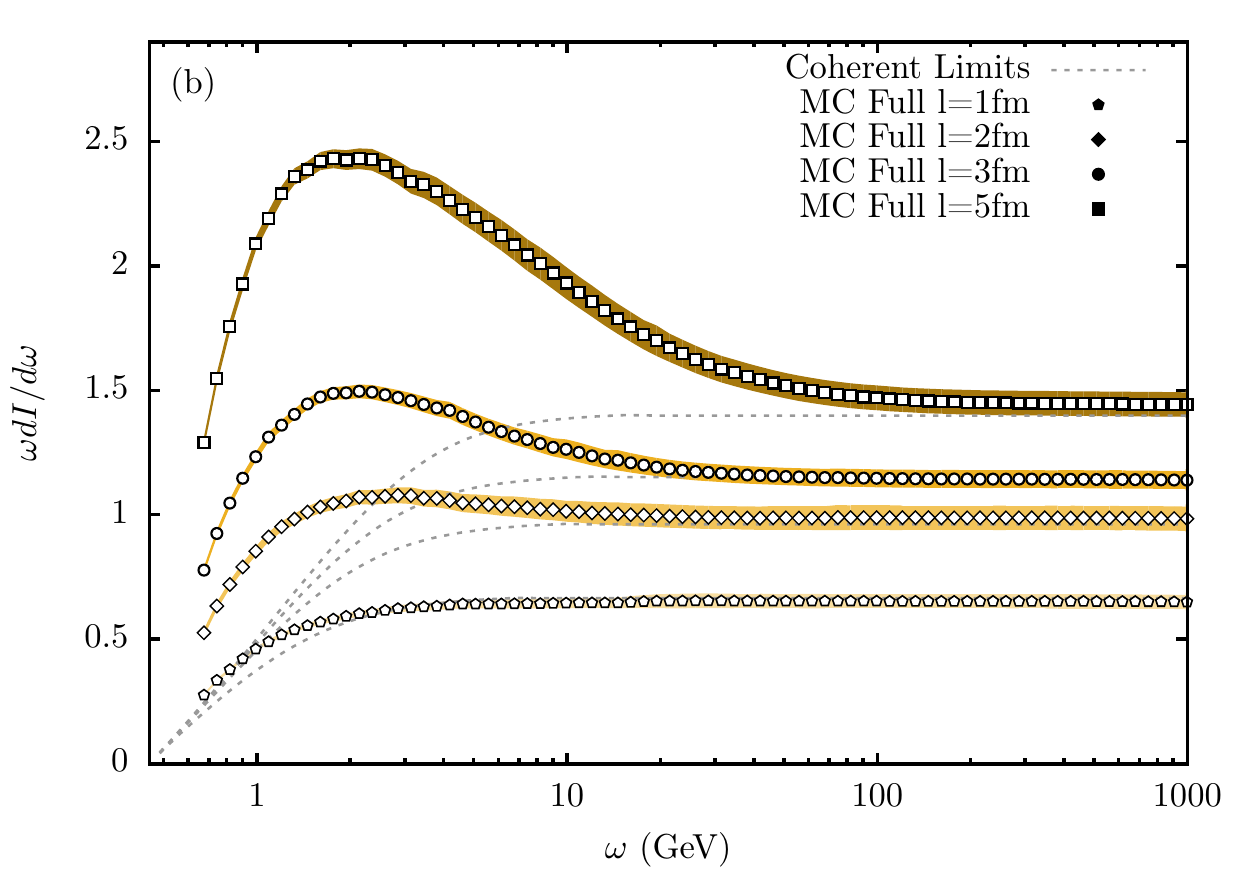}
\caption{Intensity of gluons as a function of the gluon energy
  \cite{feal2018a} emitted after traversing a medium of density $n_0T_f$ = 8
  fm$^{-3}$ corresponding to a transport parameter $\hat{q}$ = 0.98
  GeV$^2$/fm, for a gluon mass of $m_g$ = 0.45 GeV and a Debye screening mass
  of $\mu_d$ = 0.45 GeV. Coupling is set to $\alpha_s$ = 0.5. Numerical
  evaluation of \eqref{central_equation_qcd} is shown for Fokker-Planck
  approximation (a) and the Debye interaction (b) for several medium lengths,
  $l$ = 1 fm (pentagons), $l$ = 2 fm (diamonds), $l$ = 3 fm (circles) and $l$
  = 5 fm (squares). Also shown is the coherent plateau $i_0(l)$ for both
  interactions (dashed lines) and the heuristic formula \eqref{heuristic_qcd}
  for the Fokker-Planck approximation (dot-dashed line).}
\label{fig:figure_5_3}
\end{figure}

Equation \eqref{central_equation_qcd} is suitable for a numerical evaluation
under a general interaction. A Monte Carlo code has been built where the
integration in the gluon internal momenta at \eqref{central_equation_qcd} is
performed as a sum over discretized gluon paths. The gluon paths are generated
as zig-zag trajectories satisfying the effective elastic distributions
\eqref{phi_gbarg}, with a step size $\delta z$. The quark is allowed to suffer
medium interactions, but can be equally considered frozen for $\omega \ll
p_0^0$. The phases and the gluon trajectory are calculated with the exact
kinematical constraints, so that energy is always conserved in the
collisions. This property becomes relevant for gluon energies $\omega^2$
comparable to the final averaged squared momentum change $2\hat{q}l$. In a
typical run the step size is chosen as $0.01\lambda_{g\bar{g}}$ in such a way
that the largest medium $l\sim 5$ produces $\gtrsim$ 10$^4$ steps. An array of
$\sim 100$ gluon frequencies and $\sim 800$ final angles were computed for
$10^4$ discretized paths in order to guarantee that the uncertainty in all the
cases falls below the 10$\%$.

A second approach to qualitatively understand the behavior of intensity
\eqref{central_equation_qcd} would consist in an analogy with the QED
case. The classical behavior discussed at Section \ref{sec:section_3_1}
suggests interpreting intensity \eqref{central_equation_qcd} as a sum of $n$
single Bethe-Heitler amplitudes $\v{\delta}_k$ \eqref{qcd_current} of a gluon
being emitted at the point $z_k$, carrying a relative phase
\eqref{softgluon_phaseapproximation} producing interferences in the squared
emission amplitude. Each gluon appears twice, namely before or after the k-th
collision. For mediums of vanishing length gluon momentum homogeneity produces
a cancellation of the sum at \eqref{central_equation_qcd} and intensity
vanishes. For mediums of small length, satisfying $l\lesssim
\lambda_{g\bar{g}}$, the internal sum cancels and we are left with the first
and last gluons. By expanding \eqref{central_equation_qcd} in opacity for a
single layer the massive Bertsch-Gunion formula is recovered. For mediums of
larger length, compromising several collisions on average, the internal sum in
\eqref{central_equation_qcd} becomes relevant and is modulated by the
phase. In that case we can pair the sum of the single elements $\v{\delta}_k$
in groups separated by a maximum distance $\delta l=z_j-z_m$ such that their
relative phase, under the elastic distributions \eqref{phi_gbarg}, acquires a
value of the order of unity. This condition reads, using
\eqref{momentumchange_gluon_equation},
\begin{align}
\varphi^{j}_m\equiv \int^{z_j}_{z_m}dz\frac{k_\mu(z)p^\mu(0)}{p_0^0-\omega} \simeq \frac{m_g^2\delta
l}{(1+\beta_k)\omega}+\frac{\hat{q}(\delta l)^2}{2\beta_k\omega}=1.\label{coherence_lenght_1_qcd}
\end{align}
In the distance $\delta l$ the partial internal sum between $z_m$ and $z_j$ at
\eqref{central_equation_qcd} of the single amplitudes $\v{\delta}_k^n$
cancels, since their relative phase is negligible using the condition
\eqref{coherence_lenght_1_qcd}, and the first and last gluons of the group are
the only ones surviving,
\begin{align}
\sum_{k=m}^j\v{\delta}_ke^{i\varphi_k}=e^{i\varphi_m}\sum_{k=m}^j\v{\delta}_ke^{i\varphi_m^k}\simeq
e^{i\varphi_m}\sum_{k=m}^j\v{\delta}_k = e^{i\varphi_m}
\left(\frac{\v{k}_j\times\v{p}_0}{k_\mu^jp^\mu_0}-\frac{\v{k}_{k-1}\times\v{p}_{0}}{k_\mu^{k-1}p^\mu_{0}}\right),\label{auxiliary_equation_coherence}
\end{align}
where we used \eqref{qcd_current} and \eqref{coherence_lenght_1_qcd}. Then the elements in $\delta l$ coherently act
between themselves but they incoherently interfere with any other group. This
defines a coherence length, modulated by $\omega$, which by using
\eqref{coherence_lenght_1_qcd} is given by
\begin{align}
\delta l(\omega)\equiv
\frac{m_g^2}{2\hat{q}}\left(\sqrt{1+\frac{8\hat{q}\omega}{m_g^4}} -1\right),
\label{coherence_lenght_2_qcd}
\end{align}
This interference is characterized by the frequency $\omega_c$ at which the
coherence length acquires the maximum value $l$, given then by $\omega_c\simeq
(\hat{q}l+m_g^2)l$, and the characteristic frequency $\omega_s$ at which the
coherence length acquires the minimum value $\lambda_{g\bar{g}}$ required to
produce bremsstrahlung, given then by $\omega_s=m_g^4/\hat{q}$, where
$\mu_d\simeq m_g$ was assumed. Since for $\delta l(\omega)> l$ there are not
centers producing bremsstrahlung we further impose to
\eqref{coherence_lenght_2_qcd} the condition $\delta l(\omega)=l$ for $\omega>
\omega_c$. Following \eqref{auxiliary_equation_coherence} in a coherence
length the gluon is not able to resolve the scattering structure and the
matter in $\delta l(\omega)$ acts like a single and independent scatterer with
equivalent charge $n_0\delta l(\omega)$ for the distribution
\eqref{phi_gbarg}. Since in a medium of length $l$ there are $l/\delta
l(\omega)$ of these units, we write the approximated formula for the intensity
after traversing a distance $l$
\begin{align}
\omega \frac{dI(l)}{d\omega} = \frac{l}{\delta
  l(\omega)}\frac{g_s^2C_f}{(2\pi)^2}\int d\Omega_k\int\frac{d^3\v{k}_0}{(2\pi)^3}
\v{\delta}_1^2\phi_{g\bar{g}}^{(n)}(\delta\v{k}_1,\delta l(\omega)).
\end{align}
By inserting \eqref{phi_gbarg} in the above equation and integrating in the
final solid angle we get
\begin{align}
\omega \frac{dI(l)}{d\omega} =\frac{\alpha_sC_f}{\pi^2}
\frac{l}{\delta l(\omega)} \int_0^\pi
d\theta \sin(\theta) F(\theta)\Sigma_2(\delta\v{k}_1,\delta
l(\omega))\label{heuristic_qcd},
\end{align}
where the momentum change in the coherence length is given by
$|\delta\v{k}|=2\beta_k\omega\sin\theta$ and the function arising in the
angular integration is given by
\begin{align}
F(\theta)=\bigg[&\frac{1-\beta_k^2\cos\theta}{2\beta_k\sin(\theta/2)
\sqrt{1-\beta_k^2\cos^2(\theta/2)}} \nonumber \\
&\times\log\bigg[\frac{\sqrt{1-\beta_k^2
\cos^2(\theta/2)}+\beta_k\sin(\theta/2)}{\sqrt{1-\beta_k^2
\cos^2(\theta/2)}-\beta_k\sin(\theta/2)}\bigg]-1\bigg].
\end{align}
Equation \eqref{heuristic_qcd} becomes exact in the frequency interval
$\omega\ll \omega_s$ in which the gluon is able to resolve each of the
internal scatterings, which is the totally incoherent superposition of $\eta$,
the average number of collisions, single Bertsch-Gunion intensities, and in
the interval $\omega \gg \omega_c$ in which the gluon is not able to resolve
any of the internal scatterings, which is nothing but Weinberg's soft photon
theorem \cite{weinberg1965} assuming always that $p_0^0\gg\omega$ in all
cases. For $\omega\gg \omega_c$ the phase \eqref{softgluon_phaseapproximation}
vanishes and the medium acts like a single scatterer with an equivalent charge
the amount of matter contained in $l$, following a Bethe-Heitler law
$1/\omega$. The radiation in this interval grows with the quotient between the
gluon squared momentum change in $l$ and its squared mass. This medium
dependent term is a coherent plateau and dominates the radiation and the quark
energy loss. The above qualitative behavior can be made quantitative for the
Fokker-Planck approximation for \eqref{sigma_gbarg_qcd}. In that case
\eqref{heuristic_qcd} for $\omega\gg \omega_c$ has the asymptotic expressions
\begin{equation}
\lim_{l\to 0}\omega \frac{dI(l)}{d\omega} =
\frac{2}{3\pi}\alpha_sC_f\frac{2\hat{q}\delta l}{m_g^2}, \; \; \; \; \; \;
\lim_{l\to \infty}\omega \frac{dI(l)}{d\omega}=
\frac{2}{\pi}\alpha_sC_f\Bigg(\log\Bigg(\frac{2\hat{q}\delta l}{m_g^2}\Bigg)
-(\gamma+1) \Bigg).
\label{coherent_plateau_qcd_asympt}
\end{equation}
An expression interpolating between these two limits can be found
\begin{align}
i_0(l)\equiv \omega \frac{dI}{d\omega} = \frac{2}{\pi}\alpha_sC_f\frac{1+\eta}{3A+\eta}\log(1+A\eta),\label{coherent_plateau_qcd_fokker}
\end{align}
where $A=\exp(-1-\gamma)$, $\gamma$ is Euler's constant and
$\eta=2\hat{q}l/m_g^2$ is the average number of collisions in $l$. For
frequencies below $\omega_c$ the gluon starts to resolve the internal
structure of the scattering and radiation decouples into $l/\delta l(\omega)$
elements of equivalent charge the amount of matter in $\delta l(\omega)$. Then
in this regime intensity, using \eqref{coherence_lenght_2_qcd}, grows as
$l/\delta l(\omega)\simeq \sqrt{\hat{q}/\omega}$ times a Bethe-Heitler law
$1/\omega$ with a slow logarithmic charge decrease $\log(\delta
l\omega)\sim\log(\omega/\hat{q})$. This enhancement stops at $\omega_s$, where
$\delta l(\omega)$ acquires the minimum average value $\lambda_{g\bar{g}}$ to
produce a collision and thus radiation. This would consists in the totally
incoherent sum of the $l/\lambda_{g\bar{g}}$ single emitters, but suppression
due to a vanishing velocity $\beta_k$ quickly cancels this enhancement.

In Figure \ref{fig:figure_5_3} we show the angular integrated intensity of
gluons of $m_g$ = 0.45 GeV emitted from a high energy quark after performing a
multiple collision with a medium with screening $\mu_d$ = 0.45 GeV and density
$n_0T_f$ = 8 fm$^{-3}$, which corresponds to a transport parameter $\hat{q}$ =
0.98 GeV$^2$/fm. From here onwards we choose $\alpha_s $ = 0.5. The numerical
evaluation of \eqref{central_equation_qcd} is shown for the Debye interaction
\eqref{fielddefinition} and its Fokker-Planck approximation. Also shown are
the coherent plateaus $i_0(l)$ for both cases and the heuristic formula
\eqref{heuristic_qcd} in the Fokker-Planck approximation. As we can see the
coherent plateaus of the numerical evaluations match the analytical expression
\eqref{heuristic_qcd} for both interactions. The heuristic formula provides a
reasonable approximation to the LPM effect in the rest of the range. The
results are shown for medium lengths of $l$ = 1,2,3 and 5 fm, corresponding to
$\eta \sim$ 10,20,30 and 50 collisions and characteristic frequencies
$\omega_c \simeq$ 1, 4, 9 and 25 GeV, respectively. We see that although the
shortest medium compromises 10 collisions the LPM effect is negligible since
$\omega_c$ falls in the mass suppression zone. The enhancement from the
coherent plateau starts to be noticeable for larger lengths.

\begin{figure}
\centering
\includegraphics[width=0.92\textwidth]{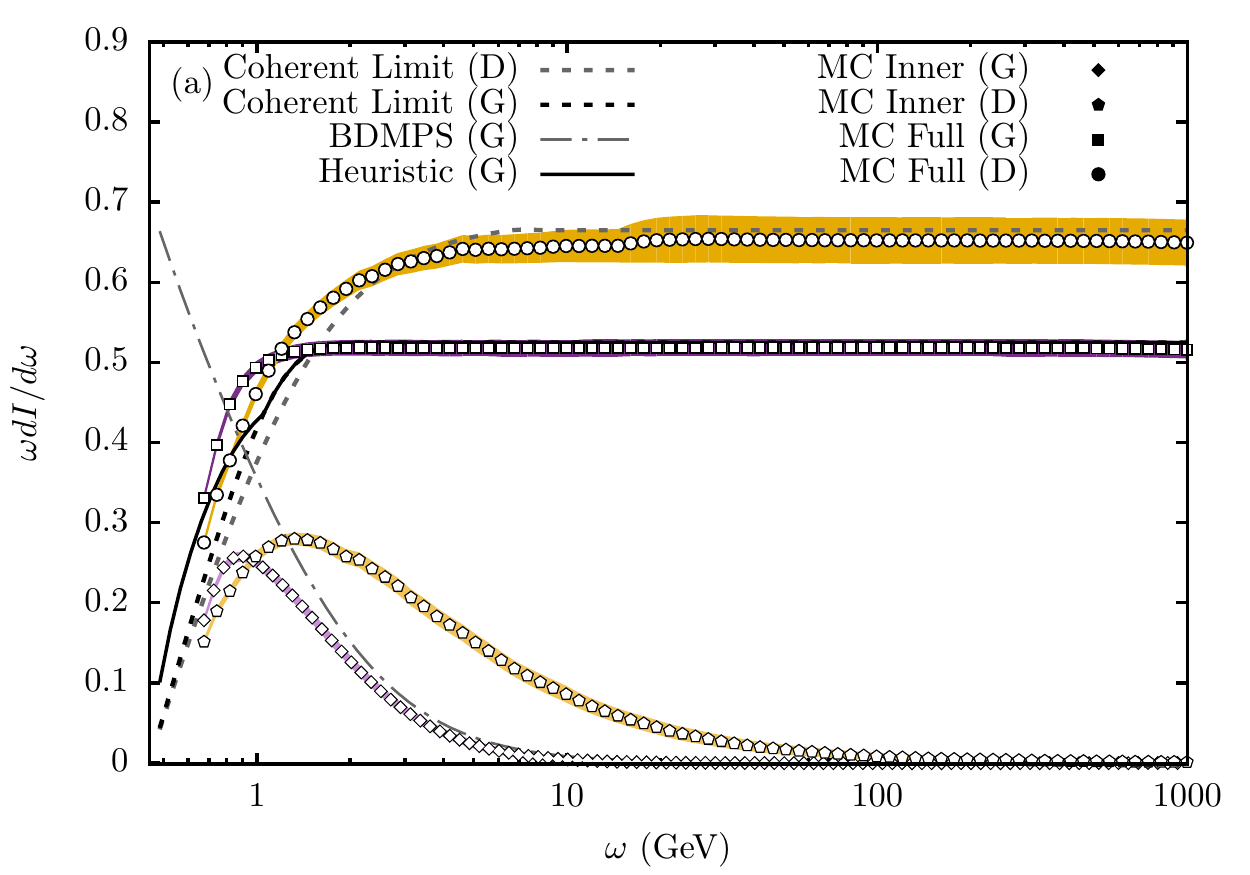}
\includegraphics[width=0.92\textwidth]{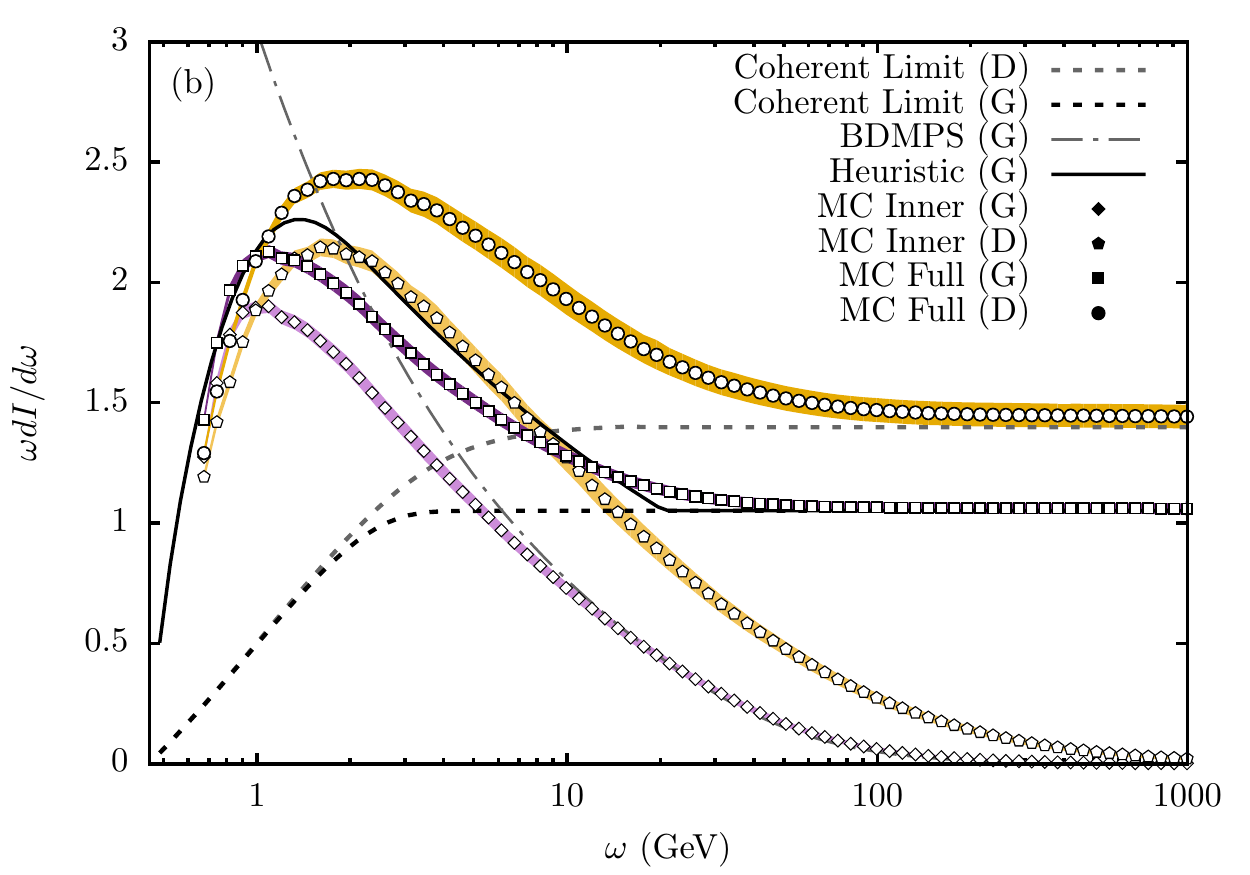}
\caption{Intensity of gluons of $m_g$ = 0.45 GeV as a function of the gluon
  energy \cite{feal2018a} for a medium of length $l$ = 1 fm (a) and $l$ = 5 fm
  (b) of density $n_0T_f$ = 8 fm$^{-3}$ corresponding to a transport parameter
  $\hat{q}$ = 0.98 GeV$^2$/fm, and medium screening mass $\mu_d$ = 0.45
  GeV. Results are shown for the Monte Carlo evaluation of
  \eqref{central_equation_qcd} both for the Debye interaction (circles) and
  the Fokker-Planck approximation (squares). Also shown is our heuristic
  formula \eqref{heuristic_formula} (solid black line) and the coherent
  plateau of the Debye interaction (grey dashed line) and of the Fokker-Planck
  approximation (black dashed line). BDMPS result is also shown (dot-dashed
  line) and the evaluation of \eqref{central_equation_qcd} only with the inner
  terms in the Debye (pentagons) and the Fokker-Planck (diamonds)
  interactions.}
\label{fig:figure_5_4}
\end{figure}

\begin{figure}
\centering
\includegraphics[width=0.92\textwidth]{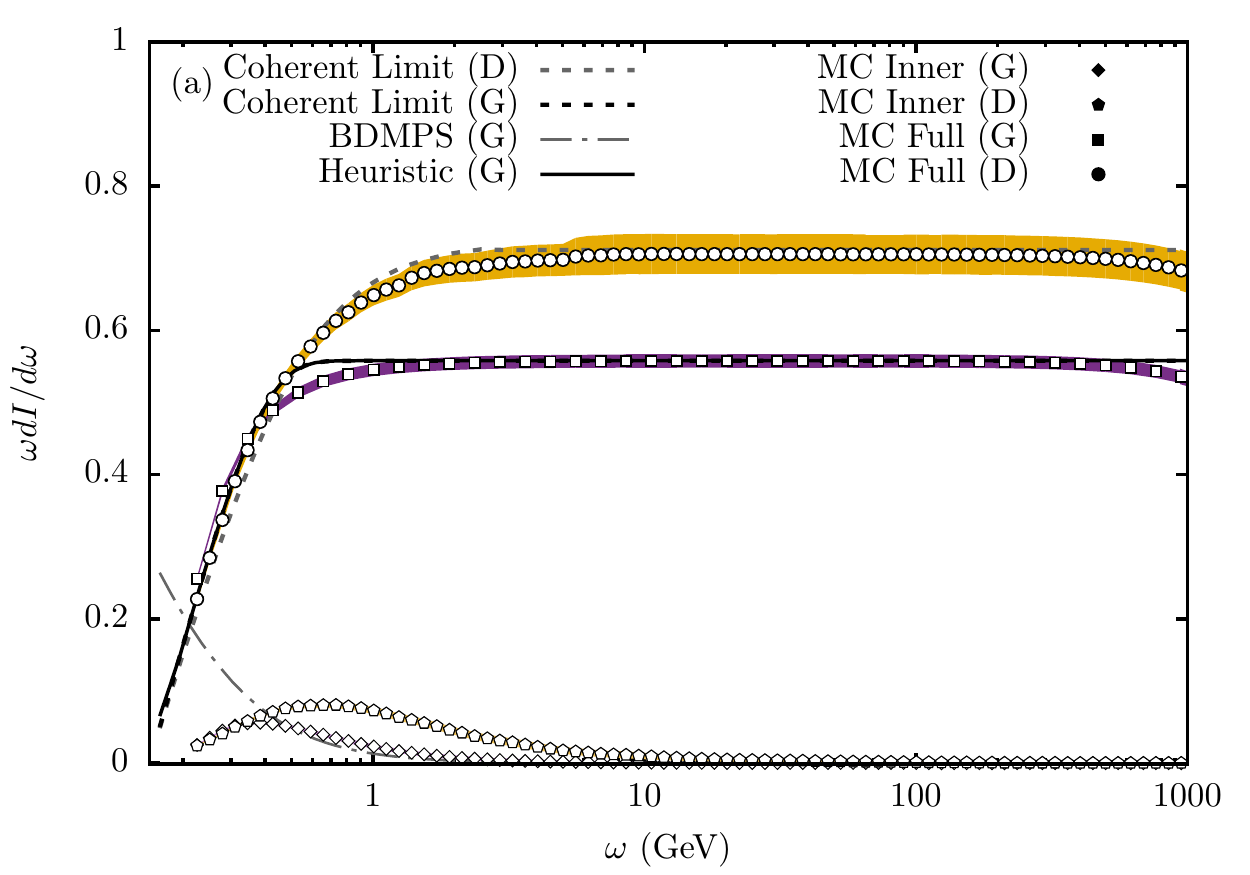}
\includegraphics[width=0.92\textwidth]{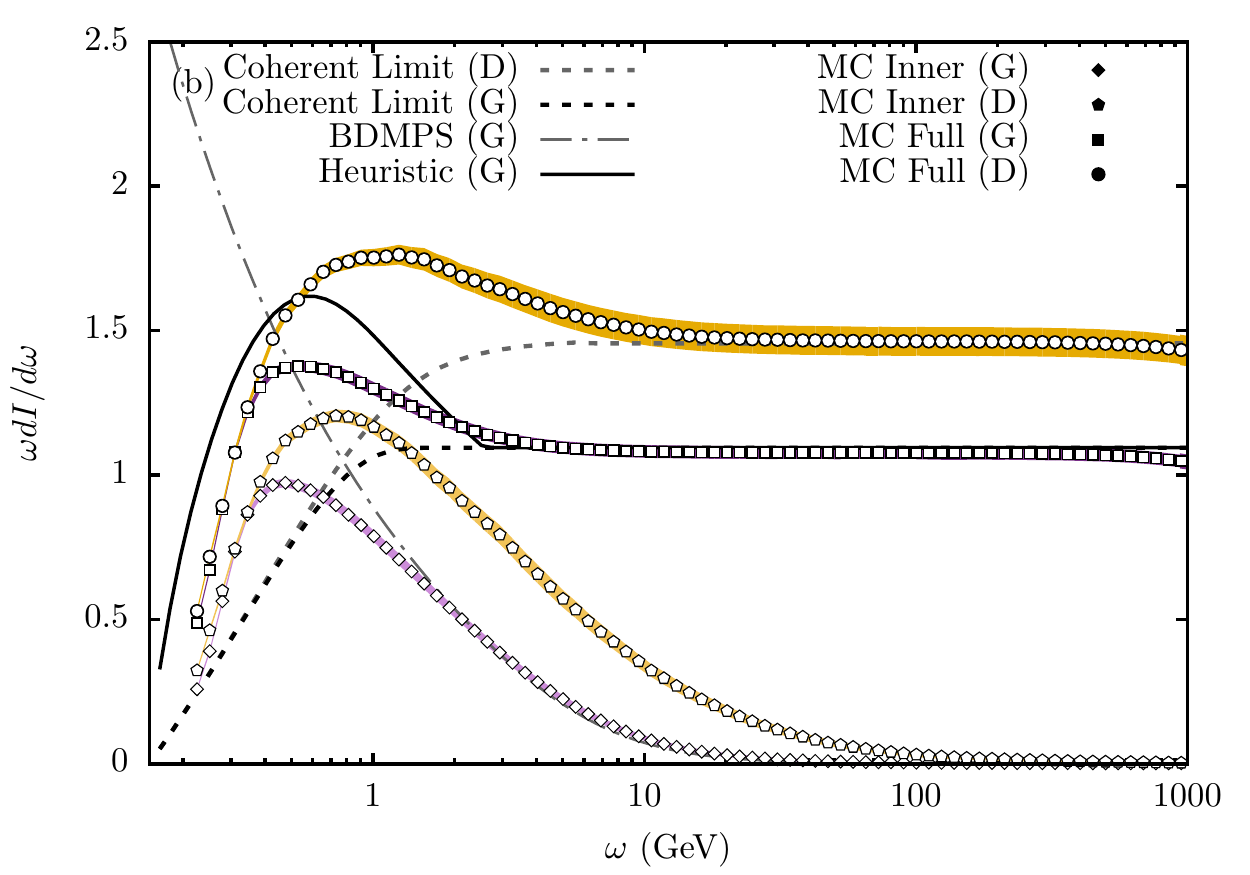}
\caption{Intensity of gluons of $m_g$ = 0.15 GeV as a function of the gluon
  energy \cite{feal2018a} for a medium of length $l$ = 1 fm (a) and $l$ = 5 fm
  (b) of density $n_0T_f$ = 1 fm$^{-3}$ corresponding to a transport
  parameter $\hat{q}$ = 0.12 GeV$^2$/fm, and a medium screening mass of
  $\mu_d$ = 0.15 GeV. Results are shown for the Monte Carlo evaluation of
  \eqref{central_equation_qcd} both for the Debye interaction (circles) and
  the Fokker-Planck approximation (squares). Also shown is our heuristic
  formula \eqref{heuristic_formula} (solid black line) and the coherent
  plateau of the Debye interaction (grey dashed line) and of the Fokker-Planck
  approximation (black dashed line). BDMPS result is also shown (dot-dashed
  line) and the evaluation of \eqref{central_equation_qcd} only with the inner
  terms in the Debye (pentagons) and the Fokker-Planck (diamonds)
  interactions.}
\label{fig:figure_5_5}
\end{figure}

In Figures \ref{fig:figure_5_4} and \ref{fig:figure_5_5} we show the results
for two medium lengths $l$ =1 and 5 fm, two medium densities $n_0T_f$ = 1 and
8 fm$^{-3}$, and two gluon masses $m_g$ = 0.15 and 0.45 GeV, for the Debye
interaction and its Fokker-Planck approximation. We see that for the same
parameters the Debye interaction produces more radiation than its
Fokker-Planck approximation. This disagreement can be cast into a redefinition
of $\hat{q}$ at the cost of making it dependent on the length $l$ and the
screening mass $\mu_d$ of the medium. We also notice that the enhancement zone
is wider in the Debye interaction than in the Fokker-Planck approximation,
starting for larger $\omega_c$. This difference can be understood as the
effect of the long-tail of the Debye interaction on $\hat{q}$ and thus on
$\omega_c$.

\begin{figure}
\centering
\includegraphics[width=0.92\textwidth]{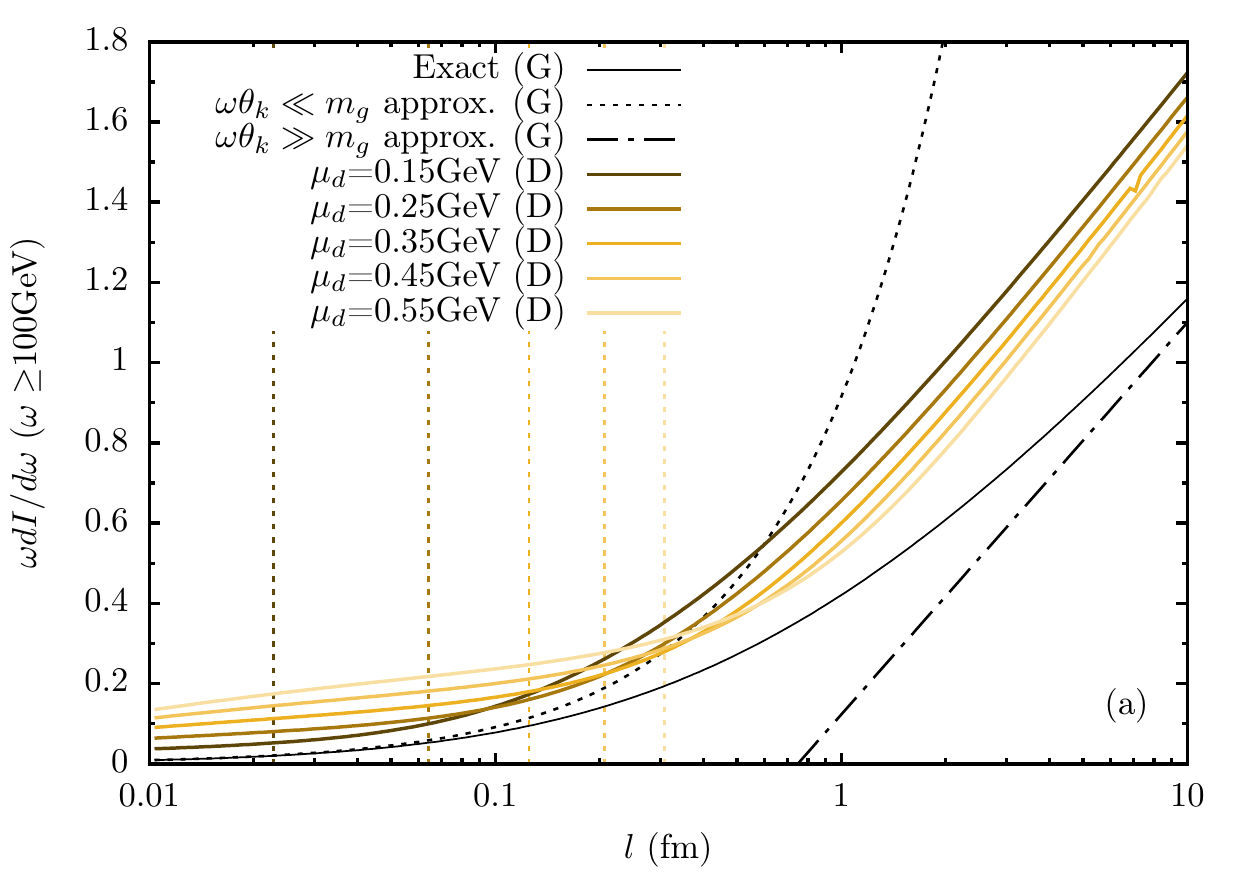}
\includegraphics[width=0.92\textwidth]{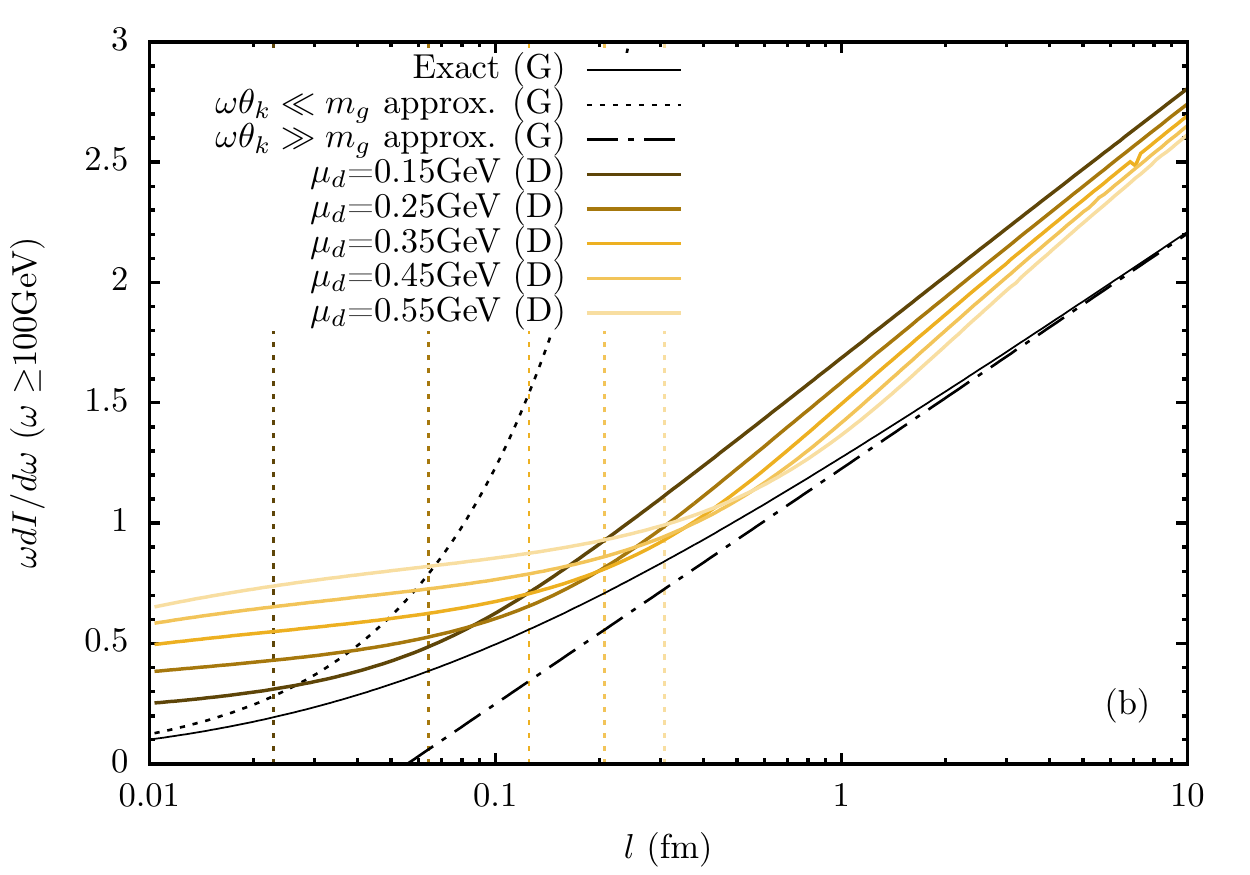}
\caption{Asymptotic emission intensity $i_0(l)$ at as a function of the medium
  length \cite{feal2018a} for a medium of density $n_0T_f$ = 8 fm$^{-3}$
  corresponding to a transport parameter of $\hat{q}$ = 0.98 GeV$^2$/fm
  ($\alpha_s$ = 0.5). Results are shown (a) for a gluon mass of $m_g$ = 0.55
  GeV and $(b)$ $m_g$ = 0.15 GeV. Debye interaction is shown in solid lines
  for different screening masses, as marked, and the Fokker-Planck
  approximation is shown in solid black line. Also shown are the small and
  large medium approximation of the Fokker-Planck approximation. Vertical
  dashed lines mark the transition between large and small media for each
  screening mass $\eta$ = 1.}
\label{fig:figure_5_6}
\end{figure}

In Figure \ref{fig:figure_5_6} we show the asymptotic emission intensity
$i_0(l)$ for large $\omega$, or coherent plateau, as a function of the medium
length for two gluon masses $m_g$ = 0.15 and 0.55 GeV and several medium
screening masses $\mu_d$, both for the Debye interaction and its Fokker-Planck
approximation keeping the transport parameter $\hat{q}$ constant. As we can
see the ratio of the two asymptotic intensities in the Debye interaction and
its Fokker-Planck approximation is not constant so that one cannot recast the
difference into a redefinition of $\hat{q}$ independently of the medium
properties.

Finally, a third approach for evaluating \eqref{central_equation_qcd} would
consist in taking the $\delta z\to 0$. In that case a Boltzmann transport
equation can be found \cite{migdal1956,bell1958,baier1996} or an equivalent path
integral in transverse coordinates
\cite{zakharov1996a,wiedemann1999,wiedemann2000a}. In both cases
the Fokker-Planck approximation produces a differential equation or a
quadratic path integral, respectively, which are solvable. While this method
leads to reasonably simple results, we have already shown that for finite
size targets the differences between the Fokker-Planck approximation and the
Debye interaction cannot be recast into a simple redefinition of the transport
parameter $\hat{q}$ independently of the medium properties and the gluon
energy. In QED the equivalence between the $\delta z\to 0$ limit of the
intensity \eqref{central_equation_qed} and the path integral formalism has
been proven at Section \ref{sec:section_3_4}. We notice that QCD intensity
\eqref{central_equation_qcd} is equivalent to the QED intensity
\eqref{central_equation_qed} if the electron role is played by a gluon moving
in the opposite direction. Indeed by comparing the soft gluon QCD phase
\eqref{softgluon_phaseapproximation} with the QED phase
\eqref{phase_difference} we have to replace $\omega/(p_0^0)^2\to 1/\omega$ so
that
\begin{align}
\medspace
\frac{\omega}{2p_0^0(p_0^0-\omega)}m_e^2\to
\frac{m_g^2}{2\omega}, \medspace\medspace\medspace\medspace\medspace
\Omega=\frac{1-i}{\sqrt{2}}\sqrt{\frac{\hat{q}\omega}{(p_0^0)^2}}\to \Omega_g=\frac{1-i}{\sqrt{2}}\sqrt{\frac{\hat{q}}{\omega}}.\label{phase_replacement}
\end{align}
Similarly the coupling of the gluon to the external field and the coupling of
the squared emission vertex has to be replaced, respectively, as
\begin{align}
(Ze^2)^2\to g_s^4T_f ,\medspace\medspace\medspace e^2\to g_s^2C_f.\label{coupling_replacement}
\end{align}
Notice that the final passage of the gluon from $z_j$ to $l$ is affected by
collisions, while the electron final momentum has been integrated out and is
the initial passage from $z_1$ to $z_k$ the one affected by
collisions. Correspondingly, the integral in space has to be reversed so that
\begin{align}
z=\infty\to z=-\infty, \medspace\medspace\medspace z=0\to
z=l \medspace\medspace\medspace z=l\to z=0 .\label{integration_replacement}
\end{align}
With these replacements the QCD equivalents of
\eqref{central_equation_pathintegral_qed} are straightforward to obtain. We
give only here the required expressions for the semi-infinite length
approximations of Migdal/Zakharov \cite{migdal1956,zakharov1996a} and the
BDMPS group \cite{baier1996} for the angle integrated intensity. Using the
replacement \eqref{integration_replacement} the Migdal/Zakharov approximation
\eqref{migdal_qed_definition} transforms to
\begin{align}
\omega \frac{dI^{(n)}_{inc}}{d\omega d\Omega_k}\equiv \int^{l}_0\int^{l}_0+\int_{l}^\infty\int_{l}^\infty=\omega \frac{dI^{(n)}_d}{d\omega d\Omega_k}+\omega \frac{dI^{(n)}_b}{d\omega d\Omega_k}.\label{migdal_qcd_definition}
\end{align}
The final result of integrating in angle and longitudinal positions
\eqref{migdal_qcd_definition} can be directly written using
\eqref{migdal_prediction} and prescriptions \eqref{phase_replacement} and
\eqref{coupling_replacement}. We obtain in the $l\to\infty$ limit
\begin{align}
 \omega \frac{dI^{(n)}_d}{d\omega}+\omega \frac{dI^{(n)}_b}{d\omega}=l
\frac{2}{\pi}g_s^2C_f&\frac{|\Omega|_g}{\sqrt{2}}\int_0^\infty dz
\exp\left(-\frac{z}{\sqrt{2}s}\right)\label{migdal_qcd_result}\\
&\times\left(\sin\left(\frac{z}{\sqrt{2}s}\right)+\cos\left(\frac{z}{\sqrt{2}{s}}\right)\right)\left(\frac{1}{z^2}-\frac{1}{\sinh^2(z)}\right),\nonumber
\end{align}
where the parameter $s$ is given for QCD by $s=2\omega\Omega_g/m_g^2$, with
$\Omega_g$ given at \eqref{phase_replacement}. Similarly using the
prescription \eqref{integration_replacement} the BDMPS approximation
\eqref{bdmps_qed_definition} transforms to
\begin{align}
\omega \frac{dI^{(n)}_{inc}}{d\omega d\Omega_k}\equiv\left|\int_{0}^\infty\right|^2-\left|\int_{l}^\infty\right|^2=\int^l_0\int^l_0+2\Re\int_{-\infty}^0\int^l_0
\equiv \omega \frac{dI^{(n)}_d}{d\omega d\Omega_k}+\omega \frac{dI^{(n)}_e}{d\omega d\Omega_k}.\label{bdmps_qcd_definition}
\end{align}
The final result of integrating in angle and longitudinal positions can be
directly written using \eqref{bdmps_qed_result} and prescriptions
\eqref{phase_replacement} and \eqref{coupling_replacement}, obtaining in the
$m_g=0$ case the BDMPS result
\begin{align}
\omega \frac{dI^{(n)}_d}{d\omega}+\omega
\frac{dI^{(n)}_e}{d\omega}=\frac{2g_s^2C_f}{\pi}\Re\left[\log\left(\cos\bigg(\Omega_g
L\bigg)\right)\right]\label{bdmps_qcd_result}
\end{align}
with $\Omega_g$ as before given at \eqref{phase_replacement}. 
\begin{figure}
\centering
\includegraphics[width=0.8\textwidth]{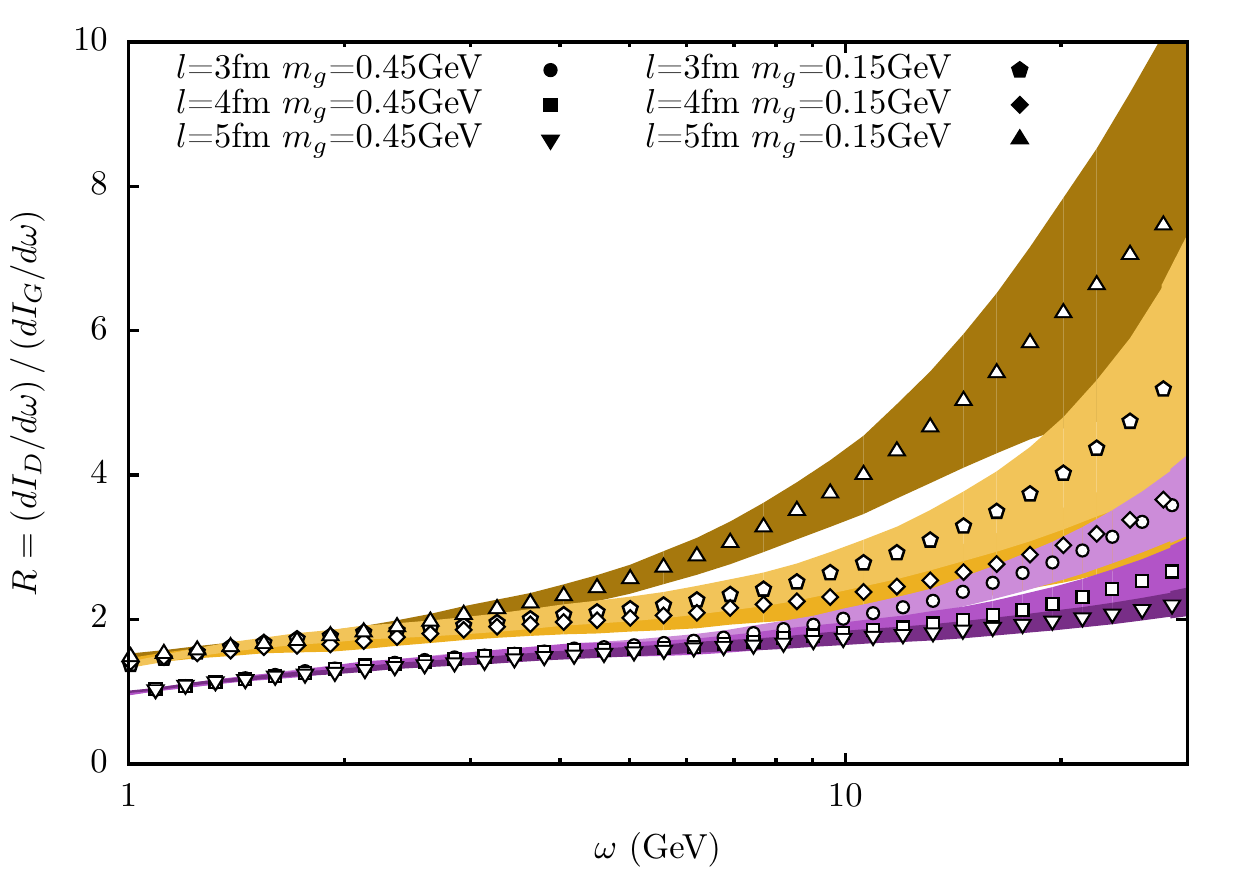}
\caption{Ratio between the inner terms corresponding to the BDMPS prescription
  of \eqref{central_equation_qed} for the Debye interaction and its
  Fokker-Planck approximation for several lengths $l$ and gluon masses $m_g$,
  as marked. Results are shown for a medium of $n_oT_f$ = 8 fm$^{-3}$
  corresponding to a transport parameter $\hat{q}$ = 0.98 GeV$^2$/fm. Debye
  screening is set to $\mu_d$ = $m_g$ and coupling $\alpha_s$ = 0.5. }
\label{fig:figure_5_11}
\end{figure}
In Figures \ref{fig:figure_5_4} and \ref{fig:figure_5_5} the BDMPS result
\eqref{bdmps_qcd_result} is shown together with our evaluation of the
intensity \eqref{central_equation_qcd} including only the inner gluons
indicated by the BDMPS definition \eqref{bdmps_qcd_definition}. Neither the
BDMPS subset for \eqref{central_equation_qcd} nor the BDMPS result itself
\eqref{bdmps_qcd_result} consider the emission in the coherent limit since the
gluon emitted in the original quark direction and the last gluon are not
included. Correspondingly, this prescription has to be considered only as an
approximation for the radiation scenario in the transverse direction after a
hard collision. Our results match the BDMPS result at high energies for the
Fokker-Planck approximation, only with differences for the small $l$ or big
$m_g$ limits, as expected. We observe that at low $\omega$ the kinematical
restriction in the integration in $\v{k}_t$ and the suppression effect due to
the gluon mass make the intensity vanish. The Debye interaction still produces
more radiation compared to its Fokker-Planck approximation, having a much
slower fall off as $\sim 1/\omega$ for $\omega\gg \omega_c$ in contrast to the
$1/\omega^2$ behavior of the BDMPS result. This can be seen in Figure
\ref{fig:figure_5_11} where we show the ratio between the intensity for the
Debye interaction and its Fokker-Planck approximation for a medium of $n_0T_f$
= 8 fm$^3$ corresponding to a transport parameter of $\hat{q}$ = 0.98
GeV$^2$/fm and two gluon masses and Debye screenings $m_g$ = $\mu_d$ 0.15 and
0.45 GeV. The ratio is not constant and strongly depends on the mass and
energy of the gluon, going from $\sim $ 1 ($\sim $ 1.5) at low energies for
$m_g$ = 0.45 GeV ($m_g$ = 0.15 GeV) to $\sim$ 2.2 ($\sim$ 8) at larger
energies. A single change in the $\hat{q}$ parameter can not fit the Debye
results in all the range. A change of $\hat{q}\to$ 3.5$\hat{q}$ fits the
result for the case $n_0T_f$ = 8 fm$^{-3}$, a gluon/Debye mass of $\mu_d$ =
0.15 GeV and large lengths above $l$ = 3 fm, with an error of 20$\%$. For a
gluon/Debye mass of $\mu_d$ = 0.45 GeV the scale factor is 2.8 instead, with
an error of 40$\%$. This enhancement of the realistic spectrum by a factor 3-4
times larger than the well known Fokker-Planck approximations
\cite{baier1995,zakharov1996a,wiedemann1999,salgado2003a} for the same medium
characteristics may suggest that single gluon total cross sections
\cite{baier2006} much larger than the leading order expectations
$\sigma_{g\bar{g}}^{(1)}(\v{0})$ $\sim$ 1.5 mb ($\sim$ 4 mb) with $\alpha_s$ =
0.3 (0.5), supporting the idea of a strong coupled nature of the formed QGP,
may not be required to obtain the medium transport parameters $\hat{q}$
favored by the data \cite{gyulassy2004,denterria2009b}.

\subsubsection{Energy loss}
For a quark coming from the infinity and going to the infinity, in a multiple
soft collision scenario, energy loss is dominated by the large $\omega$
behavior of the intensity $i_0(l)$. This term comes from the emission
amplitude of the first and last gluons emitted in the medium and produces an
energy loss proportional to the initial quark energy, $p_0^0$. Multiple gluon
emission can be taken into account by assuming independent emissions
\cite{landau1944}. Let us define the probability density $f(l,\Delta)$ of
having a total energy loss $\Delta$ at a distance $l$. The increment in the
probability $f(l,\Delta)$ after traveling a distance $\delta l$ is fed with
the number of states having lost $\Delta-\omega$ at $l$ and emptied with the
number of states already with $\Delta$ at $l$, which leads to the transport
equation
\begin{equation}
\frac{\partial f(l,\Delta)}{\partial l} = \int_0^{p_0^0} d\omega
\frac{dI(l)}{d\omega dl} \big(f(l,\Delta-\omega)-f(l,\Delta)\big).\label{energyloss_transportequation}
\end{equation}
Equation \eqref{energyloss_transportequation} can be solved by defining the
Laplace transform of $f(l,\Delta)$, we call $\varphi(l,p)$, which produces the
transformed equation
\begin{align}
\int^{\infty}_0 f(l,\Delta)e^{-p\Delta}=\varphi(l,p) \to \frac{\partial
  \varphi(l,p)}{\partial l} = \varphi(l,p)\int^{p_0^0}_0 d\omega
\frac{dI(l)}{d\omega dl}(e^{-p\omega}-1\big),
\end{align}
which leads to the inverse Laplace transform 
\begin{equation}
f(l,\Delta)=\frac{1}{2\pi i} \int^{+i\infty+\sigma}_{-i\infty+\sigma}
dp \medspace e^{p\Delta} \exp\Bigg(-\int_0^{p_0^0}
  \frac{dI(l)}{d\omega}\Big(1-e^{-p\omega}\Big)\Bigg).\label{energyloss_probability}
\end{equation}
This equation can be simplified if we assume that the spectrum can be well
approximated as constant $i_0(l)$, in which case one finds the following
expression suitable for a numerical evaluation
\begin{equation}
f(x,l)=\frac{1}{\pi}\int^{\infty}_0 ds \; \exp \Big(-i_0(l) C(s)\Big) 
\; \; \cos \Big(xs-i_0(l) S(s) \Big),
\label{fraction_density}
\end{equation}
where $x=\Delta/p_0^0$ is the total fraction of energy loss, the sum of the
single energy losses $y=\omega/p_0^0$, and the functions
\begin{equation}
C(s) = \int^s_0 du \frac{1-\cos(u)}{u}, 
\; \; \; \; \; \; \; \; \; \; 
S(s) = \int^s_0 du \frac{\sin(u)}{u},
\nonumber
\end{equation}
are related to the sine and cosine integrals. For finite quark energies,
$p_0^0$, this probability can not be correct, since the hypothesis of
independent gluon emission is not valid for energy losses $\Delta \sim p_0^0$
and \eqref{energyloss_transportequation} does not hold. On the other hand, for
small energy losses the approximation should be valid. Therefore, one can take
the probability distribution \eqref{fraction_density} and normalize it up to
$x=1$. This approximation, however, underestimates the energy loss, specially
large energy loses are strongly underestimated. In order to circumvent the
problem a different approach would be to simulate a Monte Carlo code which
exactly takes into account energy conservation at each step. A Poisson process
can be generated in this way which takes as input the single gluon intensity
given at \eqref{central_equation_qcd}.
\begin{figure}
\centering
\includegraphics[width=0.9\textwidth]{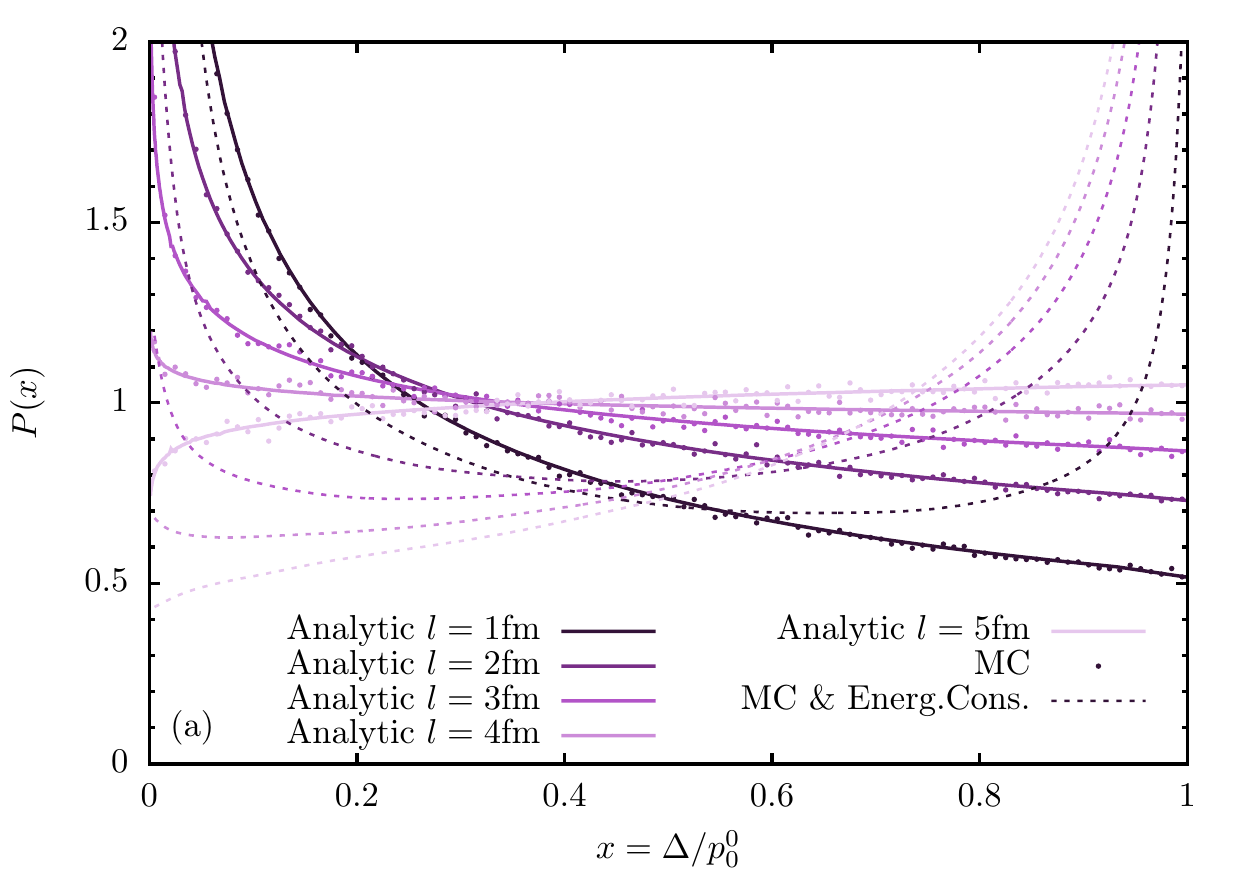}
\includegraphics[width=0.9\textwidth]{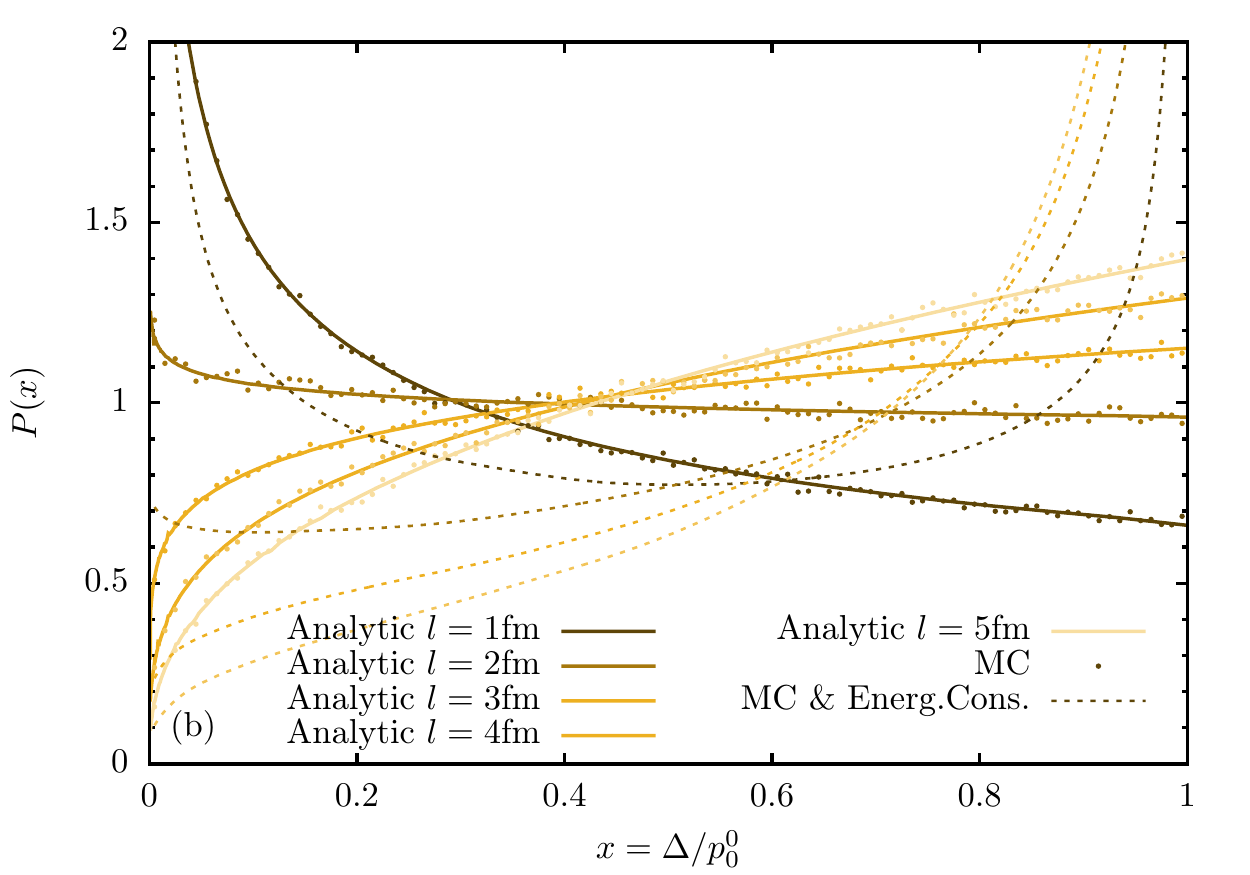}
\caption{Probability of a fraction of energy loss $x\equiv \Delta/p_0^0$ for
  several medium lengths assuming the coherent plateau spectrum
  $i_0(l)$. Results are shown for a medium of density $n_0T_f$ = 8 fm$^{-3}$,
  corresponding to a transport parameter of $\hat{q}$ = 0.98 GeV$^2$/fm, a
  gluon mass of $m_g$ = 0.45 GeV and a screening of $\mu_d$ = 0.45
  GeV. Coupling is set to $\alpha_s$ =0.5. The Fokker-Planck approximation (a)
  and the Debye interaction (b) are shown for several medium lengths, as
  marked. Evaluation of \eqref{fraction_density} is shown in solid line, while
  the Monte Carlo evaluation of the Poisson process is shown assuming
  independent emissions (dots) or assuming dependent emissions/energy
  conservation (dashed lines).}
\label{fig:figure_5_8}
\end{figure}

In Figure \ref{fig:figure_5_8} we show the probability of losing a fraction of energy
$x=\Delta/p_0^0$ assuming a constant spectrum given by the coherent plateau
$i_0(l)$ for several medium lengths under the Debye interaction and its
Fokker-Planck approximation. The evaluation of \eqref{fraction_density}
assuming single independent emissions is shown together with the Monte Carlo
evaluation of the Poisson process assuming energy conservation. In order to
check the Monte Carlo evaluation the single independent emission has also been
implemented in the simulation, exactly matching the evaluation of
\eqref{fraction_density}. For small $i_0(l)$ the multiple emission spectrum
approaches the single emission spectrum, recovering the Bethe-Heitler power
law $1/x^{\alpha}$ with $\alpha=1$. For larger $i_0(l)$ the
multiple emission possibilities gradually enhance the probability of having
larger energy losses, thus the average power $\alpha$ decreases. We observe
that energy conservation constraints further enhance the probability of having
larger energy losses.

\begin{figure}
\centering
\includegraphics[width=0.9\textwidth]{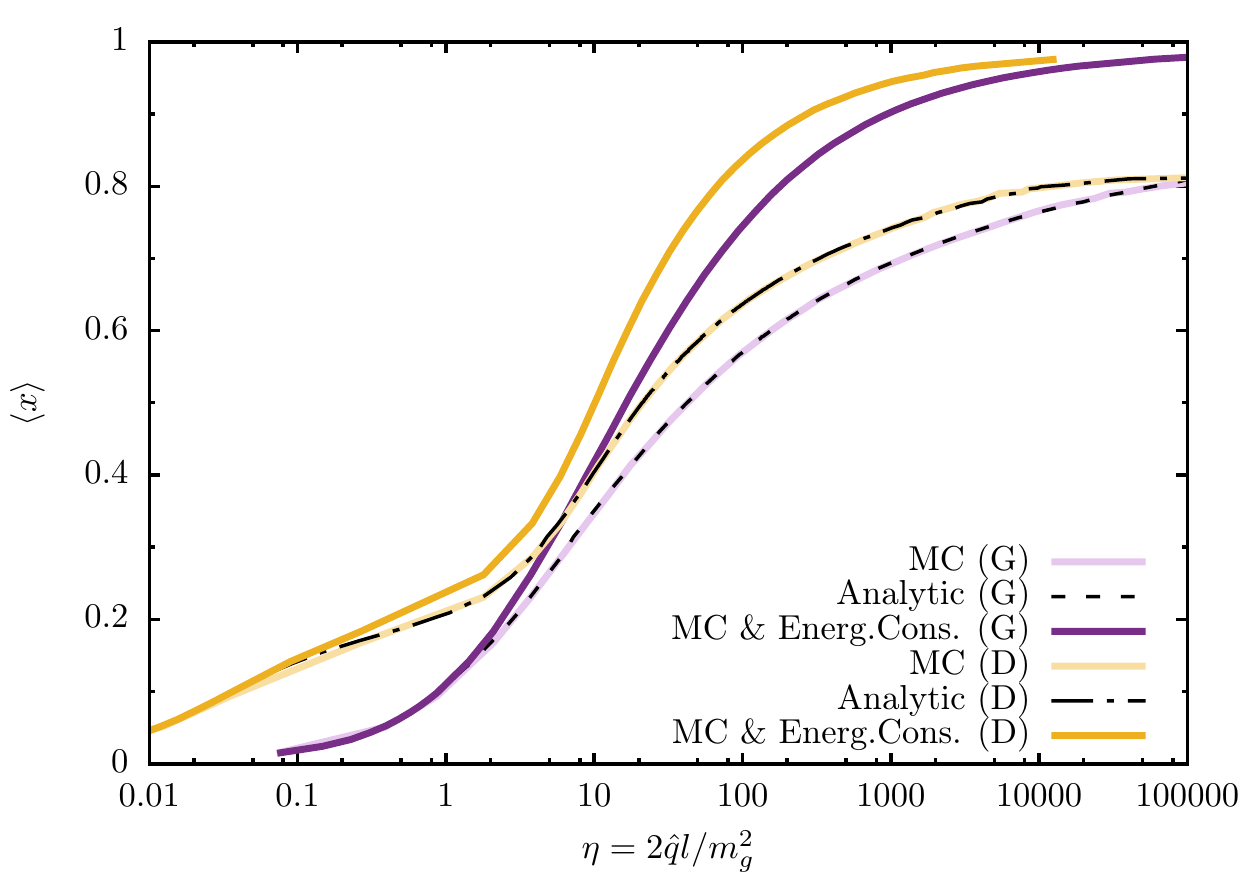}
\caption{Average fraction of energy loss as a function of the average number
  of collisions assuming a multiple emission scenario for the coherent plateau
  intensity $i_0(l)$. Results are shown for a medium of $n_0T_f$ = 8 fm$^{-3}$
  corresponding to $\hat{q}$ = 0.98 GeV$^2$/fm, $\alpha_s$ = 0.5. Gluon mass
  and Debye screening are set to $m_g$=$\mu_d$= 0.45 GeV$^2$/fm. Direct
  evaluation of the independent emission case is show for the Debye
  interaction (dot-dashed line) and its Fokker-Planck approximation (dashed
  line). Monte Carlo evaluation of the independent emission case is shown for
  the Debye interaction (light yellow) and for the Fokker-Planck approximation
  (light purple). Monte Carlo evaluation for the dependent emission scenario
  with energy conservation constraint is also shown for the Debye case
  (yellow) and the Fokker-Planck case (purple).}
\label{fig:figure_5_9}
\end{figure}

In Figure \ref{fig:figure_5_9} we show the average fraction of energy loss
assuming the coherent plateau for the spectrum $i_0(l)$, both in the Debye
interaction and its Fokker-Planck approximation, as a function of the average
number of collisions. We observe that for the same parameters the average
fraction of energy loss is larger for the Debye interaction than for the
Fokker-Planck approximation, as expected from the results at
\eqref{fig:figure_5_6}, due to the long tail of the Debye potential. We also
notice that for more than $\sim$ 10 collisions the energy loss is substantial
$\sim$ 60$\%$.

\begin{figure}
\includegraphics[width=0.5\textwidth]{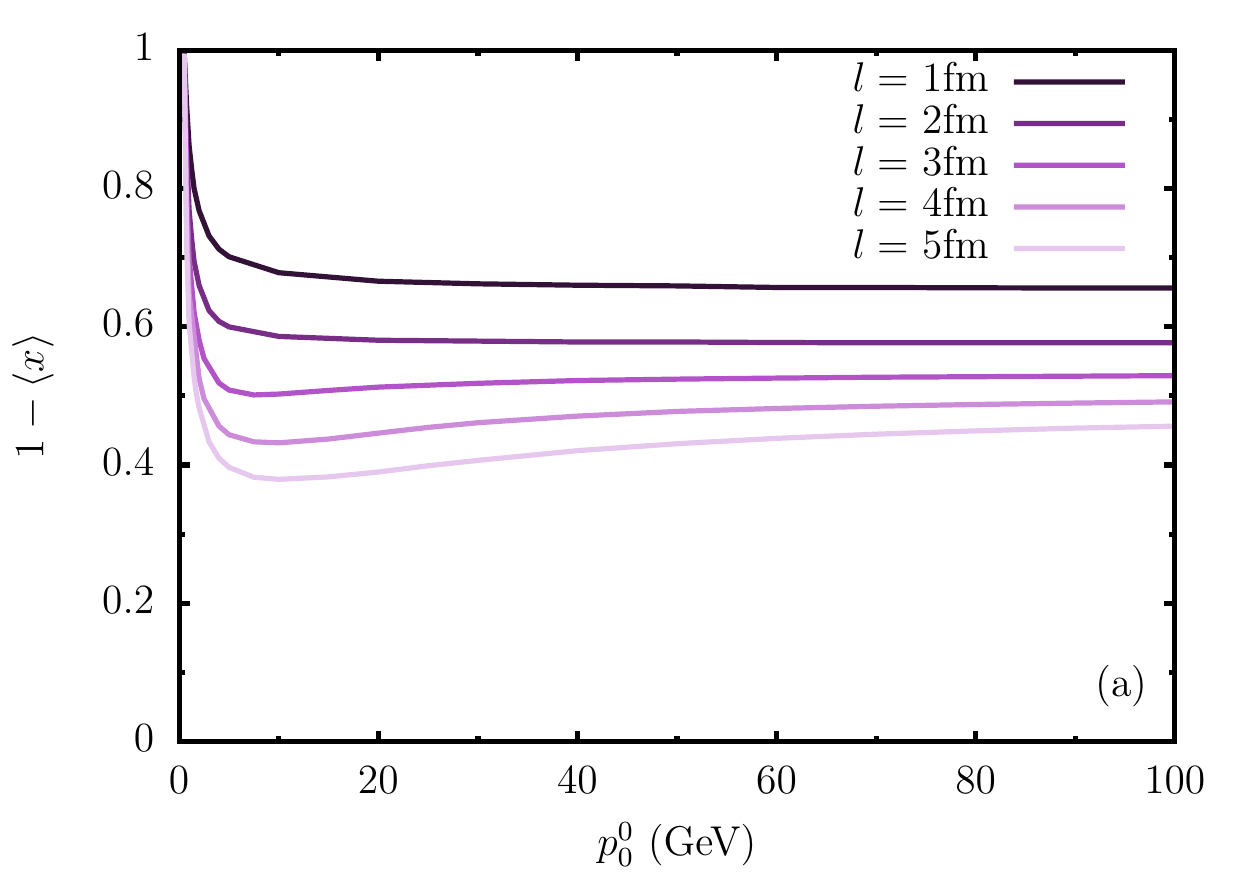}
\includegraphics[width=0.5\textwidth]{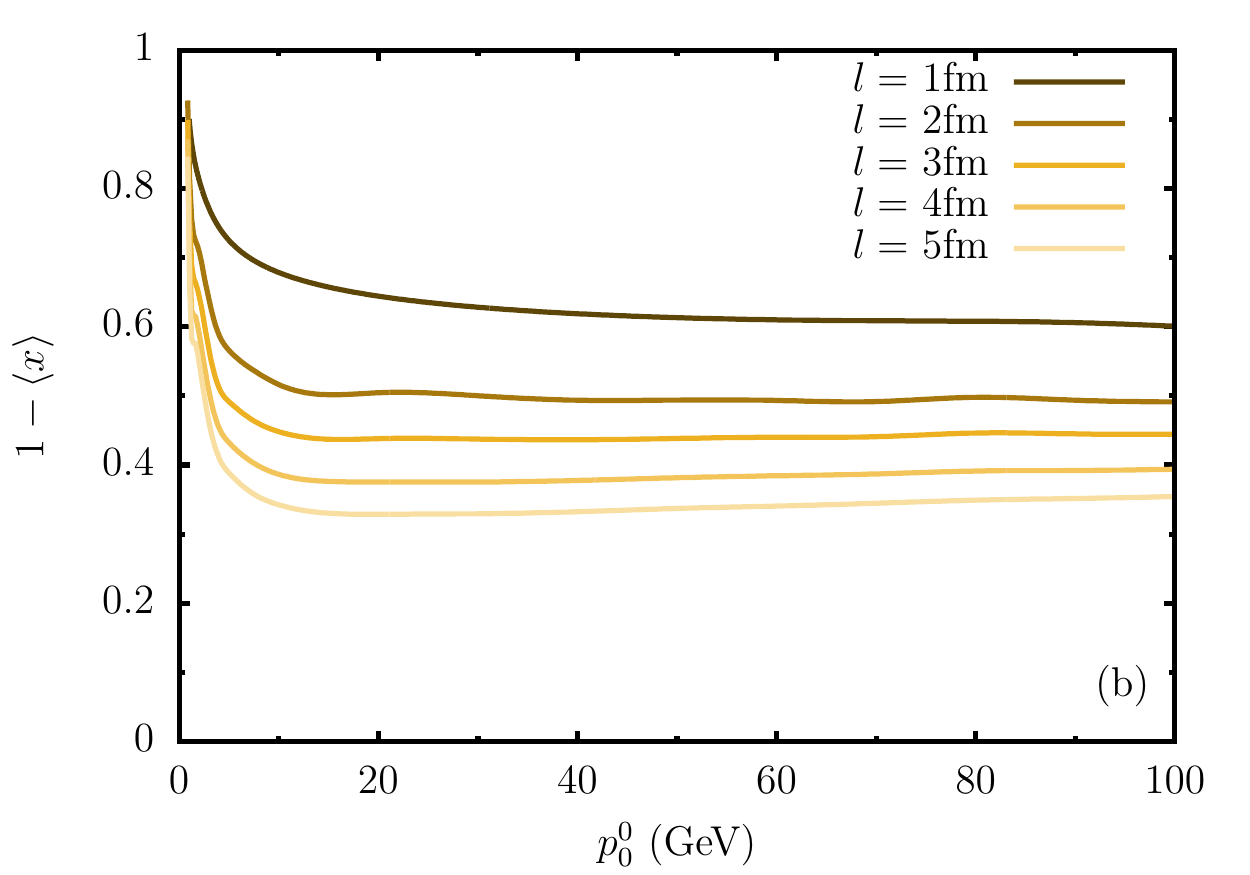}
\caption{Average fraction of remaining energy $1-\langle x \rangle$ as a
  function of the quark initial energy $p_0^0$ assuming the full spectrum
  given by \eqref{central_equation_qcd}, for a medium of density $n_0T_f$ = 8
  fm$^{-3}$ corresponding to a transport parameter of $\hat{q}$ = 0.98
  GeV$^2$/fm, a gluon mass of $m_g$ = 0.45 GeV and a Debye screening of
  $\mu_d$ = 0.45 GeV. Coupling is set to $\alpha_s$ = 0.5. Both the
  Fokker-Planck approximation (a) and in the Debye interaction (b) are shown
  for several medium lengths, as marked.}
\label{fig:figure_5_10}
\end{figure}

In Figure \ref{fig:figure_5_10} we show the average fraction of remaining
energy as a function of the initial quark energy, assuming the full intensity
\eqref{central_equation_qcd}. We observe that the LPM effect translates into a
slight deviation from the constant energy loss plateau caused by the coherent
contribution $i_0(l)$ for low energy quarks. Quarks of energy near $\omega_c$
experiment larger energy losses than high energetic quarks. For energies
approaching the gluon mass the average fraction of remaining energy approaches
to $1$ as expected due to bremsstrahlung suppression.

%% file: conclusions/conclusions.tex
\chapter*{Conclusions}

\begin{itemize}
\item 

A high energy approximation of the state of a fermion under a classical and
static external field has been written. The state becomes ordered in the
asymptotic initial direction, causing that sources placed at coordinate
$x_3<z_3$ affect the state of the particle at coordinate $z_3$. Due to this
fact, beyond eikonal amplitudes preserve the internal scattering structure and
longitudinal phases, then, cause interferences in the square of the
amplitudes.

\item

The averaged squared scattering amplitude can be always split into a
transverse coherent and a transverse incoherent contribution. The first
encodes the diffractive and quantum effect of the medium transverse boundaries
(non diagonal terms) and the second leads to the probabilistic interpretation
of the multiple scattering (diagonal terms). An evolution, and a transport
equation, can be written for both contributions. Moliere's result is recovered
for $R\to\infty$. In this limit the averaged squared momentum change becomes
additive in the traveled length for any interaction.

\item

The resulting elastic distributions for a Debye screened interaction with the
medium lead to Rutherford tails of the form $1/\v{q}^4$ which cannot be
accounted for with a Fokker-Planck/Gaussian approximation. The averaged
squared momentum change can be accounted at expenses of making the medium
transport parameter $\hat{q}$ dependent, albeit slowly, on the particle energy
and the screening.

\item

A formalism for the emission intensity in a multiple scattering scenario has
been implemented and evaluated under a realistic interaction, with the angular
dependence, and taking into account structured/finite targets and its effect
on the photon dispersion relations. Results are in good agreement with the
experimental data of SLAC and CERN, and show that Weinberg's soft photon
theorem saturates the LPM suppression.

\item
Within the Fokker-Planck approximation our result recovers Migdal Zakharov
result in the $l\to\infty$ limit and Wiedemann and Gyulassy results for finite
$l$. The Fokker-Planck approximation, however, overestimates the
angle-unintegrated spectrum at lower photon angles but underestimates it at
large angles. For the integrated spectrum, the Fokker-Planck approximation
requires a length $l$ and frequency $\omega$ dependent definition, unless the
very large limit of collisions is taken $\eta\geq$ 10$^4$. This becomes
critical for the energy loss estimations for particles of energies below
$p_0^{lpm}$, i.e. total suppression.

\item
A QCD evaluation analogous to the QED scenario is found for the multiple
scattering in a medium. The traceless color matrices lead to a higher order
in $\alpha_s$ coherent contribution. 

\item
A formalism for the intensity in a multiple scattering scenario has been
implemented within the QCD formalism which admits an evaluation for a general
interaction. The transverse coherent corrections may become relevant for
mediums of R$\leq$ 4-5 fm. For $R\to\infty$ and with the adequate
approximations for the color averaged gluon interactions, within the
Fokker-Planck approximations, Zakharov result is recovered when $l\to\infty$,
the BDMPS subset when $m_g\to 0$ and Salgado and Wiedemann results for the
general case. 

\item

Exact kinematical integration has been taken, causing corrections in the gluon
in the soft regime. Soft gluon suppression is found due to mass gluon
effects. The Fokker-Planck approximation underestimates the Debye interaction
intensity and the difference can not be cast into a single definition of the
medium transport properties through $\hat{q}$. In the BDMPS subset,
Fokker-Planck approximation underestimates the radiation by a factor $\sim$
3-4 which may suggest that larger gluon elastic cross sections, leading to the
hypothesis of strongly coupled QGP, may not be required to match the data.

\end{itemize}

%% file: appendices/appendix1.tex
\chapter{Spinors and polarizations}
\label{appendix1}
\begin{figure}[ht]
\centering
\includegraphics[scale=0.5]{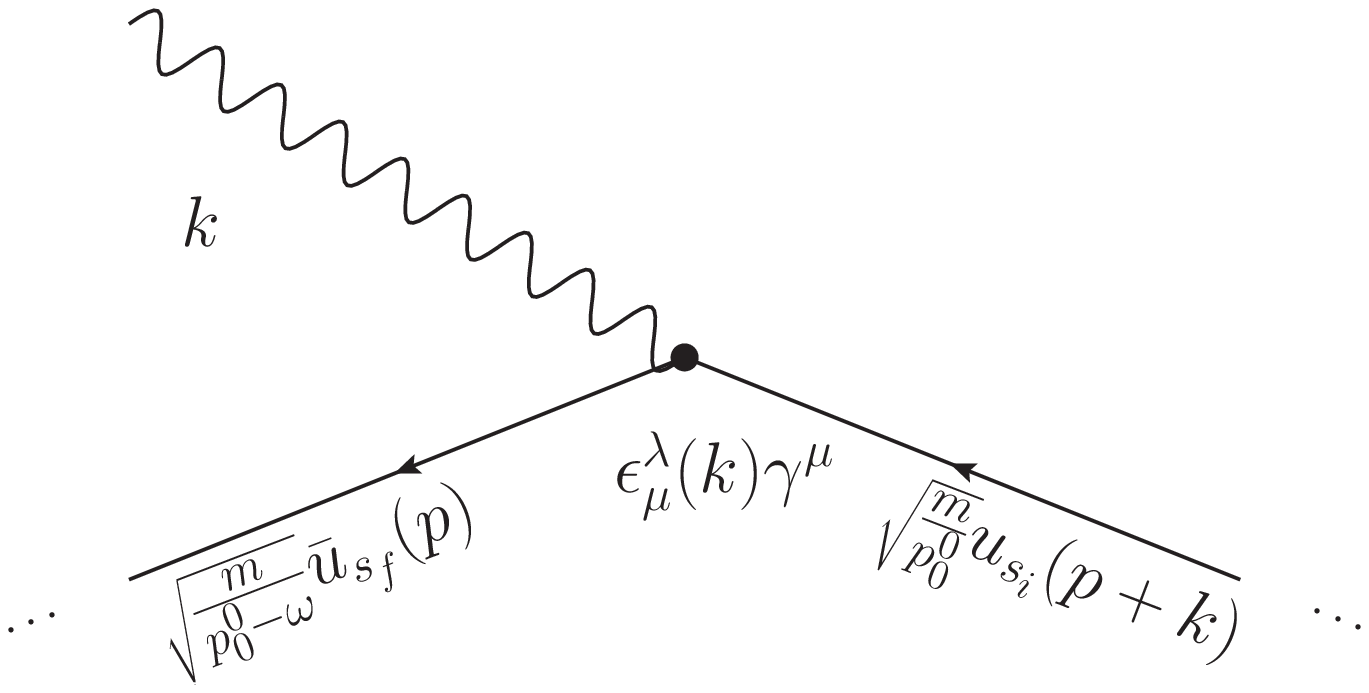}
\caption{Diagrammatic representation of the emission vertex.}
\label{fig:figure_a_1}
\end{figure}

The spinor conventions used through this work are now going to be explicitely
written, since in various cases of our interest some derivations going beyond
the usual polarization and spin sum tricks are required \cite{casimir1933a}. A
free solution to the Dirac equation is given by
\begin{equation}
\psi(x)= \mathcal{N}(p)u_s(p)e^{-ip\cdot x},
\label{free_spinor}
\end{equation}
where $\mathcal{N}(p)$ is some normalization and $u_s(p)$ a free spinor.
In order to obtain the exact form of this spinor we write for the Dirac matrices
\begin{align}
\gamma_0=\begin{bmatrix}
1 &  0 \\
0 & -1 \end{bmatrix},\medspace \gamma_i = \begin{bmatrix}
0 &  \sigma_i \\
-\sigma_i & 0 \end{bmatrix},
\end{align}
where $\v{\sigma}$ are Pauli matrices. This equation can be easily solved for
a free electron at rest. By calling $u_s(m,\v{0})$ its solution we have
\begin{align}
(p_\mu\gamma^\mu-m)u_{s}(m,\v{p})=(m\gamma^0-m)u_{s}(m,\v{0})=0.
\end{align}
We trivially obtain two orthogonal solutions, labeled with $s$=1,2 and given by
\begin{align}
u_{s}(m,\v{0})=\begin{bmatrix}\varphi_s\\0\end{bmatrix},\medspace \varphi_1
= \begin{bmatrix}1\\0\end{bmatrix},\medspace \varphi_2
  = \begin{bmatrix}0\\1\end{bmatrix}.
\end{align}
In order to construct a free solution for an arbitrary $p$ we observe
\begin{align}
(p_\mu\gamma^\mu-m)(p_\nu\gamma^\nu+m)u_{s}(m,\v{0}) =
(p^2-m^2)u_{s}(m,\v{0}) = 0,
\end{align}
so, up to an overall constant, we have found the solution for $\v{p}\ne 0$ in terms of the rest solution, indeed
\begin{align}
 u_s(m,\v{p})=(p_\mu\gamma^\mu+m)u_s(m,\v{0}) = \begin{bmatrix}
  (p_0+m)\varphi_{s} \\ (\v{\sigma}\cdot\v{p})\varphi_s\end{bmatrix}.
\end{align}
We will be using the normalization convention
\begin{align}
\delta_{s_1s_2} \equiv \bar{u}_{s_1}(p)u_{s_2}(p).
\end{align}
For our previous solutions we obtain 
\begin{align}
\bar{u}_{s_1}(m,p) u_{s_2}(m,p)&= \varphi_{s_1}^\dag \big
((p_0+m)^2-(\v{\sigma}\cdot\v{p})^2\big)\varphi_{s_2}\nonumber\\
&=\big((p_0+m)^2-\v{p}^2\big)\delta_{s_1s_2}=m(m+2p_0)\delta_{s_1s_2},
\end{align}
where we used the relations $(\v{\sigma}\cdot\v{p})^2=\v{p}^2+i(\v{p}\times\v{p})\cdot\v{\sigma}=\v{p}^2$
and $\varphi_{s_1}^\dag\varphi_{s_2}=\delta_{s_1s_2}$. This
requires us to define the spinors in \ref{free_spinor} according to 
\begin{align}
u_s(p)= \sqrt{\frac{p_0+m}{2m}} \begin{bmatrix}
  \varphi_{s}\\
  \frac{\v{\sigma}\cdot\v{p}}{p_0+m}\varphi_s\end{bmatrix} \label{free_spinor_explicit}.
\end{align}
In order to obtain $\mathcal{N}(p)$ we require current conservation, that is, 
a free spinor has to remain normalized. This condition reads
\begin{align}
\int d^4x \medspace \bar{\psi}_2(x)\gamma^0\psi_1(x) = (2\pi)^4\delta^{4}(p_2-p_1)
\mathcal{N}(p_1)\mathcal{N}(p_2)\bar{u}_{s_2}(p_2)\gamma_0 u_{s_1}(p_1).
\end{align}
From this condition, taking the $p_0\gg m$ limit in the arising normalization,
one finds
\begin{align}
\mathcal{N}(p) = \sqrt{\frac{m}{p_0}},\medspace\medspace\medspace\medspace\medspace\medspace\medspace\medspace\medspace\medspace \mathcal{N}(p)u_s(p)\simeq \sqrt{\frac{1}{2}} \begin{bmatrix}
  \varphi_{s}\\
  \frac{\v{\sigma}\cdot\v{p}}{p_0+m}\varphi_s\end{bmatrix}\label{free_spinor_explicit2}.
\end{align}
Once we explicitely know \eqref{free_spinor} we are in position to compute some
required unpolarized cross sections. 
For a photon whose polarization vector is $\epsilon_\mu^\lambda(k)$ and $\v{p}$
and $\v{u}$ of modulus $\beta (p_0^0-\omega)$, where $\beta$ is the velocity, we 
evaluate the quantity
\begin{align}
h_k(p,u)\equiv \frac{1}{2}\sum_{s_fs_i}\sum_\lambda \sqrt{\frac{m}{p_0^0-\omega}}&\bar{u}_{s_f}(p)\epsilon_\mu^\lambda(k)\gamma^\mu
u_{s_i}(p+k)\sqrt{\frac{m}{p_0^0}}\\
&\times\left(\sqrt{\frac{m}{p_0^0-\omega}}\bar{u}_{s_f}(u)\epsilon_\mu^\lambda(k)\gamma^\mu
u_{s_i}(u+k)\sqrt{\frac{m}{p_0^0}}\right)^*\nonumber.
\end{align}
where we sum over final spin and polarization and average over initial
spins. A diagrammatic representation of this vertex is shown in Figure \ref{fig:figure_a_1}.
First we note that
\begin{align}
\sum_{\lambda=1,2}
\epsilon_\mu^{\lambda}(k)\epsilon_\nu^{\lambda}(k) = \delta_{ij} -
\frac{k_ik_j}{\omega^2},\medspace\medspace\medspace\medspace\medspace i,j=1,2,3,\end{align}
so, if we place the $z$ axis in the photon direction we take advantage of a further
simplification
\begin{align}
\sum_{\lambda=1,2}
\epsilon_\mu^{\lambda}(k)\epsilon_\nu^{\lambda}(k) =
\delta_{ij}-\delta_{i3}\delta_{j3}.
\end{align}
In this reference frame we only sum over transverse directions $i=1,2$ and
then, sum over repeated spin indices assumed
\begin{align}
h_k(p,u)=  \frac{1}{2}\sum_{i=1,2}\sqrt{\frac{m}{p_0^0-\omega}}&\bar{u}_{s_f}(p)\gamma^i
u_{s_i}(p+k)\sqrt{\frac{m}{p_0^0}}\nonumber\\
&\times \sqrt{\frac{m}{p_0^0}}\bar{u}_{s_i}(u+k)\gamma^i
u_{s_f}(u)\sqrt{\frac{m}{p_0^0-\omega}}.
\end{align}
These spin sums can be done in the usual way or by directly using the explicit
forms of the spinors, in order to obtain, using \eqref{free_spinor_explicit2},
\begin{align}\sqrt{\frac{m}{p_0-\omega}}\bar{u}_{s_f}(p)\gamma^i
u_{s_i}(p+k)\sqrt{\frac{m}{p_0}} = 
\frac{1}{2}\varphi_{s_f}^\dag\left(\frac{\sigma_i\v{\sigma}\cdot(\v{p}+\v{k})}{p_0+m}+\frac{\v{\sigma}\cdot\v{p}\sigma_i}{p_0-\omega+m}\right)\medspace\varphi_{s_i},
\end{align}
and
\begin{align}
\sqrt{\frac{m}{p_0}}\bar{u}_{s_i}(u+k)\gamma^i
u_{s_f}(u)\sqrt{\frac{m}{p_0-\omega}} =
\frac{1}{2}\varphi_{s_i}^\dag\left(\frac{\sigma_i\v{\sigma}\cdot\v{u}}{p_0-\omega+m}+\frac{\v{\sigma}\cdot(\v{u}+\v{k})\sigma_i}{p_0+m}\right)\medspace\varphi_{s_f}.
\end{align}
We then find a trace, 
\begin{align}
h_k(p,u)= \frac{1}{2} \sum_{s_f,s_i}
\frac{1}{4}\varphi_{s_f}^\dag\left(\frac{\sigma_i\v{\sigma}\cdot(\v{p}+\v{k})}{p_0+m}\right.&\left.+\frac{\v{\sigma}\cdot\v{p}\sigma_i}{p_0-\omega+m}\right)\medspace\varphi_{s_i}\\
&\times\varphi_{s_i}^\dag\left(\frac{\sigma_i\v{\sigma}\cdot\v{u}}{p_0-\omega+m}+\frac{\v{\sigma}\cdot(\v{u}+\v{k})\sigma_i}{p_0+m}\right)\medspace\varphi_{s_f}\nonumber\\
= \frac{1}{8} \Tr \left\{\left(\frac{\sigma_i\v{\sigma}\cdot(\v{p}+\v{k})}{p_0+m}+\right.\right.&\left.\left.\frac{\v{\sigma}\cdot\v{p}\sigma_i}{p_0-\omega+m}\bigg)\bigg(\frac{\sigma_i\v{\sigma}\cdot\v{u}}{p_0-\omega+m}+\frac{\v{\sigma}\cdot(\v{u}+\v{k})\sigma_i}{p_0+m}\right)\right\}.\nonumber
\end{align}
In order to compute the required four traces above we use the following
relations satisfied by the Pauli matrices
\begin{align}
\Tr \big(\sigma_i\sigma_j\sigma_k\sigma_l\big) =
2(\delta_{ij}\delta_{kl}-\delta_{ik}\delta_{jl}+\delta_{il}\delta_{jk}).
\end{align}
We easily find, then for the first term
\begin{align}
\frac{1}{(p_0+m)(p_0-\omega+m)}\sum_{i=1}^2\sum_{j,k=1}^3\Tr(\sigma_i\sigma_j\sigma_i\sigma_k)u_j(p+k)_k=
\frac{-4u_3(p+k)_3}{(p_0+m)(p_0-\omega+m)},
\end{align}
for the second term,
\begin{align}
\frac{1}{(p_0+m)(p_0-\omega+m)}\sum_{i=1}^2\sum_{j,k=1}^3\Tr(\sigma_j\sigma_i\sigma_k\sigma_i)(u+k)_jp_k=
\frac{-4p_3(u+k)_3}{(p_0+m)(p_0-\omega+m)},
\end{align}
for the third term,
\begin{align}
\frac{1}{(p_0+m)^2}\sum_{i=1}^2\sum_{j,k=1}^3\Tr(\sigma_j\sigma_i\sigma_i\sigma_k)(u+k)_j(p+k)_k=
\frac{4}{(p_0+m)^2}(\v{u}+\v{k})\cdot(\v{p}+\v{k}),
\end{align}
and for the last term,
\begin{align}
\frac{1}{(p_0-\omega+m)^2}\sum_{i=1}^2\sum_{j,k=1}^3\Tr(\sigma_i\sigma_j\sigma_k\sigma_i)u_jp_k=
\frac{4}{(p_0-\omega+m)^2}\v{u}\cdot\v{p}.
\end{align}
So, by joining the four terms and expanding in transverse and longitudinal
directions we simplify to
\begin{align}
h_k(p,u)=&\frac{1}{2}\bigg(\frac{1}{p_0+m}(p+k)_3-\frac{1}{p_0-\omega+m}p_3\bigg)\bigg(\frac{1}{p_0+m}(u+k)_3-\frac{1}{p_0-\omega+m}u_3\bigg)\nonumber\\&+\frac{1}{2(p_0+m)^2}(\v{p}+\v{k})_t\cdot(\v{u}+\v{k})_t+\frac{1}{2(p_0-\omega+m)^2}\v{p}_t\cdot\v{u}_t.
\end{align}
Since the reference frame in which $\v{k}$ points in the $z$
direction has been chosen, then $\v{k}_t=0$, and $\v{p}_t$ and $\v{u}_t$
refers to the transverse components with respect to $\v{k}$. 
In a general frame these components are given by
\begin{align}
\v{p}_t=
\v{p}-\bigg(\v{p}\cdot\frac{\v{k}}{\omega}\bigg)\frac{\v{k}}{\omega},\medspace\medspace\medspace\medspace\medspace \v{u}_t=
\v{u}-\bigg(\v{u}\cdot\frac{\v{k}}{\omega}\bigg)\frac{\v{k}}{\omega},
\end{align}
so that
\begin{align}
(\v{p}+\v{k})_t\cdot(\v{u}+\v{k})_t &=
  \v{p}_t\cdot\v{u}_t=\left(\v{p}-\bigg(\v{p}\cdot\frac{\v{k}}{\omega}\bigg)\frac{\v{k}}{\omega}\right)\cdot\left(\v{u}-\bigg(\v{u}\cdot\frac{\v{k}}{\omega}\bigg)\frac{\v{k}}{\omega}\right)\nonumber\\
&=\v{p}\cdot\v{u}-\frac{1}{\omega^2}(\v{p}\cdot\v{k})(\v{u}\cdot\v{k})=\left(\v{p}\times \frac{\v{k}}{\omega}\right)\cdot \left(\v{u}\times \frac{\v{k}}{\omega}\right).
\end{align}
The transverse contribution in $h_k(p,u)$ produces, then, the soft
and classical result related then to spin no flip amplitudes
\begin{align}
\sum_{\lambda}\epsilon_\mu^\lambda(k)\epsilon_\nu^\lambda(k)p^\mu
u^\nu=(\v{p}\times\hat{\v{k}})\cdot(\v{u}\times\hat{\v{k}}),
\end{align}
 times a kinematical hard photon correction we call $h^n(y)$, where
 $y=\omega/p_0^0$ is the fraction of energy carried by the photon,
\begin{align}
\frac{1}{2}\frac{1}{(p_0+m)^2}&(\v{p}+\v{k})_t\cdot(\v{u}+\v{k})_t+\frac{1}{2}\frac{1}{(p_0-\omega+m)^2}\v{p}_t\cdot\v{u}_t\\&\simeq\frac{1}{2}\frac{p_0^2+(p_0-\omega)^2}{p_0^2(p_0-\omega)^2}
\left(\v{p}\times \frac{\v{k}}{\omega}\right)\cdot \left(\v{q}\times
\frac{\v{k}}{\omega}\right)\equiv h_n(y) \left(\hat{\v{p}}\times \frac{\v{k}}{\omega}\right)\cdot \left(\hat{\v{u}}\times
\frac{\v{k}}{\omega}\right)\nonumber,
\end{align}
and where we re-absorbed the modulus of $|\v{p}|=\beta (p_0-\omega)$ and
$|\v{u}|=\beta (p_0-\omega)$ in $h_n(y)$, and in the high energy limit
$p_0\gg m$ and $\beta$=1. Similarly for the photon longitudinal contribution,
we observe, similarly taking $\beta$=1,
\begin{align}
\frac{1}{p_0+m}(p+k)_z-\frac{1}{p_0+m-\omega}p_z\simeq
\left(1-\frac{m}{p_0}\right)-\left(1-\frac{m}{p_0-\omega}\right)=\frac{m\omega}{p_0(p_0-\omega)}
\end{align}
so that we find a contribution which corrects the classical contribution only
in the hard part of the spectrum
\begin{align}
\frac{1}{2}\left(\frac{1}{p_0^0+m}(p+k)_z\right.&-\left.\frac{1}{p_0^0-\omega+m}p_z\right)\left(\frac{1}{p_0^0+m}(u+k)_z-\frac{1}{p_0^0-\omega+m}u_z\right)\nonumber\\
&\simeq\frac{1}{2}\frac{m^2\omega^2}{p_0^2(p_0-\omega)^2}.
\end{align}
Finally, joining the two contributions we find
\begin{align}
h_k(p,u) =  h_{n}(y)\medspace\left(\hat{\v{p}}\times
\frac{\v{k}}{\omega}\right)\cdot \left(\hat{\v{u}}\times \frac{\v{k}}{\omega}\right)
+h_{s}(y), 
\end{align}
where the kinematical weights are given by
\begin{align}
h_n(y)=\frac{1}{2}\frac{(p_0^0)^2+(p_0^0-\omega)^2}{(p_0^0)^2}=1-y+\frac{1}{2}y^2, \medspace h_s(y)=\frac{1}{2}\left(\frac{m\omega}{p_0(p_0-\omega)}\right)^2\simeq\frac{1}{2}y^2,
\end{align}
where the hard limit has been taken $p_0-\omega\simeq m$ in the $h_s(y)$
function. The above form of the hard corrections to the unpolarized and spin
averaged current is the well known result \cite{bethe1934,migdal1956}.